\documentclass{dune}
\pdfoutput=1

\usepackage[pdftex,bookmarks,hidelinks]{hyperref}

\graphicspath{ {graphics/} }

\newif\ifdp
\newif\ifsp

\def\expshort{DUNE\xspace}

\def\explong{The Deep Underground Neutrino Experiment\xspace}

\def\thedocsubtitle{Deep Underground Neutrino Experiment (DUNE)} 
\def\tdrtitle{Technical Design Report}
\def\voltitleexec{Introduction to DUNE\xspace}
\def\volnumberexec{I}

\def\voltitlephysics{DUNE Physics\xspace}
\def\volnumberphysics{II}

\def\voltitletc{DUNE Far Detector Technical Coordination\xspace}
\def\volnumbertc{III}

\def\voltitlesp{The DUNE Far Detector Single-Phase Technology\xspace}
\def\volnumbersp{IV}

\def\spchinstall{Volume~\volnumbersp{}, \voltitlesp{}, Chapter~9\xspace}

\def\voltitledp{The DUNE Far Detector Dual-Phase Technology\xspace}
\def\volnumberdp{V}

\newcommand{\refsec}[2]{Volume~\csname volnumber#1\endcsname \xspace Section~#2}
\newcommand{\refch}[2]{Volume~\csname volnumber#1\endcsname \xspace Chapter~#2}
\newcommand{\refinch}[2]{#2 in Volume~\csname volnumber#1\endcsname \xspace}

\newcommand{\numu}{\ensuremath{\nu_\mu}\xspace}
\newcommand{\nue}{\ensuremath{\nu_e}\xspace}

\def\argon40{${}^{40}$Ar}       
\def\Ar39{$^{39}$Ar}
\def\Cl40{$^{40}$Cl}
\def\K40{$^{40}$K}
\def\B8{$^{8}$B}

\def\fdfiducialmass{\SI{40}{\kt}\xspace}

\def\larmass{\SI{17.5}{\kt}\xspace} 

\def\nominalmodsize{\SI{10}{kt}\xspace} 
\def\dunelifetime{\SI{20}{years}\xspace} 
 
\def\cooldown{cool-down\xspace} 

\def\spmaxfield{\SI{500}{\volt/\centi\meter}\xspace} 
\def\spmaxdrift{\SI{3.5}{\m}\xspace}
\def\tpcheight{\SI{12.0}{\meter}\xspace} 
\def\sptpclen{\SI{58.2}{\meter}\xspace} 

\def\coldbox{cold box\xspace} 

\def\dpmaxdrift{\SI{12.0}{\m}\xspace} 
\def\dpnumpmtch{\num{720}\xspace} 
\def\dpnominaldriftfield{\SI{500}{\volt/\cm}\xspace} 
\def\dptargetdriftvoltneg{\SI{-600}{\kV}\xspace} 

\newcommand{\efield}{E field\xspace}

\newcommand{\threed}{3D\xspace}
\newcommand{\twod}{2D\xspace}
\newcommand{\phel}{photoelectron\xspace} 
\newcommand{\frfour}{FR-4\xspace} 

\newcommand{\lsim}{{\;\raise0.3ex\hbox{$<$\kern-0.75em\raise-1.1ex\hbox{$\sim$}}\;}}
\newcommand{\gsim}{{\;\raise0.3ex\hbox{$>$\kern-0.75em\raise-1.1ex\hbox{$\sim$}}\;}}
\newcommand{\beq}{\begin{equation}}
\newcommand{\eeq}{\end{equation}}
\newcommand{\bea}{\begin{eqnarray}}
\newcommand{\eea}{\end{eqnarray}}

\mathchardef\minus="002D

\newcommand{\sdlwavailable}{April 2022\xspace}
\newcommand{\cucbenocc}{October 2022\xspace}
\newcommand{\accesscuccountrm}{April  2023\xspace}
\newcommand{\accesstopfirstcryo}{January 2024\xspace}
\newcommand{\startfirsttpcinstall}{August 2024\xspace}
\newcommand{\accesstopsecondcryo}{January 2025\xspace}
\newcommand{\firsttpcinstallend}{May 2025\xspace}
\newcommand{\startsecondtpcinstall}{August 2025\xspace}
\newcommand{\secondtpcinstallend}{May 2026\xspace}

\newcommand{\rrt}[1]{}

\newcommand{\microboone}{MicroBooNE\xspace} 
\newcommand{\minerva}{MINERvA\xspace} 
\newcommand{\nova}{NOvA\xspace} 

\newcommand{\lartpc}{LArTPC\xspace}

\newcommand{\docdb}{DUNE DocDB\xspace}

\newcommand{\fnal}{Fermilab\xspace} 
\newcommand{\surf}{SURF\xspace}

\newcommand{\dual}{DP\xspace}

\newcommand{\single}{SP\xspace}

\newcommand{\lar}{LAr\xspace}
\newcommand{\lntwo}{LN$_2$\xspace}  

\DeclareSIUnit \s {\second}
\DeclareSIUnit \MB {\mega\byte}
\DeclareSIUnit \GB {\giga\byte}
\DeclareSIUnit \TB {\tera\byte}
\DeclareSIUnit \PB {\peta\byte}
\DeclareSIUnit \Mbps {\mega\bit/\s}
\DeclareSIUnit \Gbps {\giga\bit/\s}
\DeclareSIUnit \Tbps {\tera\bit/\s}
\DeclareSIUnit \Pbps {\peta\bit/\s}
\DeclareSIUnit \kton {\kilo\tonne} 
\DeclareSIUnit \kt {\kilo\tonne}
\DeclareSIUnit \Mt {\mega\tonne}
\DeclareSIUnit \eV {\electronvolt}
\DeclareSIUnit \keV {\kilo\electronvolt}
\DeclareSIUnit \MeV {\mega\electronvolt}
\DeclareSIUnit \GeV {\giga\electronvolt}
\DeclareSIUnit \m {\meter}
\DeclareSIUnit \cm {\centi\meter}
\DeclareSIUnit \in {\inchcommand}
\DeclareSIUnit \km {\kilo\meter}
\DeclareSIUnit \kV {\kilo\volt}
\DeclareSIUnit \kW {\kilo\watt}
\DeclareSIUnit \MW {\mega\watt}
\DeclareSIUnit \MHz {\mega\hertz}
\DeclareSIUnit \mrad {\milli\radian}
\DeclareSIUnit \year {year}
\DeclareSIUnit \POT {POT}
\DeclareSIUnit \sig {$\sigma$}
\DeclareSIUnit\parsec{pc}
\DeclareSIUnit\lightyear{ly}
\DeclareSIUnit\foot{ft}
\DeclareSIUnit\ft{ft}
\DeclareSIUnit \ppb{ppb}
\DeclareSIUnit \ppt{ppt}
\DeclareSIUnit \samples{S}

\usepackage[toc]{glossaries}
\makeglossaries

\newcommand{\dshort}[1]{\glsentrytext{#1}}  
\newcommand{\dshorts}[1]{\glsentryshortpl{#1}}  

\newcommand{\dfirst}[1]{\glsfirst{#1}\glsunset{#1}}

\newcommand{\dword}[1]{\gls{#1}}
\newcommand{\dwords}[1]{\glspl{#1}}
\newcommand{\Dword}[1]{\Gls{#1}}
\newcommand{\Dwords}[1]{\Glspl{#1}}

\newcommand{\newduneword}[3]{
    \newglossaryentry{#1}{
        text={#2},
        long={#2},
        name={\glsentrylong{#1}},
        first={\glsentryname{#1}},
        firstplural={\glsentrylong{#1}\glspluralsuffix},
        description={#3},
        sort={#2}
    }
}

\newcommand{\newduneabbrev}[4]{
  \newglossaryentry{#1}{
    text={#2},
    long={#3},
    shortplural={{#2}\glspluralsuffix},
    longplural={{#3}\glspluralsuffix{}},
    name={\glsentrylong{#1}{} (\glsentrytext{#1}{})},
    first={#3 (#2)},
    firstplural={#3\glspluralsuffix{} (\glsentrytext{#1}\glspluralsuffix{})},
    description={#4},
    sort={#2}
  }
}

\newcommand{\newduneabbrevs}[5]{
  \newglossaryentry{#1}{
    text={#2},
    long={#3},
    plural={#4},
    shortplural={{#2}\glspluralsuffix},
    longplural={#4},
    name={\glsentrylong{#1}{} (\glsentrytext{#1}{})},
    first={#3 (#2)},
    firstplural={#4 (\glsentrytext{#1}\glspluralsuffix{})},
    description={#5},
    sort={#2}    
  }
}

\newduneword{dword}{DUNE Word}{A term in the DUNE lexicon}

\newduneword{nasa}{NASA}{U.S. National Aereonautics and Space Administration}

\newduneabbrev{nd}{ND}{near detector}{Refers to the detector(s) 
 installed close to the neutrino source at \fnal }

\newduneabbrev{fd}{FD}{far detector}{The \SI{70}{kt} total (\fdfiducialmass fiducial) mass \gls{lartpc} DUNE detector, composed of four \larmass total (\nominalmodsize fiducial) mass modules,  
  to be installed at the far site at \surf in
  Lead, SD, USA}

\newduneabbrev{sp}{SP}{single-phase}{Distinguishes one of the DUNE far detector technologies by the fact that it operates using argon in its liquid phase only}

\newduneabbrev{dp}{DP}{dual-phase}{Distinguishes one of the DUNE far detector technologies by the fact that it operates using argon 
 in both gas and liquid phases}

\newduneabbrev{pds}{PD system}{photon detection system}{The detector 
  subsystem sensitive to light produced in the \lar }

\newduneabbrev{hvs}{HVS}{high voltage system}{The detector 
  subsystem that provides the \gls{tpc} drift field}

\newduneabbrev{tpc}{TPC}{time projection chamber}{A type of particle detector that uses an \efield together with a sensitive volume of gas or liquid, e.g., \gls{lar}, to perform a \threed reconstruction of a particle trajectory or interaction. The activity is recorded by digitizing the waveforms of current
  induced on the anode as the distribution of ionization charge passes by
  or is collected on the electrode} 

\newduneabbrev{lartpc}{LArTPC}{liquid argon time-projection chamber}{A \gls{tpc} filled with liquid argon; 
the basis for the \gls{dune} \gls{fd} modules} 

\newduneabbrevs{apa}{APA}{anode plane assembly}{anode plane assemblies}{A unit of the \single
  detector module containing the elements sensitive to ionization in the \lar. 
  It contains two faces each of three planes of wires, and interfaces to the cold
  electronics and photon detection system} 

\newduneabbrev{awg}{AWG}{American wire gauge} {U.S. standard set of non-ferrous wire conductor sizes}

\newduneabbrev{ufer}{Ufer}{concrete encased electrode} {U.S. National Electrical Code grounding method refered to as Concrete Encased Electrode}

\newduneabbrev{cro}{CRO}{charge readout}{The system for detecting
  ionization charge distributions in a \dual detector module}

\newduneabbrev{lro}{LRO}{light readout}{The system for detecting
  scintillation photons in a \dual  detector module}

\newduneabbrev{shv}{SHV}{safe high voltage}{Type of bayonet mount
connector used on coaxial cables that has additional insulation 
compared to standard BNC and MHV connectors that makes it safer
for handling \gls{hv} by preventing accidental contact with the
live wire connector in an unmated connector or plug}

\newduneabbrev{fe}{FE}{front-end}{The front-end refers a point that is
  ``upstream'' of the data flow for a particular subsystem. 
  For example the \gls{sp} front-end electronics is where the cold electronics
  meet the sense wires of the TPC and the front-end \gls{daq} is where the
  \gls{daq} meets the output of the electronics}

\newduneabbrev{daqrou}{DAQ RU}{DAQ readout unit}{The first element in the data flow of the \gls{daq}}

\newduneabbrev{cots}{COTS}{commercial off-the-shelf}{Items, typically hardware such as 
computers, that may be purchased whole, without any custom design or fabrication and 
thus at normal consumer prices and availability}

\newduneabbrev{i2c}{I2C}{Inter-Integrated Circuit}{I$^2$C or I2C is a synchronous, 
multi-master, multi-slave, packet switched, single-ended, serial computer bus widely used 
for attaching lower-speed peripheral ICs to processors and microcontrollers in short-distance, 
intra-board communication} 

\newduneabbrev{spi}{SPI}{Serial Peripheral Interface}{The Serial Peripheral Interface is a 
synchronous serial communication interface specification used for short distance 
communication, primarily in embedded systems}

\newduneabbrev{miso}{MISO}{master in slave out}{The Master In Slave Out is a logic
signal on the \gls{spi} bus on which the data from the slave are transmitted once
a request from the master is received} 

\newduneabbrev{mosi}{MOSI}{master out slave in}{The Master Out Slave In is a logic
signal on the \gls{spi} bus on which the data from the master is transmitted} 

\newduneabbrev{uart}{UART}{Universal Asynchrous Receiver/Transmitter}{A universal 
asynchronous receiver-transmitter is a computer hardware device for asynchronous 
serial communication in which the data format and transmission speeds are configurable}

\newduneword{cr}{CR}{Capacitance-Resistance} 

\newduneword{dc}{DC}{direct coupling} 

\newduneword{ac}{AC}{capacitive coupling}  

\newduneabbrev{pll}{PLL}{Phase-Locked Loop}{A control system that generates an
output signal whose phase is related to the phase of an input signal}  

\newduneword{fifo}{FIFO}{First-In-First-Out} 

\newduneword{tsmc}{TSMC}{Taiwan Semiconductor Manufacturing Company}

\newduneword{saci}{SACI}{\gls{slac} \gls{asic} Control Interface}

\newduneword{om3}{OM3}{Type of multi-mode fiber optic cable, typically capable of \SI{10}{Gbps} data transmission at lengths up to \SI{300}{m}}

\newduneword{om4}{OM4}{Type of multi-mode fiber optic cable, typically capable of \SI{10}{Gbps} data transmission at lengths up to \SI{550}{m}}

\newduneword{qfp}{QFP}{Quad Flat Package} 

\newduneabbrev{ams}{AMS}{analog and mixed signal}{Verilog-AMS is a derivative of the Verilog hardware description language that includes analog and mixed-signal extensions (AMS) in order to define the behavior of analog and mixed-signal systems}

\newduneabbrev{hepa}{HEPA}{High Efficiency Particulate Air}{The High Efficiency Particulate Air filters are a type of air filter that remove \num{99.97}\% of particles that have a size greater han or equal to \SI{0.3}{$\mu$m}}  

\newduneabbrev{uvm}{UVM}{universal verification methodology}{The Universal Verification Methodology is a standardized methodology for verifying integrated circuit designs}   

\newduneword{lhc}{LHC}{Large Hadron Collider}

\newduneabbrev{lsb}{LSB}{least significant bit}{The bit with the lowest numerical value in a binary number}

\newduneabbrev{ldo}{LDO}{low-dropout regulator}{A low-dropout or LDO regulator is a \gls{dc} linear voltage regulator that can regulate the output voltage even when the supply voltage is very close to the output voltage}

\newduneabbrev{adc}{ADC}{analog-to-digital converter}{A sampling of a voltage
  resulting in a discrete integer count corresponding in some way to
  the input}

\newduneabbrev{inl}{INL}{integral non-linearity}{A commonly used measure of performance in \glspl{adc}. It is the deviation between the ideal input threshold value and the measured threshold level of a certain output code}

\newduneabbrev{dnl}{DNL}{differential non-linearity}{A commonly used measure of performance in \glspl{adc}. The DNL error is defined as the difference between an actual step width and the ideal value of one \gls{lsb}}

\newduneword{pnp}{PNP}{Type of bipolar junction transistor consistning of a
layer of N-doped semiconductor sandwiched between two layers of P-doped material}

\newduneword{spice}{SPICE}{SPICE
(``Simulation Program with Integrated Circuit Emphasis'') is a general-purpose, 
open-source analog electronic circuit simulator. It is a program used in integrated 
circuit and board-level design to check the integrity of circuit designs and to 
predict circuit behavior}

\newduneabbrev{daq}{DAQ}{data acquisition}{The data acquisition system
  accepts data from the detector \gls{fe} electronics, buffers
  the data, performs a \gls{trigdecision}, builds events from the selected
  data and delivers the result to the offline \gls{diskbuffer}}

\newduneword{detmodule}{detector module}{The entire DUNE far detector is
  segmented into four modules, each with a nominal \SI{10}{\kton}
  fiducial mass}

\newduneword{detunit}{detector unit}{A 
portion of a \gls{detmodule} may be further partitioned into a number of similar parts.   For example the \gls{sp} \gls{tpc} 
is made up of \gls{apa}  units (and other elements)}

\newduneword{diskbuffer}{secondary DAQ buffer}{A secondary
  \dshort{daq} buffer holds a small subset of the full rate as
  selected by a \gls{trigcommand}. 
  This buffer also marks the interface with the DUNE Offline}

\newduneabbrev{om}{OM}{online monitoring}{Processes that run inside
  the \gls{daq} on data ``in flight,'' specifically before landing on the
  offline disk buffer, and that provide feedback on the operation of
  the \gls{daq} itself and the general health of the data it is marshalling}

\newduneabbrev{dqm}{DQM}{data quality monitoring}{Analysis of the raw
  data to monitor the integrity of the data and the performance of the
  detectors and their electronics. This type of monitoring may be
  performed in real time, within the \gls{daq} system, or in later
  stages of processing, using disk files as input}

\newduneword{dumpbuffer}{DAQ dump buffer}{This \gls{daq} buffer
  accepts a high-rate data stream, in aggregate, from an associated
  portion of a \gls{detmodule} sufficient to collect all data likely relevant to
  a potential \gls{snb}}

\newduneabbrev{etl}{ETL}{external trigger logic}{Trigger processing
  that consumes \gls{detmodule} level \gls{trignote} information
  and other global sources of trigger input and emits
  \gls{trigcommand} information back to the \glspl{mtl}}
\newduneabbrev{daqeti}{ETI}{external trigger interface}{Interface between \glspl{mtl} and external source and sinks of relevant trigger information}

\newduneword{trignote}{trigger notification}{Information provided by
  \gls{mtl} to \gls{etl} about \gls{trigdecision} 
  processing}

\newduneword{trigprimitive}{trigger primitive}{Information derived by
  the \gls{daq} \gls{fe} hardware that describes a region of space (e.g.,
  one or several neighboring channels) and time (e.g., a contiguous set
  of \gls{adc} sample ticks) associated with some activity}

\newduneword{externtrigger}{external trigger candidate}{Information
  provided to the \gls{mtl} about events external to a
  \gls{detmodule} so that it may be considered in forming
  \glspl{trigcommand}}

\newduneabbrev{daqoob}{OOB dispatcher}{out-of-band trigger command
  dispatcher}{This component is responsible for dispatching a \gls{snb} dump
  command to all \glspl{daqfer} in the \gls{detmodule}}

\newduneabbrev{mtl}{MTL}{module trigger logic}{Trigger processing
  that consumes \gls{detunit} level \gls{trigcommand} information
  and emits \glspl{trigcommand}. 
  It provides the \gls{etl} with \glspl{trignote} and receives back any
  \glspl{externtrigger}}

\newduneword{octant}{octant}{Any of the eight parts into which 4$\pi$
  is divided by three mutually perpendicular axes. 
  In particular in referencing the value for the mixing angle
  $\theta_{23}$}

\newduneword{trigcandidate}{trigger candidate}{Summary information derived
  from the full data stream and representing a contribution toward
  forming a \gls{trigdecision}}

\newduneword{trigcommand}{trigger command}{Information derived from
  one or more \glspl{trigcandidate}  that directs elements of the
  \gls{detmodule} to read out a portion of the data stream}

\newduneabbrev{tcm}{TCM}{trigger command message}{A message flowing
  down the trigger hierarchy from global to local context.  Also see \gls{tpm}}

\newduneabbrev{mlt}{MLT}{module level trigger}{The \gls{daq} component responsible for producing a \gls{trigdecision} that will be used to command the readout of a detector module}

\newduneword{trigdecision}{trigger decision}{The process by which
  \glspl{trigcandidate} are converted into \glspl{trigcommand}}

\newduneabbrev{tpm}{TPM}{trigger primitive message}{A message flowing
  up the trigger hierarchy from local to global context.  Also see \gls{tcm}}

\newduneabbrev{ipc}{IPC}{inter-process communication}{A system for software elements to exchange information between threads, local processes or across a data network.  An IPC system is typically specified in terms of protocols  composed of message types and their associated data schema}

\newduneword{daqdispre}{discovery and presence}{As used in the context of the \gls{ipc}, a system that provides mechanisms for a node on a communication network to learn of the existence of peers and their identity (discovery) as well as determine if they are currently operational or have become unresponsive (presence)}

\newduneabbrev{pubsub}{PUB/SUB}{publish-subscribe communication pattern}{An \gls{ipc} communication pattern where one element, the publisher, sends data to all connected elements, the subscribers.  Each subscriber may connect to multiple publishers.  A variant is PUB/SUB with topics where a subscriber may register an identifier, the topic, to limit the information received to just an associated subset}

\newduneabbrev{eb}{EB}{event builder}{A software agent that executes \glspl{trigcommand}  for one  \gls{detmodule} by reading out the requested data}

\newduneabbrev{daqdfo}{DFO}{data flow orchestrator}{The process by which trigger commands are executed in parallel and asynchronous manner by the back-end output subsystem of the \gls{daq}}

\newduneabbrev{daqubi}{UBI}{upstream DAQ buffer interface}{The process which provides read-only access to data residing in the upstream \gls{daq} buffers to processes on the network}

\newduneabbrev{cob}{COB}{cluster on board}{An ATCA motherboard housing four RCEs}

\newduneabbrev{rce}{RCE}{reconfigurable computing element}{Data processor located outside of the cryostat on a \gls{cob} that contains \gls{fpga}, RAM and \gls{ssd} resources, responsible for buffering data, producing trigger primitives, responding to triggered requests for data and synching \gls{snb} dumps}

\newduneabbrev{bow}{BOW}{Bump On Wire}{A working name for the front-end readout computing elements used in the nominal \gls{daq} design to interface the \dual  crates to the \gls{daq} front-end computers}

\newduneabbrev{atca}{ATCA}{Advanced Telecommunications Computing
  Architecture}{An advanced computer architecture specification developed for the telecommunications, military, and aerospace industries that incorporates the latest trends in  high-speed interconnect technologies, next-generation processors, and improved reliability, availability and serviceability} 

\newduneabbrev{utca}{$\mu$TCA}{Micro Telecommunications Computing Architecture}{The computer architecture specification followed by the crates that house charge and light readout electronics in the \gls{dpmod}} 

\newduneabbrev{udp}{UDP}{user datagram protocol}{A simple,
  connectionless Internet protocol that supports data integrity
  checksums, requires no handshaking, and does not guarantee packet delivery}

\newduneabbrev{amc}{AMC}{advanced mezzanine card}{Holds digitizing
  electronics and lives in \gls{utca} crates}

\newduneabbrev{rf}{RF}{radio frequency}{Electromagnetic emissions
  that are within the (radio) frequency band of sensitivity of the detector
  electronics}

\newduneabbrev{fpga}{FPGA}{field programmable gate array}{An
integrated circuit technology that allows the hardware to be reconfigured to
execute different algorithms after its manufacture and deployment}

\newduneabbrev{fmc}{FMC}{FPGA mezzanine card}{Boards holding \glspl{fpga} and other integrated circuitry that attach to a motherboard}

\newduneabbrev{felix}{FELIX}{Front-End Link eXchange}{A
  high-throughput interface between \gls{fe} and trigger electronics
  and the standard PCIe computer bus}

\newduneword{daqpart}{DAQ partition}{A cohesive and
 coherent collection of \gls{daq} hardware and software working together to trigger and read out some portion of one detector module; it consists of an integral number of \glspl{daqfrag}. 
 Multiple \gls{daq} partitions may operate simultaneously, but each instance operates independently}

\newduneabbrev{fec}{DAQ FEC}{DAQ front-end computer}{The portion of one
  \gls{daqpart} that hosts the \gls{daqdr}, \gls{daqbuf} and
  \gls{daqds}.  It hosts the \gls{daqfer} and corresponding portion of the \gls{daqbuf}}

\newduneword{daqfrag}{DAQ front-end fragment}{The portion of one
  \gls{daqpart} relating to a single \gls{fec} and corresponding to an
  integral number of \glspl{detunit}.  See also \gls{datafrag}}

\newduneword{datafrag}{data fragment}{A block of data read out from a single \gls{daqfrag} that
span a contiguous period of time as requested by a \gls{trigcommand}}

\newduneabbrev{daqfer}{FER}{DAQ front-end readout}{The portion of a
  \gls{daqfrag} that accepts data from the detector electronics and
  provides it to the \gls{fec}}

\newduneabbrev{daqdr}{DDR}{DAQ data receiver}{The portion of the
  \gls{daqfrag} that accepts data from the \gls{daqfer}, emits
  trigger candidates produced from the input trigger primitives, and
  forwards the full data stream to the \gls{daqbuf}}

\newduneword{daqbuf}{DAQ primary buffer}{The portion
  of the \gls{daqfrag} that accepts full data stream from the
  corresponding \gls{detunit} and retains it sufficiently long for it
  to be available to produce a \gls{datafrag}}

\newduneword{daqds}{data selector}{The portion of the \gls{daqfrag}
  that accepts \glspl{trigcommand} and returns the corresponding
  \gls{datafrag}.  Not to be confused with \gls{daqdsn}}

\newduneword{daqdsn}{data selection}{The process of forming a trigger decision for selecting a subset of detector data for output by the \gls{daq} from the content of the detector data itself.  Not to be confused with \gls{daqds}}

\newduneabbrev{daqros}{DAQ RO}{DAQ readout subsystem}{The subsystem of the \gls{daq} for accepting and buffering data input from detector electronics}

\newduneabbrev{daqdss}{DAQ DS}{DAQ data selection subsystem}{The subsystem of the \gls{daq} responsible for forming a trigger decision based on a portion of the input data stream.  The majority subset of the \gls{daqtrs}}

\newduneabbrev{daqtrs}{DAQ TS}{DAQ trigger subsystem}{The subsystem of the \gls{daq} responsible for forming a trigger decision}

\newduneabbrev{daqbes}{DAQ BE}{DAQ back-end subsystem}{The portion of the \gls{daq} that is generally toward its output end.  It is responsible for accepting and executing trigger commands and marshaling the data they address to output storage buffers}

\newduneabbrev{daqtss}{DAQ TSS}{DAQ timing and synchronization subsystem}{The portion of the \gls{daq} that provides for timing and synchronization to various components}

\newduneabbrev{femb}{FEMB}{front-end mother board}{Refers a unit of
  the \gls{sp} \gls{ce} that contains the \gls{fe} amplifier
  and \gls{adc} \glspl{asic} covering 128 channels}

\newduneword{asic}{ASIC}{application-specific integrated circuit}

\newduneword{lv}{LV}{low voltage}

\newduneabbrev{iceberg}{ICEBERG}{ICEBERG R\&D cryostat and electronics}{Integrated Cryostat and Electronics Built for Experimental Research Goals:
a new double-walled cryostat built and installed at \gls{fnal} 
for liquid argon detector R\&D and for testing of DUNE detector components}

\newduneword{coldadc}{ColdADC}{A newly developed 16-channels \gls{asic} providing analog to digital conversion}

\newduneword{coldata}{COLDATA}{A 64-channel control and communications \gls{asic}}

\newduneword{cryo}{CRYO}{Integrated ASIC including \gls{fe} circuitry providing signal amplification and pulse shaping, analog to digital conversion, and control and communication functionalities for 64 channels}

\newduneword{larasic}{LArASIC}{A 16-channel \gls{fe} \gls{asic} that provides signal amplification and pulse shaping}

\newduneword{cmos}{CMOS}{Complementary metal-oxide-semiconductor}

\newduneabbrev{enc}{ENC}{equivalent noise charge}{The equivalent noise charge is the input charge that corresponds to a 
$\gls{snr}=1$}

\newduneword{sar}{SAR}{successive approximation register}

\newduneword{protodune}{ProtoDUNE}{Either of the two DUNE prototype detectors constructed at \gls{cern}. 
  One prototype implements \gls{sp} technology and the other \gls{dp}}
  
\newduneword{protodune2}{ProtoDUNE-2}{The second run of a \gls{protodune} detector}

\newduneword{pdsp}{ProtoDUNE-SP}{The \gls{sp} \gls{protodune} detector at \gls{cern}}

\newduneword{pddp}{ProtoDUNE-DP}{The \gls{dp} \gls{protodune} detector at \gls{cern}}

\newduneword{wa105}{WA105 DP demonstrator}{The \SI[product-units=power]{3x1x1}{m} WA105 \gls{dp} prototype detector at \gls{cern}}

\newduneword{rawevent}{DAQ event block}{The unit of data output by the
  \gls{daq}.  
  It contains trigger and detector data spanning a unique, contiguous
  time period and a subset of the detector channels}

\newduneabbrev{ssd}{SSD}{solid-state disk}{Any storage device that
  may provide sufficient write throughput to receive, both collectively and
  distributed, the sustained full rate of data from a \gls{detmodule}
  for many seconds}
\newduneabbrev{nvme}{NVMe}{Non-volatile memory express}{A specification for an interface to storage media attached via PCIe}

\newduneabbrev{hlt}{HLT}{high-level trigger}{This is actually a filter applied to data that has been triggered and aggregated in order to further reduce or characterize it}

\newduneabbrev{pid}{PID}{particle ID}{Particle identification}

\newduneword{readout window}{readout window}{A fixed, atomic and
  continuous period of time over which data from a \gls{detmodule}, in
  whole or in part, is recorded. 
  This period may differ based on the trigger that initiated the
  readout}

\newduneabbrev{zs}{ZS}{zero-suppression}{Used to delete some portion of a
  data stream that does not significantly deviate from zero or
  intrinsic noise levels. 
  It may be applied at different granularity from per-channel to per
  \gls{detunit}}

\newduneabbrev{rc}{RC}{run control}{The system for configuring,
  starting and terminating the \gls{daq}}

\newduneword{r-c}{RC}{resistive-capacitive (circuit)}

\newduneabbrev{daqccm}{CCM}{DAQ control, configuration and monitoring subsystem}{A system for controlling, configuring and monitoring other systems in particular those that make up the \gls{daq} where the CCM encompasses \gls{rc}}

\newduneword{daqrun}{DAQ run}{A period of time over which relevant data taking conditions and \gls{daq} configuration are asserted to be unchanged. 
  Multiple \gls{daq} runs may occur simultaneously when multiple \glspl{daqpart} are active. 
  This term should not be confused with DUNE experiment or beam ``runs'' that typically span many \gls{daq} runs}
\newduneword{daqrunnum}{DAQ run number}{A monotonically increasing count that uniquely and globally identifies a \gls{daqrun}}

\newduneabbrev{snb}{SNB}{supernova neutrino burst}{A prompt 
  increase in the flux of low-energy neutrinos emitted in the first few seconds of a core-collapse supernova.  It can also refer to a trigger command type that may be due to this phenomenon,
  or detector conditions that mimic its interaction signature}

\newduneabbrev{snble}{SNB/LE}{supernova neutrino burst and low
  energy}{Supernova neutrino burst and low-energy physics program}

\newduneabbrev{snews}{SNEWS}{SuperNova Early Warning System}{A global
  supernova neutrino burst trigger formed by a coincidence of \gls{snb} 
  triggers collected from participating experiments}

\newduneabbrev{pps}{1PPS signal}{one-pulse-per-second signal}{An
  electrical signal with a fast rise time and that arrives in real
  time with a precise period of one second}

\newduneabbrev{sls}{SLS}{spill location system}{A system residing at
  the DUNE far detector site that provides information, possibly
  predictive, indicating periods of time when neutrinos are being
  produced by the \fnal Main Injector beam spills}

\newduneabbrev{wib}{WIB}{warm interface board}{Digital electronics
  situated just outside the \gls{sp} cryostat that receives digital data
  from the \glspl{femb} over cold copper connections and sends it to the \gls{rce}
  \gls{fe} readout hardware}

\newduneabbrev{gps}{GPS}{Global Positioning System}{A satellite-based system that provides a highly accurate \gls{pps} that may be used to synchronize clocks and determine location}

\newduneabbrev{ntp}{NTP}{Network Time Protocol}{A networking protocol that allows synchronizing of clocks to within a few \si{\milli\second} of a time standard on a local network and within a few tens of \si{\milli\second} over the Internet} 

\newduneabbrev{ptproto}{PTP}{Precision Time Protocol}{A networking protocol that allows synchronizing of clocks to within a few \si{\micro\second} of a time standard on a local network} 

\newduneabbrev{irig}{IRIG}{inter-range instrumentation group}{A standards body that defined a time-code standard for transferring timing information}

\newduneabbrev{nic}{NIC}{network interface controller}{Hardware for controlling the interface to a communication network.  Typically, one that obeys the Ethernet protocol}

\newduneabbrev{wiec}{WIEC}{warm interface electronics crate}{Crates mounted on the signal flanges that contain the \glspl{wib}}

\newduneabbrev{ptc}{PTC}{power and timing card}{Cards that provide further processing and distribution of the signals entering and exiting the \gls{sp} cryostat}

\newduneabbrev{ptb}{PTB}{power and timing backplane}{Backplane used to connect the \gls{wib}s and the \gls{ptc}s on the \gls{wiec}. Also connects the \gls{ce} flange on the cryostat penetration}

\newduneabbrev{sipm}{SiPM}{silicon photomultiplier}{A solid-state
  avalanche photodiode sensitive to single \phel signals}

\newduneabbrev{cisc}{CISC}{cryogenic instrumentation and slow controls}{Includes equipment to monitor all detector  components and  \gls{lar} quality and behavior, and provides a control system for many of the detector components}

\newduneword{fte}{FTE}{full-time equivalent. A unit of labor
  for the project. One year of work from one person}

\newduneword{art}{art}{A software framework implementing an
  event-based execution paradigm} 

\newduneabbrev{sam}{SAM}{sequential
  access via metadata}{A data-handling system to store and retrieve
  files and associated metadata, including a complete record of the
  processing that has used the files}

\newduneword{artdaq}{artdaq}{A data acquisition toolkit for data transfer, aggregation and processing}

\newduneword{beamline}{beamline}{A sequence of control and monitoring devices used for the formation of a directed collection of particles}
\newduneabbrev{cdr}{CDR}{conceptual design report}{A formal project
  document 
   that describes the experiment
  at a conceptual level}

\newduneabbrev{cf}{CF}{conventional facilities}{Pertaining to
  construction and operation of buildings and conventional infrastructure, and for \gls{lbnf-dune}, CF includes the excavation caverns}

\newduneabbrev{cp}{CP}{charge parity}{Product of charge and parity
  transformations}

\newduneabbrev{cpt}{CPT}{charge, parity, and time reversal symmetry}{product of charge, parity
  and time-reversal transformations}

\newduneabbrev{cpv}{CPV}{charge-parity symmetry violation}{Lack of
  symmetry in a system before and after charge and parity
  transformations are applied. 
  For CP symmetry to hold,  a particle turns into its
 corresponding antiparticle under a charge transformation, and a parity
transformation inverts its space coordinates, i.e., 
produces the mirror image}

\newduneword{doe}{DOE}{U.S. Department of Energy}

\newduneabbrev{fra}{FRA}{Fermi Research Alliance}{A joint partnership of the University of Chicago and the Universities Research Association (URA) that manages and operates Fermilab on behalf of the \gls{doe}}

\newduneabbrev{dune}{DUNE}{Deep Underground Neutrino Experiment}{A leading-edge, international experiment for neutrino science and proton decay studies}

\newduneabbrev{esh}{ES\&H}{environment, safety and health}{A discipline and specialty that studies and implements practical aspects of environmental protection and safety at work} 

\newduneabbrev{ppe}{PPE}{personnel protective equipment}{Equipment worn to minimize exposure to hazards that cause serious workplace injuries and illnesses}

\newduneabbrev{odh}{ODH}{oxygen deficiency hazard}{a hazard that occurs when inert gases such as nitrogen, helium, or argon displace room air and thus reduce the percentage of oxygen below the level required for human life}

\newduneabbrev{feshm}{FESHM}{Fermilab Environment, Safety and Health Manual}{The document that contains Fermilab's policies and procedures designed to manage environment, safety, and health in all its programs}

\newduneabbrev{fscf}{FSCF}{far site conventional facilities}{The
  \gls{cf} at the DUNE far detector site, \surf}
  
\newduneabbrev{nscf}{NSCF}{near site conventional facilities}{The
  \gls{cf} at the DUNE near detector site, \fnal}

\newduneabbrevs{gut}{GUT}{grand unified theory}{grand unified theories}{A class of theories that unifies the electro-weak and strong forces}

\newduneabbrev{lar}{LAr}{liquid argon}{Argon in its liquid phase; it is a cryogenic liquid with a boiling point of $\SI{-90}{^\circ{C}}$ (\SI{87}{K}) and density of \SI{1.4}{g/ml}}

\newduneabbrev{lbl}{LBL}{long-baseline}{Refers to the distance between the 
  neutrino source  and the \gls{fd}.  It can also refer to the distance between the near and far detectors. 
  The ``long'' designation is an approximate and relative distinction. For DUNE, this distance  (between \gls{fnal} and \gls{surf}) is approximately \SI{1300}{km}}

\newduneabbrev{lbnf}{LBNF}{Long-Baseline Neutrino Facility}{The
  organizational entity responsible for developing the neutrino beam, the cryostats
  and cryogenics systems, and the conventional facilities for DUNE}
  
\newduneabbrev{lbnf-dune}{LBNF/DUNE}{LBNF and DUNE project}{The overall global project, including \gls{lbnf} and \gls{dune}}

\newduneabbrev{lbnc}{LBNC}{Long-Baseline Neutrino Committee}{The committee, composed of internationally prominent scientists with relevant expertise, charged by the \gls{fnal} director to review the scientific, technical, and managerial progress, plans and decisions associated with \gls{dune}}

\newduneabbrev{ncg}{NCG}{Neutrino Cost Group}{A group of internationally prominent scientists with relevant experience that is charged by the \gls{fnal} director to review the cost, schedule, and associated risks for the \gls{dune} experiment}

\newduneabbrev{mh}{MH}{mass hierarchy}{Describes the separation
  between the mass squared differences related to the solar and
  atmospheric neutrino problems}

\newduneabbrev{mi}{MI}{Fermilab Main Injector}{An accelerator at
  \fnal that provides a beam of high-energy protons that upon
  striking a target produce secondaries that decay to provide the
  neutrinos directed toward the DUNE far detector}

\newduneabbrev{pot}{POT}{protons on target}{Typically used as a unit
  of normalization for the number of protons striking the neutrino
  production target}

\newduneabbrev{qa}{QA}{quality assurance}{The set of actions taken to provide confidence that quality requirements are fulfilled, and to detect and correct poor results}

\newduneabbrev{qc}{QC}{quality control}{An aggregate of activities (such as design analysis and inspection for defects) performed to ensure adequate quality in manufactured products}

\newduneabbrev{sm}{SM}{standard model}{Refers to a theory describing
  the interaction of elementary particles}

\newduneabbrev{tdr}{TDR}{technical design report}{A formal project
  document 
  that describes the experiment at a technical level}

\newduneabbrev{prelimdr}{PDR}{preliminary design report}{A formal project
  document 
  that describes the experiment at a preliminary design level}

\newduneabbrev{tp}{IDR}{interim design report}{An intermediate
milestone on the path to a full \gls{tdr}} 

\newduneabbrev{ckm}{CKM matrix}{Cabibbo-Kobayashi-Maskawa
  matrix}{Refers to the matrix describing the mixing between mass and
  weak eigenstates of quarks}

\newduneabbrev{cl}{CL}{confidence level}{Refers to a probability
  used to determine the value of a random variable given its
  distribution}

\newduneabbrev{pmns}{PMNS}{Pontecorvo-Maki-Nakagawa-Sakata}{A type of matrix that describes the mixing between mass and weak eigenstates of
  the neutrino}

\newduneabbrevs{cpa}{CPA}{cathode plane assembly}{cathode plane assemblies}{The component of the \single detector module that provides the drift HV cathode}

\newduneabbrev{fc}{FC}{field cage}{The component of a \gls{lartpc} that contains and shapes the applied \efield}

\newduneword{cpafc}{CPA/FC}{A pair of \gls{cpa} panels and the top and bottom \gls{fc} portions that attach to the pair; an intermediate assembly for installation into the \gls{spmod} }

\newduneabbrev{topfc}{top FC}{top field cage}{The horizontal portions of the \gls{sp} \gls{fc}   on the top of the \gls{tpc}}

\newduneabbrev{botfc}{bottom FC}{bottom field cage}{The horizontal portions of the \gls{sp} \gls{fc} on the bottom of the \gls{tpc}}

\newduneabbrev{ewfc}{endwall FC}{endwall field cage}{The vertical portions of the \gls{sp} \gls{fc} near the wall}

\newduneabbrev{gp}{GP}{ground plane}{An electrode held electrically neutral relative to Earth ground voltage; it is mounted on the \gls{fc} in a \gls{spmod} to protect the cryostat wall}

  \newduneword{gg}{ground grid}{An electrode held electrically neutral relative to Earth ground voltage; it is installed between the cathode and the \glspl{pd} in a \gls{dpmod} to protect the \glspl{pmt}, maintaining high transparency to light}

\newduneabbrev{alara}{ALARA}{as low as reasonably
  achievable}{Typically used with regard management of radiation
  exposure but may be used more generally. It means making every
  reasonable effort to maintain e.g., exposures, to as far below the
  limits as practical, consistent with the purpose for that the
  activity is undertaken}

\newduneabbrev{ecal}{ECAL}{electromagnetic calorimeter}{A detector
  component that measures energy deposition of traversing particles (in the near detector conceptual design)}

\newduneabbrev{hv}{HV}{high voltage}{Generally describes a voltage
  applied to drive the motion of free electrons through some media, e.g., LAr}

\newduneword{spmod}{SP module}{single-phase DUNE \gls{fd} module}
\newduneword{dpmod}{DP module}{dual-phase DUNE \gls{fd} module}

\newduneabbrev{tcoord}{TC}{technical coordinator}{A member of the \gls{dune} management team responsible for organizing the technical aspects of the project effort; is head of \gls{tc}}

\newduneabbrev{rcoord}{RC}{resource coordinator}{A member of the \gls{dune} management team responsible for coordinating the financial resources of the project effort}

\newduneword{tc}{technical coordination}{The DUNE organization responsible for overall integration 
of the detector elements and successful execution of the detector
construction project; areas of responsibility include 
general project oversight, systems engineering, \gls{qa} 
and safety}

\newduneabbrev{exb}{EB}{executive board}{The highest level DUNE
  decision-making body for the collaboration}

\newduneabbrev{tb}{TB}{technical board}{The DUNE organization responsible for
  evaluating technical decisions}

\newduneabbrev{rrb}{RRB}{Resources Review Board}{A part of \gls{dune}'s international project governance structure, composed of representatives of all funding agencies that sponsor the project, and of  \gls{fnal} management, established to provide coordination among funding partners and oversight of \gls{dune}}

\newduneabbrev{inc}{INC}{International Neutrino Council}{A highest-level international advisory body to the U.S. \gls{doe} and the  \gls{fnal} directorate on matters related to the  \gls{lbnf} and the  \gls{pip2} projects. This council is composed of representatives from the international funding agencies and  \gls{cern} that make major contributions the infrastructure}

\newduneabbrev{cc}{CC}{charged current}{Refers to an interaction
  between elementary particles where a charged weak force carrier
  ($W^+$ or $W^-$) is exchanged}

\newduneabbrev{dis}{DIS}{deep inelastic scattering}{Refers to the 
  interaction of an elementary charged particle with a nucleus in an
  energy range where the interaction can be modeled as taking place with
  individual nucleons}

\newduneabbrev{fsi}{FSI}{final-state interactions}{Refers to
  interactions between elementary or composite particles subsequent to
  the initial, fundamental particle interaction, such as may occur as
  the products exit a nucleus}

\newduneword{geant4}{Geant4}{A
  software toolkit for the simulation of the passage of particles
  through matter using \gls{mc} methods}

\newduneabbrev{genie}{GENIE}{Generates Events for Neutrino Interaction
  Experiments}{Software providing an object-oriented neutrino
  interaction simulation resulting in kinematics of the products of
  the interaction}

\newduneabbrev{mc}{MC}{Monte Carlo}{Refers to a method of numerical
  integration that entails the statistical sampling of the integrand
  function. 
  Forms the basis for some types of detector and physics simulations}

\newduneabbrev{qe}{QE}{quasi-elastic}{Refers to interaction between
  elementary particles and a nucleus in an energy range where the
  interaction can be modeled as occurring between constituent quarks
  of one nucleon and resulting in no bulk recoil of the resulting
  nucleus}

\newduneabbrev{mou}{MoU}{memorandum of understanding}{A document
  summarizing an agreement between two or more parties}

\newduneabbrev{pip2}{PIP-II}{Proton Improvement Plan II}{A \gls{fnal} project for
  improving the protons on target delivered delivered by the \gls{lbnf} neutrino production beam. 
  This is version two of this plan and it is planned to be followed by a PIP-III}
  
\newduneabbrev{sdsta}{SDSTA}{South Dakota Science and Technology
  Authority}{The legal entity that manages \gls{surf}, in Lead, S.D}
  
\newduneabbrev{sdsd}{SDSD}{Fermilab South Dakota Services Division}{A Fermilab division responsible providing host laboratory functions at SURF in South Dakota}

\newduneabbrev{firus}{FIRUS}{Facility Information Reporting Utility System}
 {The safety system at \surf}

\newduneabbrev{bsi}{BSI}{building and site infrastructure}
 {The work package for outfitting of the \gls{lbnf} underground infrastructure}

\newduneabbrev{wbs}{WBS}{work breakdown structure}{An organizational
  project management tool by which the tasks to be performed are
  partitioned in a hierarchical manner}

\newduneabbrev{br}{BR}{branching ratio}{A fractional probability for a
  decay of a composite particle to occur into some specified set or
  sets of products}
\newduneword{bsm}{BSM}{beyond the standard model}

\newduneabbrev{dm}{DM}{dark matter}{The term given to the unknown
  matter or force that explains measurements of galaxy motion 
  that are otherwise inconsistent with the amount of mass associated
  with the observed amount of photon production}
  
  \newduneabbrev{bdm}{BDM}{boosted dark matter}{A new model that describes a relativistic dark matter particle boosted by the annihilation of heavier dark matter participles in the galactic center or the sun}

\newduneabbrev{cern}{CERN}{European Organization for Nuclear
Research}{The leading particle physics laboratory in Europe and home to the ProtoDUNEs. (In French, the Organisation Europ\'{e}enne pour la Recherche Nucl\'{e}aire, derived from Conseil Europ\'{e}en pour la Recherche Nucl\'{e}aire}

\newduneabbrev{dsnb}{DSNB}{diffuse supernova neutrino background}{The
  term describing the pervasive, constant flux of neutrinos due to all
  past supernova neutrino bursts}

\newduneabbrev{espp}{ESPP}{European Strategy for Particle Physics}{The
cornerstone of Europe's
decision-making process for the long-term future of the
field. Mandated by the \gls{cern} Council, it is formed through a broad
consultation of the grass-roots particle physics community, it
actively solicits the opinions of physicists from around the world,
and it is developed in close coordination with similar processes in
the USA and Japan in order to ensure coordination between regions and
optimal use of resources globally}

\newduneabbrev{gar}{GAr}{gaseous argon}{argon in its gas phase}
\newduneabbrev{gartpc}{GArTPC}{gaseous argon time-projection chamber}{A \gls{tpc} filled with gaseous argon; a possible technology choice for the \gls{nd}}

\newduneabbrev{globes}{GLoBES}{General Long-Baseline Experiment
  Simulator}{A software package for simulating energy spectra of
  neutrino flux, interactions, and energy spectra measured after application of some
  model of a detector response)}

\newduneabbrev{snowglobes}{SNOwGLoBES}{SuperNova
Observatories with GLoBES} {From the official description~\cite{snowglobes}: 
SNOwGLoBES is public software for computing interaction rates and distributions of observed quantities for \gls{snb} neutrinos in common detector materials} 

\newduneword{l/e}{L/E}{length-to-energy ratio}
\newduneword{lri}{LRI}{long-range interactions}

\newduneabbrev{nc}{NC}{neutral current}{Refers to an interaction
  between elementary particles where a neutrally charged weak force carrier
  ($Z^0$) is exchanged}

\newduneabbrev{nh}{NH}{normal hierarchy}{Refers to the neutrino mass
  eigenstate ordering whereby the sign of the mass squared difference
  associated with the atmospheric neutrino problem is positive}

\newduneabbrev{ih}{IH}{inverted hierarchy}{Refers to the neutrino mass
  eigenstate ordering whereby the sign of the mass squared difference
  associated with the atmospheric neutrino problem is negative}

\newduneabbrev{no}{NO}{normal ordering}{Refers to the neutrino mass
  eigenstate ordering whereby the sign of the mass squared difference
  associated with the atmospheric neutrino problem is positive}

\newduneabbrev{io}{IO}{inverted ordering}{Refers to the neutrino mass
  eigenstate ordering whereby the sign of the mass squared difference
  associated with the atmospheric neutrino problem is negative}

\newduneabbrev{msw}{MSW}{Mikheyev-Smirnov-Wolfenstein effect}{Explains
  the oscillatory behavior of neutrinos produced inside the sun as
  they traverse the solar matter}

\newduneabbrev{nsi}{NSI}{nonstandard interaction}{A general class of
  theory of elementary particles other than the Standard Model}

\newduneabbrev{pfive}{P5}{Particle Physics Project Prioritization
Panel}{The Particle Physics Project Prioritization Panel (P5) was a
subpanel of the High Energy Physics Advisory Panel (HEPAP). It completed
its Report, a ten-year strategic plan for high energy physics in the
U.S., in 2014. This report included a recommendation that ``host a world-leading neutrino
program that will have an optimized set of short- and long-baseline neutrino oscillation experiments, and its long-term focus
is a reformulated venture referred to here as the Long Baseline
Neutrino Facility (LBNF)''}

\newduneabbrev{sme}{SME}{standard-model extension}{an effective field theory that contains the \gls{sm}, general relativity, and all possible operators that break Lorentz symmetry (Wikipedia)}

\newduneabbrev{susy}{SUSY}{supersymmetry}{Theoretical symmetry between a fermion and a boson}

\newduneabbrev{wimp}{WIMP}{weakly-interacting massive particle}{A
  hypothesized particle that may be a component of dark matter}

\newduneabbrev{ce}{CE}{cold electronics}{Analog and digital readout electronics that operate at cryogenic temperatures}

\newduneabbrev{crp}{CRP}{charge-readout plane}{In the \gls{dp} technology, a  collection of
  electrodes in a planar arrangement placed at a particular voltage
  relative to some applied \efield such that drifting electrons
  may be collected and their number and time may be measured}

\newduneabbrev{dram}{DRAM}{dynamic random access memory}{A computer memory technology}

\newduneabbrev{fnal}{Fermilab}
{Fermi National Accelerator Laboratory}{U.S. national laboratory in Batavia, IL. It is the laboratory that hosts \gls{dune} and serves as its near site}

\newduneabbrev{bnl}{BNL}{Brookhaven National Laboratory}{US national laboratory in Upton, NY}

\newduneabbrev{slac}{SLAC}{SLAC National Accelerator Laboratory}{US national laboratory in Menlo Park, CA}

\newduneabbrev{lbnl}{LBNL}{Lawrence Berkeley National Laboratory}{US national laboratory in Berkeley, CA}

\newduneabbrev{anl}{ANL}{Argonne National Laboratory}{US national laboratory in Lemont, IL}

\newduneabbrev{lanl}{LANL}{Los Alamos National Laboratory}{US national laboratory in Los Alamos, NM}

\newduneabbrev{fs}{FS}{full stream}{Relates to a data stream that has not undergone selection, compression or other form of reduction}

\newduneabbrev{lem}{LEM}{large electron multiplier}{A micro-pattern detector suitable for use in ultra-pure argon vapor; LEMs consist of copper-clad PCB boards with sub-millimeter-size holes through which electrons undergo amplification}

\newduneabbrev{lng}{LNG}{liquefied natural gas}{Pertaining to natural gas in its liquid phase}

\newduneabbrev{mip}{MIP}{minimum ionizing particle}{Refers to a
  particle traversing some medium such that the particle's mean energy loss is  
  near the minimum}

\newduneabbrev{pd}{PD}{photon detector}{The detector
  elements involved in measurement of the number and arrival times of
  optical photons produced in a detector module} 

\newduneabbrev{pmt}{PMT}{photomultiplier tube}{A device that makes use
  of the photoelectric effect to produce an electrical signal from the
  arrival of optical photons}

\newduneabbrev{ppm}{ppm}{parts per million}{A concentration equal to one part in $10^{-6}$}
\newduneabbrev{ppb}{ppb}{parts per billion}{A concentration equal to one part in $10^{-9}$}
\newduneabbrev{ppt}{ppt}{parts per trillion}{A concentration equal to one part in $10^{-12}$}

\newduneword{rio}{RIO}{reconfigurable input output}

\newduneabbrev{s/n}{S/N}{signal-to-noise}{signal-to-noise ratio}
\newduneword{snr}{\mbox{S/N}}{signal-to-noise ratio}

\newduneword{ssp}{SSP}{SiPM signal processor}

\newduneabbrev{sbn}{SBN}{Short-Baseline Neutrino}{A \gls{fnal} program consisting of three collaborations, \gls{microboone}, \gls{sbnd}, and \gls{icarus}, to perform sensitive searches for $\nue$ appearance and $\numu$ disappearance in the Booster Neutrino Beam}

\newduneword{stt}{STT}{straw tube tracker}

\newduneword{wire board}{wire board}{At the head end of the APA in the \single TPC, stacks of electronics boards referred to as ``wire boards'' are arrayed to anchor the wires.  They also provide the connection between the wires and the cold electronics} 

\newduneabbrev{wls}{WLS}{wavelength-shifting}{A material or process by
  which incident photons are absorbed by a material and photons are
  emitted at a different, typically longer, wavelength}
  
\newduneabbrev{tpb}{TPB}{tetra-phenyl butadiene}{A 
\gls{wls} material}

\newduneabbrev{ptp}{PTP}{p-terphenyl}{A 
\gls{wls} material}

\newduneabbrev{sft}{SFT}{signal feedthrough}{A cryostat penetration allowing for the passage of cables or other extended parts}
\newduneabbrev{sftchimney}{SFT chimney}{signal feedthrough chimney}{In the \dual technology, a volume above the cryostat penetration used for a signal feedthrough}

\newduneabbrev{catiroc}{CATIROC}{charge and time integrated readout chip}{A complete read-out chip manufactured in AustriaMicroSystem designed to read arrays of 16 photomultipliers}

\newduneabbrev{wr}{WR}{White Rabbit}{A component of the timing system that forwards clock signal and time-of-day reference data to the master timing unit}

\newduneabbrev{mch}{MCH}{MicroTCA Carrier Hub}{An network switching device}

\newduneabbrev{wrmch}{WR-MCH}{White Rabbit \gls{utca} Carrier Hub}{A card mounted in \gls{utca} crate that recieves time syncronization information and trigger data packets over \gls{wr} network and disributes them to the \gls{amc} over \gls{utca} backplane} 

\newduneabbrev{wrtsn}{WR-TSN}{White Rabbit TimeStamping Node}{A unit on the \gls{wr} network that timestamps the trigger signals and sends out trigger data packets to \gls{wrmch}}

\newduneword{cmp}{CMP}{configuration management plan}
\newduneword{qap}{QAP}{quality assurance plan} 
\newduneword{ieshp}{IESHP}{integrated environmental, safety and health plan}
\newduneword{dmp}{DMP}{data management plan} 
\newduneword{qam}{QAM}{quality assurance manager} 

\newduneabbrev{dss}{DSS}{detector support system}{The system used to support a \gls{sp} \gls{detmodule} within its cryostat}

\newduneabbrev{ddss}{DDSS}{DUNE detector safety system}{The system used to manage key aspects of detector safety}

\newduneabbrev{lc}{LC}{logistics center}{A facility where \gls{lbnf} and \gls{dune} components will be received and transhipped to \gls{surf}}

\newduneabbrev{tco}{TCO}{temporary construction opening}{An opening in the side of a cryostat through which detector elements are brought into the cryostat; utilized during construction and installation}

\newduneabbrev{surf}{SURF}{Sanford Underground Research Facility}{The laboratory in South Dakota where the \gls{lbnf} \gls{fscf} will be constructed and the \gls{dune} \gls{fd} will be installed and operated}

\newduneabbrev{sit}{SIT}{surface installation team}{An organizational unit responsible for logistics and integration in South Dakota}

\newduneabbrev{uit}{UIT}{underground installation team}{An organizational unit responsible for installation in the underground area at the \gls{surf} site}

\newduneabbrev{cmgc}{CMGC}{construction manager/general contractor}{The organizational unit responsible for management of the construction of conventional facilities at the underground area at the \surf site}

\newduneword{cdrev}{conceptual design review}{A project management device by which a conceptual design is reviewed} 
\newduneword{pdr}{preliminary design review}{A project management device by which an early design is reviewed} 
\newduneword{fdr}{final design review}{A project management device by which a final design is reviewed}
\newduneword{prr}{production readiness review}{A project management device by which the production readiness is reviewed}
\newduneword{irr}{installation readiness review}{A project management device by which the plan for installation is reviewed}
\newduneword{orr}{operational readiness review}{A project management device by which the operational readiness is reviewed}
\newduneword{ppr}{production progress review}{A project management device by which the progress of production is reviewed} 
\newduneabbrev{edms}{EDMS}{engineering document management system}{A computerized document management system developed and supported at \gls{cern} in which some DUNE documents, drawings and engineering models are managed}
\newduneabbrev{ecr}{ECR}{engineering change request}{The first step in the change control process in which a proposed change is described}
\newduneabbrev{docdb}{DocDB}{Document DataBase}{A computerized document management system developed and supported at \gls{fnal} in which virtually all LBNF and most DUNE documents are managed}

\newduneword{wrgm}{WR grandmaster}{White Rabbit grandmaster}

\newduneabbrev{larsoft}{LArSoft}{Liquid Argon Software}{A shared base of physics software across \lartpc experiments}
\newduneword{nova}{NOvA}{The \nova off-axis neutrino oscillation experiment at \gls{fnal}}
\newduneword{minerva}{MINERvA}{The \minerva neutrino cross sections experiment at  \gls{fnal}}
\newduneword{microboone}{MicroBooNE}{The \lartpc-based \microboone neutrino oscillation experiment at  \gls{fnal}}
\newduneword{sbnd}{SBND}{The Short-Baseline Near Detector experiment at  \gls{fnal}}
\newduneabbrev{nexo}{nEXO}{Enriched Xenon Observatory}{Experiment at Lawrence Livermore National Laboratory (U.S. national lab in Livermore, CA)searching for new physics with neutrinoless double-beta decay}
\newduneword{argoneut}{ArgoNeuT}{The ArgoNeuT test-beam experiment and \gls{lartpc} prototype at  \gls{fnal}}
\newduneword{icarus}{ICARUS}{A neutrino experiment that was located at the Laboratori Nazionali del Gran Sasso (LNGS) in Italy, then refurbished at \gls{cern} for re-use in the same neutrino beam from \gls{fnal} used by the MiniBooNE, \gls{microboone} and \gls{sbnd} experiments. The ICARUS detector is being reassembled at \gls{fnal}}
\newduneword{atlas}{ATLAS}{One of two general-purpose detectors at the \gls{lhc}. It investigates a wide range of physics, from the search for the Higgs boson to extra dimensions and particles that could make up \gls{dm}}

\newduneword{lbne}{LBNE}{Long Baseline Neutrino Experiment (a terminated US project that was reformulated in 2014 under the auspices of the new \gls{dune} collaboration, an internationally coordinated and internationally funded program, with \gls{fnal} as host)}

\newduneabbrev{lbno}{LBNO}{Long Baseline Neutrino Observatory} {A terminated European project that, during its six-year duration, assessed the feasibility of a next-generation deep underground neutrino observatory in Europe)}

\newduneword{wirecell}{Wire-Cell}{A tomographic automated \threed neutrino event reconstruction method for \lartpc{}s}
\newduneabbrev{wct}{WCT}{Wire-Cell Toolkit}{A software toolkit with data flow processing components for \lartpc noise and signal simulation, noise filtering, signal processing, and tomographic \threed ionization activity imaging}
\newduneword{ftslite}{F-FTS-lite}{Light-weight version of the \fnal File Transfer system used for rapid data transfers out of the online systems}
\newduneabbrev{fts}{FTS}{File Transfer System}{A file transfer system developed at \fnal to catalog and move data to permanent storage}

\newduneword{35t}{35 ton prototype}{A prototype cryostat and \gls{sp} detector built at \fnal before the \gls{protodune} detectors}

\newduneabbrev{mcr}{MCR}{main communications room}{Space at the \gls{fd} site for cyber infrastructure}

\newduneabbrev{cuc}{CUC}{central utility cavern}{The utility cavern at the 4850L of \gls{surf} located between the two detector caverns. It contains utilities such as central cryogenics and other systems, and the underground data center and control room}

\newduneabbrev{cfd}{CFD}{computational fluid dynamics}{High performance computer-assisted modeling of fluid dynamical systems}
\newduneword{vuv}{VUV}{vacuum ultra-violet}
\newduneword{tallbo}{TallBo}{A cylindrical cryostat at \gls{fnal} primarily used for developing scintillation light collection technologies for \gls{lartpc} detectors}

\newduneword{root}{ROOT}{A modular scientific software toolkit. It provides all the functionalities needed to deal with big data processing, statistical analysis, visualisation and storage. It is mainly written in C++ but integrated with other languages such as Python and R}

\newduneabbrev{eos}{EOS}{EOS}{The XRootD-based distributed file system developed by CERN}
\newduneabbrev{ehn1}{EHN1}{Experiment Hall North One}{Location at CERN of the ProtoDUNE experiments}
\newduneword{led}{LED}{Light-emitting diode}
\newduneabbrev{rtd}{RTD}{resistance temperature detector}{A temperature sensor consisting of a material with an accurate and reproducible resistance/temperature relationship}
\newduneword{swc}{SWC}{Software \& Computing}
\newduneabbrev{las}{LAS}{LEM-anode Sandwich}{In the \dual technology, a \gls{lem} and its corresponding anode are mounted together in a module called a LEM-anode sandwich}

\newduneword{roi}{ROI}{region of interest}
\newduneabbrev{hpc}{HPC}{high-performance computing}{high-performance computing facilities; generally computing facilities emphasizing parallel computing with aggregate power of more than a teraflop}

\newduneword{comfund}{common fund}{The shared resources of the collaboration}
\newduneabbrev{ims}{IMS}{integrated master schedule}{A project management device consisting of linked tasks and milestones}

\newduneword{hvdb}{HVDB}{HV divider board}
\newduneword{sas}{SAS}{Another term for the materials airlock; a pass-through chamber used to ensure safe transfer of materials into a clean room, avoiding contamination in both directions}

\newduneabbrev{fea}{FEA}{finite element analysis}{Simulation of a physical phenomenon using the numerical technique called Finite Element Method (FEM), a numerical method for solving problems of engineering and mathematical physics}

\newduneword{fss}{FSS}{field shaping strips}
\newduneword{lvds}{LVDS}{low-voltage differential signaling}

\newduneword{esd}{ESD}{electrostatic discharge}

\newduneabbrev{rp}{RP}{resistive panel}{Resistive panels form the constant potential surfaces for a \gls{spmod} \gls{cpa}; they are composed of a thin layer of carbon-impregnated Kapton and laminated to both sides of a \frfour sheet}

\newduneword{uhmwpe}{UHMWPE}{ultra-high molecular weight polyethylene}

\newduneword{cts}{CTS}{Cryogenic Test System}
\newduneword{plc}{PLC}{programmable logic controller}

\newduneword{mppc}{MPPC}{\SI{6}{mm}$\times$\SI{6}{mm} Multi-Pixel Photon Counters produced by Hamamatsu\texttrademark{} Photonics K.K}

\newduneabbrev{sfp}{SFP}{small form-factor pluggable}{a particular standard for optical transceivers}

\newduneabbrev{minipod}{MiniPOD}{miniature parallel optical device}{a family of types of multi-channel optical transceivers}

\newduneword{ccc}{CCC}{configuration change command}
\newduneword{act}{ACT}{activation time stamp}
\newduneword{lcm}{LCM}{light calibration module}
\newduneword{lpm}{LPM}{light pulser module}
\newduneword{dac}{DAC}{digital-to-analog converter}
\newduneword{arapuca}{ARAPUCA}{A \gls{pds} design that consists of a light trap that captures wavelength-shifted photons inside boxes with highly reflective internal surfaces until they are eventually detected by \gls{sipm} detectors or are lost}
\newduneword{sarapu}{S-ARAPUCA}{Standard \gls{arapuca} design with different \gls{wls} coatings on both faces of the dichroic filter window(s) of the cell}
\newduneword{xarapu}{X-ARAPUCA}{Extended \gls{arapuca} design with \gls{wls} coating on only the external face of the dichroic filter window(s) but with a \gls{wls} doped plate inside the cell}
\newduneword{feb}{FEB}{front-end board}

\newduneabbrev{lsnd}{LSND}{Liquid Scintilator Neutrino Detector}{A scintillation detector and associated experiment located at Los Alamos National Laboratory}

\newduneabbrev{cvn}{CVN}{convolutional visual network}{An algorithm for identifying neutrino interactions based on their topology and without the need for detailed reconstruction algorithms}

\newduneword{pandora}{Pandora}{The Pandora multi-algorithm approach to pattern recognition} 

\newduneabbrev{pma}{PMA}{Projection Matching Algorithm}{A reconstruction algorithm that combines \twod reconstructed objects to form a \threed representation}
\newduneabbrev{bdt}{BDT}{boosted decision tree}{A method of multivariate analysis}
\newduneabbrev{cnn}{CNN}{convolutional neural network}{A deep learning technique most commonly applied to analyzing visual imagery}
\newduneword{pdg}{PDG}{Particle Data Group}

\newduneword{pci}{PCI}{Peripheral Component Interconnect}

\newduneword{labview}{LabVIEW}{Laboratory Virtual Instrument Engineering Workbench is a system-design platform and development environment for a visual programming language from National Instruments}

\newduneword{pcb}{PCB}{printed circuit board}

\newduneword{crio}{cRIO}{Compact Reconfigurable Input Output}

\newduneword{dcs}{DCS}{Distributed Communications System}

\newduneword{opc-ua}{OPC-UA}{OPC  Unified Architecture is a machine to machine communication protocol for industrial automation developed by the OPC Foundation. OPC stands for Object Linking and Embedding for Process Control}

\newduneword{cabangle}{Cabibbo angle}{A quark mixing parameter that governs the coupling of up quarks to strange quarks}
\newduneword{valor}{VALOR}{A neutrino oscillation fitting framework that is used by \gls{t2k}; the name stands for VALencia-Oxford-Rutherford, the original three institutions that developed it}
\newduneword{cafana}{CAFAna}{Common Analysis File Analysis}
\newduneabbrev{pca}{PCA}{principal component analysis}{A statistical procedure that uses an orthogonal transformation to convert a set of observations of possibly correlated variables into a set of values of linearly uncorrelated variables called principal components (Wikipedia)}
\newduneword{numi}{NuMI}{a set of facilities at \fnal, collectively called ``Neutrinos at the Main Injector.''  The NuMI neutrino beamline target system converts an intense proton beam into a focused neutrino beam}
\newduneword{gibuu}{GiBUU}{Giessen Boltzmann-Uehling-Uhlenback Project; a unified theory and transport framework in the MeV and GeV energy regimes for elementary reactions on nuclei }
\newduneabbrev{rpa}{RPA}{random phase approximation} {an approximation method commonly used for describing the dynamic linear electronic response of electron systems (Wikipedia)}
\newduneword{t2k}{T2K}{T2K (Tokai to Kamioka) is a long-baseline neutrino experiment in Japan studying neutrino oscillations}
\newduneword{mptdet}{MPT detector}{multipurpose tracking detector}

\newduneword{lariat}{LArIAT}{The repurposed ArgoNeuT \gls{lartpc}, modified for use in a charged particle beam, dedicated to the calibration and precise characterization of the output response of these detectors}

\newduneword{captain}{CAPTAIN}{Experimental program sited at \gls{lanl} that is designed to make measurements of scientific importance to \gls{lbl} neutrino physics and physics topics that will be explored by large underground detectors}

\newduneword{dayabay}{Daya Bay}{a neutrino-oscillation experiment in Daya Bay, China, designed to measure the mixing angle $\Theta_{13}$  using antineutrinos produced by the reactors of the Daya Bay and Ling Ao nuclear power plants}

\newduneword{nuwro}{NuWro}{neutrino interaction generator}

\newduneabbrev{neut}{NEUT}{neutrino interaction generator}{A neutrino interaction simulation program library for the studies of atmospheric accelerator neutrinos}

\newduneword{minos}{MINOS}{A long-baseline neutrino experiment, with a near detector at \gls{fnal} and a far detector in the Soudan mine in Minnesota, designed to observe the phenomena of neutrino oscillations (ended data runs in 2012)}

\newduneabbrev{efig}{EFIG}{Experimental Facilities Interface Group}{The body responsible for the required high-level coordination between the \gls{lbnf} and \gls{dune} projects}
\newduneword{ashriver}{Ash River}{The Ash River, Minnesota, USA \gls{nova} experiment far site, used as an assembly test site for \gls{dune}} 

\newduneword{ipd}{project integration director}{Responsible for integration and installation of \gls{lbnf} and \gls{dune} deliverables in South Dakota. Manages the \gls{integoff}}

\newduneabbrev{jpo}{JPO}{Joint Project Office}{The framework through which team members from the LBNF project office, \gls{integoff}, and DUNE \gls{tc} work together to provide coherence in project support functions across the global enterprise. 
Its functions include global project configuration and integration, installation planning and coordination, scheduling, safety assurance, technical review planning and oversight, development of partner agreements, and financial reporting}

\newduneword{ifbeam}{IFbeam}{Database that stores beamline information 
indexed by timestamp}

\newduneabbrev{marley}{MARLEY}{Model of Argon Reaction Low Energy
Yields}{Developed at UC Davis, MARLEY is the first realistic model of
neutrino electron interactions on argon for enegies less than \SI{50}{MeV}. This includes the energy range important for \gls{snb}
neutrinos and also solar 8--boron neutrinos}

\newduneabbrev{es}{ES}{elastic scattering}{Events in which a neutrino
elastically scatters off of another particle}

\newduneabbrev{cno}{CNO}{carbon nitrogen oxygen}{The CNO cycle (for carbon-nitrogen-oxygen) is one of the two known sets of fusion reactions by which stars convert
hydrogen to helium, the other being the proton-proton chain reaction
(pp-chain reaction). In the CNO cycle, four protons fuse, using
carbon, nitrogen, and oxygen isotopes as catalysts, to produce one
alpha particle, two positrons and two electron neutrinos}

\newduneabbrev{sdwf}{SDWF}{South Dakota Warehouse Facility}{Warehousing operations in South Dakota responsible for receiving LBNF and DUNE goods and coordinating shipments to the Ross shaft at \gls{surf}}

\newduneabbrev{wms}{WMS}{warehouse management system}{Commercial software package used to track shipments and interface to freight forwarders. This includes a database for shipping}

\newduneabbrev{dcdb}{DCDB}{DUNE construction database}{Database used by DUNE to track the history and testing of all parts of each \gls{detmodule}}

\newduneabbrev{aup}{AUP}{acceptance for use and possession}{Required for beneficial occupancy of the underground areas at SURF for LBNF and DUNE}

\newduneabbrev{bms}{BMS}{building management system}{Part of the safety system at \gls{surf} that includes the fire and life safety system}
\newduneabbrev{fls}{FLS}{fire and life safety system}{Part of the safety system at \gls{surf}}

\newduneabbrev{sno}{SNO}{Sudbury Neutrino Observatory}{The Sudbury
Neutrino Observatory was a detector built 6800 feet under ground, in
INCO's Creighton mine near Sudbury, Ontario, Canada. SNO was a
heavy-water Cherenkov detector designed to detect neutrinos produced
by fusion reactions in the sun}

\newduneword{sk}{Super-Kamiokande}{Experiment sited in the Kamioka-mine, Hida-city, Gifu, Japan that uses a large water Cherenkov detector to study neutrino properties through the observation of solar neutrinos, atmospheric neutrinos and man-made neutrinos}

\newduneabbrev{id}{ID}{inner diameter}{Inner diameter of a tube}

\newduneabbrev{od}{OD}{outer diameter}{Outer diameter of a tube}

\newduneabbrev{rms}{RMS}{root mean square}{The square root of the arithmetic mean of the squares of a set of values, used as a measure of the typical magnitude of a set of numbers, regardless of their sign}

\newduneabbrev{orc}{ORC}{operational readiness clearance}{Final safety approval prior to the start of operation}

\newduneabbrev{gsc}{GSC group}{global safety coordination group}{DUNE group that evaluates applicable codes and standards, including international code equivalency, for the design, assembly, and installation of the \gls{fd}}

\newduneabbrev{ha}{HA}{hazard analysis}{A first step in a process to assess risk; the result of hazard analysis is the identification of the hazards present for a task or process}
\newduneword{har}{HAR}{hazard analysis report}

\newduneabbrev{tap}{TAP}{trip action plan}{A document required for any trip by a worker to the underground area at \gls{surf}, per that site's access control program; 
it describes the work to be accomplished during the trip} 

\newduneword{em}{EM}{emergency management}
\newduneword{ert}{ERT}{emergency response team}

\newduneabbrev{ndk}{NDK}{nucleon decay}{The hypothetical, baryon number violating decay of a proton or a bound neutron into lighter particles}

\newduneabbrev{emi}{EMI}{electromagnetic interference}{Disturbance generated by an external source that affects an electrical circuit by electromagnetic induction, electrostatic coupling, or conduction}

\newduneabbrev{pe}{PE}{photoelectron}{An electron ejected from the surface of a material by the photoelectric effect}

\newduneabbrev{spe}{SPE}{single photoelectron}{A single photoelectron}

\newduneabbrev{fwhm}{FWHM}{full width at half maximum}{Width of a distribution measured between those points at which the distribution is equal to half of its maximum amplitude}

\newduneabbrev{gdml}{GDML}{geometry description markup language}{An application-indepedent, geometry-description format based on XML}

\newduneabbrev{xml}{XML}{extensible markup language}{A markup language that defines a set of rules for encoding documents in a format that is both human-readable and machine-readable}

\newduneabbrev{crt}{CRT}{cosmic ray tagger}{Detector external to the TPC designed to tag TPC-traversing cosmic ray particles}

\newduneabbrev{sn}{SN}{supernova}{Event that occurs upon the death of certain types of stars}

\newduneabbrev{wg}{WG}{working group}{A group of persons working together to achieve specified goals}

\newduneabbrev{ctsf}{CTSF}{coating, testing and storage facility}{A facility where the the \dual photon detectors will be coated, tested, and stored}

\newduneword{rucio}{Rucio}{Data management system originally developed
by \gls{atlas} but now open-source and shared across HEP}
\newduneabbrev{doma}{DOMA}{data organization, management, and
access}{data organization, management, and access efforts through the
HEP Software Foundation}

\newduneabbrev{hsf}{HSC}{HEP Software Foundation Collaboration}{A foundation that facilitates cooperation and common efforts in high energy physics software and computing internationally}

\newduneabbrev{wlcg}{WLCG}{Worldwide LHC Computing Grid}{Worldwide LHC
Computing Grid}
\newduneabbrev{osg}{OSG}{Open Science Grid}{Open Science Grid}
\newduneabbrev{sci}{SCI}{Scientific Computing Infrastructure}{Proposed
extension of the infrastructure component of \gls{wlcg} to other
experiments}
\newduneabbrev{csc}{CSC}{computing and software consortium}{DUNE
computing and software consortium}

\newduneword{dirac}{DIRAC}{Computing workflow management designed for
LHCb and now used by many HEP experiments}

\newduneword{frp}{FRP}{fiber-reinforced plastic}
\newduneabbrev{hdpe}{HDPE}{high-density polyethylene}{High-density polyethylene plastic}
\newduneword{hvps}{HVPS}{\gls{hv} power supply}
\newduneword{aisi}{AISI}{American Iron and Steel Institute}
\newduneword{ific}{IFIC}{Instituto de Fisica Corpuscular (in Valencia, Spain)}
\newduneabbrev{rsds}{RSDS}{radioactive source deployment system}{Proposed calibration system based on the deployment of
radioactive sources inside the \gls{dune} cryostat}
\newduneword{2p2h}{2p2h}{two particle, two hole}
\newduneabbrev{duneprism}{DUNE-PRISM}{\gls{dune} Precision Reaction-Independent Spectrum Measurement}{a mobile near detector that can perform measurements over a range of angles off-axis from the neutrino beam direction in order to sample many different neutrino energy distributions}
\newduneword{arcube}{ArgonCube}{The name of the core part of the \gls{dune} \gls{nd}, a \gls{lartpc}}

\newduneabbrev{citf}{CITF}{cryogenic instrumentation test facility}{A facility at \fnal with small ($<\,\SI{1}{ton}$) to intermediate ($\sim\,\SI{1}{ton}$) volumes of instrumented, purified TPC-grade \lar, used for testing devices intended for use in \gls{dune}}

\newduneabbrev{3dst}{3DST}{3D scintillator tracker}{The core part of the \threed projection scintillator tracker spectrometer in the near detector conceptual design}
\newduneabbrev{3dsts}{3DST-S}{3D scintillator tracker spectrometer}{The \threed projection scintillator tracker spectrometer  in the near detector conceptual design}
\newduneabbrev{mpd}{MPD}{multi-purpose detector}{A component of the near detector conceptual design; it is a magnetized system consisting of a \gls{hpgtpc} and a surrounding \gls{ecal}}
\newduneabbrev{hpg}{HPG}{high-pressure gas}{gas at high pressure to be used in a \gls{hpgtpc}} 
\newduneabbrev{hpgtpc}{HPgTPC}{high-pressure gaseous argon TPC}{A \gls{tpc} filled with gaseous argon; a possible component of the \gls{dune} \gls{nd}}

\newduneword{src}{SRC}{short-range correlated nucleon-nucleon interactions}
\newduneword{larpix}{LArPix}{ \gls{asic} pixelated charge readout for a \gls{tpc} }
\newduneword{arclt}{ArCLight}{a light detector \gls{arcube} effort}
\newduneword{fhc}{FHC}{forward horn current ($\numu$ mode)}
\newduneword{rhc}{RHC}{reverse horn current ($\overline{\nu}_{\mu}$ mode)}
\newduneword{mwpc}{MWPC}{multi-wire proportional chamber}
\newduneword{na61}{NA61}{CERN hadron production experiment}
\newduneword{pdnd}{ProtoDUNE-ND}{a prototype \gls{dune} \gls{nd}}
\newduneword{ccqe}{CCQE}{charged current quasielastic interaction} 
\newduneabbrev{roc}{ROC}{readout chamber}{readout chamber for gaseous argon \gls{tpc}}
\newduneabbrev{iroc}{IROC}{inner readout chamber}{inner (radial) readout chamber for gaseous argon \gls{tpc}}
\newduneabbrev{oroc}{OROC}{outer readout chamber}{outer (radial) readout chamber for gaseous argon \gls{tpc}}

\newduneword{lux}{LUX}{Large Underground Xenon (LUX) dark matter detector at \gls{surf} }

\newduneword{mjdemo}{Majorana Demonstrator}{Experiment sited at \gls{surf} that  seeks to determine whether neutrinos are their own antiparticles}

\newduneword{lz}{LZ}{Experiment sited at \gls{surf} that  seeks to detect faint interactions between galactic dark matter and regular matter}

\newduneword{mu2e}{Mu2e}{An experiment sited at \gls{fnal} that searches for charged-lepton flavor violation and seeks to discover physics beyond the \gls{sm}}

\newduneword{pdsp2}{ProtoDUNE-SP-2}{A second test run in the singe-phase
ProtoDUNE test stand at CERN, acting as a validation of the final
single-phase detector design}

\newduneword{osha}{OSHA}{Occupational Safety and Health Administration (USA Department of Labor) formed by the Occupational Safety and Health Act of 1970}
\newduneabbrev{pns}{PNS}{pulsed neutron source}{Calibration system based
on neutron capture gamma showers spread out in the whole detector}

\newduneabbrev{fv}{FV}{fiducial volume}{The detector volume within the \gls{tpc} 
that is selected for physics analysis through cuts on reconstructed event position}

\newduneword{p6}{P6}{framework used to plan and status the resource-loaded schedule of activities associated with the USA contributions to \gls{lbnf} and \gls{dune} }
\newduneabbrev{evms}{EVMS}{earned value management system}{Earned Value Management is a systematic approach to the integration and measurement of cost, schedule, and technical (scope) accomplishments on a project or task. It provides both the government and contractors the ability to examine detailed schedule information, critical program and technical milestones, and cost data (text from the US DOE); the EVMS is a system that implements this approach}

\newduneword{core}{CORE}{CORE contributions are in either monetary units or labor hours. They can be technical components for the facility or experiment and the effort of the staff needed to produce, install, and test them;  major facilities for the experiment; or other products and services relevant for the completion of the facility or experiment} 

\newduneabbrev{ahj}{AHJ}{Authority Having Jurisdiction}{An organization, office, or individual responsible for enforcing the requirements of a code or standard, or for approving equipment, materials, an installation, or a procedure (OSHA)}
\newduneword{cte}{CTE}{coefficient of thermal expansion}

\newduneabbrev{opc}{OPC}{open platform communications}{Open platform communications is a series of standards and specifications for industrial telecommunication} 
\newduneword{scada}{SCADA}{supervisory control and data acquisition}
\newduneword{ln}{LN$_2$}{liquid nitrogen}
\newduneabbrev{lapd}{LAPD}{Liquid Argon Purity Demonstrator}{Cryostat at Fermilab for long-term studies requiring a large volume of argon}

\newduneabbrev{pab}{PAB}{Proton Assembly Building}{Home of several \gls{lar} facilities at Fermilab}
\newduneword{hep}{HEP}{high energy physics}
\newduneword{sc}{SC}{scientific computing}  
\newduneword{cms}{CMS}{Compact Muon Solenoid experiment at CERN}
\newduneword{alice}{ALICE}{A Large Ion Collider Experiment, at CERN}
\newduneword{gpib}{GPIB}{general purpose interface bus}

\newduneabbrev{pfparticle}{PFParticle}{particle flow particle}{Each of the individual reconstructed particles in the hierarchy (or particle flow) describing the reconstructed event interaction}

\newduneabbrev{mcparticle}{MCParticle}{Monte Carlo Particle}{Individual true simulated particle}
\newduneword{au}{AU}{astronomical unit}
\newduneword{nufit}{NuFIT 4.0}{The NuFIT 4.0 global fit to neutrino oscillation data}

\newduneabbrev{sgft}{SGFT}{term}{add def (DP install)}
\newduneword{uhv}{UHV}{ultra high vacuum}
\newduneword{lps}{LPS}{laser positioning system}

\newduneword{unicamp}{UNICAMP}{University of Campinas, Sao Paulo, Brazil}
 
\newduneabbrev{fbk}{FBK}{Fondazione Bruno Kessler}{FBK is a research non-profit entity in Trento, Italy that partners in the development of technology with applications in various fields including High Energy Physics}

\newduneword{fft}{FFT}{fast Fourier transform}
\newduneabbrev{enob}{ENOB}{effective number of bits}{The effective number of bits is a measure of the dynamic range of an \gls{adc} and its associated circuitry. The resolution of an \gls{adc} is specified by the number of bits used to represent the analog value, in principle giving 2N signal levels for an N-bit signal. However, all real \gls{adc} circuits introduce noise and distortion. ENOB specifies the resolution of an ideal \gls{adc} circuit that would have the same resolution as the circuit under consideration}
\newduneabbrev{sndr}{SNDR}{signal to noise and distortion ratio}{Also known as SINAD. Ratio of the \gls{rms} signal amplitude to the mean value of the root-sum-square of all other spectral components, including harmonics, but excluding \gls{dc} levels. It is a good indication of the overall dynamic performance of an \gls{adc} because it includes all components which make up noise and distortion}
\newduneabbrev{sfdr}{SFDR}{spurious free dynamic range}{Spurious free dynamic range is the ratio of the \gls{rms} value of the signal to the \gls{rms} value of the worst spurious signal regardless of where it falls in the frequency spectrum. The worst spur may or may not be a harmonic of the original signal}
\newduneabbrev{thd}{THD}{total harmonic distortion}{Total harmonic distortion is the ratio of the \gls{rms} value of the fundamental signal to the mean value of the root-sum-square of its harmonics} 
\newduneword{tvs}{TVS}{transient voltage suppression}

\newduneword{riskprob}{risk probabilities}{The risk probability, after taking into account the planned mitigation activities, is ranked as 
L (low $<\,$\SI{10}{\%}), 
M (medium \SIrange{10}{25}{\%}), or 
H (high $>\,$\SI{25}{\%}). 
The cost and schedule impacts are ranked as 
L (cost increase $<\,$\SI{5}{\%}, schedule delay $<\,$\num{2} months), 
M (\SIrange{5}{25}{\%} and 2--6 months, respectively) and 
H ($>\,$\SI{20}{\%} and $>\,$2 months, respectively)}

\newduneabbrev{lbls}{LBLS}{laser beam location system}
{Auxiliary calibration system providing an independent location measurement of the ionization laser beams direction}

\newduneabbrev{lsst}{LSST}{Large Synoptic Survey Telescope}{8.4 m telescope with 3.2G-pixel camera that will start taking data in 2023}
\newduneabbrev{ska}{SKA}{Square Kilometer Array}{International radio telescope array planned to start data-taking in 2027}
\newduneabbrev{hyperk}{HyperK}{Hyper Kamiokande}{260 kt water Cerenkov neutrino detector to begin construction at Kamiokande in 2020}
\newduneword{lhcb}{LHCb}{LHC experiment dedicated to forward physics}
\newduneword{belleii}{Belle II}{B-factory experiment now running at KEK}

 \newduneabbrev{ldm}{LDM}{light-mass dark matter}{Refers to dark matter particles with mass values much lower than the electroweak scale, specifically below the 1~GeV level}
 
\newduneabbrev{bnv}{BNV}{baryon-number violating}{Describing an interaction where \gls{baryonnumber} is not conserved}

\newduneword{bugey}{Bugey}{Neutrino experiment that operated at the Bugey nuclear power plant in France}

\newduneword{minosplus}{MINOS$+$}{The successor to the \gls{minos} experiment, utilizing the same detectors and beam line, but operating at higher beam energy tune than \gls{minos}, parasitic with \gls{nova}}

\newduneword{baryonnumber}{baryon number}{A quantity expressing the total number of baryons in a system minus the number of antibaryons}

\newduneword{np04}{NP04}{CERN North Area hadron beamline used for the \gls{sp} test beam run}

\newduneword{ua1}{UA1}{UA1 (Underground Area 1) was a particle detector at \gls{cern}'s  Super Proton Synchrotron (SPS). It ran from 1981 until 1990, when the SPS was used as a proton-antiproton collider, searching for traces of W and Z particles in collisions. (CERN) The UA1 dipole magnet was reused in the NOMAD experiment and currently provides the magnetic field for the \gls{t2k} ND280 detector}

\newduneword{ssc}{SSC}{The Superconducting Super Collider was to be a huge underground ring complex beneath the area near Waxahachie, Texas, USA, that would have been the world’s most energetic particle accelerator. It was begun in 1990, but canceled by the U.S. Congress in 1993 (scientificamerican.com Oct 2013)}

\newduneword{daphne}{DAPHNE}{Detector electronics for Acquiring PHotons from NEutrinos is a custom-developed warm front-end waveform digitizing electronics module derived from the readout system developed at Fermilab for the Mu2e experiment}
 
\newduneword{nersc}{NERSC}{National Energy Research Computing Facility at \gls{lbnl}}

  \newduneword{integoff}{integration office}{The office that incorporates the onsite team responsible for coordinating integration and installation activities at SURF}

\newduneabbrev{sma}{SMA}{SubMiniature version A}{Connector interface for coaxial cables
with a screw-type coupling mechanism}

\newduneword{kloe}{KLOE}{KLOE is a $e^+ e^-$ collider detector spectrometer operated at DAFNE, the $\phi$-meson factory at Frascati, Rome. In DUNE it will consist of a \SI{26}{cm} Pb+scintillating fiber ECAL surrounding a cylindrical open detector region that is  \SI{4.00}{m} in diameter and \SI{4.30}{m} long. The ECAL and detector region are embedded in a \SI{0.6}{T} magnetic field created by a \SI{4.86}{m} diameter superconducting coil and a \SI{475}{tonne} iron yoke}

\newduneword{ro}{review office}{An office within the \gls{integoff} that organizes reviews }

\newduneabbrev{doecd}{CD}{critical decision}{The U.S. DOE's Order 413.3B outlines a series of staged project approvals, each of which is referred to as a critical decision (CD)}

\newduneabbrev{lbnfspac}{LBNF SPAC}{LBNF Strategic Project Advisory Committee}{A committee charged by the host laboratory director to provide expert, independent advice on significant issues and strategies related to LBNF project organization, management, and risks}

\newduneabbrev{sand}{SAND}{System for on-Axis Neutrino Detection}{The beam monitor component of the near detector that remains on-axis at all times and serves as a dedicated neutrino spectrum monitor}

\newduneword{4850l}{4850L}{The depth in feet (1480 m) of the top of the cryostats underground at SURF; used more generally to refer to the DUNE underground area. Called the ``4850 level'' or ``4850L''}

\hypersetup{
    pdftitle={\expshort TDR \thedocsubtitle},
    pdfauthor={\expshort Collaboration},
    final=true,
    colorlinks=false,
    linktocpage=true,
    linkbordercolor=blue,
    citebordercolor=green,
    urlbordercolor=magenta,
    filecolor=black,
    pdfpagemode=UseOutlines,
    pdfborderstyle={/S/U},  
}

\renewcommand\thedoctitle{\voltitletc} 
\newcommand\thevolumenumber{\volnumbertc} 

\begin{document}

\pagestyle{titlepage}
\includepdf[pages={-}]{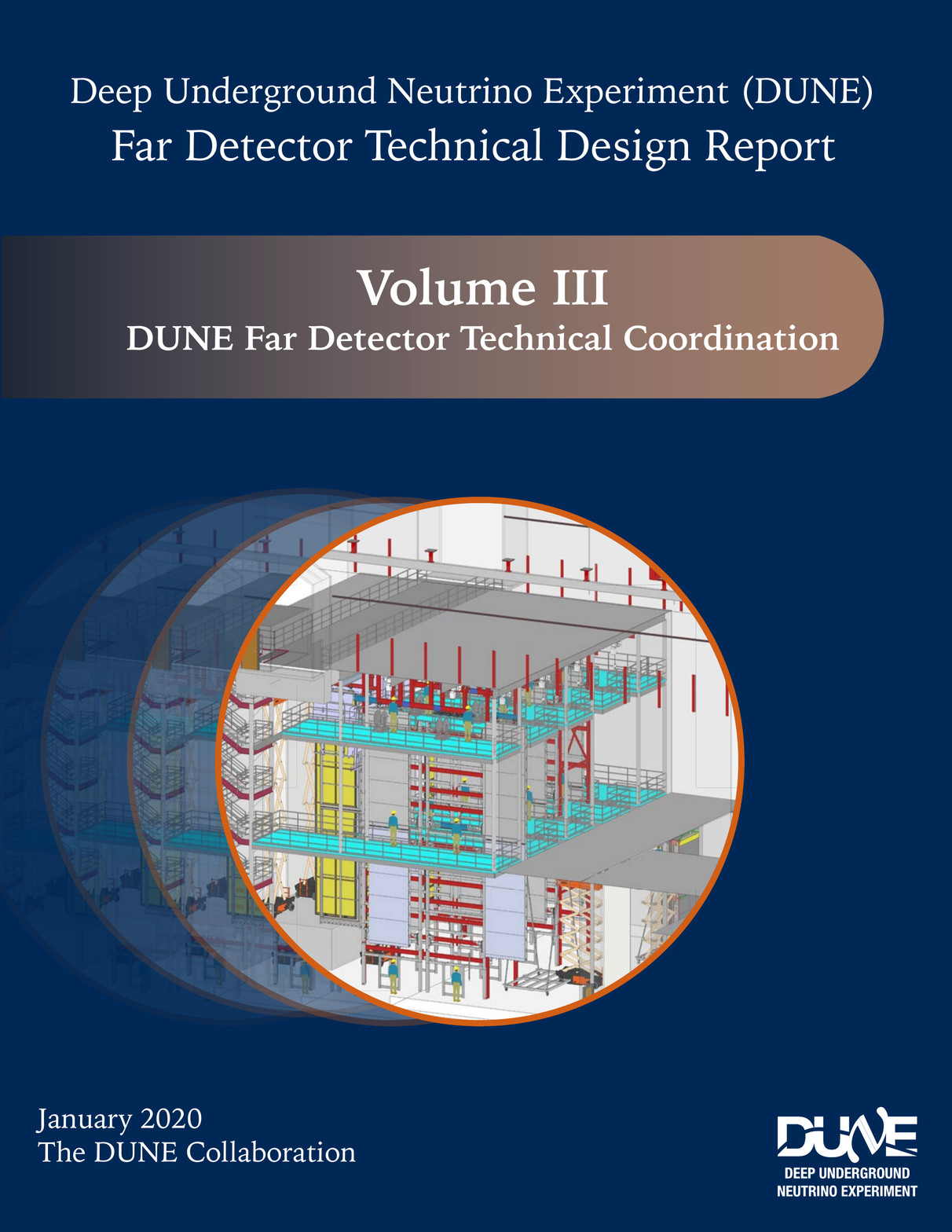}
\cleardoublepage

\cleardoublepage
\vspace*{16cm} 
  {\small  This document was prepared by the DUNE collaboration using the resources of the Fermi National Accelerator Laboratory (Fermilab), a U.S. Department of Energy, Office of Science, HEP User Facility. Fermilab is managed by Fermi Research Alliance, LLC (FRA), acting under Contract No. DE-AC02-07CH11359.
  
The DUNE collaboration also acknowledges the international, national, and regional funding agencies supporting the institutions who have contributed to completing this Technical Design Report.  
  }
\includepdf[pages={-}]{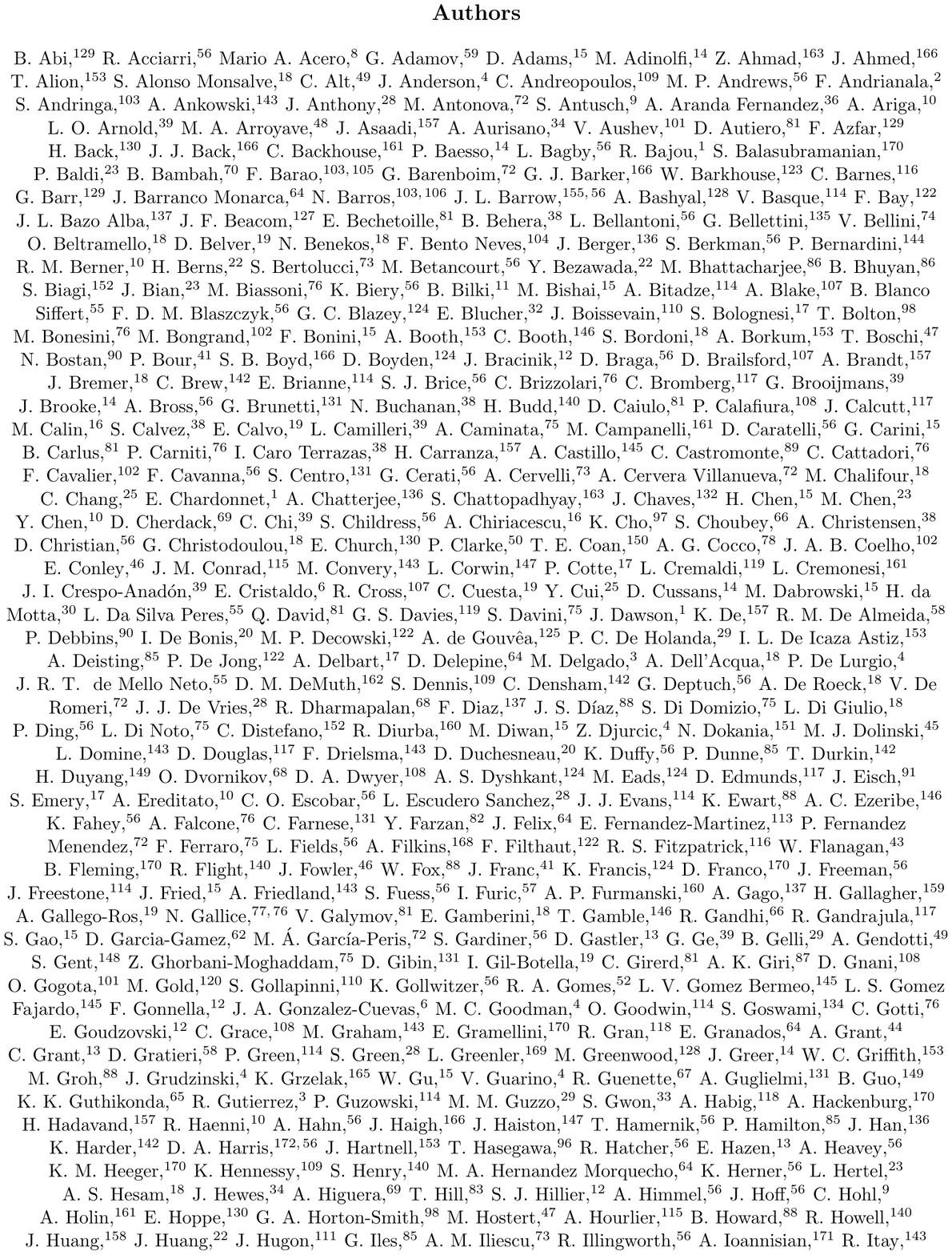}

\renewcommand{\familydefault}{\sfdefault}
\renewcommand{\thepage}{\roman{page}}
\setcounter{page}{0}

\pagestyle{plain}

\textsf{\tableofcontents}

\textsf{\listoffigures}

\textsf{\listoftables}
  \vspace{4mm}
  \addcontentsline{toc}{chapter}{A Roadmap of the DUNE Technical Design Report}

\iffinal\else
\textsf{\listoftodos}
\clearpage
\fi

\renewcommand{\thepage}{\arabic{page}}
\setcounter{page}{1}

\pagestyle{fancy}

\renewcommand{\chaptermark}[1]{%
\markboth{Chapter \thechapter:\ #1}{}}
\fancyhead{}
\fancyhead[RO,LE]{\textsf{\footnotesize \thechapter--\thepage}}
\fancyhead[LO,RE]{\textsf{\footnotesize \leftmark}}

\fancyfoot{}
\fancyfoot[RO]{\textsf{\footnotesize The DUNE Technical Design Report}}
\fancyfoot[LO]{\textsf{\footnotesize \thedoctitle}}
\fancypagestyle{plain}{}

\renewcommand{\headrule}{\vspace{-4mm}\color[gray]{0.5}{\rule{\headwidth}{0.5pt}}}



\cleardoublepage
\chapter*{A Roadmap of the DUNE Technical Design Report}

The \dword{dune} \dword{fd} \dword{tdr} describes the proposed physics program,  
detector designs, and management structures and procedures at the technical design stage.  

The TDR is composed of five volumes, as follows:

\begin{itemize}
\item Volume~\volnumberexec{} (\voltitleexec{}) provides an overview of all of DUNE for science policy professionals.

\item Volume~\volnumberphysics{} (\voltitlephysics{}) describes the DUNE physics program.

\item Volume~\volnumbertc{} (\voltitletc{}) outlines DUNE management structures, methodologies, procedures, requirements, and risks. 

\item Volume~\volnumbersp{} (\voltitlesp{}) and Volume~\volnumberdp{} (\voltitledp{}) describe the two \dword{fd} \dword{lartpc} technologies.

\end{itemize}

The text includes terms that hyperlink to definitions in a volume-specific glossary. These terms  appear underlined in some online browsers, if enabled in the browser's settings.

\cleardoublepage

\cleardoublepage

\chapter{Executive Summary}

\label{vl:tc-execsum}

This volume describes how the activities required to design,
construct, fabricate, install, and commission the \dword{dune}
\dword{fd} modules are organized and managed. The \dword{fd} modules
are hosted at the \dword{lbnf} site at the \dword{surf}. The
\dword{dune} \dword{fd} construction project is one piece of the
global \dword{lbnf-dune}, which encompasses all of the facilities,
supporting infrastructure, and detector elements required to carry out
the \dword{dune} science program at \dword{surf}.
      
The \dword{dune} collaboration has  responsibility for the design 
and construction of the \dword{dune} detectors.  Groups of collaboration 
institutions, referred to as consortia, assume responsibility for 
the different detector subsystems.  The activities of the consortia are 
overseen and coordinated through the \dword{dune} \dword{tc} organization 
headed by the \dword{dune} \dword{tcoord}.  The \dword{tc} organization 
provides project support functions such as safety coordination, 
engineering integration, change control, document management, scheduling, 
risk management, and technical review planning.  \dword{dune} \dword{tc} 
manages internal, subsystem-to-subsystem interfaces, and is responsible 
for ensuring the proper integration of the different subsystems.   

A \dword{jpo} establishes the global engineering
and documentation requirements adhered to within the \dword{dune} 
\dword{fd} construction project, manages external \dword{dune} detector 
interfaces with \dword{lbnf}, and is responsible for ensuring proper 
integration of the \dword{dune} detector elements within the facilities 
and supporting infrastructure.
\dword{dune} \dword{tc} works closely with the support teams of its 
\dword{lbnf-dune} partners within the framework of the \dword{jpo} to 
ensure coherence in project support functions across the entire global 
enterprise.  To ensure consistency of the \dword{dune} \dword{esh} 
and \dword{qa} programs with those across \dword{lbnf-dune}, the 
\dword{lbnf-dune} \dword{esh} and \dword{qa} managers, who sit within 
the \dword{jpo}, are embedded within the \dword{dune} \dword{tc} 
organization.  

The \dword{lbnf}/\dword{dune} \dword{integoff} under the 
direction of the \dword{ipd} incorporates the on site team responsible 
for coordinating integration and installation activities at \dword{surf}.
Detector integration and installation activities are supported by the
\dword{dune} consortia, which maintain responsibility for ensuring
proper installation and commissioning of their subsystems.  External
\dword{dune} interfaces with the on site integration and installation
activities are managed through the \dword{jpo}. Support services are
provided by the \dword{sdsd} and \dword{surf}.

The ordering of the subsequent chapters is chosen to provide first,  
additional detail regarding the organizational structures summarized 
here; second, overviews of the facilities, supporting infrastructure, 
and detectors for context; and third, information on project-related 
functions and methodologies used by \dword{dune} \dword{tcoord} 
focusing on the areas of integration engineering, technical reviews, 
\dword{qa}, and safety oversight.  Because of their more advanced stage 
of development, functional examples presented here focus primarily on 
the \dword{sp} \dword{detmodule}.

\cleardoublepage

\chapter{Global Project Organization}
\label{vl:tc-global}

\section{Global Project Partners}
\label{sec:partners}

The \dword{lbnf} project is responsible for providing both the
\dword{cf} and supporting infrastructure (cryostats and
cryogenics systems) that house the \dword{dune} \dword{fd}
modules. \dword{lbnf} is a U.S. \dword{doe} project incorporating
contributions from international partners and is headed by the
\dword{lbnf} project director who also serves as the \dword{fnal}
deputy director for \dword{lbnf}.  
The international \dword{dune}
collaboration under the direction of its management team is
responsible for the detector components.  The \dword{dune} \dword{fd}
construction project encompasses all activities required for designing
and fabricating the detector elements and incorporates contributions
from a number of international partners.  The organization of the
global \dword{lbnf-dune}, which encompasses both project elements, is
shown in Figure~\ref{fig:DUNE_global}.
\begin{dunefigure}[Global project organization]{fig:DUNE_global}
  {\dword{lbnf-dune} organization.}
  \includegraphics[width=0.95\textwidth]{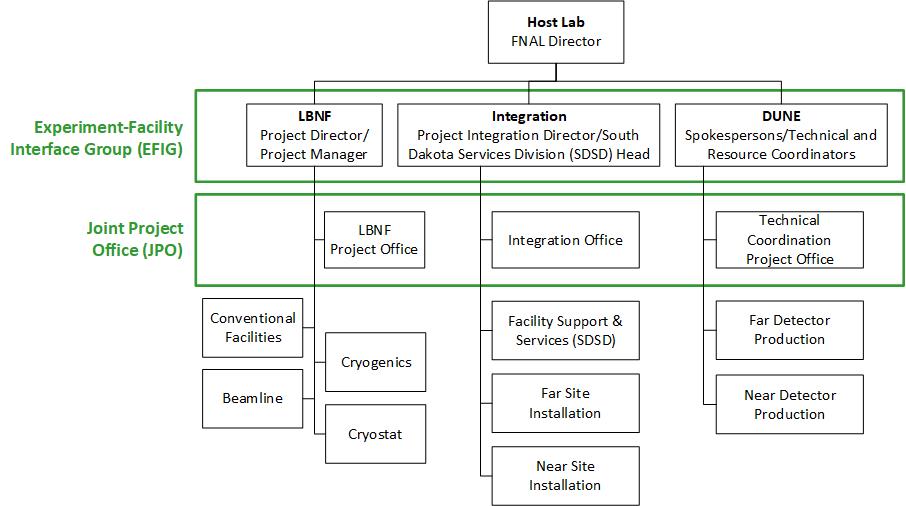}
\end{dunefigure}

In addition to the \dword{lbnf} and \dword{dune} pieces, the overall
coordination of installation activities in the underground caverns 
is managed as a separate element of \dword{lbnf-dune} under the
responsibility of the \dword{ipd}, who is appointed by and reports
to the \dword{fnal} director.  To ensure coordination across
all elements of \dword{lbnf-dune}, the \dword{ipd} connects to both
the facilities and detector construction projects through ex-officio
positions on the \dword{lbnf} Project Management Board and
\dword{dune} \dword{exb}, respectively.  In carrying out these
responsibilities, the \dword{ipd} receives support from the \dword{sdsd},
a \dword{fnal} division established to 
provide the necessary supporting infrastructure for installation, commissioning, and operation 
of the \dword{dune} detector.

The \dword{ipd} works closely with the \dword{lbnf} and \dword{dune}
teams in advance of these activities to coordinate planning
and ensure that detector elements are properly integrated within the 
supporting infrastructure.  Although the \dword{ipd} is responsible 
for overall coordination of  installation activities
at \dword{surf}, the \dword{dune} consortia maintain responsibility 
for the installation and commissioning of their detector subsystems
and support these activities by providing dedicated personnel and
equipment resources.  Likewise, \dword{lbnf} retains responsibility
for the installation and commissioning of supporting infrastructure
items and provides dedicated resources to support these activities.          

\section{Experimental Facilities Interface Group}
\label{sec:efig}

The \dword{efig} is the body responsible for the required high-level
coordination between the \dword{lbnf} and \dword{dune} construction 
projects.  The \dword{lbnf} project director and \dword{ipd} 
co-chair the \dword{efig}.  \dword{efig} leadership also incorporates 
the four members of the \dword{dune} collaboration management 
team (co-spokespersons, \dword{tcoord}, and \dword{rcoord}).  
The \dword{efig} is responsible for steering the integration and   
installation of the \dword{lbnf-dune} deliverables and operates via the 
consensus of its leadership team.  If issues arise for which consensus 
cannot be achieved, decision-making responsibility is passed to the 
\dword{fnal} director.

\section{Joint Project Office}
\label{sec:jpo}

The \dword{efig} is augmented by a \dword{jpo} that supports both 
the \dword{lbnf} and \dword{dune} projects as well as the integration
effort that connects the two together. The \dword{jpo} combines
project support functions that exist within the different elements 
of the global project to ensure proper coordination across the entire 
\dword{lbnf-dune} enterprise.  Project functions coordinated globally 
through the \dword{jpo} are shown in Figure~\ref{fig:DUNE_jpo} along 
with the personnel currently supporting those functions.  The team 
members who support these functions within the \dword{jpo} framework
are drawn from the \dword{lbnf} project office, \dword{dune} \dword{tc}, 
and \dword{lbnf}/\dword{dune} \dword{integoff} personnel.  
\begin{dunefigure}[JPO functions]{fig:DUNE_jpo}
  {\dword{jpo} global support functions and teams}
  \includegraphics[width=0.85\textwidth]{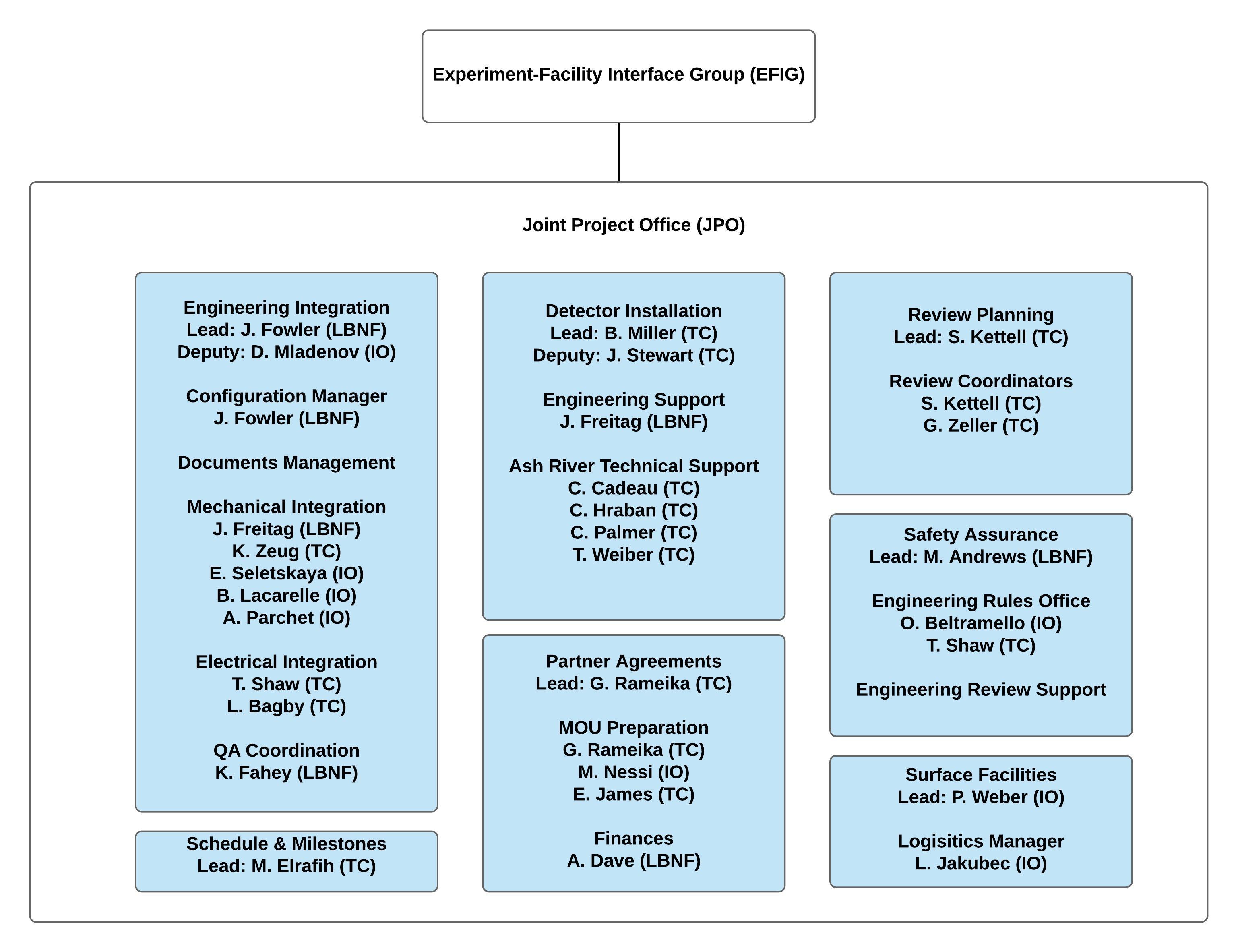}
\end{dunefigure}
Team members focusing on specific project activities and functions 
within the \dword{jpo} are typically those carrying equivalent 
responsibilities within their home organization.  For example, 
the \dword{jpo} team responsible for building the fully integrated 
\threed CAD model of the detector within its supporting infrastructure 
and surrounding facility includes the members of the \dword{lbnf} and 
\dword{dune} project teams responsible for integrating the individual 
elements.

\section{Coordinated Global Project Functions}
\label{sec:global_project}

Project support functions requiring \dword{jpo} coordination include
safety, engineering integration, change control and document 
management, scheduling, review planning and oversight, and development 
of partner agreements.  Additional detail is provided in the subsequent 
sections regarding the \dword{jpo} role in coordinating these support 
functions.

Planning activities related to detector installation and the provision 
of surface facilities are also currently embedded within the framework 
of the \dword{jpo} to ensure that all project elements are properly 
incorporated.  At the time when \dword{lbnf} \dword{fscf} delivers 
\dword{aup} of the underground detector caverns at \dword{surf}, the 
coordination of onsite activities associated with detector installation 
and the operation of surface facilities will be fully embedded within 
the \dword{lbnf}/\dword{dune} \dword{integoff} under the direction of the \dword{ipd}.  
Some current members of the \dword{lbnf} project office and \dword{dune}
\dword{tc} are expected to be moved into the \dword{integoff} at that point in time.  
The \dword{integoff} team required to coordinate post-excavation activities at 
\dword{surf} is described in Chapter~\ref{ch:tc-jpo}.  

\subsection{Safety}
\label{sec:dune_safety}

To ensure a consistent approach to safety across \dword{lbnf-dune},
there is a single \dword{lbnf-dune} \dword{esh} manager who reports 
to the \dword{lbnf} project director, \dword{ipd}, and \dword{dune}
management (via the \dword{dune} \dword{tcoord}).  This individual
directs separate safety teams responsible for implementing the
\dword{lbnf-dune} \dword{esh} program within both the \dword{lbnf} 
and \dword{dune} projects as well as the \dword{lbnf}/\dword{dune}
installation activities at \dword{surf}. The safety organization 
is shown in Figure~\ref{fig:dune_esh} and is described further in
Sections~\ref{sec:tc_safety} and~\ref{sec:far_site_safety} and in 
Chapter~\ref{vl:tc-ESH}.
\begin{dunefigure}[\dshort{lbnf-dune} \dshort{esh}]{fig:dune_esh}
  {High level \dword{lbnf-dune} \dword{esh} organization.}
  \includegraphics[width=0.85\textwidth]{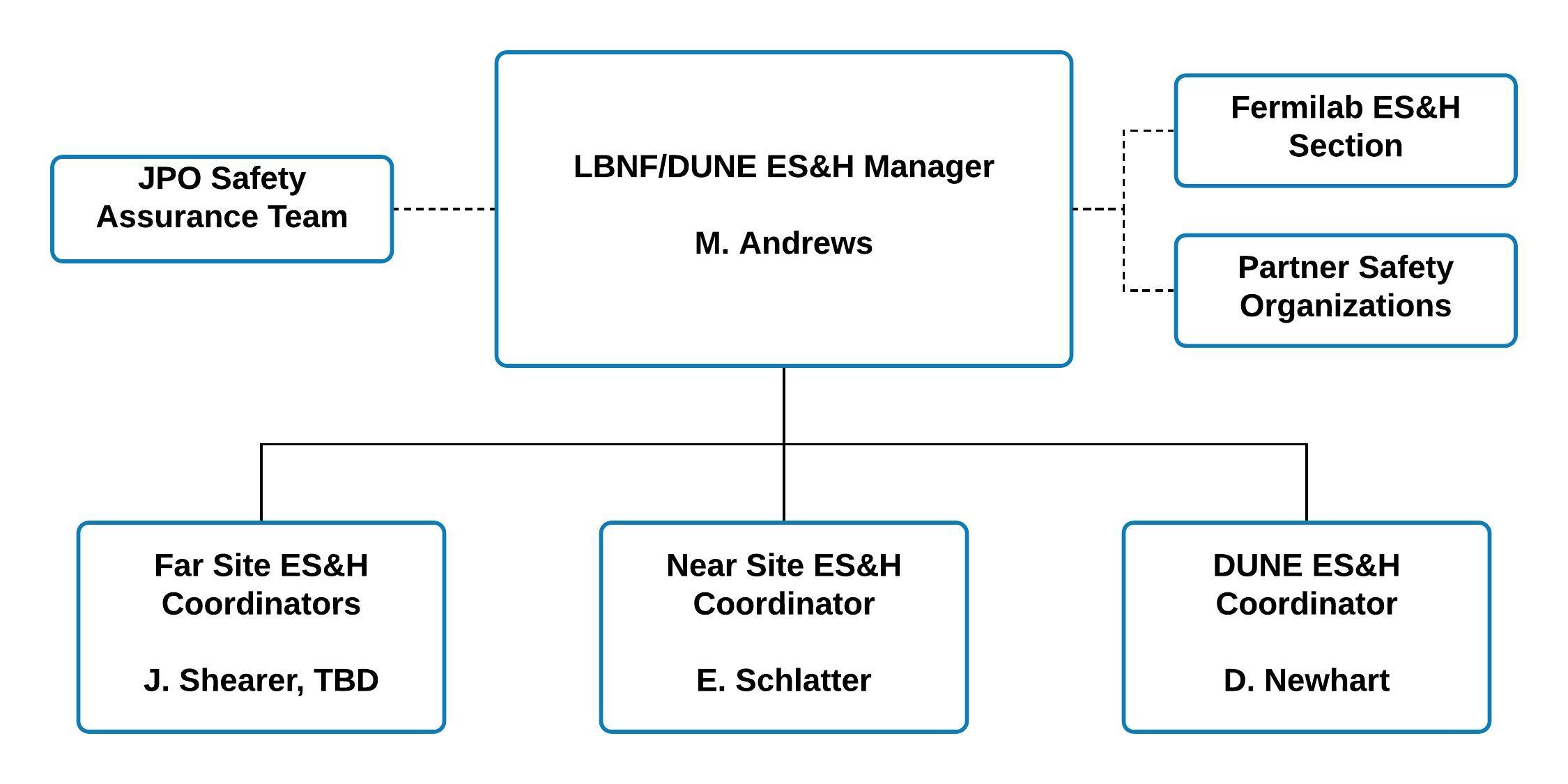}
\end{dunefigure}
The \dword{lbnf-dune} \dword{esh} manager works with the \dword{fnal} 
and \dword{surf} safety organizations to ensure that all project-related 
activities comply with the rules and regulations of the host 
organizations.  For example, the \dword{lbnf-dune} \dword{esh} manager 
works with the host safety organizations to develop the rules and 
regulations governing work in the underground areas at \dword{surf}, 
which are then consistently applied across all underground project 
activities.

The \dword{jpo} engineering safety assurance team defines a common 
set of design and construction rules (mechanical and electrical) to 
ensure consistent application of engineering standards and engineering 
documentation requirements across \dword{lbnf-dune}.  This team works 
with the \dword{lbnf-dune} \dword{esh} manager to develop equivalencies 
in codes and standards across the international project as needed.  
Following on lessons learned from the processes used for the 
\dword{protodune} detectors, an important mandate of the engineering 
safety assurance team is to ensure that safety issues related to 
component handling and installation are incorporated within the 
earliest stages of the design review process.  The \dword{jpo} team 
incorporates engineering resources to perform independent validation 
of required mechanical analyses that ensure the structural integrity 
of detector components through all stages of construction, installation, 
and operation.

\subsection{Engineering Integration}
\label{sec:dune_engineering}

A central \dword{jpo} engineering team is responsible for building 
an integrated model of the detectors within their supporting
infrastructure and the \dword{fscf} that house them.  The team
builds and maintains a full \threed CAD model of everything in the
underground detector caverns from the models of the individual
components provided by the \dword{lbnf} and \dword{dune} design 
teams.  Starting from the latest, approved version of the full CAD 
model, the \dword{jpo} team incorporates approved changes as they 
are received and checks to ensure that no errors or space conflicts 
are introduced into the model.  As part of this process, \twod control 
drawings are produced from the \threed CAD model to validate adherence 
with critical component-to-component clearances within the integrated
model.  The updated working model is passed back to the individual
\dword{lbnf} and \dword{dune} design teams to validate that their
design modifications have been properly incorporated within the 
global model.  After receiving the appropriate sign-offs from all 
parties, the \dword{jpo} team tags a new frozen release of the model 
and makes it available to the design teams as the current release 
against which the next set of design changes will be generated.

Electrical engineers are incorporated within the central \dword{jpo} 
team to ensure proper integration of the detector electrical 
components.  This team is responsible for ensuring that detector 
grounding and shielding requirements, the maintenance of which are 
critical for detector performance, are strictly adhered to.  The 
team oversees the layout of electronics racks and cable trays both 
on the top of the cryostats and within the \dword{cuc} counting room 
that hosts the \dword{daq} electronics.  It also oversees the design 
of the power and cooling distribution systems that are required to 
support the electronics infrastructure.

The \dword{jpo} engineering team is responsible for documenting and
controlling the interfaces between the \dword{lbnf} and \dword{dune} 
projects as well as the interfaces between these projects and the 
\dword{lbnf}/\dword{dune}  installation activities at \dword{surf}.  
To define these interfaces, the \dword{jpo} team develops formal 
documents which, subsequent to the approval of the relevant managers, 
are placed under signature and versioning control.  These documents 
are monitored regularly to ensure that no missing scope or technical 
incompatibilities are introduced at the boundaries between the project elements.

\subsection{Change Control and Document Management}
\label{sec:dune_changecontrol}

The \dword{lbnf-dune} project partners have agreed to adopt 
the formal change control process developed previously for the 
\dword{lbnf} project.  The change control process applies to 
proposed modifications of requirements, technical designs, 
schedule, overall project scope, and assigned responsibilities 
for individual scope items.  The formal \dword{lbnf-dune} 
change control process is described in~\citedocdb{82}.  The 
process includes separate decision paths for items affecting 
only \dword{dune} or \dword{lbnf} and incorporates an additional 
pathway for items affecting both projects.  A hierarchy of 
decision-making layers is built into each pathway based on 
pre-determined thresholds related to the extent of the proposed 
change.  The lowest-level change control body for modifications 
affecting both \dword{lbnf} and \dword{dune} is the \dword{efig}.  
The \dword{jpo} incorporates a configuration manager within the 
engineering integration team who is responsible for formally 
implementing changes that are approved through this process.  
After technical changes are incorporated into the global \threed 
CAD models, the engineering integration team is responsible for 
checking production drawings and verifying that no potential 
space conflicts have been introduced.  Under the direction of 
the configuration manager, all project changes are documented 
in detail and approved by the appropriate project partners using 
the \dword{lbnf} change control tool.

The configuration manager oversees a document management team
responsible for 
administering and managing the \dword{lbnf-dune}
document management system, which is hosted in the \dword{edms}.  
All technical documents and drawings are stored in the \dword{edms} system
under formal signature and versioning control.  A product breakdown 
structure (PBS) database will be maintained to track the history of
each detector component (and supporting infrastructure item) through
construction, assembly, testing, transport, and installation.  The
\dword{lbnf-dune} \dword{qa} manager is embedded within the
\dword{jpo} engineering integration team as the \dword{qa} coordinator
and has responsibility for ensuring that all necessary documents and
testing results used to validate component quality are stored within
the PBS database.

\subsection{Scheduling}
\label{sec:dune_schedule}

The \dword{jpo} team is responsible for creating a single project
schedule for \dword{lbnf-dune} that incorporates all \dword{lbnf} and
\dword{dune} activities together with the installation activities at
\dword{surf}, incorporating all interdependencies. A brief discussion
is provided in Section~\ref{sec:fdsp-coord-controls}. This schedule
will be used to track the status of the global enterprise.  The
project partners have agreed that the \dword{lbnf-dune} schedule 
will be managed within the same Primavera \dword{p6} framework used 
to plan and set the status of the resource-loaded schedule of activities required 
for  \dword{doe} contributions to \dword{lbnf} and \dword{dune}.
Activities falling under the responsibility of other international
partners are included and linked within the \dword{p6} schedule but 
do not incorporate associated resource information required for  
\dword{doe} activities.  Non-\dword{doe} activities will not be 
tracked using the formal \dword{evms} procedures required for the 
\dword{doe} project activities, but rather through regular assessments 
of progress towards completion by the management teams responsible 
for those activities.  A substantial number of milestones will be 
embedded within the schedule at an appropriate level of granularity 
to allow for high-level tracking of the project progress towards its 
completion.

\subsection{Review Planning and Oversight}
\label{sec:dune_review}

As described in Chapter~\ref{vl:tc-review}, all reviews conducted 
across the \dword{lbnf-dune} enterprise are coordinated through 
the \dword{jpo} review planning team to ensure coherence in the 
review process.  \dword{dune} collaboration management via the 
\dword{tcoord} has responsibility for design (\dword{pdr} and 
\dword{fdr}) and production (\dword{prr} and \dword{ppr}) 
reviews focusing on the different detector elements.  Similarly, 
\dword{lbnf} project management has responsibility for design 
and production reviews covering the supporting infrastructure 
pieces within its scope.  Installation readiness reviews and \dwords{orr}, on the 
other hand, are the responsibility of the \dword{ipd}.  Central 
coordination of the review process through the \dword{jpo} review 
planning team ensures that issues related to installation and 
operation are incorporated within all stages of the review process.  
Safety issues related to handling and installation of components 
are addressed starting from the earliest design reviews -- with the 
development of detailed engineering notes containing the required 
structural analysis -- through installation and operations reviews 
with detailed hazard analyses.  The \dword{jpo} team also takes 
responsibility for tracking review recommendations and closing 
them as appropriate, based on resulting actions.

\subsection{Development of Partner Agreements}
\label{sec:dune_agreements}

Partner contributions to all project elements will be detailed 
in a series of written agreements.  In the case of \dword{lbnf}, 
these contributions will be spelled out in bilateral agreements 
between \dword{doe} and each of the contributing partners.  In 
the case of \dword{dune}, there will be a \dword{mou} 
detailing the contributions of all participating partners.  The 
\dword{mou} will detail the deliverables being provided by each 
partner and summarize required contributions to common items, 
for which the collaboration assumes shared responsibility.  
A series of more technical agreements describing the exact 
boundaries between partner contributions and the terms and 
conditions under which they will be delivered will lie just 
beneath the primary agreements.  The \dword{jpo} team focusing 
on partner agreements will coordinate the process for drafting 
written agreements and work to obtain the appropriate partner 
approvals on each.  

\section{\dshort{protodune} Experience}
\label{sec:dune_protodune}

The global structure of \dword{lbnf-dune} is based heavily on 
the organization that successfully executed the construction,
installation, commissioning, and operation of the \dword{protodune}
detectors at the \dword{cern}.  The onsite team at \dword{cern} 
responsible for the overall installation of detector and 
infrastructure components within the test beam facility played a 
critical role in the successful execution of the \dword{protodune} 
program.  The separate projects responsible for the construction 
of the detector and infrastructure components interacted effectively 
with the central, onsite team to minimize the issues encountered 
during the installation and commissioning process.  In cases where 
issues did arise, construction project team members interacted 
effectively with their counterparts on the onsite team to reach 
quick resolutions.

Some lessons learned from the \dword{protodune} experience have 
been applied in creating the \dword{lbnf-dune} organization for 
the \dword{dune} \dword{fd}.  The integration of installation 
safety issues into the early stages of the design review process 
is one such example.  Delays were encountered in getting approvals 
for the installation of some \dword{pdsp} components stemming from 
the absence of a coordinated approach in the review process for 
these items.  The creation of the \dword{jpo} review planning 
team charged with organizing a coherent review process across 
\dword{lbnf-dune} is meant to address this issue.  In general, 
the successful implementation of the \dword{protodune} detectors 
demonstrates the capacity of the organizational structures to 
safely execute the project and meet performance requirements, as 
was seen in \dword{pdsp}.

The team that led the installation of \dword{protodune} at
\dword{cern} also led the installation of \dword{minos} in the 
Soudan mine in Minnesota, and that experience extrapolates to 
the upcoming installation at \dword{surf}. \dword{fnal}  
established the \dword{sdsd} to work with \dword{surf} to 
ensure that appropriate site infrastructure and support mechanisms 
needed to execute onsite project activities are in place.

\cleardoublepage

\chapter{Detector Design and Construction Organization}
\label{vl:tc-overview}

The \dword{dune} \dword{fd} construction project refers collectively 
to the activities associated with the design and construction of 
necessary detector components.  \dword{dune} collaboration management 
is responsible for overseeing this portion of the \dword{lbnf-dune} and 
ensuring its successful execution.  The high-level \dword{dune} 
collaboration management team consisting of the co-spokespersons, 
\dword{tcoord}, and \dword{rcoord} is responsible for the
management of the project.  

\section{\dshort{dune} Consortia}
\label{sec:consortia}

Construction of the \dword{dune} far \dwords{detmodule} is carried out by 
consortia of collaboration institutions who assume responsibility 
for detector subsystems.  Each consortium plans and executes the 
construction, installation, and commissioning of its subsystem.

Management of each consortium is through an overall consortium leader 
and a technical lead.  The consortium leader chairs an institutional 
board composed of one representative from each of the collaborating 
institutions contributing to the activities of the consortium.  Major 
consortium decisions such as technology selections and assignment of 
responsibilities within the institutions should pass 
through its institutional board.  These decisions are then passed 
as recommendations to the \dword{dune} \dword{exb}, as described in 
greater detail below, for formal collaboration approval.

Figure~\ref{fig:DUNE_consortia_org} shows an example consortium 
organizational chart incorporating the basic structures mandated 
by \dword{dune} collaboration management.  In addition to the pieces 
described above, consortia in most cases need to manage the design 
and construction of subsystem deliverables that are supported by 
multiple funding agencies.  In the sample case illustrated here, 
responsibilities for subsystem deliverables are shared between the 
USA, UK, and Switzerland (CH), where each of the funding agencies 
is expected to manage its own internal projects with responsibility 
for different sets of assigned deliverables.  To ensure coordination 
between the separate internal projects contributing to the consortia, 
technical leads are responsible for chairing consortium project 
management boards incorporating separate managers from each of 
the internal projects.   
\begin{dunefigure}[\dshort{dune} internal consortium structure]{fig:DUNE_consortia_org}
  {Sample \dword{dune} internal consortium structure}
  \includegraphics[width=0.99\textwidth]{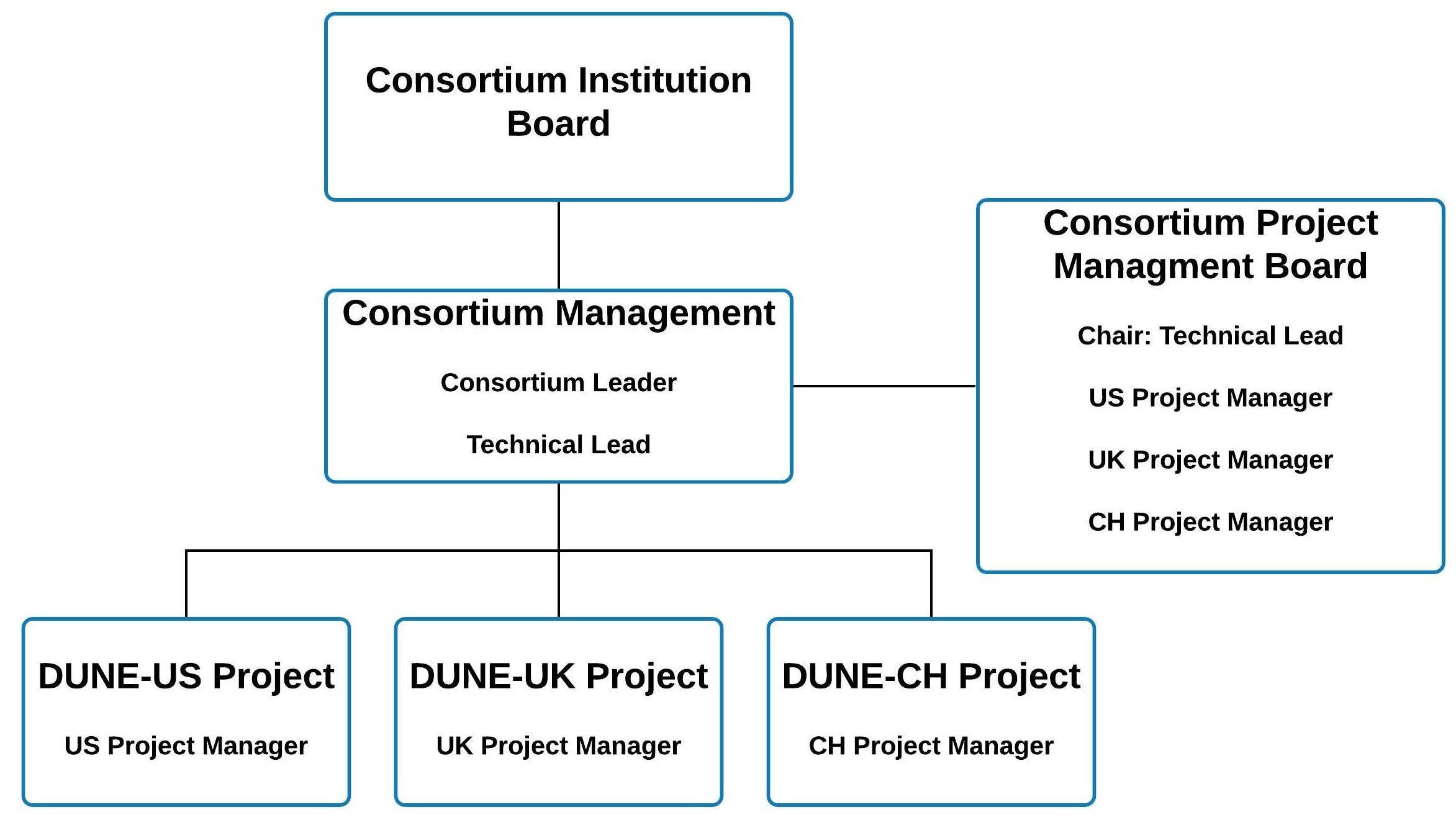}
\end{dunefigure}

In addition to the mandated organizational pieces described here, 
the consortia incorporate additional internal structures as needed 
to deliver their assigned subsystems.  For example, working groups 
with convenors are typically appointed to focus on specific consortium 
activities, and steering committees are in many cases formed to help 
guide technical and strategic decisions within the consortia.  Each 
consortium is also expected to appoint both safety and \dword{qa} 
representatives as well as a representative with responsibility for 
integration and installation.  These individuals are charged 
with interacting  with the appropriate project support team 
personnel to ensure coordination in these areas across the consortia.        

\section{\dshort{dune} Collaboration Management}
\label{sec:dune_mgmt}

The high-level \dword{dune} collaboration management structure is 
shown in Figure~\ref{fig:DUNE_org}.  The \dword{dune} \dword{exb} is 
the primary collaboration decision-making body and as such includes 
representatives from all major areas of activity within the 
collaboration.
\begin{dunefigure}[\dshort{dune} org chart]{fig:DUNE_org}
  {\dword{dune} Organizational Chart}
  \includegraphics[width=0.9\textwidth]{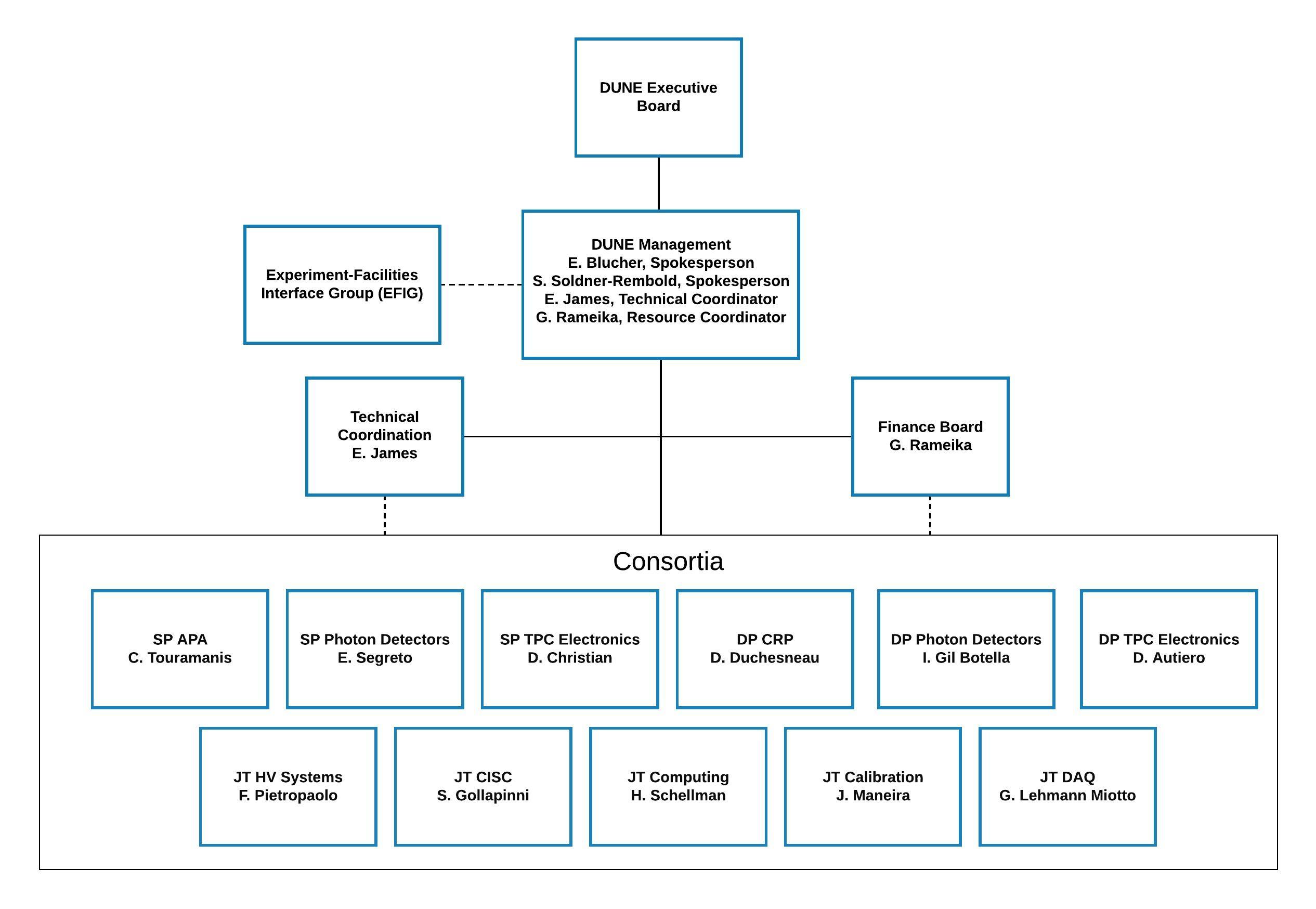}
\end{dunefigure}

Each consortium is represented on the \dword{dune} \dword{exb} by its 
consortium leader.  All collaboration decisions, especially those with 
potential impacts on the \dword{dune} scientific program or connected 
with the assignment of institutional responsibilities, pass through the 
\dword{exb}.  \dword{exb} decisions are made by  
consensus.  In cases where consensus cannot be obtained, decision-making 
responsibility passes to the co-spokespersons.

\section{Technical Coordination}
\label{sec:tc}

Because the consortia operate as self-managed entities, a strong
\dword{tc} organization is required to ensure overall integration 
of the detector elements and successful execution of the detector
construction project.  \Dword{tc} areas of responsibility include 
general project oversight, systems engineering, \dword{qa}, and 
safety.  \Dword{tc} also supports the planning and execution 
of integration and installation activities at \dword{surf} (see 
Chapter~\ref{ch:tc-jpo}).  

\Dword{tc} is headed by the \dword{tcoord},  a \dword{fnal}
employee appointed jointly by the \dword{fnal} director 
and the \dword{dune} co-spokespersons. 
A deputy \dword{tcoord}  selected from within the collaboration will assist the \dword{tcoord}.

The \dword{tcoord} manages the overall detector construction 
project through regular technical and project board meetings with 
the consortium leadership teams and members of the \dword{tc} 
organization (see Section~\ref{sec:tco}).  These board meetings 
are used to identify and resolve issues and serve as the primary 
fora for required interactions between the consortia.

Technical board meetings are used to evaluate consortia design
decisions with potential impacts on overall detector performance,
ensure that interfaces between the different subsystems are well
understood and documented, and monitor the overall construction
project to identify and address both technical and interface 
issues as they arise.

Project board meetings are used to ensure that the scopes of 
each consortium are fully documented with assigned institutional
responsibilities, develop and manage risks held within a global
project registry, review and manage project change requests, and
monitor the status of the overall detector construction schedule.

Any decisions generated through these board meetings are passed to 
the \dword{dune} \dword{exb} as recommendations for formal approval.
Depending on the agenda items to be discussed at a specific board
meeting, the \dword{tcoord} will invite additional members of the
collaboration with specific knowledge or particular expertise to
participate.  In addition, for major decisions, the \dword{tcoord}
will officially appoint internal collaboration referees with no associated 
conflicts of interest to 
assist in evaluating the 
 technical issues behind these major decisions.

\section{Technical Coordination Organization}
\label{sec:tco}

The \dword{tcoord} heads an organization that supports the work of 
the consortia and has responsibility for a number of major project 
support functions prior to the delivery of detector components to 
\dword{surf} including
\begin{itemize}
\item ensuring that each consortium has a well defined and complete
  scope, that interactions between consortia are sufficiently 
  well defined, and that any missing scope outside of the 
  consortia is provided through other sources such as collaboration
  common funds;
\item defining and documenting scope boundaries and technical 
  interfaces both between consortia and with \dshort{lbnf};  
\item developing an overall schedule with appropriate dependencies
  between activities covering all phases of the project; 
\item ensuring that appropriate engineering and safety standards 
  are developed, understood, and agreed to by all key stakeholders 
  and that these standards are conveyed to and understood by each
  consortium;
\item ensuring that all \dword{dune} requirements on \dword{lbnf} 
  for \dword{fscf}, cryostat, and cryogenics are clearly defined and 
  agreed to by each consortium;
\item ensuring that each consortium has well developed and reviewed
  component designs, construction plans, \dword{qc} processes, and 
  safety programs; and
\item monitoring the overall project schedule and the progress of 
  each consortium towards delivering its assigned scope. 
\end{itemize}

The \dword{dune} \dword{tc} organizational structure is shown 
in Figure~\ref{fig:DUNE_tc}.  The structure incorporates teams 
with responsibilities for project coordination, engineering 
support, and installation interfaces.  Many \dword{tc} team 
members also contribute to the activities of the \dword{jpo} 
teams (shown in Figure~\ref{fig:DUNE_jpo}) in order to ensure 
coherence in project support functions across \dword{lbnf-dune}.
\begin{dunefigure}[\dshort{dune} technical coordination org chart]{fig:DUNE_tc}
  {\dword{dune} \dword{tc} organizational chart}
  \includegraphics[width=0.99\textwidth]{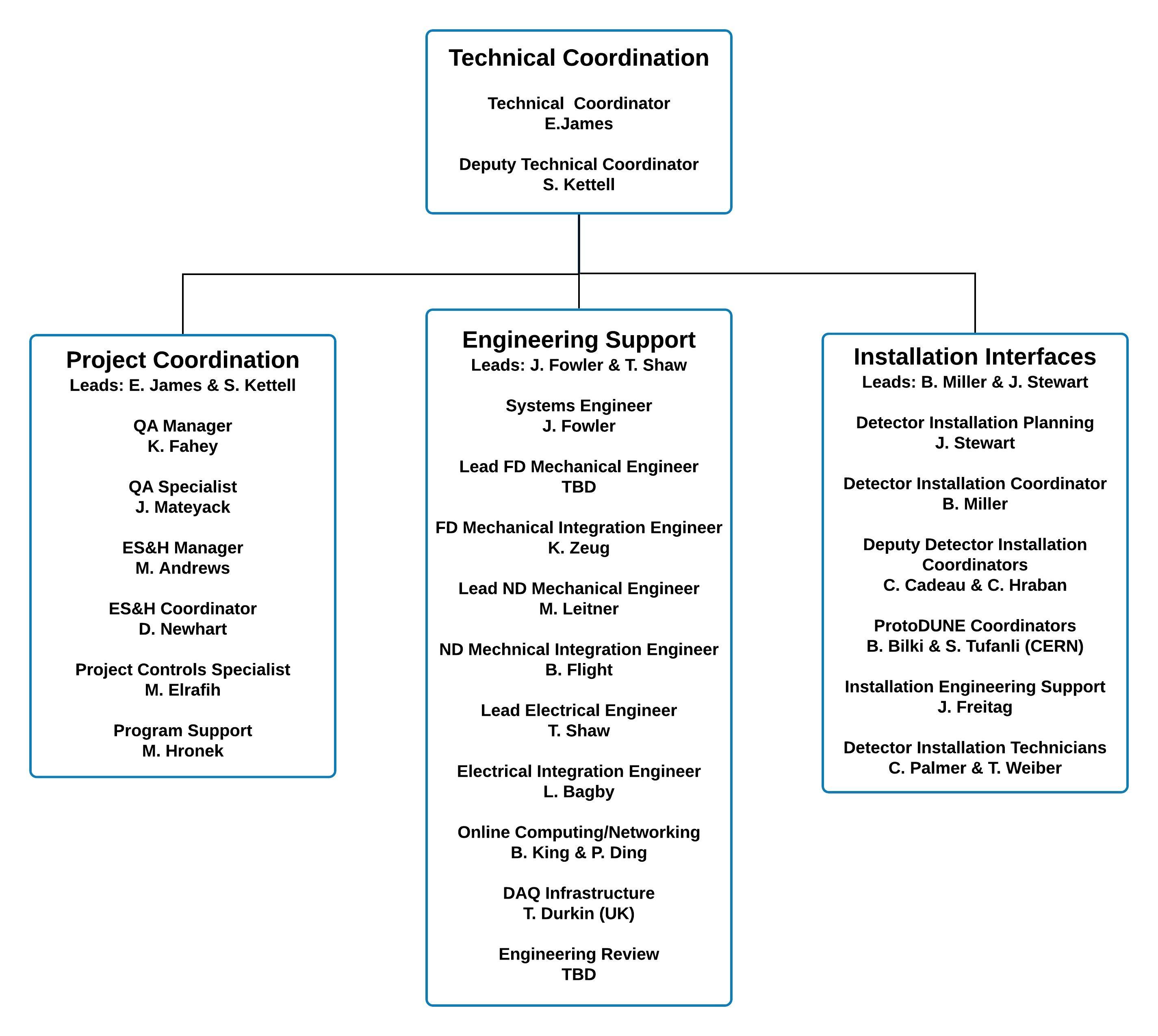}
\end{dunefigure}
The \dword{tc} project coordination team incorporates \dword{esh}, 
\dword{qa}, and project controls specialists.  Overall integration 
of the detector elements is coordinated through the \dword{tc} 
engineering support team headed by the \dword{lbnf-dune} systems 
engineer and lead \dword{dune} electrical engineer.  Planning 
coordinators for integration and installation activities at 
\dword{surf} sitting within the \dword{lbnf}/\dword{dune} \dword{integoff} also head the \dword{tc} installation interfaces team.  
The dual placement of these individuals facilitates the required 
coordination of integration and installation planning efforts between 
the core team directing these activities and the \dword{dune} 
consortia, which maintain primary responsibility for the individual 
detector subsystems.  Members of the \dword{tc} organization meet 
weekly to review project progress and discuss technical issues. 
     
Within the framework of the \dword{dune} \dword{fd} construction 
project, \dword{tc} project support functions associated with its 
coordination role include safety, engineering integration, change control, document management, scheduling, risk management, conducting reviews, and workflow. These functions are described in the following sections. 

\subsection{Safety}
\label{sec:tc_safety}

The \dword{tcoord} is responsible for implementing the safety program
covering the \dword{dune} construction project.  The \dword{tcoord} is
supported in this role by the \dword{lbnf-dune} \dword{esh}
manager.  A dedicated \dword{dune} \dword{esh} coordinator sits within
the \dword{tc} organization and guides the \dword{dune} safety program
under the direction of the \dword{lbnf-dune} \dword{esh} manager. The
safety organization for the \dword{dune} construction project is shown
in Figure~\ref{fig:dune_esh_construction} and is further described in
Chapter~\ref{vl:tc-ESH}.
\begin{dunefigure}[\dshort{dune} construction \dshort{esh}]{fig:dune_esh_construction}
  {High level \dword{dune} construction \dword{esh} organization.}
  \includegraphics[width=0.85\textwidth]{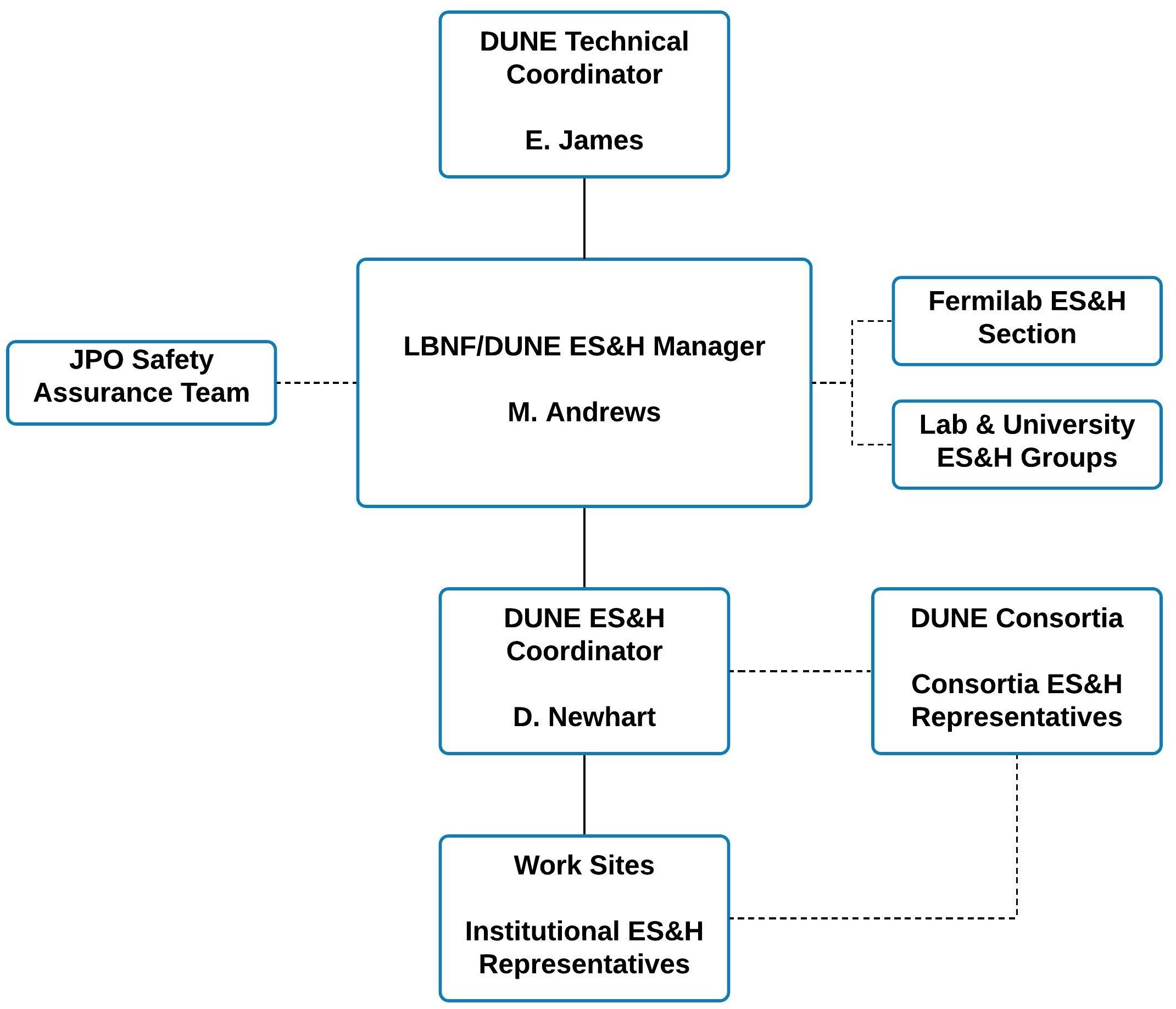}
\end{dunefigure}

The \dword{dune} construction project is carried out at many different
institutions in many different countries.  Participating institutions
sign a \dword{mou} in which they agree to abide by the requirements of
the \dword{dune} safety program.  Each of the participating
institutions assumes primary responsibility for the safe execution of
their assigned construction activities.  The \dword{dune} \dword{esh}
coordinator interacts with all participating institutions to
ensure that their programs comply with \dword{dune} safety
requirements.  Prior to the start of any construction activities, the
\dword{dune} \dword{esh} coordinator participates in the \dword{prr}
incorporating on site visits to confirm that approved safety controls
are in place.  Follow-up on site visits by the \dword{dune} \dword{esh}
coordinator, including \dwords{ppr}, over the course of the
construction period are used to validate continuing compliance with
program requirements.

\subsection{Engineering Integration}

\dword{dune} \dword{tc} works with the collaboration to 
collect and validate requirements associated with the \dword{fd} 
modules.  Each detector module has its own set of high-level 
requirements with potential effects on the \dword{dune} physics 
program, which are owned by the \dword{dune} \dword{exb}.  The
\dword{exb} must approve any proposed changes to these 
requirements.  Lower-level technical requirements associated 
with each detector subsystem are developed by the consortia 
under the guidance of the \dword{tc} engineering support team.  
Appendix~\ref{sec:fdsp-coord-requirements} contains tables 
summarizing the high-level requirements associated with each of 
the \dword{fd} modules.

Starting from these requirements, the \dword{tc} engineering team
responsible for detector integration works with the
\dword{dune} consortia to build and validate integrated detector
models from the designs of the individual subsystems.  The team
ensures that the detector subsystems fit together properly and 
that the fully assembled detectors meet structural requirements 
associated with all operational conditions (both warm and cold), as 
is discussed further in Chapter~\ref{sec:fdsp-coord-integ-sysengr}.  
The team is responsible for validating that the integrated detector 
designs satisfy the defined requirements needed to meet the goals 
of the \dword{dune} physics program.

Integration of the full detector models within the global model
encompassing the detector caverns and supporting infrastructure
is the responsibility of the central \dword{jpo} engineering 
integration team.  The \dword{tc} engineering team is responsible 
for validating interfaces within the combined model between 
\dword{dune} detector components and supporting infrastructure 
pieces.  All proposed changes to the global model require approval 
from the \dword{tcoord} based on guidance received from the lead 
engineers embedded within the \dword{tc} detector integration team.

As part of these efforts, the engineering team works 
with consortium leadership teams to develop controlled 
documents describing interfaces between the different detector 
subsystems.  These documents are placed under signature control 
within the \dword{lbnf-dune} document management system.  
Proposed changes to interface documents must be approved by 
the consortium leadership teams on both sides of the interface 
as well as by the lead engineers in the \dword{tc} 
detector integration team.

The \dword{tc} engineering team works with the
\dword{lbnf-dune} systems engineer, who heads the \dword{jpo} 
configuration and integration team, to develop required documents 
detailing interfaces between the \dword{lbnf} and \dword{dune} 
\dword{fd} construction projects and the interfaces of these 
projects with \dword{lbnf-dune} installation and integration
activities at \dword{surf}.  These \dword{lbnf-dune} interface 
documents are placed under signature control within the 
\dword{lbnf-dune} document management system and proposed 
changes require approvals from both the lead \dword{lbnf-dune} 
systems engineer and the responsible individuals associated 
with each branch of the global project (\dword{lbnf} project 
manager, \dword{dune} \dword {tcoord}, and \dword{ipd}).

Appendix~\ref{sec:fdsp-coord-interface} contains a table 
cataloging the \dword{dune} interface documents 
and providing web links for accessing approved versions 
 in place at the time of the release of this document.
 
\subsection{Change Control and Document Management}
\label{sec:tc_change control}

The \dword{dune} project follows the \dword{lbnf-dune} change control
process described in Section~\ref{sec:fdsp-change} and
Section~\ref{sec:change-control}.  The decision path for changes not
impacting the \dword{lbnf} project or \dword{lbnf-dune} 
installation activities at \dword{surf} is self-contained within
the \dword{dune} collaboration management structure.  A hierarchy of
decision-making levels is defined based on pre-determined thresholds
related to the extent of the proposed change with the most significant
changes requiring \dword{dune} \dword{exb} approval.

For document management, the \dword{dune} construction project 
relies on the \dword{lbnf-dune} document management system 
administrated by the \dword{jpo} engineering integration team 
configuration manager.  \Dword{tc} works with the \dword{lbnf-dune} 
\dword{qa} manager to ensure that the information needed to track 
the history of each detector component through construction, 
assembly, and testing is properly captured within the PBS 
database.  A dedicated \dword{dune} \dword{qa} specialist 
sits within the \dword{tc} organization and coordinates the 
\dword{dune} \dword{qa} program under the direction of the
\dword{lbnf-dune} \dword{qa} manager.  The \dword{dune} 
\dword{qa} program is described in much greater detail in 
Chapter~\ref{vl:tc-QA}.
 
\subsection{Schedule}

The lead project controls specialist within the \dword{tc} team works
with the \dword{dune} consortia to build schedules covering
the design, testing, and construction activities associated with their
subsystems and incorporate these within the \dword{lbnf-dune}
schedule.  The project controls specialist communicates with consortia
technical leads on a monthy basis to track the status of activities
and update the \dword{lbnf-dune} schedule accordingly.  Milestones
positioned at regular intervals within the subsystem construction
schedules are incorporated to enable high-level tracking of these
efforts.  Section~\ref{sec:fdsp-coord-controls} (in
Appendix~\ref{ch:tc-sp-project}) contains a table summarizing key
detector milestones around which the individual consortia
schedules are constructed.

\subsection{Risk Management}

\dword{dune} \dword{tc} maintains a global registry containing both
subsystem-specific risks identified by the consortia and self-held
risks associated with the overall coordination of the \dword{dune}
construction project as discussed in
Sections~\ref{sec:fdsp-coord-risks}~and~\ref{sec:fdsp-app-risk}.  The
\dword{tcoord} uses Project Board Meetings to regularly review the
risk registery with the consortium leadership teams and define
mitigation actions as necessary to prevent identified risks from being
realized.  The \dword{tcoord} does not have control over
contingency funds held by the internal projects of the participating
funding agencies.  In cases of identified need, the \dword{tcoord}
works with the consortium leadership teams to implement risk
reduction strategies.  Identified issues that cross consortia
boundaries are discussed at project board meetings and brought to the
\dword{dune} \dword{exb} if they need to be addressed at a higher
level.

Appendix~\ref{sec:fdsp-coord-risks} contains tables summarizing the 
highest-level identified risks within the \dword{tc} risk registry.

\subsection{Review Process}

The \dword{tcoord} has primary responsibility for conducting design 
and \dwords{prr} covering each detector subsystem.  As 
described in Section~\ref{sec:dune_review} and Chapter~\ref{vl:tc-review}, 
reviews are coordinated through the \dword{jpo} review planning team to 
ensure coherence in the review process across the entire \dword{lbnf-dune} 
enterprise.  The deputy \dword{tcoord} is the \dword{jpo} 
team member responsible for organizing the reviews.  The full review process is 
described in greater detail in Chapter~\ref{vl:tc-review}.

\section{\dshort{dune} Work Flow}
\label{sec:workflow}

Table~\ref{tab:responsibility} lists the time-ordered set of activities 
required to realize the \dword{dune} \dword{fd} modules, from the design 
of individual detector components through operation of the fully-assembled 
modules.  Primary responsibility for the detector subsystems is held 
by the \dword{dune} consortia over the full course of these activities.
\dword{dune} \dword{tc} under the direction of the \dword{tcoord} is 
responsible for coordinating the consortia efforts related to the design, 
prototyping, fabrication, and transport (to South Dakota) of the required 
detector elements.  Efforts related to the receipt (in South Dakota), 
processing, installation, and check-out of detector components are 
coordinated through the \dword{integoff} under the direction of the \dword{ipd} as 
described in Chapter~\ref{ch:tc-jpo}.  The \dword{integoff} also coordinates efforts 
related to the installation and commissioning of the supporting cryogenic 
infrastructure, for which the \dword{lbnf} project has 
responsibility.  
The \dword{dune} collaboration takes responsibility for coordinating 
activities occuring after the cryostats are filled with liquid argon 
beginning with final commissioning of the detectors and continuing into 
long-term detector operation.
\begin{dunetable}
  [DUNE responsibility matrix]
  {p{0.125\linewidth}p{0.125\linewidth}p{0.125\linewidth}p{0.125\linewidth}p{0.125\linewidth}p{0.125\linewidth}}
  {tab:responsibility}
  {Responsibility matrix for activities occuring during each phase of
   the process for implementing the \dword{dune} \dword{fd} modules.}
               & \dword{dune} & \dword{dune} & \dword{dune}  &             &              \\ 
\rowtitlestyle  Phase        & Consortia    & \dword{tc}   & Collaboration & \dword{lbnf-dune} IO & \dword{lbnf} \\ \toprowrule
  Design       & Lead         & Coordinate   & Support       & Support     &              \\ \colhline
  Prototype    & Lead         & Coordinate   & Support       & Support     &              \\ \colhline
  Fabricate    & Lead         & Coordinate   & Support       & Support     &              \\ \colhline
  Ship         & Lead         & Coordinate   & Support       & Support     &              \\ \colhline
  Receive      & Lead         & Support      & Support       & Coordinate  &              \\ \colhline
  Integrate    & Lead         & Support      & Support       & Coordinate  &              \\ \colhline
  Install      & Lead         & Support      & Support       & Coordinate  & Lead         \\ 
               & (Detector)   &              &               &             & (Cryogenics) \\ \colhline
  Commission   &              &              & Support       & Coordinate  & Lead         \\ 
  (Cryogenics) &              &              &               &             &              \\ \colhline
  Commission   & Lead         &              & Coordinate    & Support     & Support      \\ 
  (Detector)   &              &              &               &             &              \\ \colhline
  Operate      & Lead         &              & Coordinate    &             &              \\ 
\end{dunetable}

Although the organizations responsible for coordinating activities during 
each stage of the time-ordered process required to bring the \dword{fd} 
modules online are clearly delineated in Table~\ref{tab:responsibility},
additional, important support roles are connected to each stage.  
The \dword{integoff} interacts with \dword{dune} \dword{tc} during detector design, 
prototyping, and construction to ensure that detector elements integrate 
properly within the supporting infrastructure.  Participation of the 
consortia in the planning process for integration and installation
activities at \dword{surf} is facilitated through \dword{dune} \dword{tc}.
The \dword{dune} collaboration also provides 
support throughout 
the entire process.  The collaboration is responsibile for defining the 
\dword{dune} science program and performing the physics studies used to 
define detector requirements.  The collaboration also provides resources 
that support acquisition of common detector infrastructure items sitting 
outside the scope of the consortia as well as necessary personnel and 
equipment to support integration and installation efforts at \dword{surf}.
\begin{dunetable}
  [DUNE decision-making matrix]
  {p{0.2\linewidth}p{0.2\linewidth}}
  {tab:responsibility2}
  {Responsibility matrix for high-level decision-making occuring during each 
phase of the process for implementing the \dword{dune} \dword{fd} modules}
  & \\
  \rowtitlestyle  Phase             & Decision-making          \\ \toprowrule
  Design            & DUNE EB \\ \colhline
  Prototype         & DUNE EB  \\ \colhline
  Fabricate         & DUNE EB \\ \colhline
  Ship              & DUNE EB  \\ \colhline
  Receive           & EFIG            \\ \colhline
  Integrate         & EFIG     \\ \colhline
  Install           & EFIG      \\ \colhline
  Commission - Cryo & EFIG     \\ \colhline
  Commission - Det  & DUNE EB  \\ \colhline
  Operate           & DUNE EB  \\ 
\end{dunetable}

During each stage of the time-ordered process for implementing the 
\dword{fd} modules, issues may arise requiring high-level decisions 
on steps required for moving forward.  Table~\ref{tab:responsibility2} 
summarizes which body within the global project structure (\dword{dune}
\dword{exb} or \dword{efig}) takes responsibility for high-level 
decision-making during each stage of the \dword{dune} \dword{fd} 
construction project.  
         
The \dword{dune} project has already completed an initial round of design 
and prototyping activities culminating in the construction and operation 
of the \dword{protodune} detectors.  Moving forward, the project is 
updating detector component designs to account for lessons learned from 
the \dword{protodune} experience. Once the designs are final, the 
project will construct first production versions of all components that 
will be installed and operated in a second phase of \dword{protodune} 
operations prior to the start of full-scale production.  The operation 
of the \dword{protodune2} detectors will follow roughly two years after
the end of operations for the corresponding \dword{protodune} detectors.
In a few cases, the production of long lead-time components will need to 
be started in parallel with the operation of first production components 
in \dword{protodune2}.

\cleardoublepage

\chapter{Detector Installation and Commissioning Organization}
\label{ch:tc-jpo}

As discussed in Chapter~\ref{vl:tc-global}, the \dword{ipd} has
responsibility for coordinating the planning and execution of 
the \dshort{lbnf-dune} installation activities, both 
in the underground detector caverns at \dword{surf} and in 
nearby surface facilities.  The \dword{dune} consortia maintain 
responsibility for their subsystems over the course of these 
activities and provide the expert personnel and specialized 
equipment necessary to integrate, install, and commission their 
detector components.  Likewise, \dword{lbnf} has responsibility 
for activities associated with the installation 
of supporting infrastructure items, which are coordinated under 
the direction of the \dword{ipd}.       

The \dword{lbnf-dune} \dword{integoff} will evolve over 
time to incorporate the team in South Dakota responsible for the 
overall coordination of on site installation activities.  In the 
meantime, the installation planning team within the \dword{integoff} works with 
the \dword{dune} consortia and \dword{lbnf} project team members 
to plan these activities.  

The \dword{integoff} installation planning team is responsible for specification 
and procurement of common infrastructure items associated with 
installation of the detectors that are not included within 
the scope of the \dword{dune} consortia.  Some of these items 
are detector components such as racks, cable trays, cryostat flanges, 
and mechanical structures for supporting the detectors within 
the cryostats.  Others are general items required for detector 
installation such as clean rooms, cranes, scaffolding, and 
personnel lifts.

The on site \dword{integoff} team includes rigging teams responsible for moving 
materials in and out of the shaft, through the underground drifts, 
and within the detector caverns.  It includes personnel responsible 
for overseeing safety and logistics planning.  These team members 
are anticipated to sit within the \dword{sdsd}, an organization 
formed to provide \dword{fnal} support services in South Dakota.    

\section{Far Site Safety}
\label{sec:far_site_safety}

The foundation of a credible installation plan is an \dword{esh}
program that ensures the safety of team members and
equipment supporting the program, as well as protection of the environment at
the \dword{surf} site.  The \dword{ipd} has responsibility for
implementing the \dword{lbnf-dune} \dword{esh} program for 
installation activities in South Dakota.  The \dword{lbnf-dune}
\dword{esh} manager heads the on site safety organization and reports
to the \dword{ipd} to support the execution of this
responsibility.

The far site \dword{esh} coordinators sitting under the
\dword{lbnf-dune} \dword{esh} manager oversee the day-to-day execution
of the installation work as shown in
Figure~\ref{fig:dune_esh_installation} and described further in
Chapter~\ref{vl:tc-ESH}.
\begin{dunefigure}[\dshort{dune} installation \dshort{esh}]{fig:dune_esh_installation}
  {High level \dword{dune} installation \dword{esh} organization.}
  \includegraphics[width=0.85\textwidth]{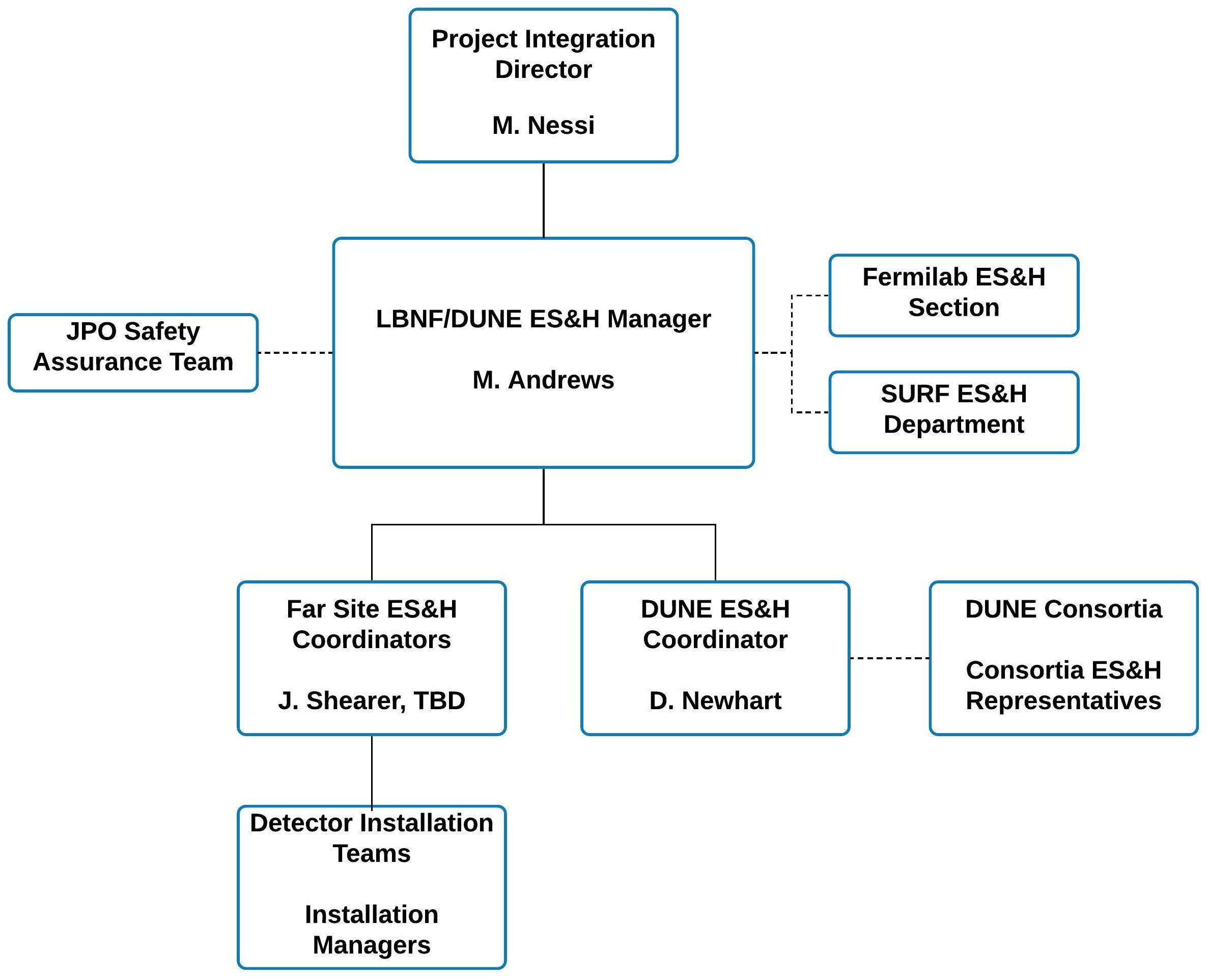}
\end{dunefigure}
As we move to two-shift installation activity, additional far site \dword{esh} 
coordinators will be assigned to each work shift.  The safety coordinator 
assigned to a particular shift is responsible for leading safety discussions 
during  toolbox meetings and for ensuring that all workers on that shift, 
including those from the consortia or contractors, are properly trained.  The 
reporting chain for safety incidents goes through the on site safety team to 
the \dword{lbnf-dune} \dword{esh} manager to minimize any potential conflicts 
of interest.  All \dword{integoff} installation team members as well as \dword{dune} 
consortia personnel and \dword{lbnf} project team members have the right to 
stop work for any safety issues.

Operation of all equipment used for installation activities such as
cranes, power tools, and personnel lifts is restricted to team members
who have been properly trained and certified for use of that
equipment.  The safety coordinator for each shift is responsible for
ensuring that all team personnel are properly trained and that safety
documentation and work procedures are up-to-date and stored within the
\dword{edms}.

Documentation, including accident reports, near misses, weekly
reports, equipment inspection, and training records is an important
component of the \dword{lbnf-dune} \dword{esh} program. The work
planning and \dword{ha} program utilizes detailed work plan documents,
\dword{ha} reports, equipment documentation, safety data sheets,
\dword{ppe}, and job task training to mimimize work place hazards and
maximize efficiency.  Sample documentation is developed through the
\dword{ashriver} trial assembly process, which maps out the step by
step procedures and brings together the documentation needed for
approving the work plan.  The sample documentation is modified to
account for differences required for performing work underground and
the updated procedures are provided to the review process (as
discussed in Chapter~\ref{vl:tc-review}) for \dwords{irr} and
\dwords{orr}.

\section{Integration Office Management}
\label{vl:tc-facility_mgmt}

The \dword{ipd} is responsible for coordinating all installation
activities at \dword{surf} including those that fall under the 
responsibility of \dword{lbnf} and the \dword{dune} consortia.  
The coordinators of this activity and crucial technical support
staff sit within the \dword{integoff}.  The organization of this on site team is 
shown in Figure~\ref{fig:io-org-chart}.
\begin{dunefigure}[Integration office installation team org chart]{fig:io-org-chart}
  {\Dword{integoff} installation team organization chart}
  \includegraphics[width=0.95\textwidth]{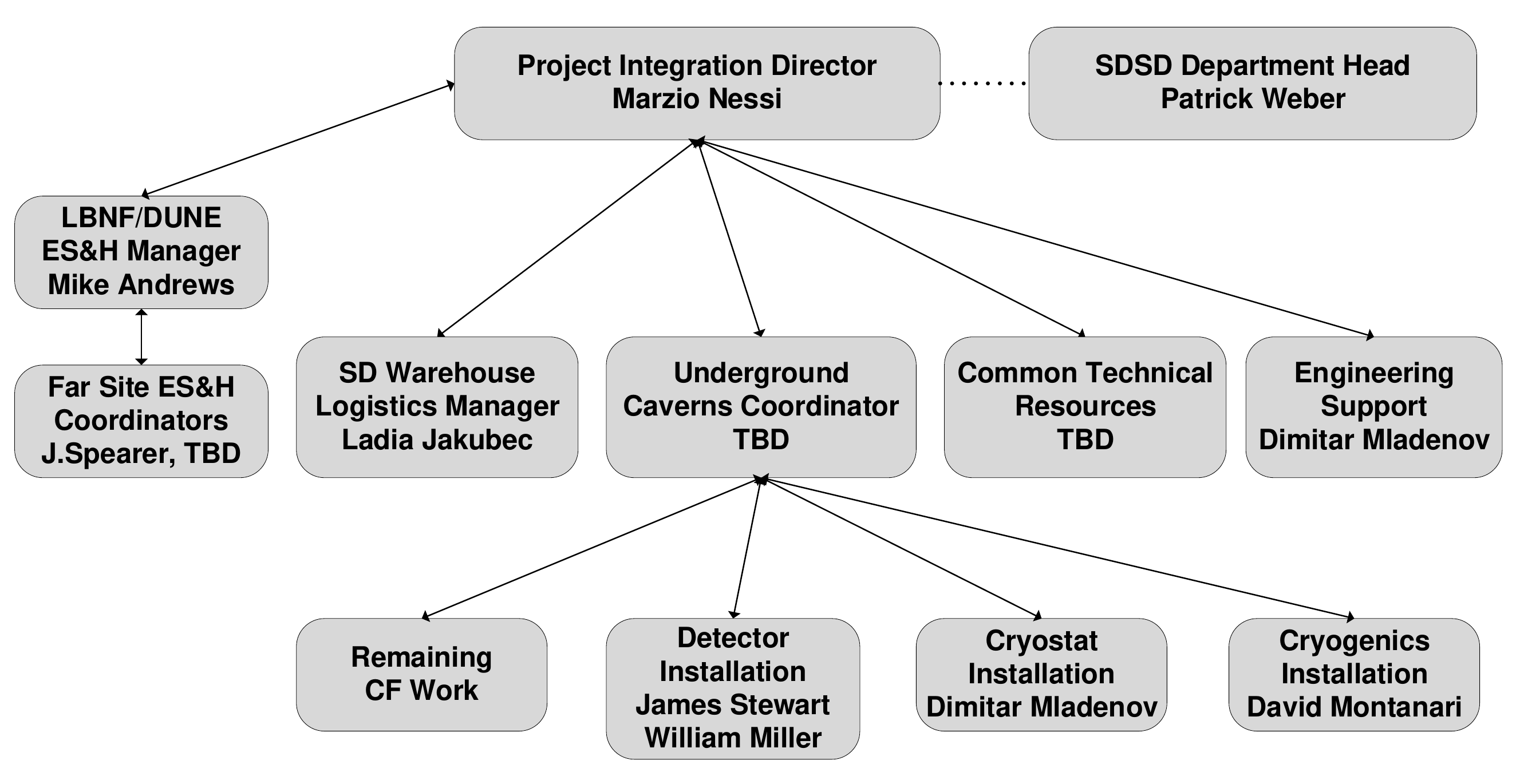}
\end{dunefigure}
 
As discussed in Section~\ref{sec:far_site_safety}, the on site safety
organization including the far site \dword{esh} coordinators
working under the direction of the \dword{lbnf-dune} \dword{esh}
manager oversee all on site activities and report 
to the \dword{ipd}.

Since \dword{surf} lacks space on the surface, a separate
warehousing facility in the vicinity of \dword{surf} is required to
receive and store materials in advance of their delivery to the
undergound area, as discussed in Section~\ref{sec:sdwf}.  Warehouse
operations are coordinated by the \dword{lbnf-dune} logistics manager
who is tasked with determining the exact sequence in which materials
are delivered into the underground areas.

The underground cavern coordinator is responsible for managing all 
activities in the two undergound detector caverns, as well as the
\dword{cuc}, including contracted workers.  Work within the
detector caverns follows a time-ordered sequence that includes
installation of the cryostats (warm and cold), cryogenic systems, and
the detectors themselves.  Work in the \dword{cuc} includes
installation of major cryogenic system pieces and the detector
\dword{daq} electronics.  The underground cavern coordinator relies on
separate installation teams focusing on cryostats, cryogenic systems,
and the detectors.  The cryostat and cryogenics system installation
teams are contracted resources provided by \dword{lbnf}.  For this
reason, coordinators of these activities are jointly placed within 
both the \dword{lbnf} project team and the \dword{integoff}.  
The detector
installation teams incoporate a substantial number of scientific and
technical personnel from the \dword{dune} consortia.  \Dword{integoff} coordinators 
of the detector installation effort are jointly placed within 
\dword{dune} \dword{tc} to facilitate consortia involvement in  
detector installation activities.  Any modifications to the facilities 
occuring after \dword{aup} are managed by the underground cavern 
coordinator under the direction of the \dword{ipd}.

The \dword{ipd} manages common technical and engineering resources to
support installation activities.  Technical resources include the
support crews needed for rigging materials on and off the hoist at the
top and bottom of the shaft, transporting materials to the underground
caverns from the bottom of the shaft, and rigging the detector and
infrastructure pieces within the underground caverns during the
installation process.  
Welders and survey teams are used in all installations, as are on-call electricians and plumbers. Electricians and plumbers are also needed for operational issues that may arise. They are provided through \dword{sdsd}, which employs full-time staff and contracts with support staff for necessary functions.

Engineering resources for installation activities sit within the \dword{integoff}.  
The engineering team, which includes both mechanical and electrical 
engineers, resolves last minute issues associated with component 
handling and detector grounding that arise over the course of the 
installation process.  Other required engineering functions include 
procurement support, configuration management, and particpation in 
the safety review process.

\subsection{South Dakota Warehouse Facility}
\label{sec:sdwf}

The \dword{sdwf} is a leased 5000m$^2$ facility hosted by 
\dword{sdsd}.  Approximately six months before \dword{aup}
of the underground detector caverns is received, the \dword{sdwf} 
must be in place for receiving cryostat and detector 
components.  Laydown space near the Ross headframe is extremely 
limited.  For this reason, the transportation of materials from 
the \dword{sdwf} to the top of the Ross shaft requires careful 
coordination. The \dword{lbnf-dune} logistics manager works 
with the \dword{cmgc} through the end of excavation activities  
and with other members of the \dword{integoff} team to coordinate transport 
of materials into the underground areas.  Since no materials or 
equipment can be shipped directly to the Ross or Yates headframes, 
the \dword{sdwf} is used for both short and long-term storage, as 
well as for any re-packaging of items required prior to transport 
into the underground areas. 

A small number of \dword{dune} consortia members work at the
\dword{sdwf} to check received components for potential damage
incurred during shipment and to track all materials coming in and out of
the facility, using the inventory management system.  In some cases,
re-packaging of materials is required for lowering them down the shaft
into the underground areas.  The \dword{dune} consortia take
responsibility for these efforts.

\subsection{Underground Caverns}

The installation 
process in the underground detector caverns and \dword{cuc} can 
be broken into a time-sequenced set of activities, coordinated 
through the \dword{integoff}.  In the detector caverns, installation 
of the warm and cold cryostat strucutres is followed by (with 
some overlap) installation of the cryogenic infrastructure and 
detectors.  In the \dword{cuc}, installation of the \dword{daq}  
infrastructure and detector readout components proceeds in 
parallel with that of the cryogenic infrastructure.  A high-level 
schedule showing the inter-dependencies between these activities 
is shown in Figure~\ref{fig:underground_schedule}.
\begin{dunefigure}[Underground summary schedule]{fig:underground_schedule}
  {Summary schedule of the different phases of work underground}
  \includegraphics[width=0.95\textwidth]{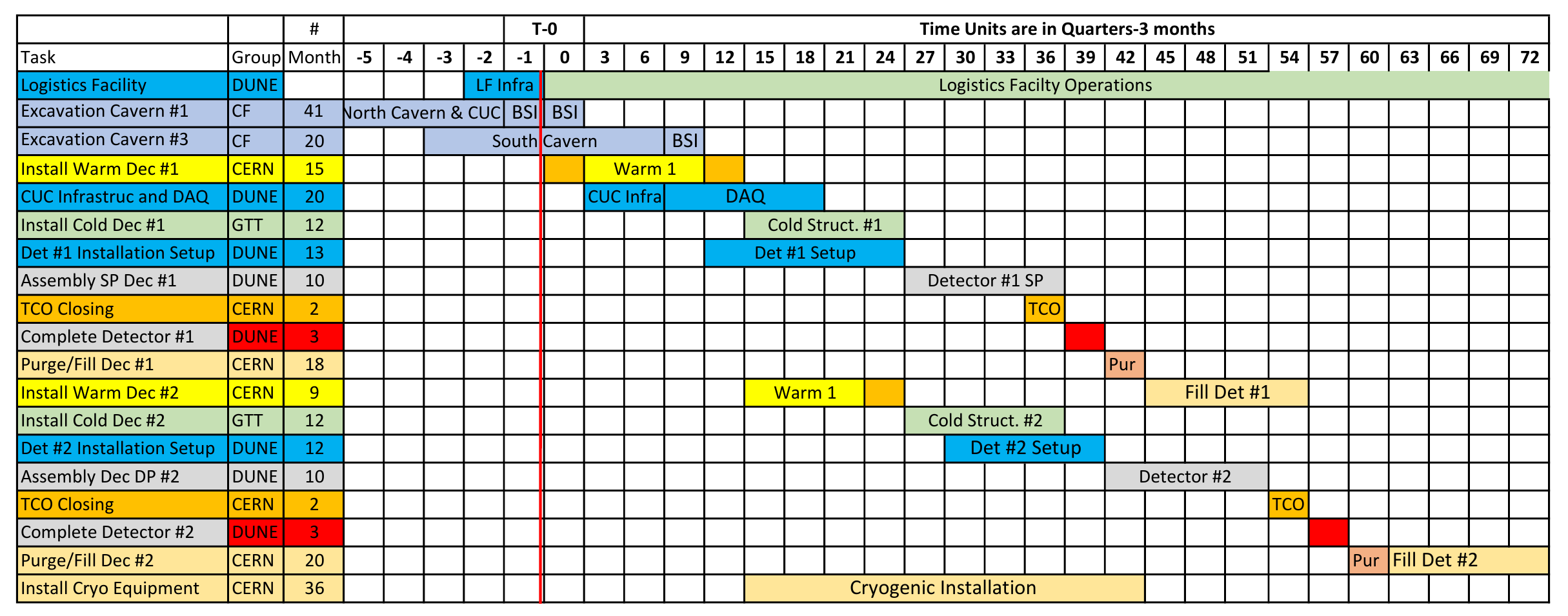}
\end{dunefigure}

The ability of \dword{lbnf-dune} to meet this schedule depends
critically on its ability to work within limitations on the total
number of people allowed in the underground areas at a given time
(144 people) as well as occupancy limits on work in the cryostat.  In order
to satisfy these limitations, careful balancing of the numbers of
workers assigned to different concurrent tasks taking place within the
different underground caverns is required.  This is a particular
challenge during the excavation period for the second detector cavern,
which runs in parallel with cryostat installation in the first
detector cavern.  The \dword{integoff} works with its \dword{lbnf-dune} project 
partners to manage and optimize the underground work schedule so that 
interferences between concurrent work efforts are minimized.

The underground cavern coordinator manages the contributions of 
the technical team supporting  installation activities.
The size of the technical support team is anticipated to evolve  
over time to meet the needs of the specific installation tasks 
taking place.  The functions provided by the technical team 
supporting the work in the underground caverns include the
following:
\begin{itemize}
  \item {material transport:} The transport team shown in
    Figure~\ref{fig:ctr_orgchart} is responsible for unloading of
    materials from the trucks arriving from the \dword{sdwf}, loading
    or rigging of materials at top of Ross Shaft, unloading or rigging
    of materials at bottom of Ross Shaft, and delivery of materials
    from the bottom of shaft to the underground caverns. This does not
    include operation of the hoists, which is performed by
    \dword{surf}.
  \item {cavern rigging operations:}  Storage of 
        components within the available spaces 
        in the cavern and movement of materials as required to 
        execute the installation process (three rigging stages
        for detector installation are moving components into 
        clean room, integrating components within clean room, 
        and installing integrated elements inside cryostat).
  \item {installation technicians:}  General technician 
        support for specific installation activities. 
  \item {welders and survey crews:}  Perform 
        specific tasks incorporated within each of the different 
        installation efforts.
  \item {electrical technicians and plumbers:} On-call support staff
    to modify systems as work transitions from one stage to the next
    and to address issues as they arise.
\end{itemize}   
    
The organization responsible for managing contributions of 
the technical support team to the installation 
activities taking place in the underground caverns is shown 
in Figure~\ref{fig:ctr_orgchart}.  The structure is illustrated
for the case of the largest anticipated workforce (approaching 
roughly 60 team members in total covering multiple shifts) 
for the periods with ongoing detector installation efforts.
These personnel support two 10-hour shifts on Mondays through 
Thursdays and a day shift on the remaining days to cover 
activities occurring over weekends. 
\begin{dunefigure}[Common technical resources]{fig:ctr_orgchart}
  {Summary of the \dshort{integoff}/\dword{sdsd} common technical resources}
  \includegraphics[width=0.95\textwidth]{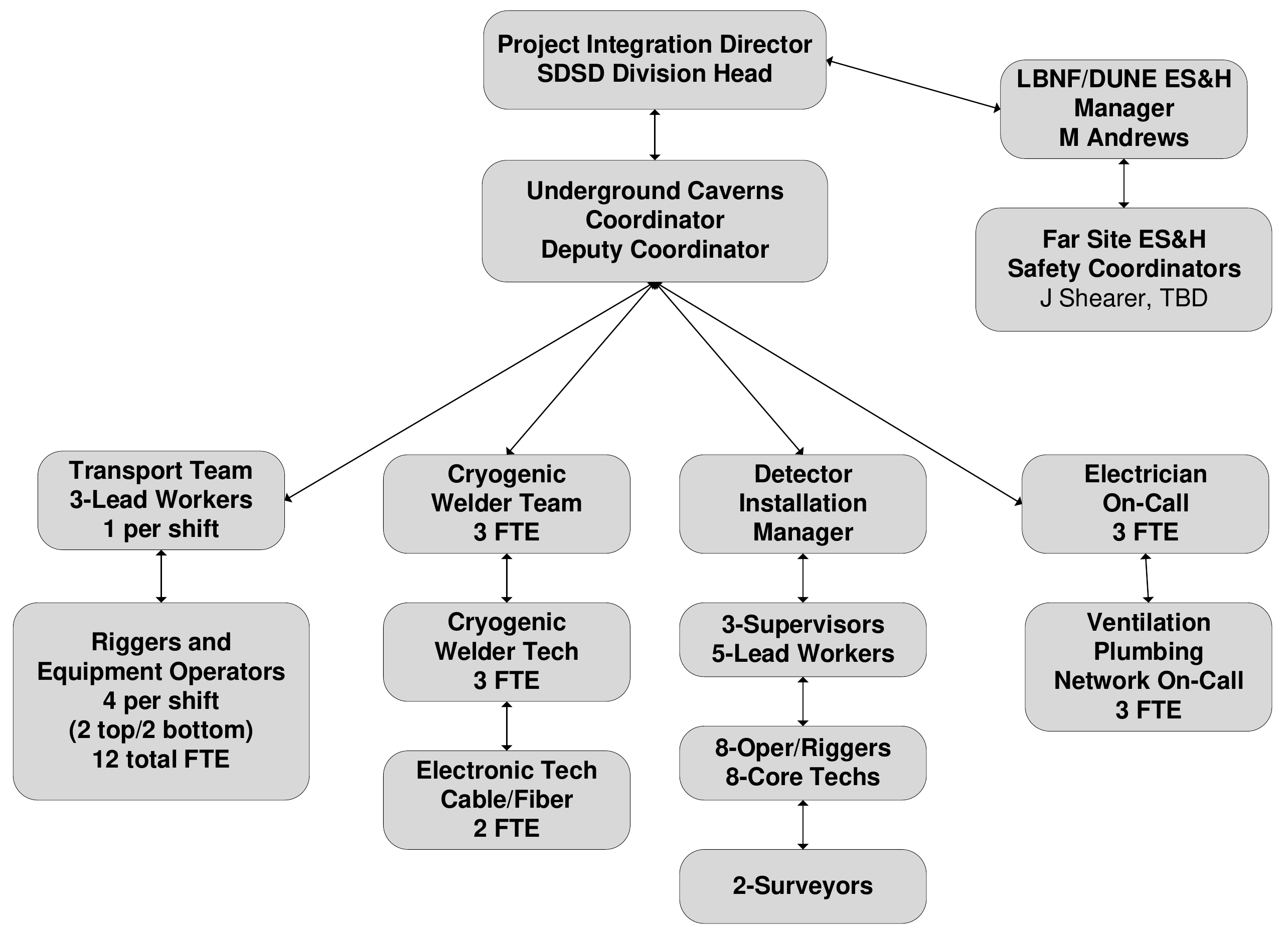}
\end{dunefigure}

On the surface at the start of each shift, there is a toolbox safety
meeting and work assignment update. An hour separates the two shifts,
in which the lead workers, safety coordinators, and other management
team members are paid overtime to overlap with each other and
transfer information from one shift to the next. The safety
coordinator for each shift is responsible for conducting the safety
discussion at the meeting and ensuring that all workers assigned to
that shift have the proper training.

The team responsible for detector installation incorporates 
members of the technical support team described above 
and includes scientific and technical personnel from 
the \dword{dune} consortia.  The team is led by the detector
installation manager who has three shift supervisors working 
with them to provide on site coverage for every shift.
The management team works with the underground cavern
coordinator to ensure that required technical support team 
members are available as needed and that required materials 
are delivered to the detector caverns on a schedule to keep
the installation effort moving forward.         

The management team supervises technical resources assigned to 
the detector installation effort and works with consortia 
team members to maximize the overall efficiency of the installation 
process.  The organizational structure to manage the detector 
installation activities is shown in Figure~\ref{fig:uit_orgchart}.
\begin{dunefigure}[Underground detector installation team]{fig:uit_orgchart}
  {The \dword{uit}}
  \includegraphics[width=0.95\textwidth]{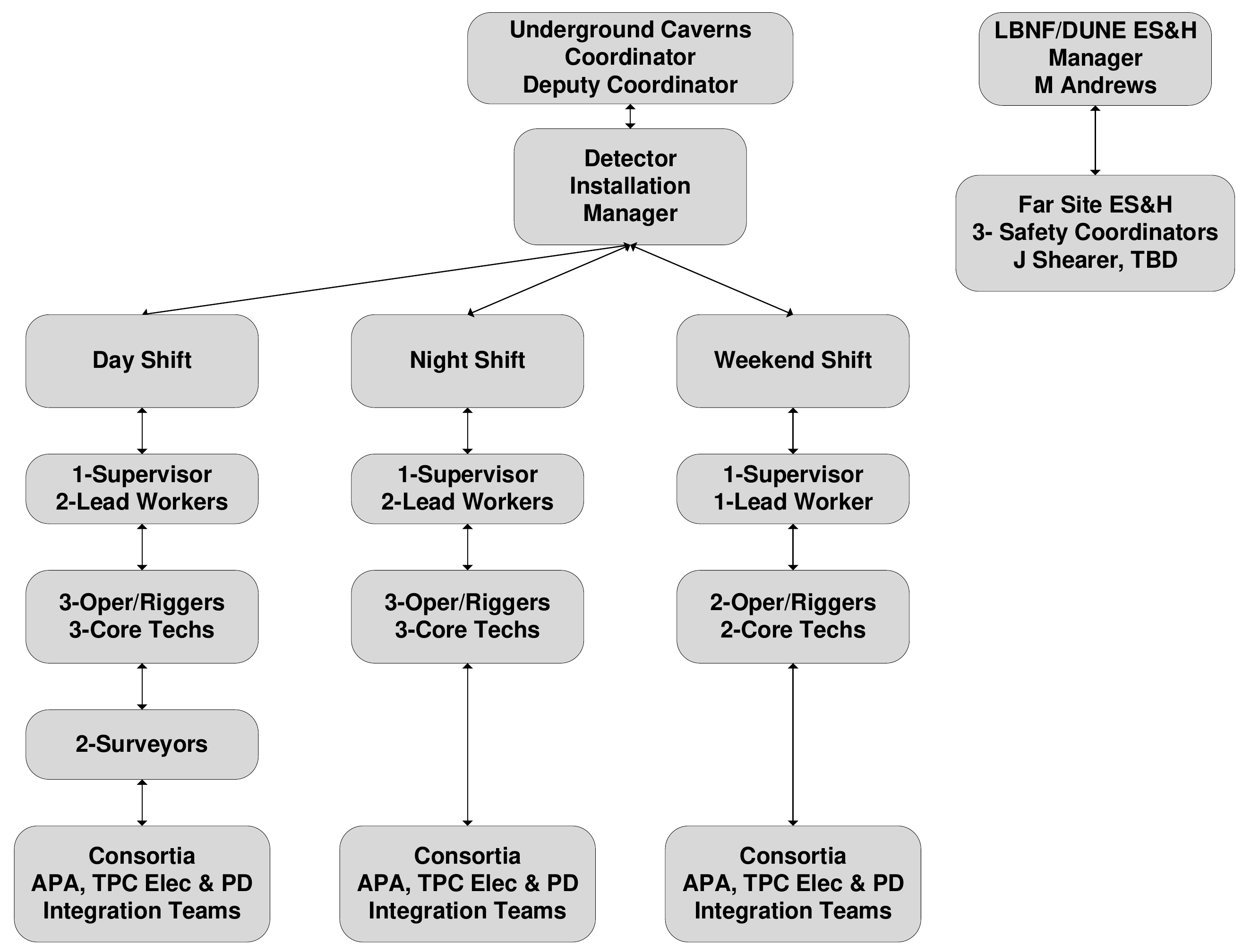}
\end{dunefigure}

The detector installation manager oversees all shifts and serves as
the supervisor for specific shifts as needed.  They serve as the
contact with the underground cavern coordinator for obtaining
required technical support team members and organizing the delivery of
needed materials into the detector caverns.  The detector installation
manager attends all high-level meetings with the underground cavern
coordinator and is tasked with submitting weekly progress reports.
They work with the \dword{dune} consortia to manage the overall work
schedule and ensure that the correct resources are in the right place
at the right time.
    
The installation supervisors are working managers, trained as riggers
and equipment operators to fill in as needed on their shifts.  They
are fully trained in all installation procedures and work with the
consortia shift team members to keep the installation effort on
schedule.  Installation supervisors fill in for their lead workers as
needed and are the primary points of contact for information exchange
between shifts.  Lead workers direct the technical support personnel
assigned for their shift.  The lead workers are trained in all
installation procedures and provide assistance to the consortia work
teams as needed.

\subsection{Trial Assembly at \dshort{ashriver}}

The trial assembly work at \dword{ashriver}, site of the \dword{nova}
far detector, focuses on mechanical tests of the installation
process for \dword{dune}. This effort is critical to confirm final
detector component designs, including modifications originating from
\dword{protodune}, confirming and practicing installation techniques
for both the cleanroom and cryostat.  The \dword{nova} far site
detector hall in Ash River, Minnesota, has facilities that
match \dword{dune} needs, including a \SI{16.75}{m} deep pit with
$\sim$\SI{300}{m$^2$} of floor space available for testing full-scale
\dword{dune} detector components and a capable workforce that is
needed for \dword{nova} operations and can be leveraged in a cost effective
manner for \dword{dune}.  The \dword{nova} far detector laboratory is
managed by the University of Minnesota (UMN) and is partially funded
through an operations contract from \dword{fnal}.  Work performed at
the \dword{ashriver} site follows university safety regulations and
any \dword{dune} safety requirements. University code officials approve
all building permits, which include engineered drawings signed by an
engineer registered in Minnesota. All hazard analyses and work
procedure documents are approved by the joint \dword{dune}/UMN safety
committee with members drawn from both the University of Minnesota
(UMN) and \dword{dune} that includes specialists as needed.

The work at \dword{ashriver} has five main goals:
\begin{itemize}
  \item use prototype \dword{dune} components to verify that the
    detector can be installed in a safe and efficient
    manner,
    \item test installation equipment needed to install the
      \dword{dune} detector at \dword{surf},
  \item validate mechanical design changes made to the detector
    elements subsequent to \dword{protodune} operation,
  \item complete a set of reviewed engineering and procedural
    documents that will serve as the basis for work to be performed
    underground at \dword{surf}, and
  \item serve as a training center for personnel who will 
    contribute to \dword{dune}  installation at \dword{surf}.
\end{itemize}

The full time staff of five people at \dword{ashriver}
includes a manager, deputy manager, and three experienced technicians
that all participated in \dword{pdsp} installation at \dword{cern} and
the \dword{pdsp} trial assembly at \dword{ashriver}.  The staff
oversees operations of the \dword{nova} detector and performs trial
assembly studies of the \dword{dune} detector components.  One of the
three technicians also serves as the site safety officer and
chairperson of the joint \dword{dune}/UMN safety committee.  Two additional staff
members will be added in the near future to handle the additional
workload associated with preparations for the \dword{protodune2}
installation effort.
\begin{dunefigure}
  [Phase 1 \dshort{apa} installation frame being installed on
    the \dshort{apa} assembly tower at \dshort{ashriver}]{fig:ashriver1}
  {Phase 1 \dword{apa} installation frame (in red) being installed on the
  \dword{apa} assembly tower at \dword{ashriver}. In the foreground is
  the \dword{pdsp} trial assembly structure.}
   \includegraphics[width=0.65\textwidth]{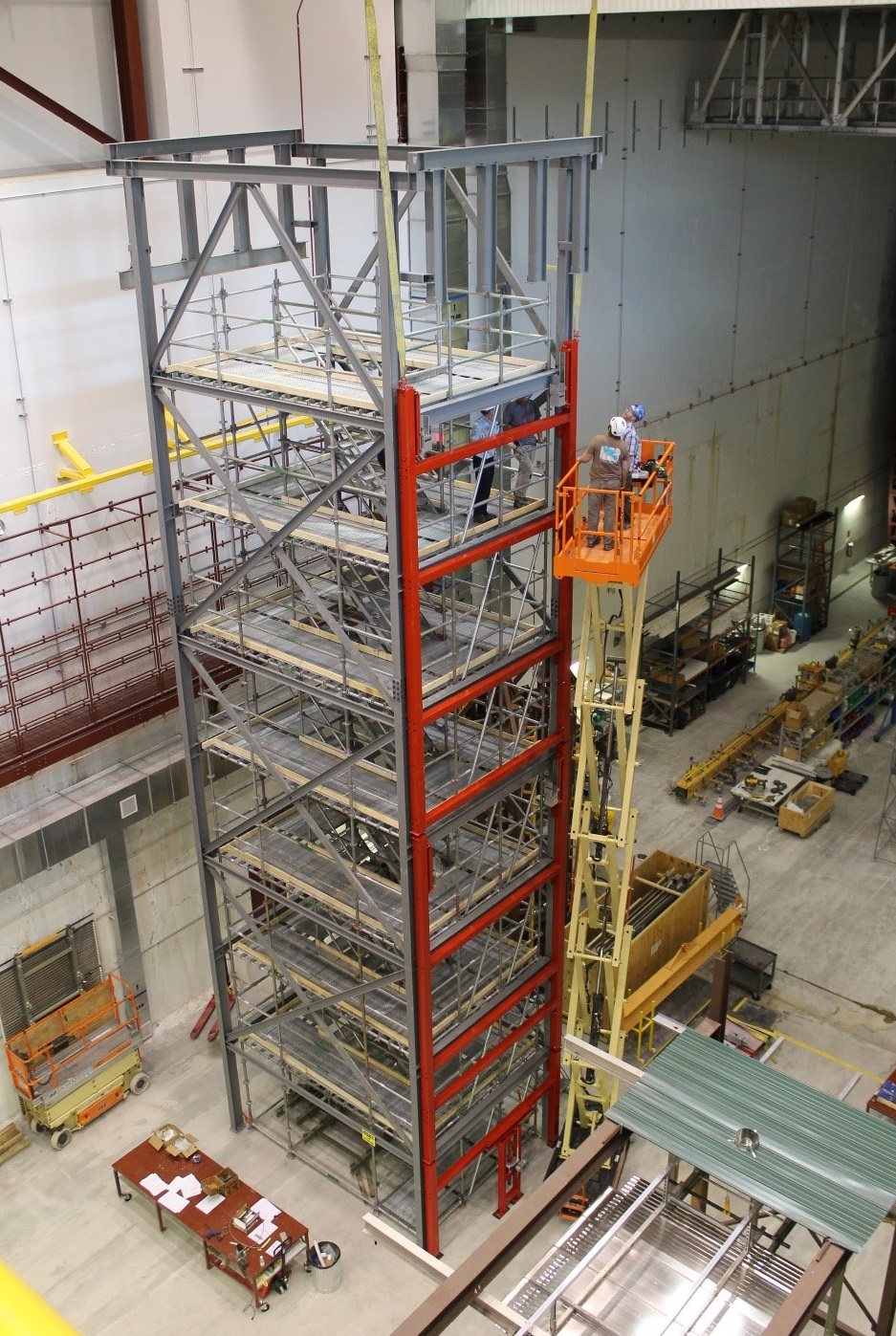}
\end{dunefigure}

The work at \dword{ashriver} is divided in three major phases:
\begin{itemize}
  \item {Phase 0:} A vertical cabling test using two full-scale 
         \dword{apa} side tubes connected top to bottom and mounted 
         against a vertical column in the detector hall.  Using this 
         setup the proposed cable bundles have been run through the 
         tubes to see how well the designed conduit system functions.
         This work has led to several proposed modifications to the 
         designs that are currently being considered.  The older 
         \dword{protodune} trial assembly structure is concurrently 
         being used to perform mechanical tests of \dword{protodune2} 
         components. 
  \item {Phase 1:} A prototype of the \dword{dune} \dword{apa} 
         assembly tower using a steel frame large enough to hold a 
         commercial stair scaffold within its mid-section, as shown 
         in Figure~\ref{fig:ashriver1}, was constructed and 
        used to test the process for connecting top 
         and bottom \dword{apa} pairs together and installing the 
         required cable bundles.  The next step will be to add a
         \dword{cpa} assembly station and test assembly procedures 
         for the updated \dword{cpa} designs.  A prototype 
         \dword{apa} shipping frame is also being constructed to 
         test the mechanical features of the shipping container 
         design.  
  \item {Phase 2:} A more complex steel structure will be 
         designed and fabricated to mock up the network of rails 
         and support structures used to install the \dword{dune}
         \dword{fd} modules including pieces of the \dword{dss} that 
         sits inside the cryostat.  This structure, as 
         illustrated in Figure~\ref{fig:ashriver2}, will provide 
         a platform for performing more detailed tests of the 
         proposed detector installation plan.  Installation steps 
         to be tested include \dword{dss} installation, transfer 
         of \dword{tpc} components through the \dword{tco}, 
         installation of the \dword{tpc} end walls, cabling 
         through the cryostat penetrations, movement of the 
         \dword{apa} and  \dword{cpa} pairs into their final 
         positions, and deployment of the top and bottom field 
         cage modules.
\end{itemize}
\begin{dunefigure}[Phase 2 trial assembly at \dshort{ashriver}]{fig:ashriver2}
  {Phase 2 trial assembly at \dword{ashriver}.}
  \includegraphics[width=0.65\textwidth]{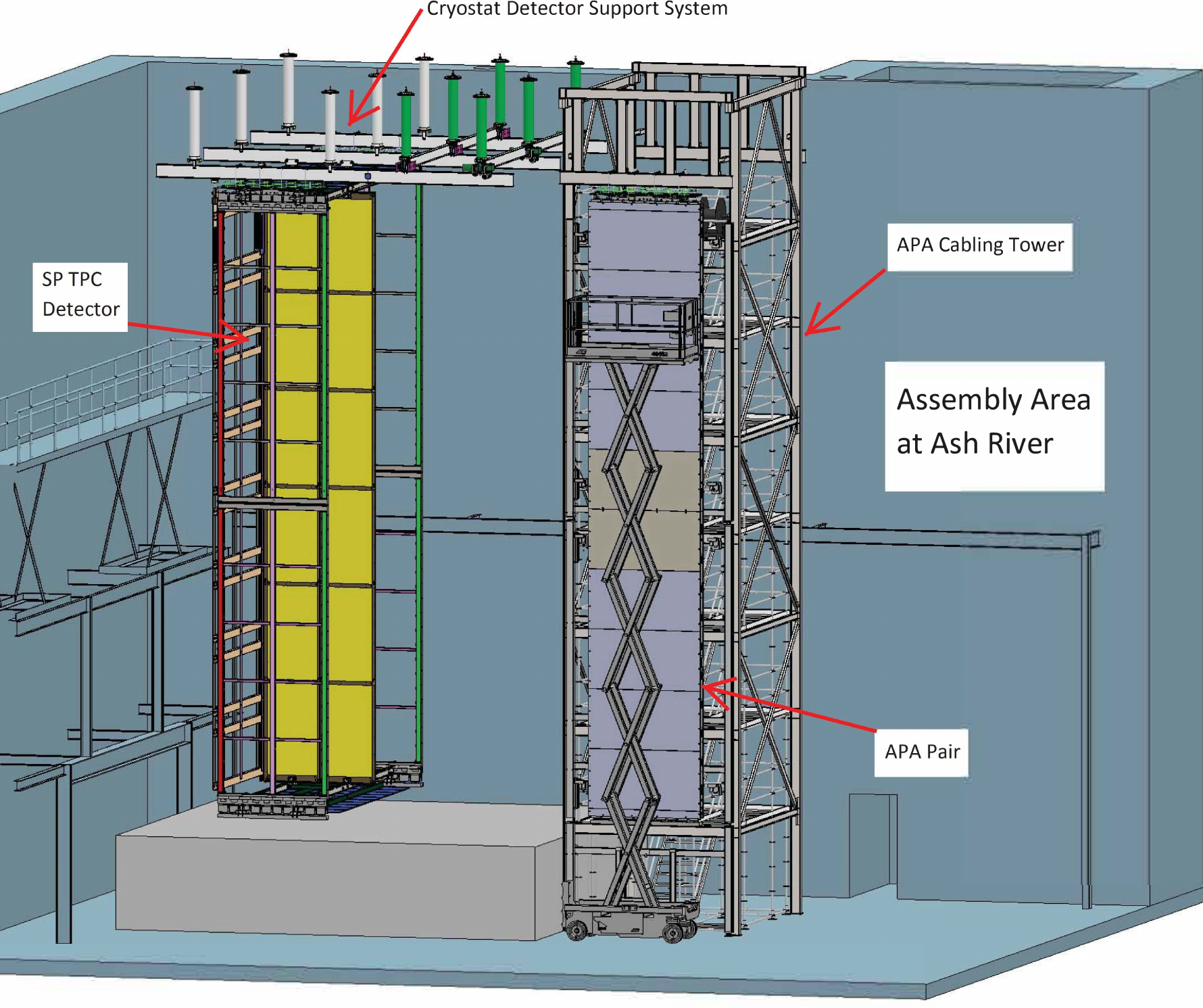}
\end{dunefigure}

\section{South Dakota Services Division}
\label{sec:fdsp-coord-host_facility_services}

\dword{surf} is operated by the \dword{sdsta} through a cooperative 
agreement with the \dword{doe}.  Prior to this agreement going forward, 
\dword{sdsta} signed onto a \dword{mou} with the Fermi Research Alliance 
(FRA) detailing facility support services to be provided by \dword{sdsta}
in support of \dword{lbnf-dune}.  The \dword{mou} establishes a Joint 
Coordination Team with regularly scheduled meetings to ensure that all  
\dword{surf} support functions necessary for achieving \dword{lbnf-dune} 
objectives are provided.             

\dword{fnal} has established the \dword{sdsd} to support integration
and installation activities in South Dakota. \dword{sdsd} will support
these activities, which are the responsibility of the \dword{ipd}, by
providing access to the required technical resources.  These resources
include dedicated \dword{fnal} personnel sitting within the division
and contracted labor provided through the division.  Some examples of
\dword{sdsd} support staff include rigging teams to support activities 
in underground caverns and at the headframe and bottom of the shaft,
transport crews for moving materials between the warehouse and site 
and from the bottom of the shaft into the underground caverns, and a 
core group of technicians for performing maintenance and installation
activities.  \dword{sdsd} will also assist \dword{lbnf-dune} partners 
in understanding work requirements at \dword{surf} and ensuring that 
appropriate provisions are incorporated into partner contracts with 
external contractors.  The \dword{sdsd} will have its own procurement 
team to assist the \dword{ipd} in acquiring the common infrastructure 
items required for the installation effort.  This team will also be 
responsible for handling any contracts associated with further work 
on the facilities subsequent to departure of the \dword{lbnf} \dword{cmgc}.         

Much like the \dword{ashriver} site, where University of Minesota 
officials are responsible for any building permits, \dword{sdsd} 
is responsible for any electrical or building permits required 
for the leased spaces at \dword{surf}.  \dword{sdsd} also takes 
responsibility for badging personnel requiring access to the 
leased areas at \dword{surf} in coordination with the \dword{fnal} 
Global Services Office and the \dword{surf} Administrative Services 
Office. This includes providing and coordinating the trainings 
required to access surface and underground areas.  To maintain safe 
working conditions within the leased areas, \dword{sdsd} performs 
regular inspections and maintenance of all \dword{lbnf-dune} 
equipment operating at \dword{surf} including lifts, conveyances, 
networking equipment, cooling and ventillation equipment, rigging 
equipment, electrical power installations, life safety systems, 
and controlled access equipment.

\cleardoublepage

\chapter{Facility Description}
\label{vl:tc-facility}

The \dword{dune} detectors are located in the main underground campus
at \dword{surf}. The main campus is located at the 4850 foot level (\dword{4850l}), between
the Ross and Yates Shafts. This campus and associated surface
facilities are being developed by \dword{lbnf} through excavation
(EXC) and outfitting (\dword{bsi}) contracts with the civil
contractor. Once the contractor has delivered and \dword{aup} has
concluded, cryostat construction will commence. Further infrastructure
will be delivered by \dword{sdsd}.  The following sections describe
the facilites as they are related to the \dword{dune} detectors.

\section{Underground Facilities and Infrastructure}
\label{sec:fdsp-coord-uderground-excavation}

The \dword{dune} underground campus at the \dword{surf} \dword{4850l} is shown in
Figure~\ref{fig:dune-underground}.
\begin{dunefigure}[Underground campus]{fig:dune-underground}
  {Underground campus at the \dword{4850l}.}
  \includegraphics[width=0.75\textwidth]{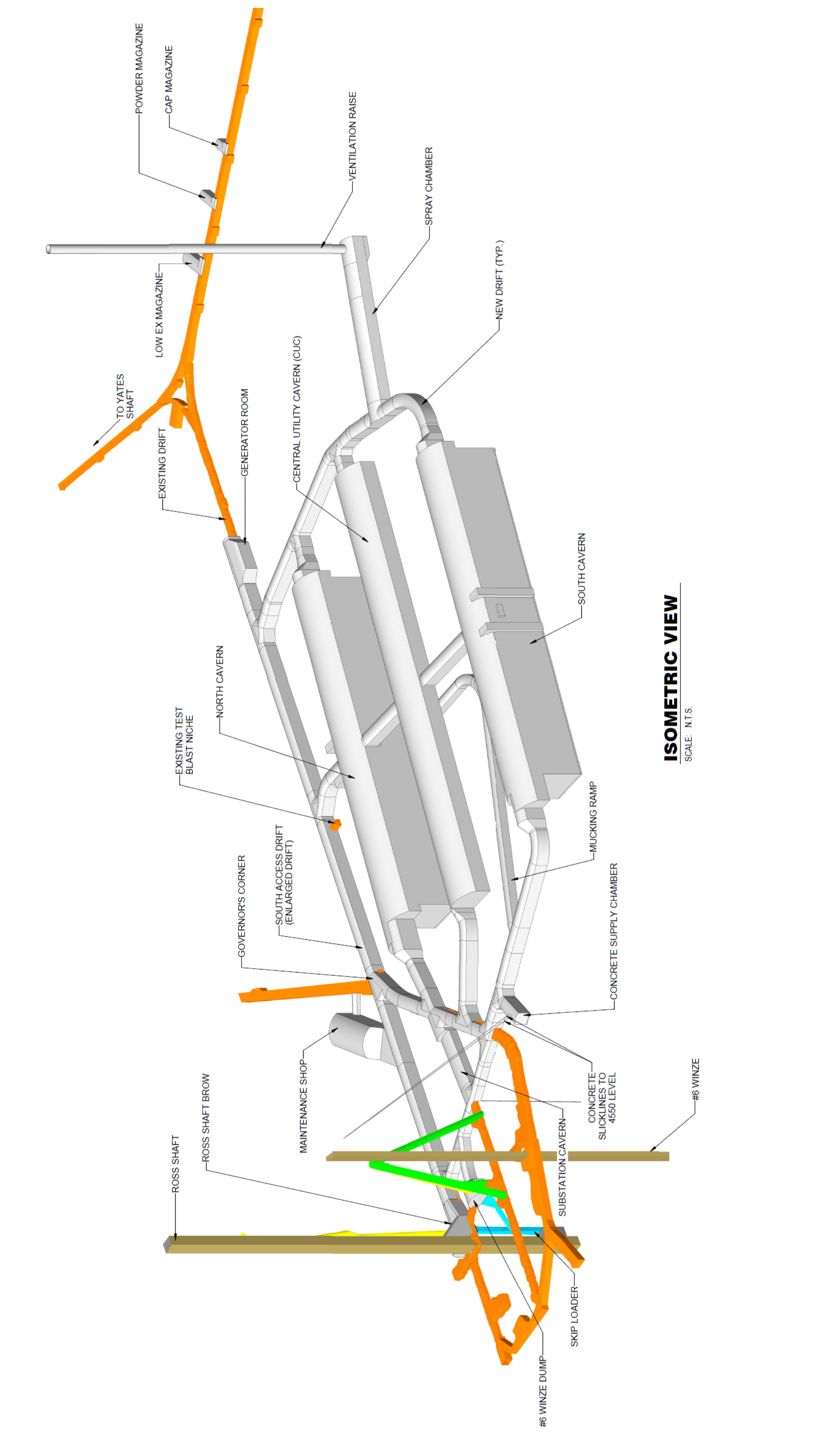}
\end{dunefigure}
\dword{lbnf} will provide facilities and services, on the surface and
underground, to support the \dword{dune} detectors.  This includes
logistical, cryogenic, electrical, mechanical, cyber, and environmental
facilities and services.  All of these facilities are provided for the
safe and productive operation of the \dwords{detmodule}.

The primary path for both personnel and material access to the
underground excavations is through the Ross Shaft. On the surface, the
Ross Shaft and Ross Headframe are undergoing major upgrades. The
structure of the headframe that supports the conveyances is being
reinforced and renovated to make the work flow more efficient and
safer.  All of the wood timber from the 100+ year old shaft has been
removed and replaced with steel sets to improve the time required to
traverse the \dword{4850l} to the underground spaces.  The brakes, drive,
clutches, and controls for the conveyance are being completely
overhauled and updated.  In addition to all of these improvements, a
new cage is being designed and fabricated for the Ross Shaft.  The new
cage incorporates features to allow the transport of larger items and
includes rigging points underneath for slung loads without the removal
of the cage.  Details on the Ross Cage design and size constraints for
material and slung loads can be found in~\citedocdb{3582}.

On the surface, a new compressor building is being constructed
adjacent to the headframe.  This building will house the cryogenics
systems for receiving cryogenic fluids and preparing them for delivery
down the Ross Shaft.  New piping is being installed down the Ross
shaft compartment to transport \dword{gar} and N$_2$ underground.

New transformers are being installed in the Ross substation on the
surface to support the underground power needs.  New power cables are
being installed down the Ross Shaft to transmit the power underground
to a new substation.

A portion of the Ross Dry basement is being refurbished to house the
surface cyber infrastructure (\dword{mcr}) required for data and other
underground information.  Redundant fiber optic cables will leave the
\dword{mcr} to travel down both the Ross and Yates shafts to newly
excavated underground communication distribution room (CDR), which is
located near the west entrance drift of the north cavern.  From the
CDR, the fibers branch out to the \dword{cuc} and detector caverns to
support detector data, cryogenic and detector safety systems and will
be tied into the \dword{bms}.  The \dword{bms} controls the facilities
\dword{fls}.  All \dword{fls} signals from the detector or cryogenic
safety systems will be tied into the facilities \dword{bms} for
communications to personnel underground and on the surface.

\section{Detector Caverns}
\label{sec:fdsp-coord-faci-caverns}

Underground spaces are being excavated to support the four
\dword{dune} detectors and infrastructure.  Two large detector caverns
are being excavated.  Each of these caverns will support two
\larmass cryostats.  These caverns, labeled north and
south, are \SI{144.5}{\meter} long, \SI{19.8}{\meter} wide,  and 
\SI{28.0}{\meter} high. The tops of the cryostats are approximately
aligned with the \dword{4850l} of \dword{surf} with the cryostats resting
at the 4910L.  A \SI{12}{\meter} space between the cryostats will
be used as part of the detector installation process, placement of
cryogenic pumps and valves, and for access to the 4910L.  The
\dword{cuc}, between the north and south caverns, is \SI{190}{\meter}
long, \SI{19.3}{\meter} wide, and \SI{10.95}{\meter} high.  The
\dword{cuc} will house infrastructure items, such as cryogenic
equipment, nitrogen dewars and compressors, data acquisition racks and
electronics, chillers for the underground cooling, and electrical
services to support the underground space and detectors.  These three
main caverns will be connected via a series of drifts at the \dword{4850l} as shown in
Figure~\ref{fig:dune-underground}. Access tunnels lie at the east and
west ends of each cavern, the north and south sides of the north cavern, and
the north side of the south cavern. Additionally, the cross section of the
existing drifts where experimental components will be transported is
being increased to allow passage.  Other ancillary spaces being
provided underground include a maintenance shop, electrical
substation, concrete supply chamber, compressor room, and  a spray chamber
to house the cooling tower for heat rejection to the mine air for
exhaust through a new 1200 foot$\times$12 foot diameter bore hole up
to the 3650L.  This bore hole will provide additional ventilation
to support excavation.

The first detector will be built in the east side of the north cavern and
the second detector will be built in the east side of the south cavern, as
shown in Figure~\ref{fig:dune-full_assmebly}.
\begin{dunefigure}[\dshort{dune} cryostat]{fig:dune-full_assmebly}
  {Placement of detectors in the north and south caverns}
  \includegraphics[width=0.85\textwidth]{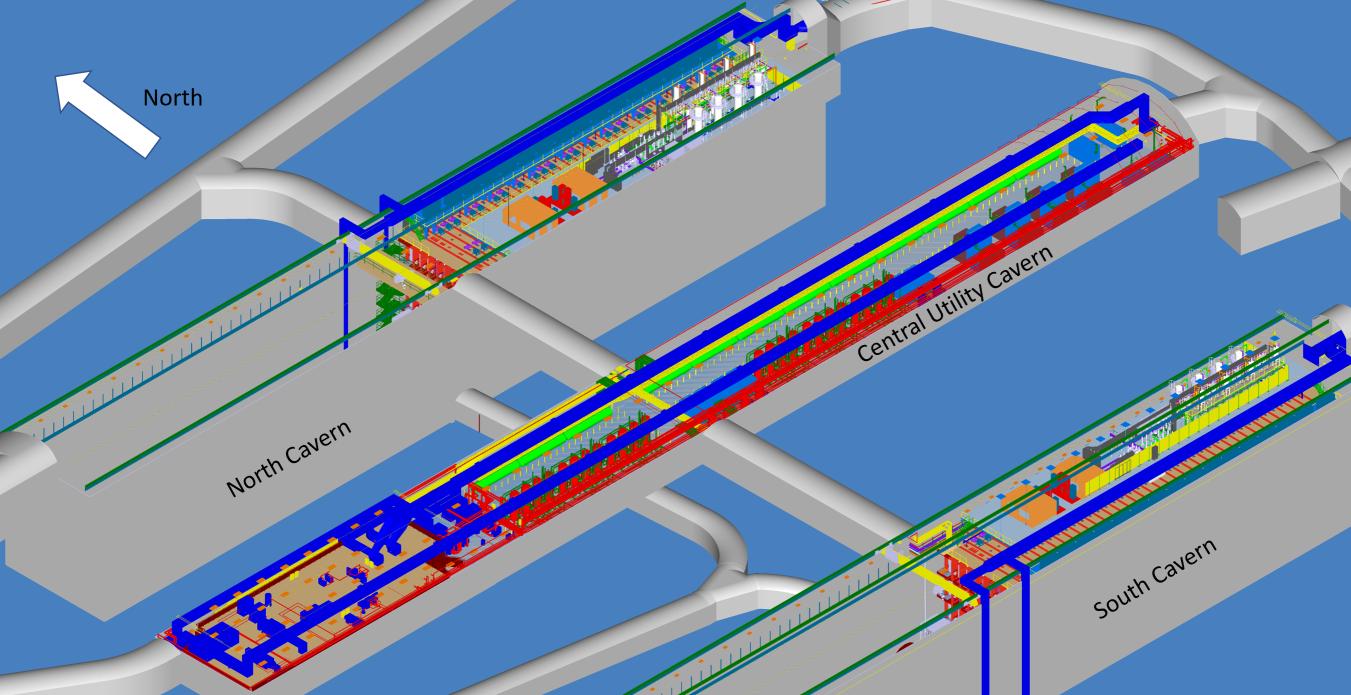}
\end{dunefigure}
The primary reasons for constructing the first detectors in the east
sides of two caverns are:
\begin{itemize}
\item {\bf Personnel safety during construction and filling}: It is
  planned that the first detector will be in the filling phase while the second detector
  is in the construction phase. Therefore, construction of the 
  second detector has to take place in a separate cavern from
  filling of the first detector. In addition, the airflow in both
  caverns is from west to east; therefore, construction
  personnel who are primarily on the west side of the detectors will
  be upstream of the cryogens.
\item{\bf Access during construction}: The main access for bringing
  large items into both caverns is from the west side. It is
  advantageous for construction of the second detector in each cavern
  that items that enter from the west are directly lowered and do not
  have to pass over the detector construction site.
\end{itemize}
Both reasons drive the plan for detectors one and two to be built on
the east side of the two separate caverns.  It is therefore planned
that both caverns are ready for \dword{aup} at the start of respective
detector construction phases.  It should be noted that both detectors
will be built at the 4910L. However, with the exception of the
\dword{lar} circulation pumps, all services are at the \dword{4850l} to
facilitate access during the operations phase.

To support the electrical requirements new 35~MVA power feeds will be
routed from the surface substation down the Ross Shaft to the
underground electrical substation and from the substation to the
electrical room in the \dword{cuc}.  The electrical room will have
multiple transformers from which to distribute power for various
functions such as lighting, HVAC, cryogenics equipment, and detector  
power.  Each detector cavern and the \dword{daq} room will have
dedicated feeds from the electrical room.  Details of this are
described in the electrical section.  There are multiple electrical
panels planned in each cavern to provide power for the different
phases of construction.  These phases include the installation
of the cryostat warm structure and cold membrane, detector
installation and detector operations. During detector operations, 
separate panels are available for detector (clean) and building
(dirty) power as is discussed further in
Section~\ref{sec:fdsp-coord-faci-power}.

To support underground cooling requirements, four 400-ton chillers
will be located in the \dword{cuc} adjacent to the electrical room and
cryogenics equipment.  These chillers are designed to provide
\SI{400}{\kilo\watt} of cooling for the various cryogenics systems,
\SI{500}{\kilo\watt} of cooling for the electronics on each of the
detectors and \SI{750}{\kilo\watt} of cooling to the \dword{daq} room
in the \dword{cuc}.  These chillers will also provide chilled water to
the HVAC systems underground to maintain the ambient temperature and
humidity.

Fire protection in the underground spaces will be determined by zone,
hazard type, and requirements.  Details of this can be found on ARUP
drawing U1-FD-E-651 in the Underground Electrical package, which is shown
in Figure~\ref{fig:lbnf-fire}.
\begin{dunefigure}[Underground fire zones]{fig:lbnf-fire}
  {Underground fire zones}
 \includegraphics[width=0.85\textwidth]{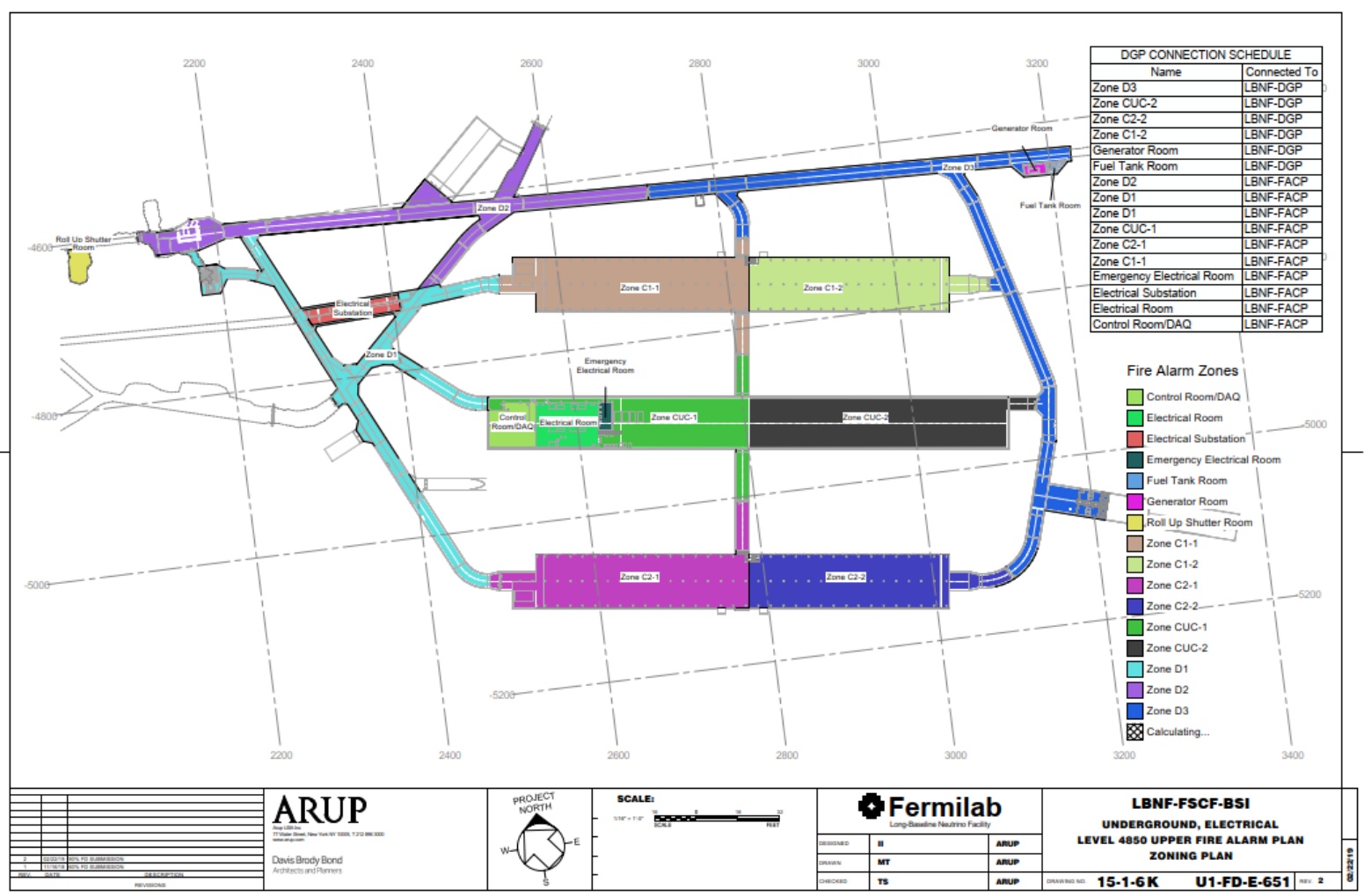}
\end{dunefigure}
The drifts will be outfitted with normal wet type sprinklers and
broken into three zones.  The north and south caverns will be outfitted
with a pre-action type sprinkler system to protect sensitive detector
electronics from water damage from accidental release.  A pre-action
system requires two signals to activate.  These are the detection of
smoke by one of the sensors and the fusing of the sprinkler
head.  This type of system was the most economical choice to reduce the
risk of unnecessary discharge of water over the detector electronics.
The DDSS will interface with the BMS fire system to turn off power
to the racks before water is introduced and reduce the impact of
the water on the system and reduce the risk of electrical shock.  A
pre-action system will also be installed in the CUC over the cryogenic
equipment as shown in the figure.  In the electrical substation,
electrical room, \dword{daq} room, and CDR, a clean-agent type system will be
employed.  Clean-agent systems use either inert gas or chemical agents
to extinguish a fire and are typically used in areas that contain
sensitive electronics or data/power centers.

In addition to the above services, systems are being installed to
provide compressed air, industrial water, internet, and a configurable
security access system.

\section{Cryostat}
\label{sec:fdsp-coord-cryostat}

Each detector will be housed inside a cryostat designed to hold the
liquid argon (\dword{lar}), cryogenic piping, and the detector as shown in
Figure~\ref{fig:dune-cryostat}.
\begin{dunefigure}[\dshort{dune} cryostat]{fig:dune-cryostat}
  {Overall construction and dimensions of the \dword{dune} cryostat.}
  \includegraphics[width=0.85\textwidth]{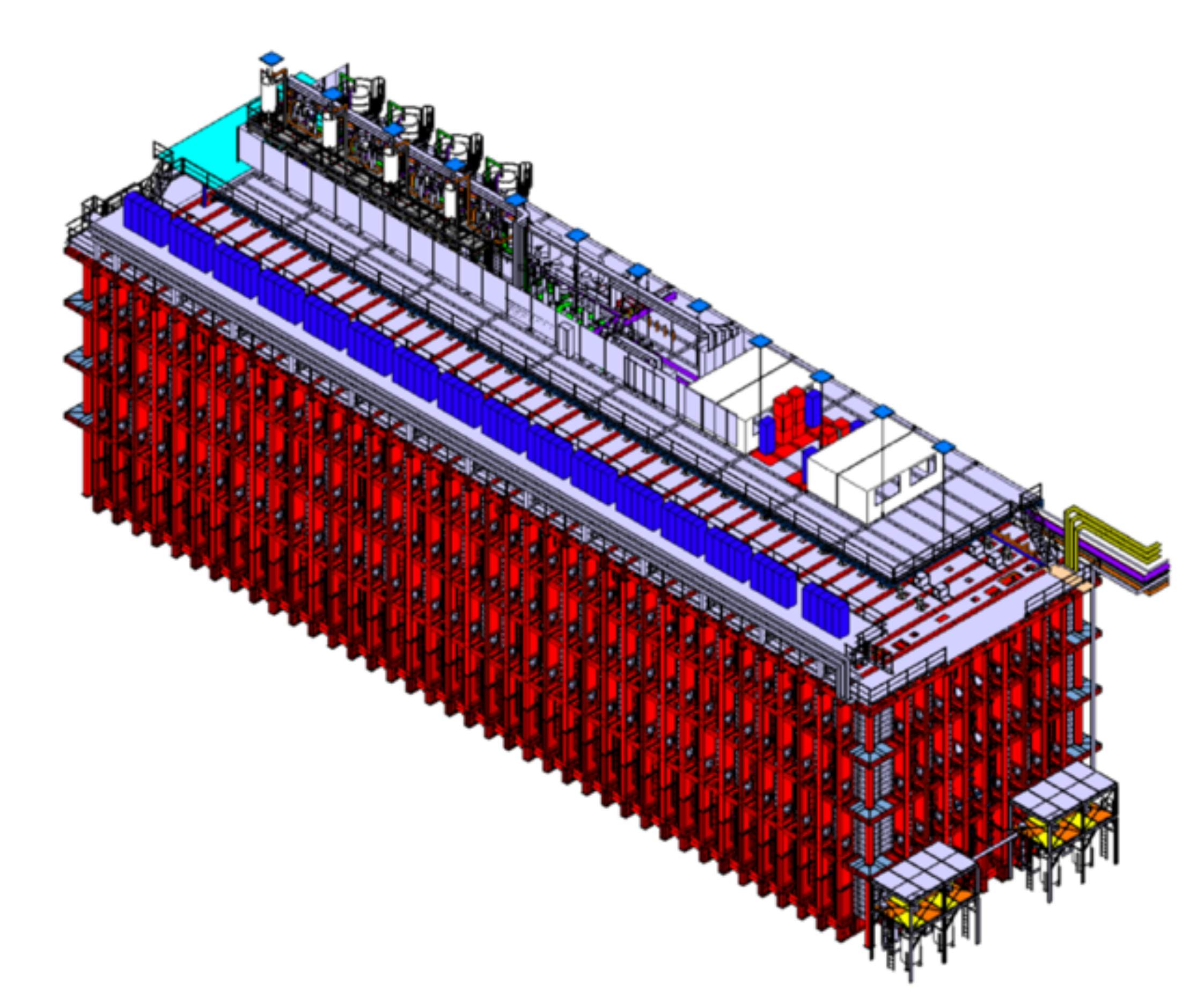}
\end{dunefigure}
Each cryostat consists of a warm structure that provides structural
support, insulating layers that maintain the temperature of the fluid
inside, and the inner membrane that provides a compliant inner surface
that helps to maintain the \dword{lar} purity and serves as primary
containment.

The external warm structure measures \SI{65.84}{\meter} long,
\SI{18.94}{\meter} wide, and \SI{17.84}{\meter} high, and its internal dimensions are 
\SI{63.6}{\meter} long, \SI{16.7}{\meter} wide, and \SI{15.6}{\meter}
high.  It is a welded and bolted structure constructed of
I-beam elements and a \SI{12}{\mm} inner steel plate reinforced with
ribs.  It is designed to support the
weight of the structure itself, 
insulation, inner membrane, liquid, detector, and any equipment placed
on the structure.  It also supports the hydrostatic pressure of the
fluid inside and resists the small gas overpressure in the ullage.
The structure is positioned on the concrete slab provided at the 4910L
of the north and south caverns.  The \SI{12}{\mm} inner steel plate of the
warm structure also serves as the tertiary layer of containment.

The insulation layer and inner membrane will be installed inside the
warm structure.  The technology for this comes from the shipping
vessels used for transporting liquid natural gas. The insulating
layers are made from prefabricated panels of reinforced polyurethane
foam.  There are two layers of the foam panels that have a flexible
membrane in between that serves as a secondary containment membrane.
The two foam layers are installed in an overlapping pattern to reduce
heat introduced into the fluid.  The total thickness of the two foam
layers is \SI{0.8}{\meter} with a density of approximately
\SI{90}{mg/cm$^3$} and a residual heat input of
\SI{6.3}{W/\meter$^2$}.  A stainless steel corrugated membrane will be
installed inside the insulating layers to create the primary
containment membrane for the \dword{lar}.  This is constructed from \SI{1.2}{\mm}
thick stainless steel panels that overlap and are welded together.
The entire inner surface of the cryostat will be tiled with these
panels to create the inner surface.  These panels need to be
corrugated to accommodate the shrinkage of the stainless steel from
ambient temperature to \dword{lar} temperature.  With the addition of the
insulation and inner membrane, the internal dimensions of the finished
vessel are \SI{62}{\meter} long, \SI{15.1}{\meter} wide, and
\SI{14}{\meter} high.

The top of the cryostat will have penetrations provided based on
drawings developed by LBNF and \dword{dune} to accommodate detector support,
electronic and data cables, cryogenic pipes, and connections and other
devices.

Each cryostat will have a vertical \dword{tco} on one short end.
These openings will be approximately \SI{13.43}{\meter} high and
\SI{2.68}{\meter} wide and will be used to move the detector elements
into the vessel during installation.  Once most of the detector
material is inside the cryostat, the \dword{tco} will be closed and
leak tested.  After the \dword{tco} closing, the last of the detector
installation will be completed. After final detector installation and
equipment removal through the roof openings, the cryostat will be
closed for purge, \cooldown{}, and filling.

At the \dword{tco} end of the cryostat, there will be four penetrations for
the cryogenic fluid pumps to be connected.  A normally closed valve
will be installed at each penetration to prevent any loss of fluid
with pump maintenance or damage.

\section{Cryogenics}
\label{sec:fdsp-coord-cryogenics}

The detector cryogenics system supplies \dword{lar} and provides
circulation, re-condensation, and purification. The cryogenic system
components are housed inside the \dword{cuc}, on top of each detector module
and between the two detector modules in each cavern. The cryogenics system comprises 
\begin{itemize}
\item {\bf Infrastructure Cryogenics}: This includes \dword{lar} and \dword{ln} receiving
  facilities on the surface, nitrogen refrigeration systems (both
  above ground and underground), \dword{ln} buffer storage
  underground, piping to interconnect equipment (\dword{ln}, GN$_2$, and \dword{gar}),
  components in the detector cavern, and the \dword{cuc} and process control/support
  equipment.
\item {\bf Proximity cryogenics}: This includes reliquefaction 
  and purification subsystems for the argon (both gas and liquid), associated
  instrumentation and monitoring equipment and \dword{lar} piping to
  interconnect equipment and components in the detector cavern and the
  \dword{cuc}. The proximity cryogenics are split into three areas: in the
  \dword{cuc}; on top of the mezzanine, as shown in Figure~\ref{fig:detector_mezzanines};
  and on the side of the cryostat where \dword{lar} circulation pumps are installed.
\item {\bf Internal cryogenics}: This includes \dword{lar} and \dword{gar} distribution
  systems inside the cryostat, as well as features to cool the
  cryostat and the detector uniformly.
\end{itemize}

Figure~\ref{fig:dune-cryogenics} shows the process flow diagram of the
\dword{lbnf} cryogenic system. For convenience, only one cryostat is shown. The three main areas are 
\begin{enumerate}
  \item Surface (on the left), with the receiving facilities and the
    recycle compressors of the nitrogen system.  All these items are
    part of the infrastructure cryogenics.
  \item
  \dword{cuc} (in the center), with the \dword{lar}/\dword{gar}
    purification and regeneration systems (part of the proximity
    cryogenics) and the \coldbox{}es, expanders, and \dword{ln} storage (part of
    the infrastructure cryogenics).
\item Detector cavern (on the right), with the argon condensing system
  and the distribution of argon from the purification system in the
  \dword{cuc} to the cryostat and vice versa.
\end{enumerate}
Argon and nitrogen are received and stored on the surface in the
liquid phase.  They are vaporized and transferred underground in the
gas phase, the nitrogen as part of the nitrogen system, the argon separately.
\begin{dunefigure}[Cryogenics system]{fig:dune-cryogenics}
  {Overall process flow diagram of the cryogenic system showing one
    cryostat only; other cryostats are the same.}
  \includegraphics[width=0.85\textwidth]{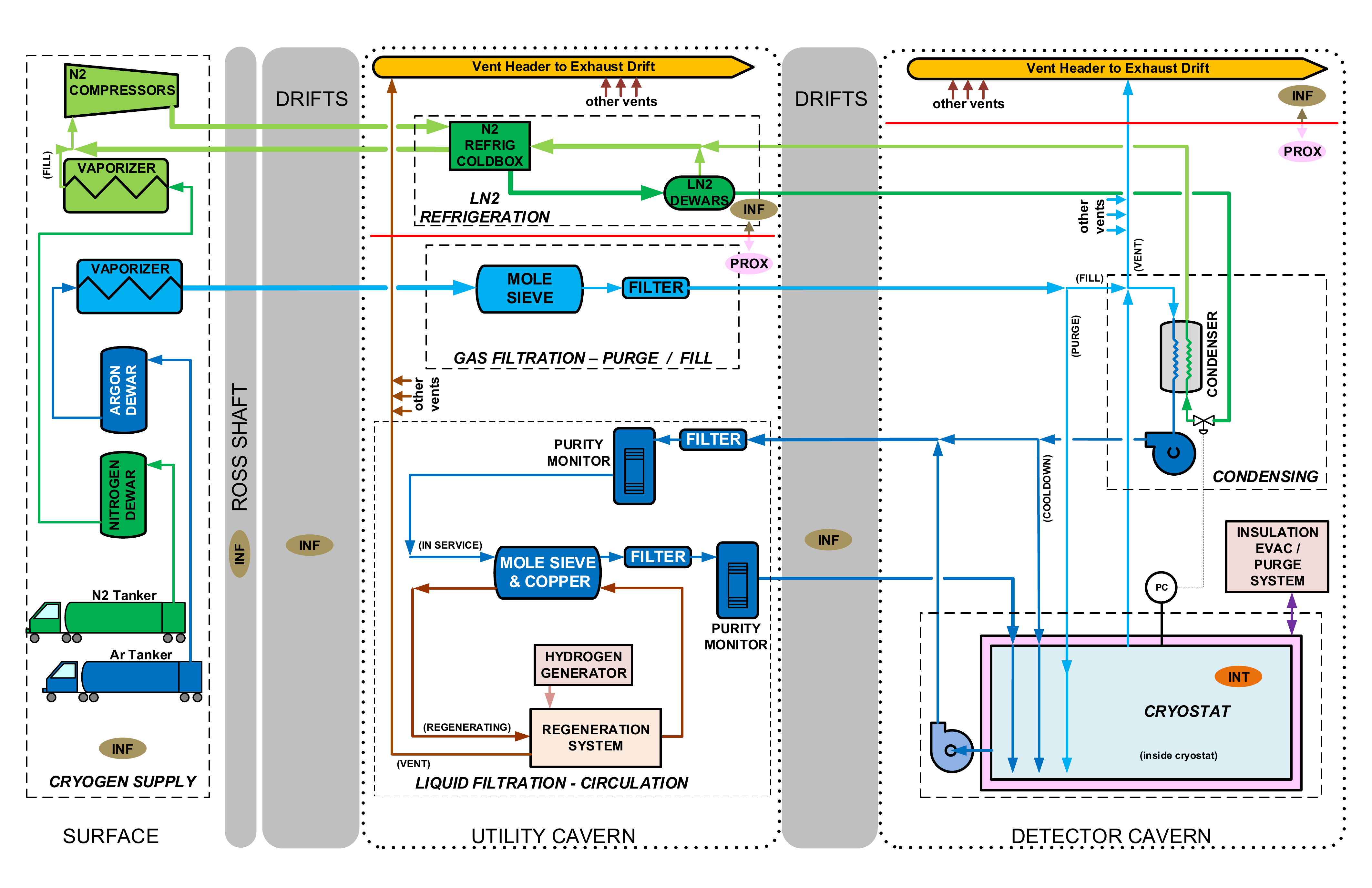}
\end{dunefigure}

The cryogenics system fulfills the following modes of operations:
\begin{itemize}
  \item {\bf Gaseous argon purge}: Initially, each cryostat is filled
    with air, which must be removed by means of a slow \dword{gar} piston
    purge.  A slow flow of argon is introduced from the bottom and
    displaces the air by pushing it to the top of the cryostat where
    it is vented.  Once the impurities, primarily nitrogen, oxygen, and 
    water, drop below the parts per million (ppm) level, the argon
    exhausted at the top of the cryostat is circulated in closed loop
    through the gas purification system and re-injected at the
    bottom. Once the contaminants drop below the ppm level, the
    cool-down can commence. The \dword{gar} for the purge comes directly from
    above ground, passes through the \dword{gar} purification and then it is
    injected at the bottom of the cryostat by means of a \dword{gar}
    distribution system.
  \item {\bf Cryostat and detector cool-down}: The detector elements
    must be cooled in a controlled and uniform manner. Purified \dword{lar}
    flows into sprayers that atomize it. A second set of sprayers
    flowing purified \dword{gar} moves the mist of argon around to achieve a
    uniform cooling. The cooling power required to recondense argon in
    the condensers outside the cryostat is supplied by the
    vaporization of nitrogen from the nitrogen system. Once the
    detector elements reach about 90 K, the filling can commence. The
    \dword{gar} for the cool-down comes directly from above ground, passes
    through the \dword{gar} purification,  then is condensed in the
    condensers before being injected into the sprayers. The \dword{gar} moving
    the mist of argon around only goes through the \dword{gar} purification
    and not the condensers.
  \item {\bf Cryostat filling}:Argon is vaporized and transferred
    underground as a gas from the receiving facilities on the surface.
    It first flows through the \dword{gar} purification system and is
    recondensed in the argon condensers by means of vaporization of
    \dword{ln}.  It then flows through the \dword{lar} purification and is introduced
    in the cryostat. The filling of each cryostat varies in duration,
    from 8 to 15 months, depending on the available cooling power at
    each stage. With the full refrigeration system available, the filling of the fourth
cryostat  will take approximately 15 months. 
  \item{\bf Steady state operations}: The \dword{lar} contained inside each
    cryostat is continuously purified through the \dword{lar} purification
    system using the main external \dword{lar} circulation pumps. The boil-off
    \dword{gar} is recondensed in the argon condensers and purified as liquid
    in the same \dword{lar} purification system as the bulk of the \dword{lar}.
  \item{\bf Cryostat emptying}: At the conclusion of the experiment,
    each cryostat is emptied and the \dword{lar} is removed from the system.
\end{itemize}
Each detector has its own stand-alone process controls system, which is
redundant and independent of the others. It resides locally in each
cavern.  PLC racks are located on the mezzanine, in the pit over the
protective structure of the main \dword{lar} circulation pump, in the \dword{cuc} and
on the surface. 
A workspace on the mezzanine and a desk with two stations in the \dword{cuc}  are available
during installation and commissioning.

Before argon is offloaded from each truck into the receiving tanks, a
sample of the \dword{lar} is analyzed locally to ensure compliance with the
requirements. If the specifications are met, the truck driver is given permission
 to offload the truck. The process is automated to reduce
human error. The purity is measured before and after the purification
system by custom-made purity monitors to verify correct
functioning of the system.

\section{Detector and Cavern Integration}
\label{sec:fdsp-coord-det-cav-integ}

Figure~\ref{fig:detector_ew_elevation} shows the north
elevation view of the detector in the cavern. The services from the
\dword{cuc} enter the cavern through a passage visible on the left.
\begin{dunefigure}[North elevation view of detector]{fig:detector_ew_elevation}
  {North elevation view of one detector module in the cavern.}
  \includegraphics[width=0.75\textwidth]{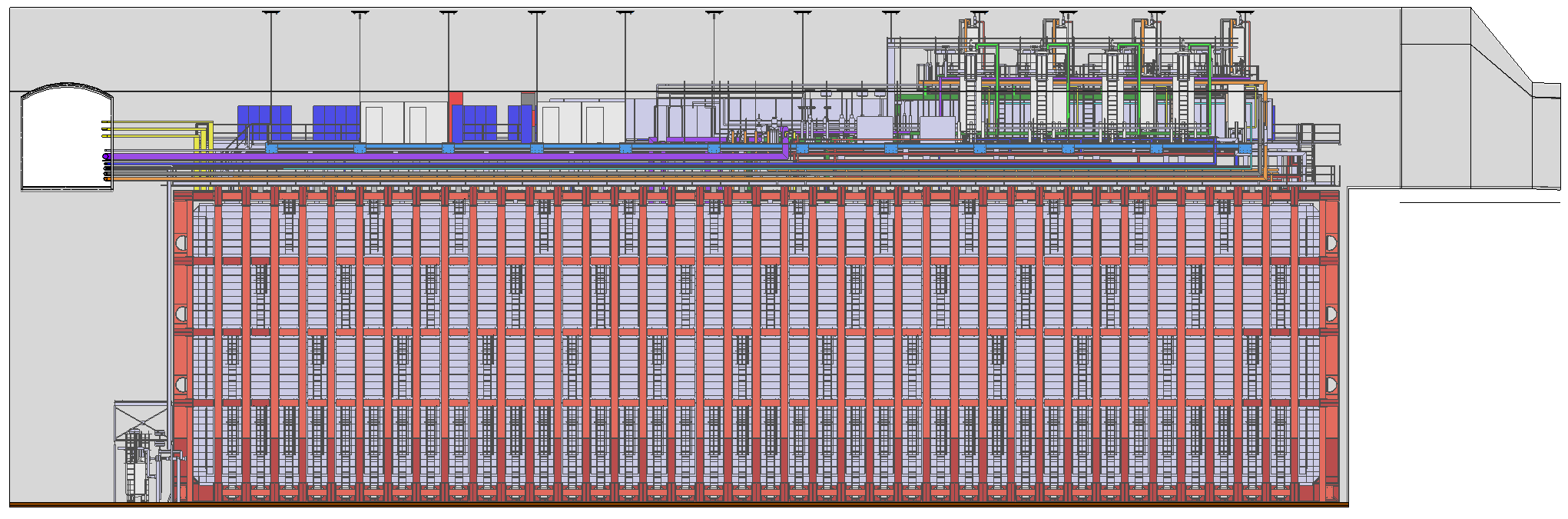}
\end{dunefigure}

Figure~\ref{fig:dune-cryostat} shows one detector module in the
cavern. In this figure, the cryogenics equipment and racks on top of
the detector are visible. The \dword{lar} recirculation pumps can also be seen
on the lower level.
Figure~\ref{fig:detector_ns_elevation} shows the west
elevation view of the detector in the cavern. The services entering
from the \dword{cuc} are visible on the right.
\begin{dunefigure}[West elevation view of detector]{fig:detector_ns_elevation}
  {West elevation view of one detector in the cavern with overall dimensions.}
  \includegraphics[width=0.9\textwidth]{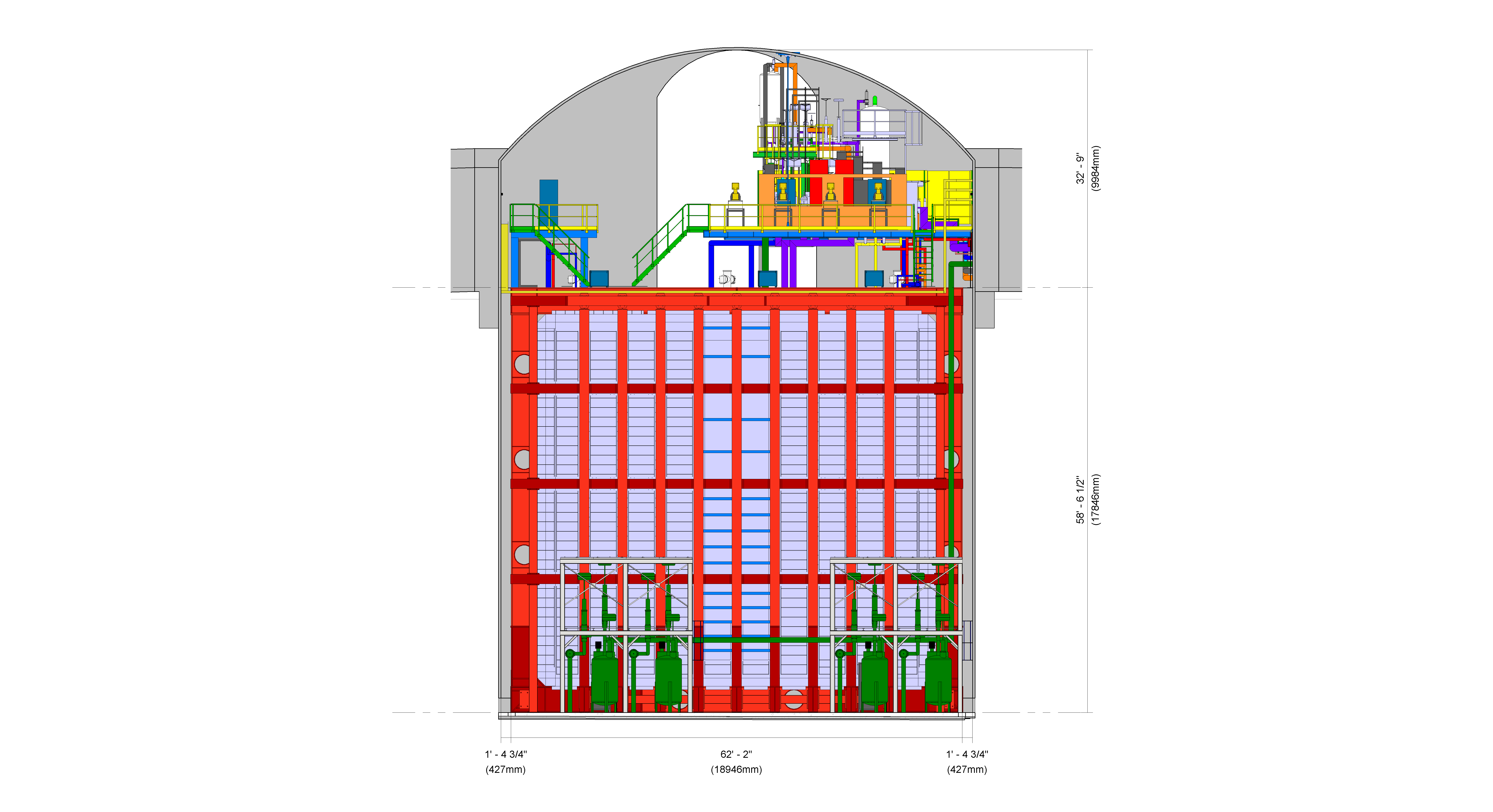}
\end{dunefigure}
Figure~\ref{fig:detector_mezzanines} shows the elevation view of the
top of cryostat showing mezzanines, cryogenic equipment, and
electronic racks.
The cryogenics are installed on a mezzanine supported from
the cavern roof and cavern wall. Cryogenic distribution lines are
routed under the mezzanine. Local control rooms for the
cryogenic equipment are on the mezzanine.
\begin{dunefigure}[Elevation view of top of cryostat]{fig:detector_mezzanines}
  {Elevation view of top of cryostat showing mezzanines, cryogenics
    equipment, and electronic racks.}
  \includegraphics[width=0.85\textwidth]{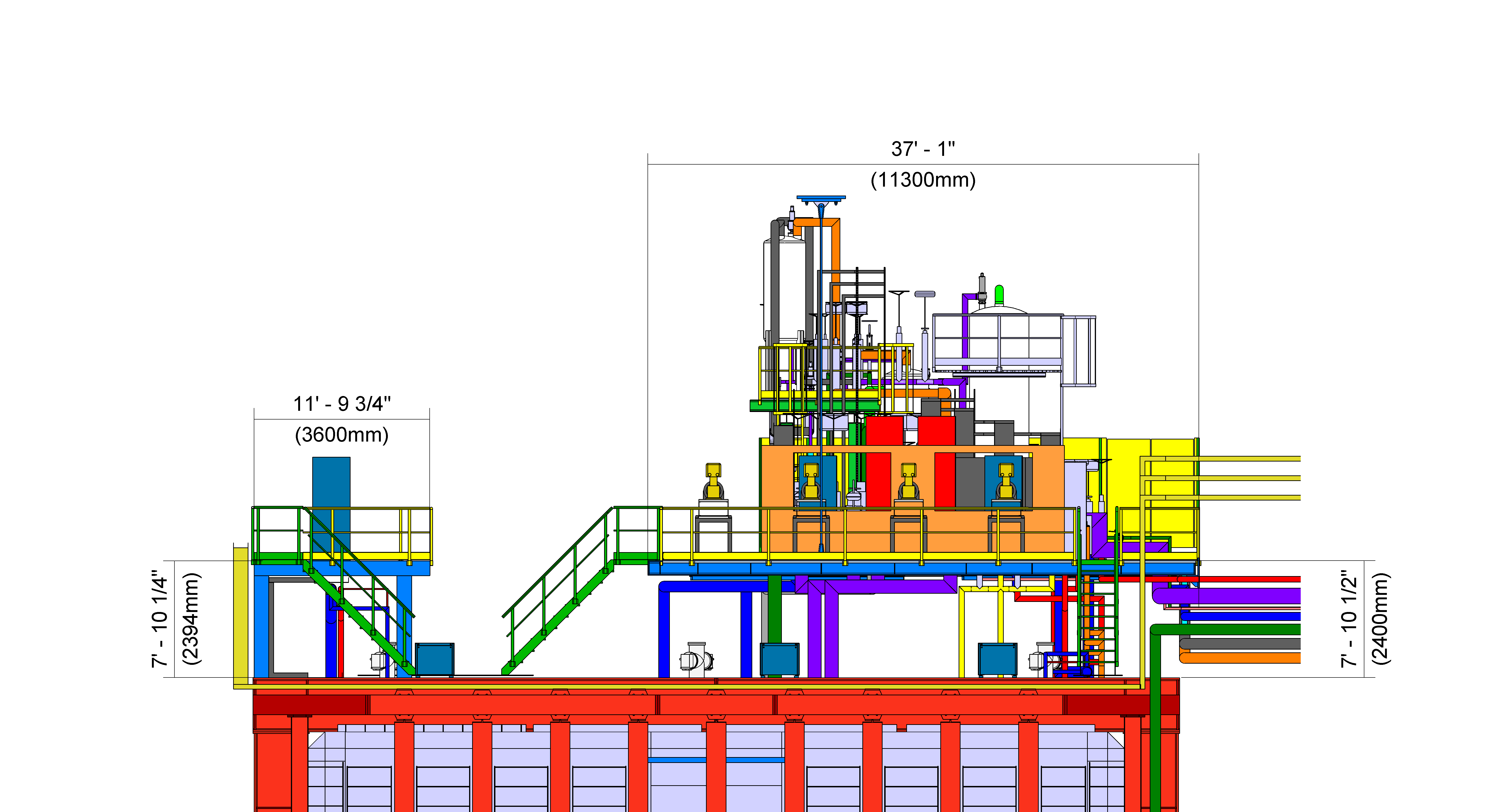}
\end{dunefigure}

Detector electronics are installed in short racks close to
feedthroughs and in taller racks installed on a separate electronics
mezzanine shown on the left of Figure~\ref{fig:detector_mezzanines}.
This will allow easy access for maintenance and reduce complexity on
top of the detector.

\section{Detector Grounding}
\label{sec:fdsp-coord-faci-grounding}

The grounding strategy provides the detectors with independent
isolated grounds to minimize any environmental electrical noise that
could couple into detector readout electronics either conductively or
through emitted electromagnetic interference.

The detectors will be placed at the 4910L of \surf. The
electrical characteristics of the various rock masses are unknown, but
should have extremely poor and inconsistent conductive
properties. Ensuring adequate sensitivity of the detectors requires a
special ground system that will isolate the detectors from all other
electrical systems and equipment, minimize the influence of inductive
and capacitive coupling, and eliminate ground loops. The grounding
infrastructure should reduce or eliminate ground currents through the
detector that would affect detector sensitivity, maintain a low
impedance current path for equipment short circuit and ground fault
currents, and ensure personnel safety by limiting any potential for
equipment-to-equipment and equipment-to-ground contact.

The infrastructure grounding plan of the underground facilities is
fully described elsewhere~\cite{bib:cernedms2095975}. 
We have planned a separate detector ground, isolated from the rest of the facility, for each of the four \dwords{detmodule}.    
The detector ground will primarily 
comprise the steel containment vessel, cryostat membrane, and
connected readout electronics.  The facility ground is constructed out
of two interconnected grounding structures; these are the cavern
ground and the \dword{ufer} grounds which are described below.  For safety
reasons, a saturable inductor will connect the detector ground to the
facility ground.

Figure~\ref{fig:dune-grounding} shows the areas of construction for
the cavern and \dword{ufer} grounds.
\begin{dunefigure}[Overall \dshort{dune} grounding structure]{fig:dune-grounding}
  {Overall \dshort{dune} facility grounding structure incorporated in cavern.}
  \includegraphics[width=0.85\textwidth]{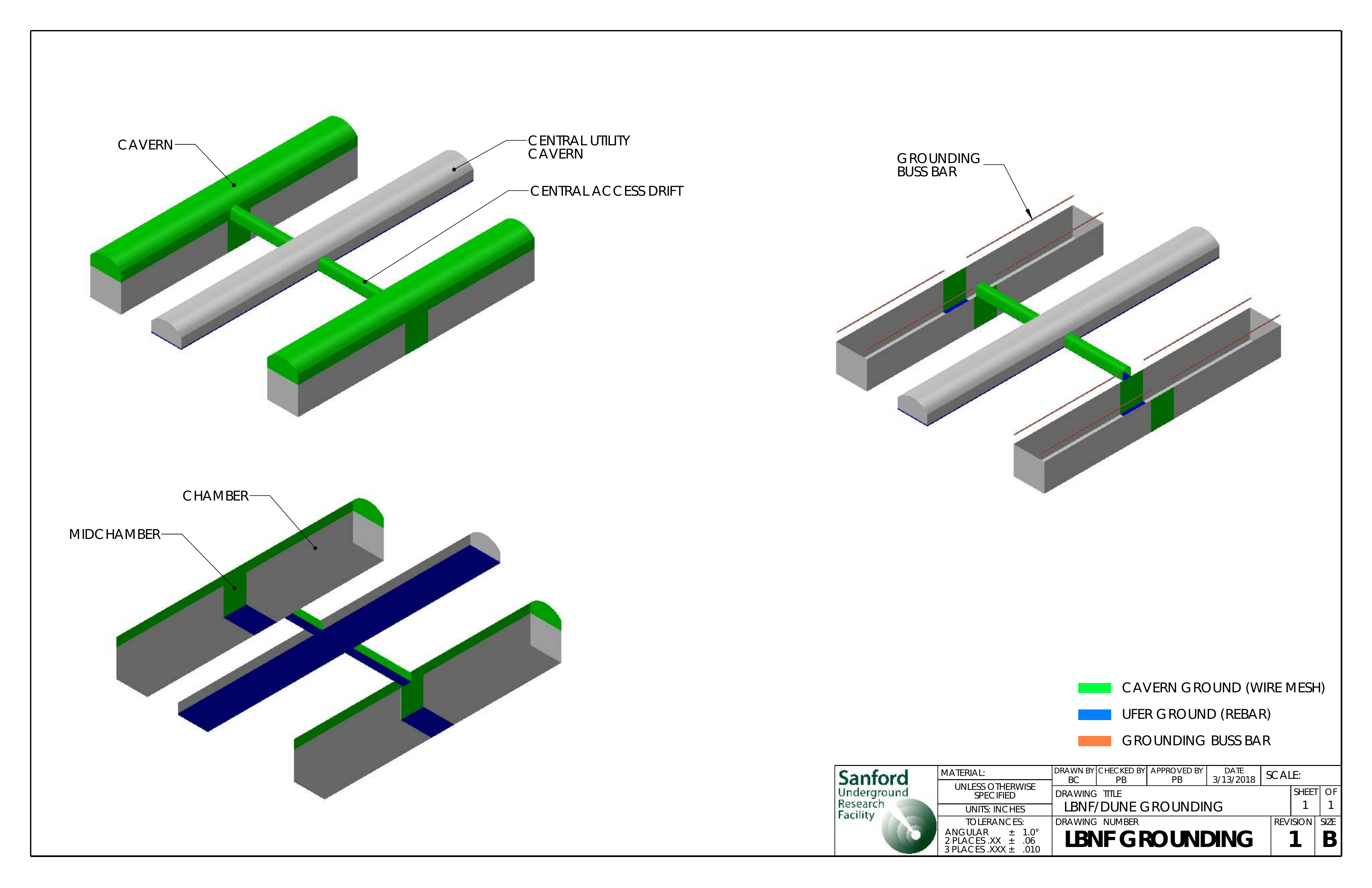}
\end{dunefigure}
Grounding structure definitions include 
\begin{enumerate}
 \item Cavern ground consisting of overlapping welded wire mesh
   supported by rock bolts and covered with shotcrete. The
   \dword{lbnf}/\dword{dune} cavern ground includes all walls and
   crown areas above the \dword{4850l} in the north and south detector
   caverns and their associated central access drifts, as well as tin-plated
   copper bus bars that run the length of the detector vessels
   on each side along the cavern walls and are mounted external to the
   shotcrete.  The cavern ground structure
\begin{enumerate}
 \item spans the full length of the cavern from the west end access
   drift entrance through the mid-chamber to the east end access drift
   entrance;
 \item spans the full width of the cavern from the \dword{4850l} sill
   (top of the detector vessels and mid-chamber floor) on both sides
   up and across the crown of the cavern;
 \item includes mid chamber walls to the 4910L; and 
 \item includes the east and west end walls of the cavern, from the
   \dword{4850l} to the crown.
\end{enumerate}
 \item \dword{ufer} ground consisting of metal rebar embedded in
   concrete floors. The \dword{lbnf}/\dword{dune} \dword{ufer} ground
   system includes the concrete floors in the cavern mid-chambers,
   center access drifts, and \dword{cuc}. The cavern and
   \dword{ufer} grounds will be well bonded electrically to construct
   a single facility ground isolated from detector ground.
 \item Detector ground consisting of the steel containment vessel
   enclosing the cryostat and all metal structures attached to or
   supported by the detector vessel.
\end{enumerate}

To ensure safety, a safety ground with one or more saturable inductors
will be installed between the detector ground and the electrically
bonded \dword{ufer} and cavern grounds that form the facility ground.
Figure~\ref{fig:dune-grounding_figure} illustrates the use of the
safety ground. The safety ground inductors saturate with flux under
low-frequency high currents, presenting minimal impedance to these
currents.  Thus, an AC power fault current would be shunted to the
facility ground and provide a safe grounding design. At higher
frequencies and lower currents, such as coupled noise currents, the
inductor provides high impedance, restricting current flow
between grounded metal structures. The desired total impedance between
the detector ground structure and the cavern/\dword{ufer} ground
structure should be a minimum of \SI{10}{Ohms} at \SI{10}{MHz}.
\begin{dunefigure}[Simplified detector grounding]{fig:dune-grounding_figure}
  {Simplified detector grounding scheme.}
  \includegraphics[width=0.5\textwidth]{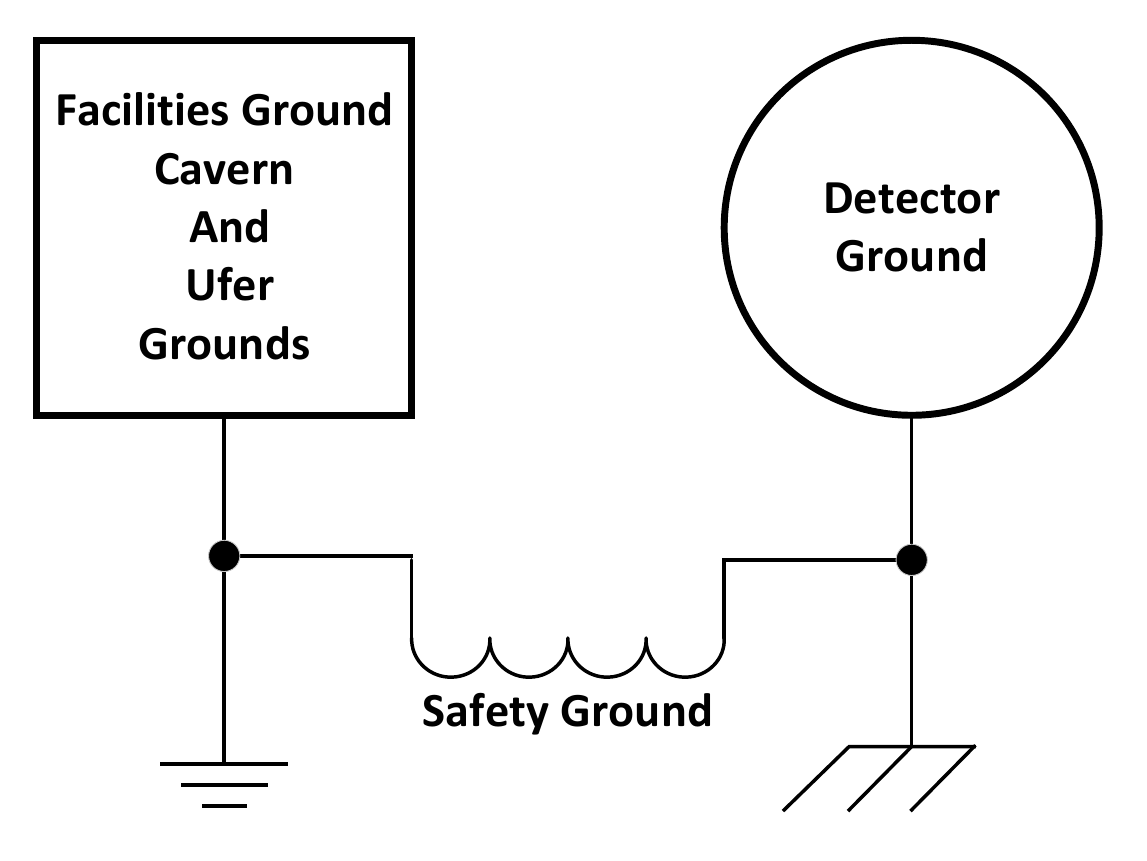}
\end{dunefigure}

As stated above, the detector ground exists only in the area of the
steel containment vessel enclosing the cryostat and all metal
structures attached to or supported by the detector vessel.  All
signal cables that run between the detector and the \dword{daq} underground
processing room in the \dword{cuc} will be fiber optic.
All connections to the cryogenics plant on the facility
ground will be isolated from the cryostat with dielectric breaks.  A
conceptual drawing showing the isolation of the cryostat is presented
in Figure~\ref{fig:dune-grounding_scheme}.
\begin{dunefigure}[Detector grounding schematic]{fig:dune-grounding_scheme}
  {Schematic of detector grounding system.}
  \includegraphics[width=0.99\textheight,angle=90]{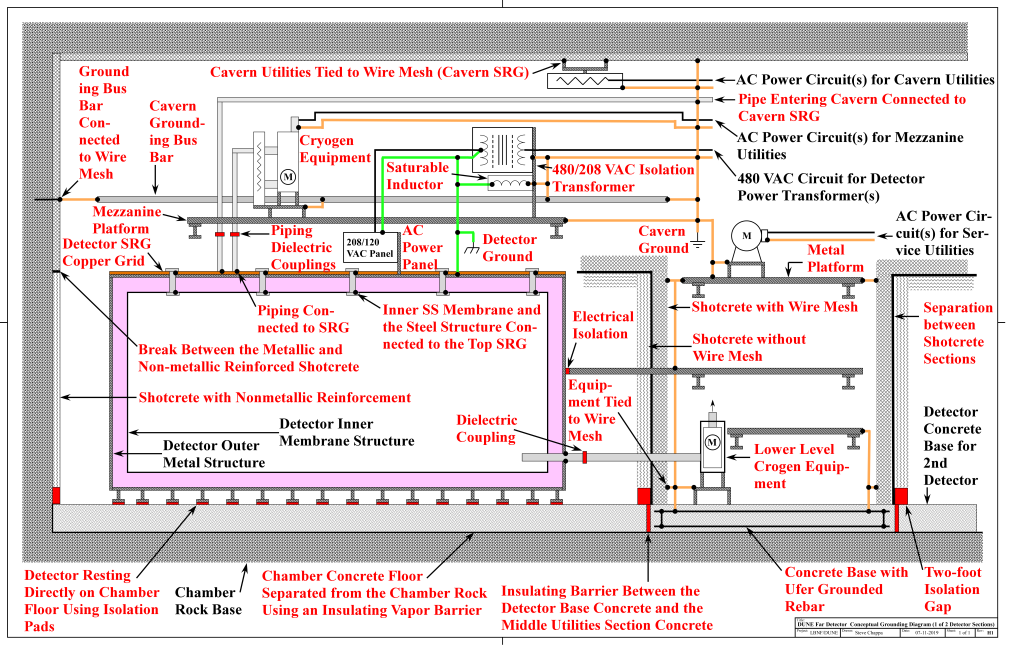}
\end{dunefigure}

The construction of the facility ground provides a low impedance path
for return currents of the facility services, such as cryogenic pumps,
and noise coupling from facility services will be greatly reduced or
eliminated.  The experiment has also been carefully designed such that
facility return currents will not flow under the cryostats.

The cryostat itself is treated as a Faraday cage.  Any connections
coming from facilities outside of actual detector electronics are
electrically isolated from the cryostat.  For detector electronics,
specific rules for signal and power cables penetrating the cryostat
exist~\cite{bib:cernedms2095958}. 

\section{Detector Power}
\label{sec:fdsp-coord-faci-power}

After requirements were given to \dword{fscf}, 
limits were established on the size of the cavern excavation, cooling
capabilities, and electronics power consumption.   
The  \dword{dune} \dwords{detmodule} must stay within these limits. Of the available 360~kW per module, the \dword{spmod} (\dword{dpmod}) will only use
216(253)~kW,  
leaving a margin of 40\%(30\%).

The \dword{fscf} will supply a 1000~kVA transformer for
each cavern.  Each cavern will host two \dword{dune}  \dwords{detmodule}.  Power from
this initial transformer will be de-rated with no more than 75\%
of total power available at the electrical distribution panels.  We
plan for a maximum consumption of 360~kW per  \dword{detmodule}.

Figure~\ref{fig:power} summarizes estimated detector loads of the 
\dword{dune} electronics located in the detector caverns.  The \dword{daq} power is 
described at the end of this section. 
\begin{dunefigure}[\dshort{dune} estimated power consumption.]
{fig:power}
{\dword{dune} estimated power consumption.}
  \includegraphics[width=0.8\textwidth]{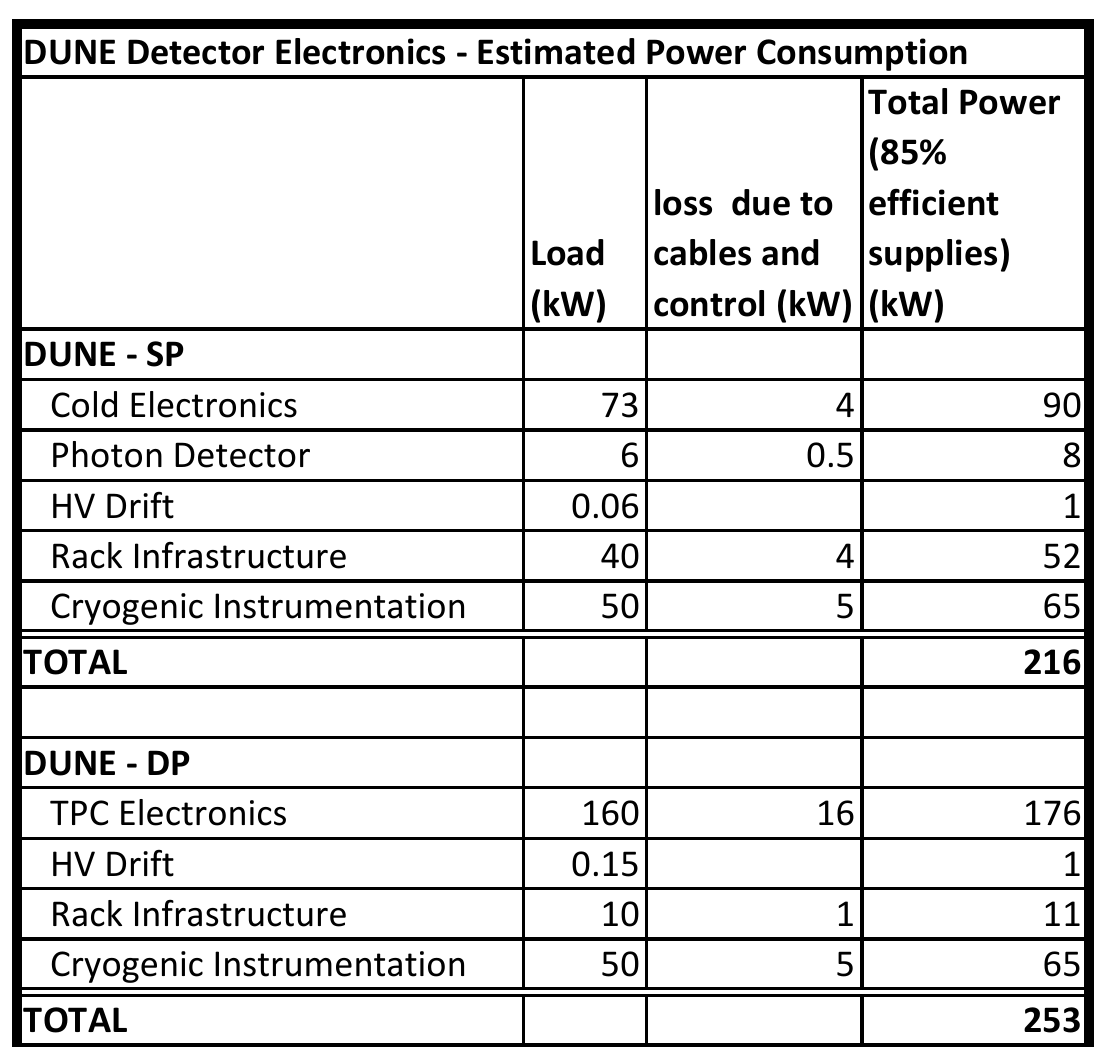}
\end{dunefigure}

The TPC electronics readout dissipates an estimated maximum of \SI{360}{W}
per \dword{apa}. An additional \SI{20}{W} will be consumed due to power loss in the
cables.  The low-voltage power supplies have a controller that adds
approximately \SI{35}{W} per \dword{apa}, and supplies have an efficiency of
approximately 85\%. This adds up to about \SI{488}{W} per \dword{apa}, or a
total load of  \SI{73}{kW} per detector module. The \dword{apa} wire-bias power
supplies have a maximum load of \SI{465}{W} per set of six \dword{apa}s, for a total
budget of around \SI{12}{kW}. Cooling fans and heaters near the feedthroughs
will use a 
small amount of power, so the overall power budget for
TPC electronics is expected to be less than \SI{90}{kW}.

The \dword{pds} electronics is based on the \dword{mu2e}
    (\dword{daphne}) electronics, from which we estimate a total power
    budget of approximately 6~kW. \dword{dune} plans a slightly higher
    power budget of 8~kW to account for cable and power supply
    inefficiencies.  The \dword{pds} electronics presents a significantly
    lower power load than the \dword{ssp} alternative solution used in 
    \dword{pdsp}, which requires a power budget of approximately 72~kW per 
    detector module.

Each of the approximately 80 detector racks will have fan units,
Ethernet switches, rack protection, and slow controls modules, adding
a load of about 500~W per rack, bringing the total to 40~kW.

Twenty-five racks are reserved for cryogenics instrumentation with a
per-rack load conservatively estimated at 2~kW, for a total of 50~kW.

The SP detector module will thus use an estimated 216~kW of power. The
higher-load SSP alternative for the PDS would increase this to
280~kW. This higher estimate represents approximately 60--78\% of
our available power.

The \dword{dp} electronics estimate is
approximately 253~kW and also fits well within the planned maximum of
360~kW.

For the \dword{spmod}, the power will be largely distributed to a number
of detector racks that will sit on a detector rack mezzanine above
the cryostat.  Each of the racks will receive a \SI{30}{A} \SI{120}{V} 
service, with a maximum of 80 racks.

The other area where \dword{dune} requires power underground is in the \dword{daq}
room.  There the power budget is determined by the available 750~kVA
transformer.  The available power must be de-rated to 80\% at
the electrical distribution panel and another 80\% for equipment
efficiency.  Thus, 480~kW of power will be distributed to a maximum of
60 racks.  Each water-cooled rack will have approximately 8~kW
available for computing power.

A minimal level of UPS power will be provided to the \dword{daq} equipment to
allow for powering down servers.  Cryogenic controls and any
critical safety interlocks will have access to long term UPS back-up.
At this time, no UPS power is being proposed for detector readout
electronics.

\section{Data Fibers}
\label{sec:fdsp-coord-faci-fibers}

The \dword{dune} experiment requires a number of fiber optic pairs to
run between the surface and the \dword{4850l}.  A total of 96 fiber
pairs, which accommodates both \dword{dune} and \dword{lbnf} needs, will
be supplied through redundant paths with bundles of 96 pairs coming
down both the Ross and Yates shafts.  The individual fibers are
specified to allow for transmission of 100 Gbps.  A schematic view of
the fiber paths from surface to underground is shown in
Figure~\ref{fig:dune-fiber_path}. From the surface main communications room (MCR, the surface \dword{daq}
room), we will connect to the WAN and ESnet as described in
Section~\ref{sec:fdsp-coord-surf-rooms}, which is being designed by
\dword{fscf} and the \dword{fnal} networking groups.

\begin{dunefigure}[Detector fiber path]{fig:dune-fiber_path}
  {Schematic of detector fiber path.}
  \includegraphics[width=0.95\textwidth]{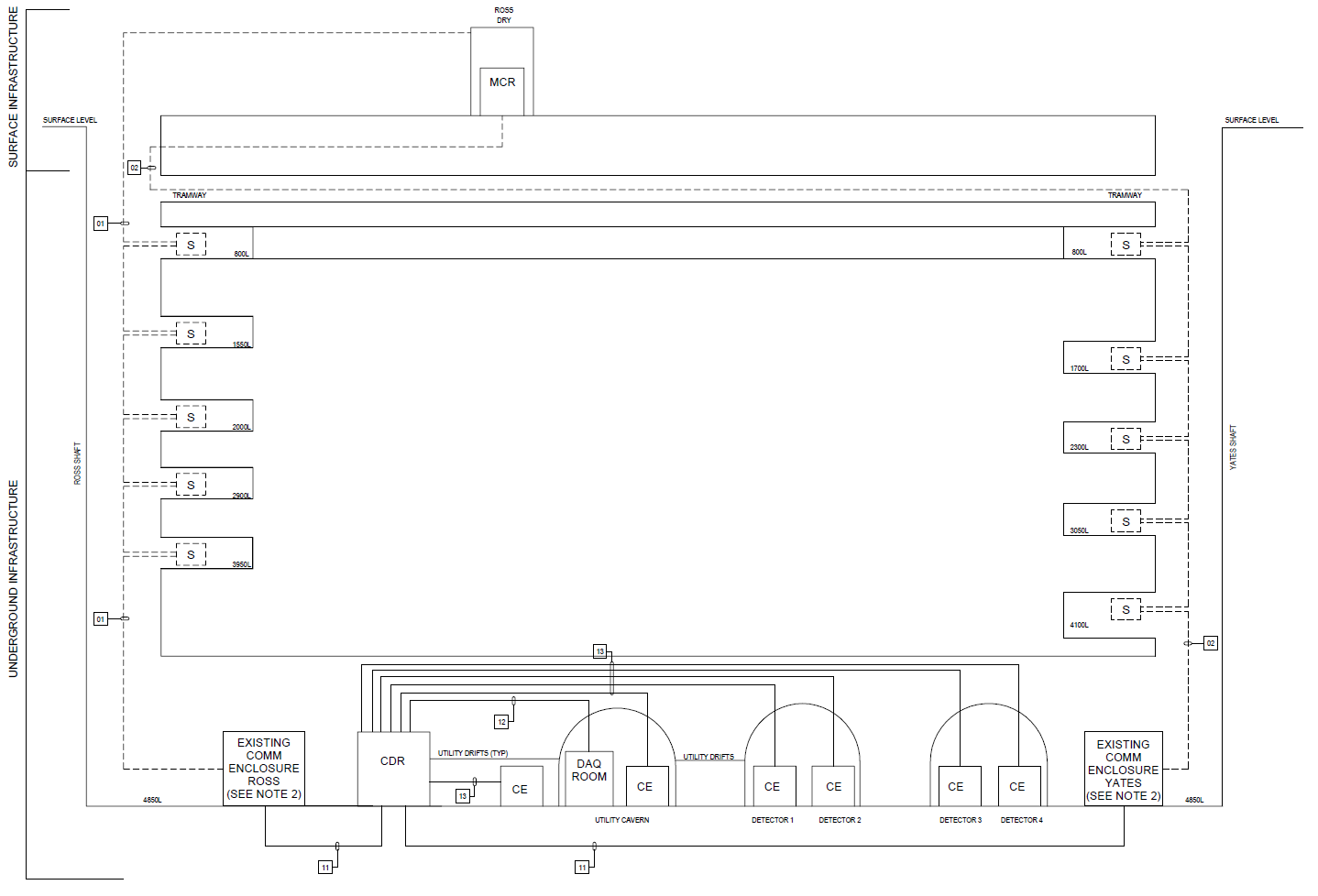}
\end{dunefigure}

The redundant fiber cable runs of 96 fiber pairs are received in
existing communications enclosures at the \dword{4850l} entrances of the Ross
and Yates shafts.  The two 96 fiber pair bundles are next routed to
the CDR.  The fibers will be received and terminated in an optical
fiber rack. Two additional network racks will be used to route the
fiber data between the surface and underground.

The plan is to use only the Ross Shaft set of 96 fiber pairs with the
Yates set being redundant.  The network switches will allow for
switchover in case of catastrophic failure of fibers in the Ross
shaft. 
There is a very low risk of catastrophic failure, but it could occur
if, for instance, rock fell and damaged the fibers.  The Yates path is
viewed as a hot spare. Plans are being formed to periodically test
the redundant Yates path and verify its viability.

From the CDR, fibers designated for \dword{fscf} are routed to provide
general network connections to the detector caverns,  \dword{cuc}, and the underground \dword{daq} room.

A total of 96 fiber pairs are routed to the underground \dword{daq} room for
use by the \dword{dune} experiment and LBNF. The fibers are reserved as
follows:
\begin{itemize}
  \item 15 pairs for \dword{dune} data per detector-total 60 pairs,
\item 1 pair for slow controls per detector-total 4 pairs,
\item 2 pairs reserved for \dword{gps},
\item 6 pairs for \dword{fscf},
\item 4 pairs for \dword{lar} cryogenics,
\item 4 pairs for \dword{ln} cryogenics, and
  \item 16 pairs reserved as spares.
\end{itemize}
The set of reserved fiber pairs total to 80.

\section{Central Utility Cavern Control and \dshort{daq} Rooms}
\label{sec:fdsp-coord-cuc-daq}

The \dword{cuc} contains various cryogenic equipment and the
\dword{daq} and Control Room for the \dword{dune} experiment.  The
cryogenic system and areas are described in
Section~\ref{sec:fdsp-coord-cryogenics}. Both the control and \dword{daq}
rooms are at the west end of the \dword{cuc} (see
Figure~\ref{fig:dune-cuc}).
\begin{dunefigure}[\dshort{daq} and control room in CUC]{fig:dune-cuc}
  {Location of underground \dword{daq} and control room in the CUC.}
  \includegraphics[width=0.85\textwidth]{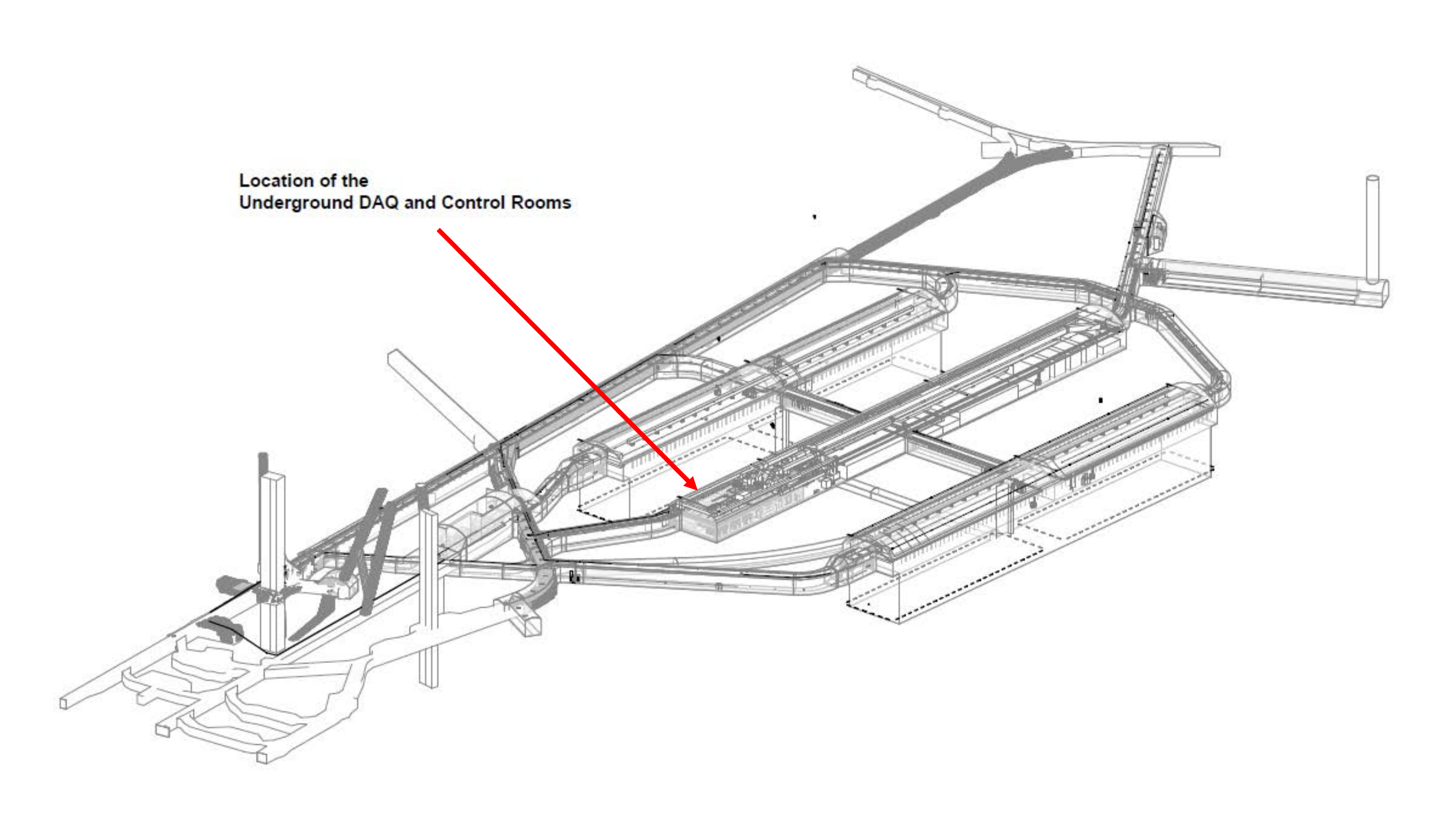}
\end{dunefigure}
\begin{dunefigure}[\dshort{daq} and control room]{fig:dune-daq}
  {Underground \dword{daq} and control room layout.}
  \includegraphics[width=0.85\textwidth]{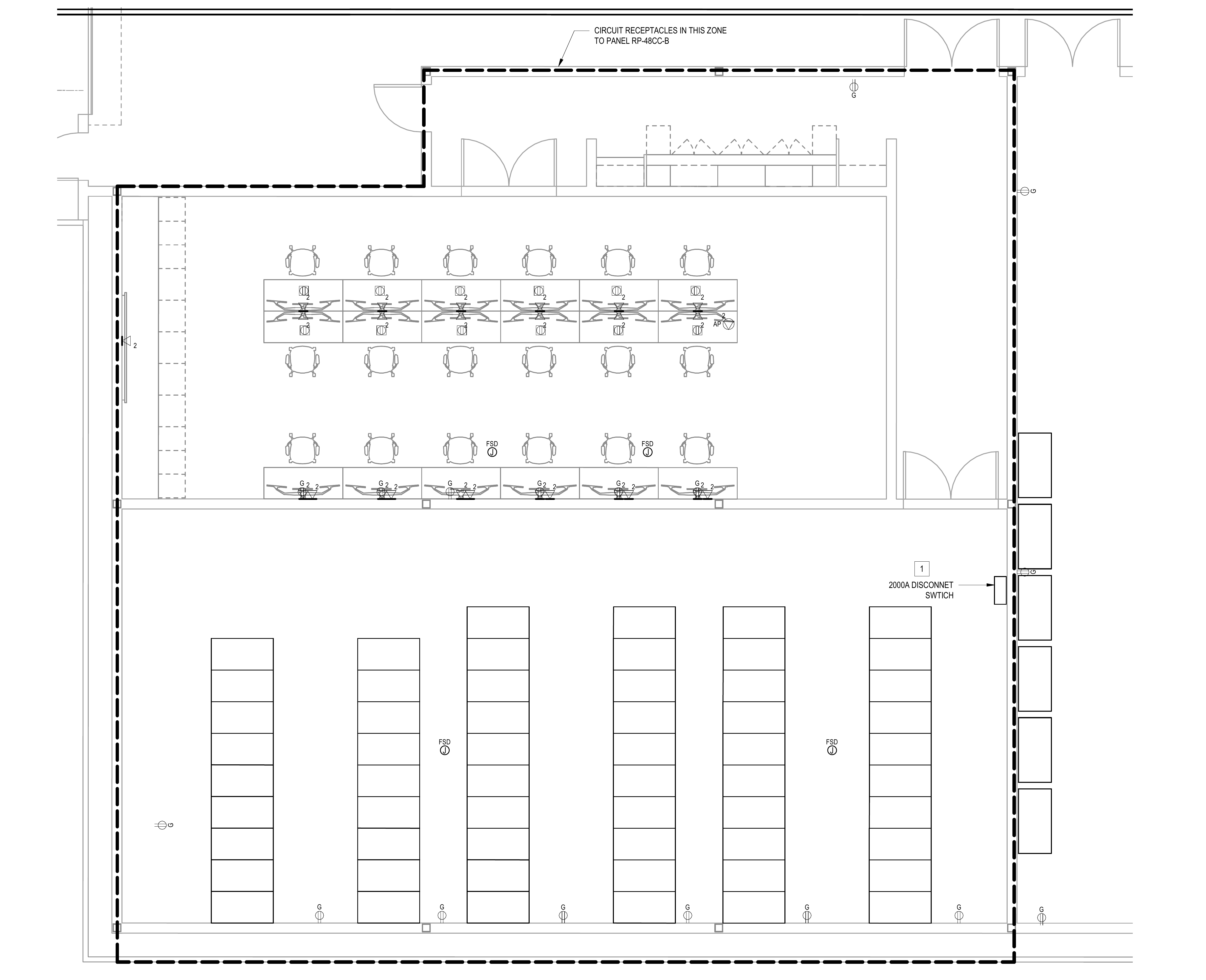}
\end{dunefigure}

The control room is an underground space of approximately 18 x 48 feet
and serves multiple purposes.  It provides a meeting or work space
during system commissioning. It provides easy access to \dword{daq}
equipment for debugging and service during commissioning.  During
experimental operations, the \dword{dune} experiment will have a remote
control room located at Fermilab's main campus.

Figure~\ref{fig:dune-daq} shows the layout and suggested outfitting of
the rooms. Additionally, the control room provides the required
workstation for monitoring of Fire and Life Safety and the building
management system.  These are facility services for which \dword{dune} is not
directly responsible, but the experiment will need to
interface with these systems.  One example of this interface would be
the reporting of any smoke detected within a detector rack.

Lastly, the cryogenic team requires a space allocation within the
Control Room of two racks and two work benches for the technicians who
monitor the cryogenic systems. This space is needed only during
commissioning with remote operation to follow. Additional space for
cryogenics commissioning may be available on the mezzanines.
       
The \dword{daq} room is approximately 26 $\times$ 56 square feet and will
contain 52--60 racks that will be used for fiber optic cable
distribution, networking, \dword{dune} \dword{daq} and two or three
racks for conventional facilities.  The current design, shown in
Figure~\ref{fig:dune-DAQ_layout}, shows a total of 60 racks 
possible.
\begin{dunefigure}[\dshort{daq} room layout]{fig:dune-DAQ_layout}
  {Proposed rack layout in \dword{daq} room.}
  \includegraphics[width=0.85\textwidth]{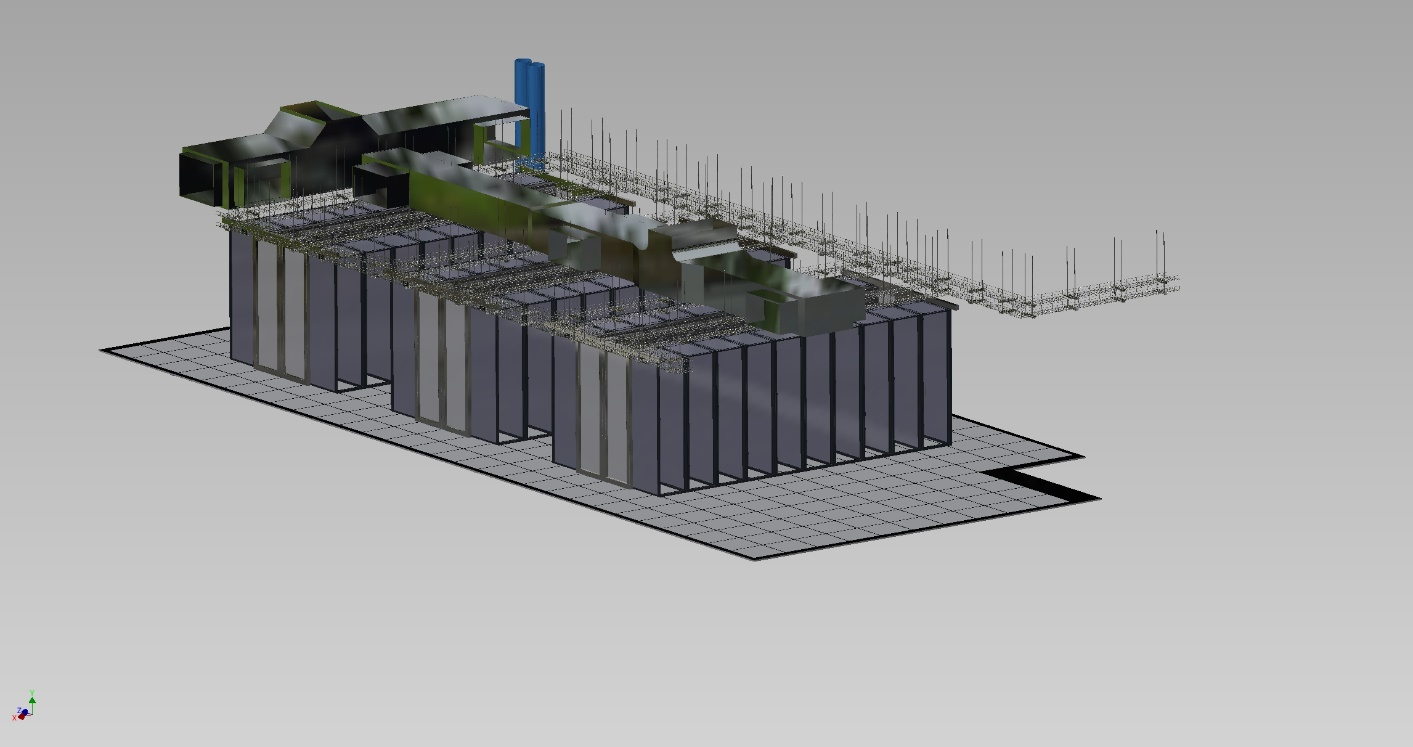}
\end{dunefigure}

A quantity of 48 racks are reserved exclusively for \dword{daq}.  Two
additional racks are required for optical fiber distribution and
network connection to the surface.

The \dword{fscf} will supply the \dword{daq} room with cooling
water, a \SI{46}{cm} (18~inch) raised floor, lighting, HVAC, dry fire protection
and a dedicated 750~kVA transformer.  
The \dword{integoff} has responsibility for installing the remaining infrastructure which includes water cooled racks, the piping required to distribute water to the racks,  the electrical distribution system required to provide AC power for the racks, and supporting cable trays.

\section{Surface Rooms}
\label{sec:fdsp-coord-surf-rooms}

The \dword{dune} experiment requires space on the surface for a small
number of \dword{daq}, networking, and fiber optic distribution racks.  Space
is also allocated for cryogenics.  The surface cryogen building and
operations is described in 
Section~\ref{sec:fdsp-coord-cryogenics}.

The \dword{daq} consortium requires a surface computer room with eight
racks and a minimum of 50~kVA of power.  \dword{daq} also requires connection
to the optical fibers running to the \dword{4850l} via the Ross and Yates
shafts as well as to the Energy Sciences Network (ESnet).

The surface \dword{daq}, networking equipment, and fiber distribution racks
will be placed in a new main communications room (MCR) in the Ross Dry
building.  The MCR is approximately 628 square feet and will be
completed as part of the LBNF project, with seven racks installed for
the conventional facilities and space allocated for eight racks
provided by the experiment.  The seven racks allocated for
conventional facilities will include networking and fiber optic
distribution.  The eight racks allocated to the experiment will
contain computer servers, disk buffer, and some network connections.
Power and cooling will be provided as part of the \dword{lbnf} project.
\begin{dunefigure}[Surface rooms layout]{fig:dune-surface_layout}
  {Ross area surface rooms layout.}
  \includegraphics[width=0.95\textwidth]{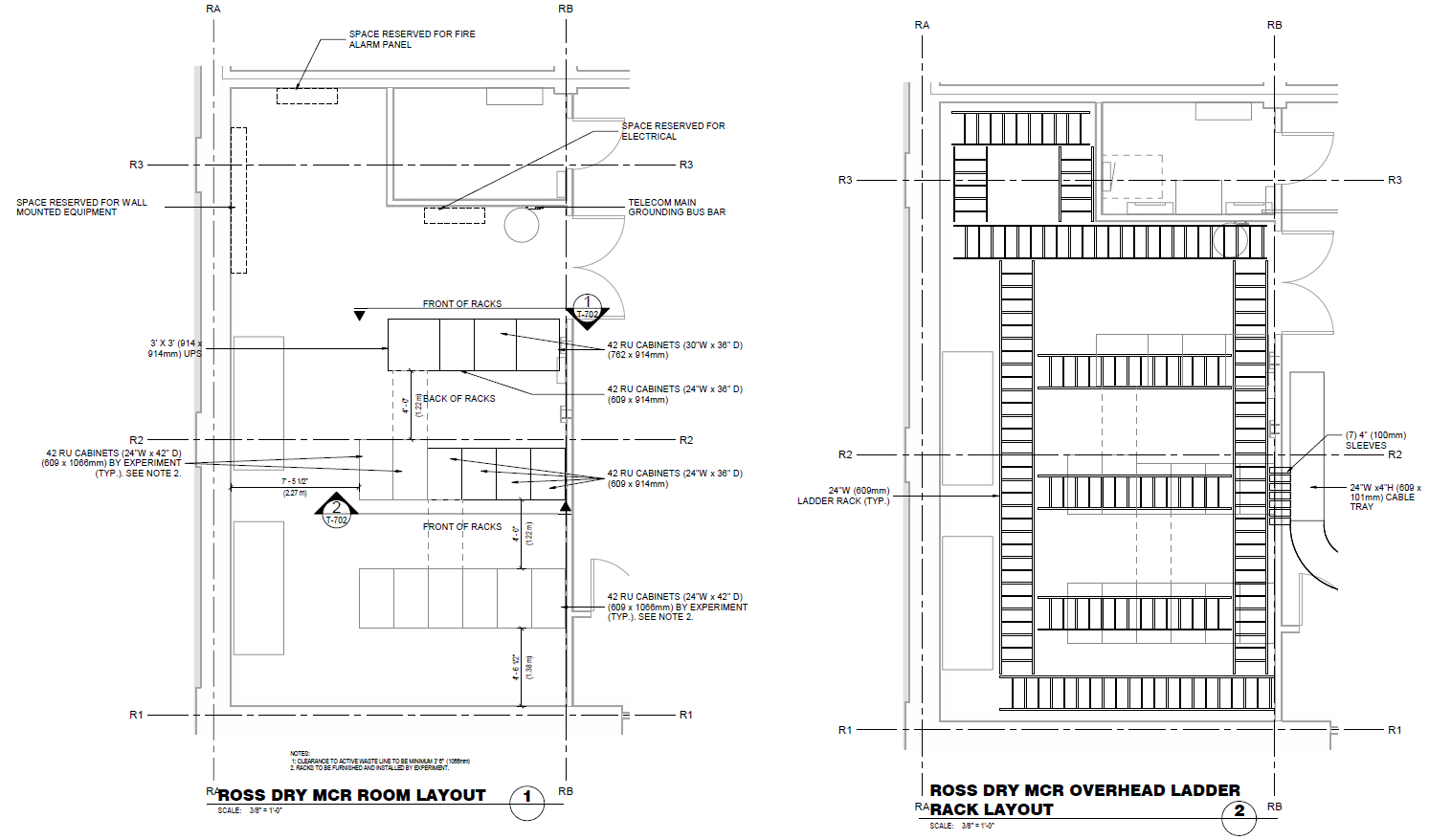}
\end{dunefigure}

\section{DUNE Detector Safety System}
\label{sec:fdsp-coord-det-safety}

The \dword{ddss} functions to protect experimental equipment.  The
system must detect abnormal and potentially harmful operating
conditions.  It must recognize when conditions are not within the
bounds of normal operating parameters and automatically take
pre-defined protective actions to protect equipment. Protective
actions are hardware or \dword{plc} driven.

\dword{dune} \dword{tc} works with the consortia to identify equipment
hazards and ensure that harmful operating conditions can be detected
and mitigated.  These hazards include smoke detected in racks, leak(s)
detected in water cooling areas, \dword{odh} detection, a drop in the
cryostat \dword{lar} liquid level, laser or radiation hazards in
calibration systems, over or under voltage conditions, and others to
be determined.  Some hazards will be unique to the actual detector
being implemented.

The slow controls system plays an active role and collects, archives,
and displays data from a broad variety of sources and provides
real-time status, alarms, warnings, and hardware interlock status for
the \dword{ddss} and detector operators. Slow controls 
monitor operating parameters for items such as HV systems, TPC electronics,
and \dword{pd} systems. Data is acquired via network interfaces, and status and
alarm levels will be sent to the \dword{ddss}.  Safety-critical issues that
require a hardware interlock, such as smoke detection or a drop in the
\dword{lar} level, which could cause \dword{hv} damage to components, are monitored by
slow controls; interlock status is provided to the \dword{ddss}.  The
protective action of a safety critical issue is done through hardware
interlocks and does not require the action of an operator, software, or
\dword{plc}.

The \dword{ddss} will provide input to the \dword{4850l}
fire alarm system.  The \dword{4850l} fire alarm system will provide life
safety and play an integral role in detecting and responding to an
event as well as notifying occupants and emergency responders.  This
system is the responsibility of the host laboratory and \dword{sdsta}.  The
fire alarm system is described in the BSI design~\cite{bib:cernedms2093229}.  
The \dword{4850l} fire alarm system is connected to the surface incident command vault in the
second floor of the Yates Administration building. Limits on the
number of occupants and egress paths are discussed. The table of
triggering inputs and Fire Alarm Sequence of Operation is documented
in the set of BSI underground electrical drawings, sheet U1-FD-E-308,
which is also found in~\cite{bib:cernedms2093229}. 
The
experiment will be adding a list of initiating inputs, such as smoke
detected in electronic racks or water leaks detected in the
\dword{daq} room to this sheet as designs reach a higher level of
maturity.  

The \dword{ddss} must communicate to the \dword{dune} slow controls
system as well as the \dword{4850l} fire alarm system.  The \dword{dune} slow
controls system monitors and records detector
status.  Working together through communication links, the three
systems will (1) monitor the status of the experiment (slow controls),
(2) protect equipment (\dword{ddss}),  and (3) and provide life safety (\dword{4850l} fire alarm system). Figure~\ref{fig:dune-DDSS} indicates how these systems
interact. 
\begin{dunefigure}[Examples of \dshort{dune} detector safety system information flow]{fig:dune-DDSS}
  {Sample \dword{ddss} information flow.}
  \includegraphics[width=0.85\textwidth]{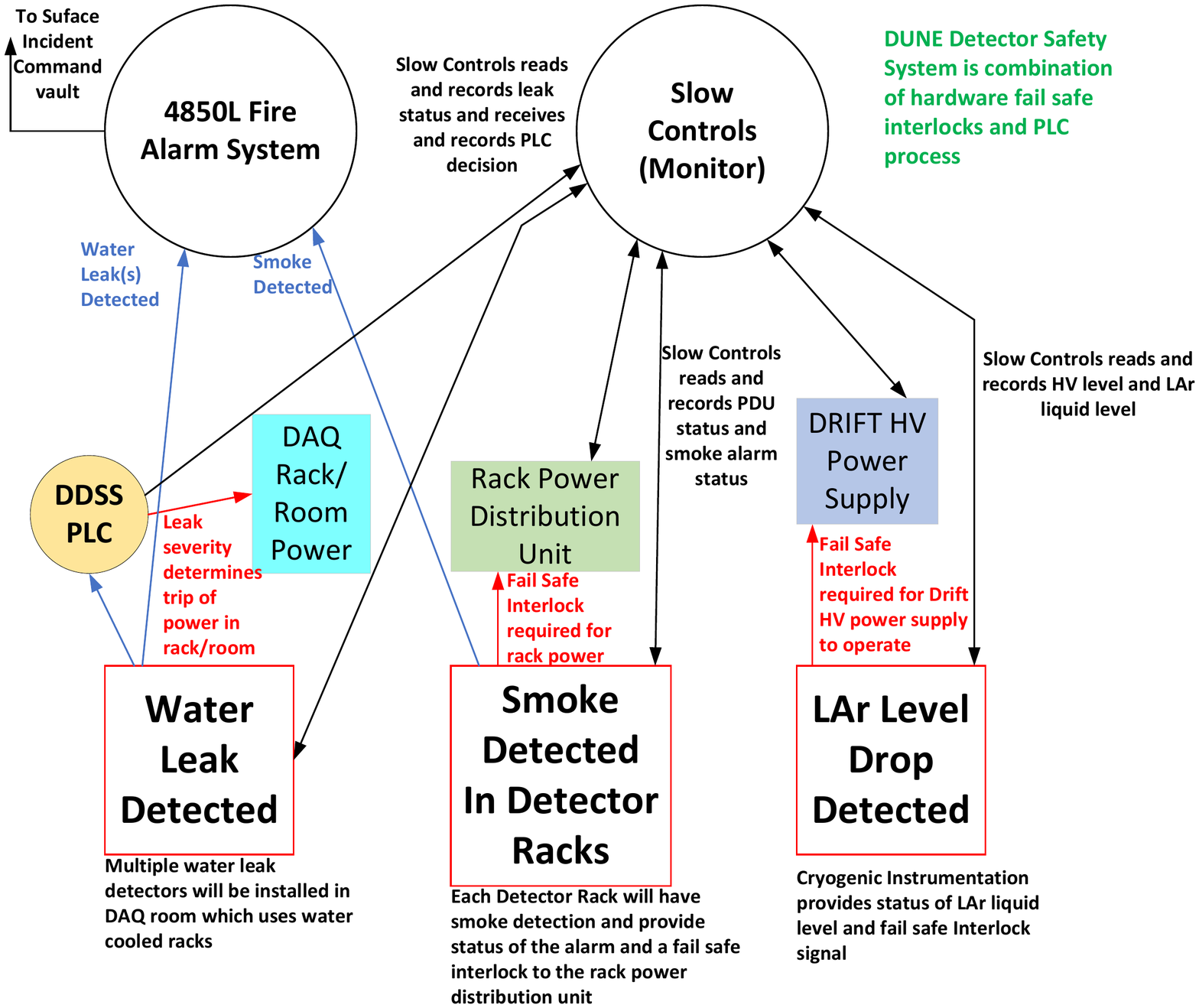}
\end{dunefigure}

The \dword{ddss} will be implemented through robust sensors feeding
information to redundant \dwords{plc} that activate hardware
interlocks. The selection of the \dword{plc} hardware platform is
still an open decision.  Listed below are some of the general
\dword{dune} experimental conditions that require intervention of the
\dword{ddss}:
\begin{enumerate}
 \item A drop in the \dword{lar} level.  This condition requires a hardware
   interlock on the liquid level.  If the level drops below a
   pre-determined level, the drift \dword{hv}  must automatically be 
   shut off to prevent equipment damage due to \dword{hv} discharge.  Slow controls would be
   alerted through normal monitoring and record the status of the detector.
 \item Smoke or a temperature/humidity increase above normal operating
   levels. This could be detected inside a rack or near an instrumented
   feedthrough.  If any of these conditions are detected, local
   power must be automatically switched off. If smoke is detected, a
   dedicated line will alert \dword{4850l} fire alarm system.
 \item A water leak detected near energized equipment in the \dword{daq}
   underground data processing room.  Water leak detectors 
   report to the \dword{ddss} \dword{plc} and a decision will be made to either
   issue an alert or immediately shut down power to the room, depending
   on the detected magnitude of the leak.  This condition would also be reported
   to the \dword{4850l} fire alarm system.
\end{enumerate}


\cleardoublepage

\chapter{DUNE Detector Construction Management}
\label{vl:tc-dune_overview}

This chapter provides an overview of
the \dword{dune} \dword{fd} modules and their construction
management. The \dword{fd} will have approximately \SI{70}{kt} of  \dword{lar}  mass divided into four cryostats. Each \dword{detmodule} is contained in
its own   \larmass{} cryostat, of which at least \nominalmodsize of the \dword{lar} is active (fiducial). \dword{dune} has two detector designs:
\dword{sp} and \dword{dp}.  Full descriptions can be found in
 Volume~\volnumbersp\ and Volume~\volnumberdp\
of this \dword{fd} \dword{tdr}.

\section{DUNE Single-Phase Far Detector Module}
\label{sec:fdsp-SP-module}

The \dword{sp} \dword{lartpc} is a \nominalmodsize module,
contributing to the full \SI{40}{\kilo\tonne} \dword{fd} fiducial
mass.  One \nominalmodsize \dword{spmod} is shown in
Figure~\ref{fig:DUNE_SP_model}.
\begin{dunefigure}[A \nominalmodsize SP module.]
{fig:DUNE_SP_model} 
{A \nominalmodsize DUNE FD SP module, showing alternating APAs,
    CPAs, FC and ground planes, detector support system, cryostat
    and cryogenics distribution.}
  \includegraphics[width=0.99\textwidth]{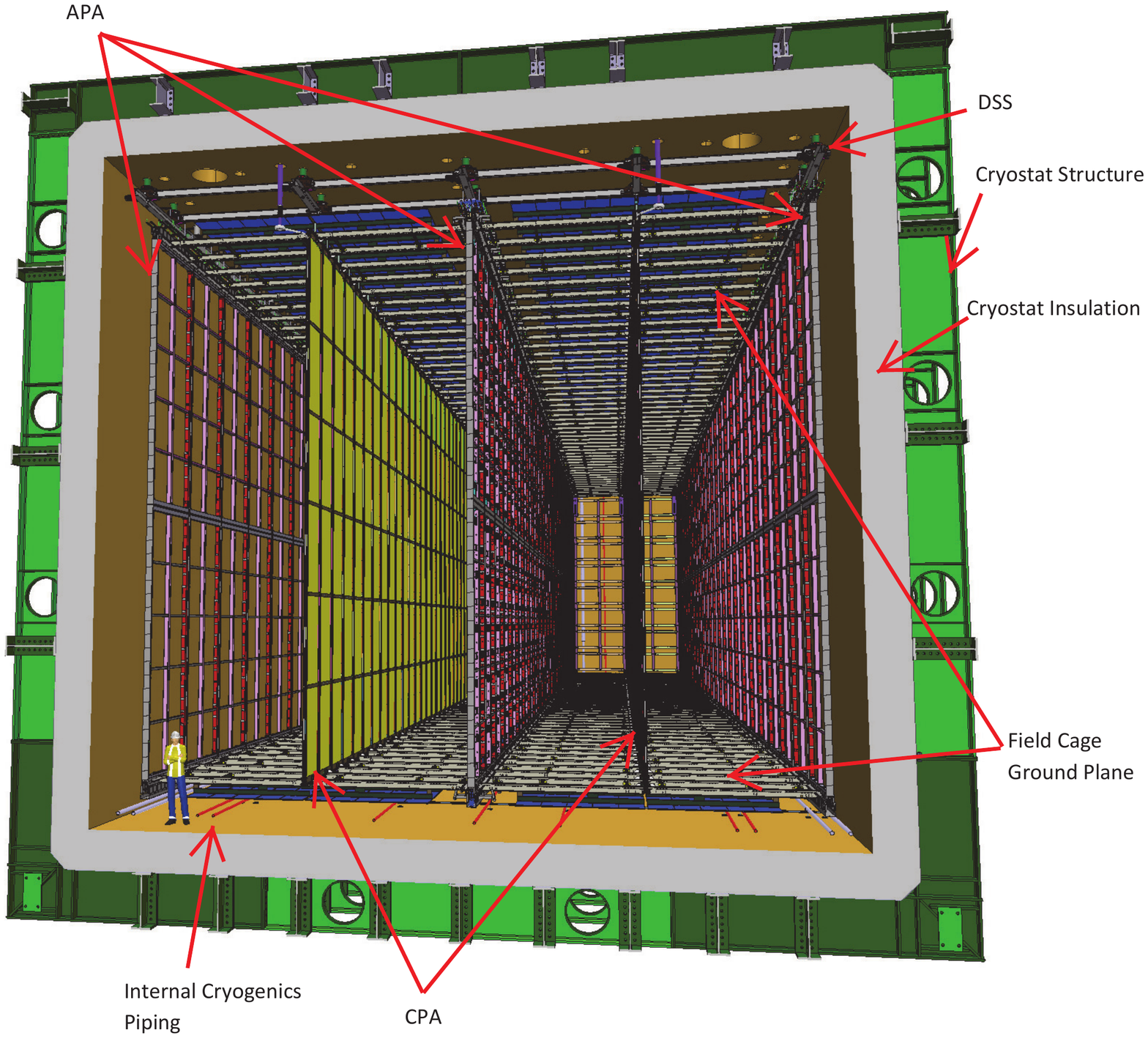}
\end{dunefigure} 

The module is contained in a cryostat as shown in
Figure~\ref{fig:dune-cryostat}.  Four drift volumes are created
between alternating \dword{apa} and \dword{cpa} walls on sides,
\dwords{fc} and \dwords{gp} on top and bottom, and two \dwords{ewfc}.
Each drift volume is \sptpclen long, \spmaxdrift wide and \tpcheight
high.  The entire assembly is supported from the roof of the cryostat
with \dword{dss}.

Each of the three \dword{apa} walls consists of an array of \num{25}
wide and \num{2} high individual \dword{apa}s, each with dimensions of
\SI{6.2}{\meter} high and \SI{2.32}{\meter} wide. There are \num{150}
\dword{apa}s in one module. Each \dword{apa} contains \num{10}
\dwords{pd}. Each of the two \dword{cpa} walls consists of an array of
\num{25} wide and \num{2} high \dword{cpa} panels. Two panels are
stacked to form planes, each with dimensions of \SI{12.1}{\meter} high
and \SI{1.16}{\meter} wide. There are \num{100} \dword{cpa}s in one
module.  \dword{cpa} walls are held at $-$\SI{180}{\kilo\volt}. With
\dword{apa} walls held close to ground, the result is a
\SI{511}{\volt/\centi\meter} gradient across the drift volume. The
drift in the \dword{spmod} is in the horizontal direction. The
\dword{lar} level is above the top set of \dwords{gp} and just below
the \dword{dss}.

Readout electronics are mounted on the \dword{apa}s. Cables from the
readout electronics and \dwords{pds} are routed to the cryostat roof
where they exit through a set of \num{75} feedthroughs. The cables are
connected to \dwords{wib}, which are contained in \num{150}
\dwords{wiec} for \dword{apa}s and \num{75} crates for \dwords{pds}. Data
fibers from the \dwords{wiec} carry the data to \dword{daq} racks
inside the \dword{cuc}.

There are four \dword{hv} feedthroughs on top of the cryostat, two on
each end. In addition, there are feedthroughs for calibration,
instrumentation and cryogenics distribution. Power supplies, and
controls are located on top of the cryostat on a dedicated
mezzanine. Cryogenics equipment is also installed on a separate
mezzanine on top of the cryostat.

\section{DUNE Dual-Phase Far Detector Module}
\label{sec:fdsp-DP-module}

Each \dword{dpmod} is a \nominalmodsize \dword{lartpc}, contributing
to the full \SI{40}{\kilo\tonne} \dword{fd} fiducial mass.  One
\nominalmodsize \dword{dpmod} is shown in
Figure~\ref{fig:DUNE_DP_model}.
\begin{dunefigure}[A \nominalmodsize DP module]{fig:DUNE_DP_model} {A \nominalmodsize DUNE   \dword{dpmod}.}
  \includegraphics[width=0.8\textwidth]{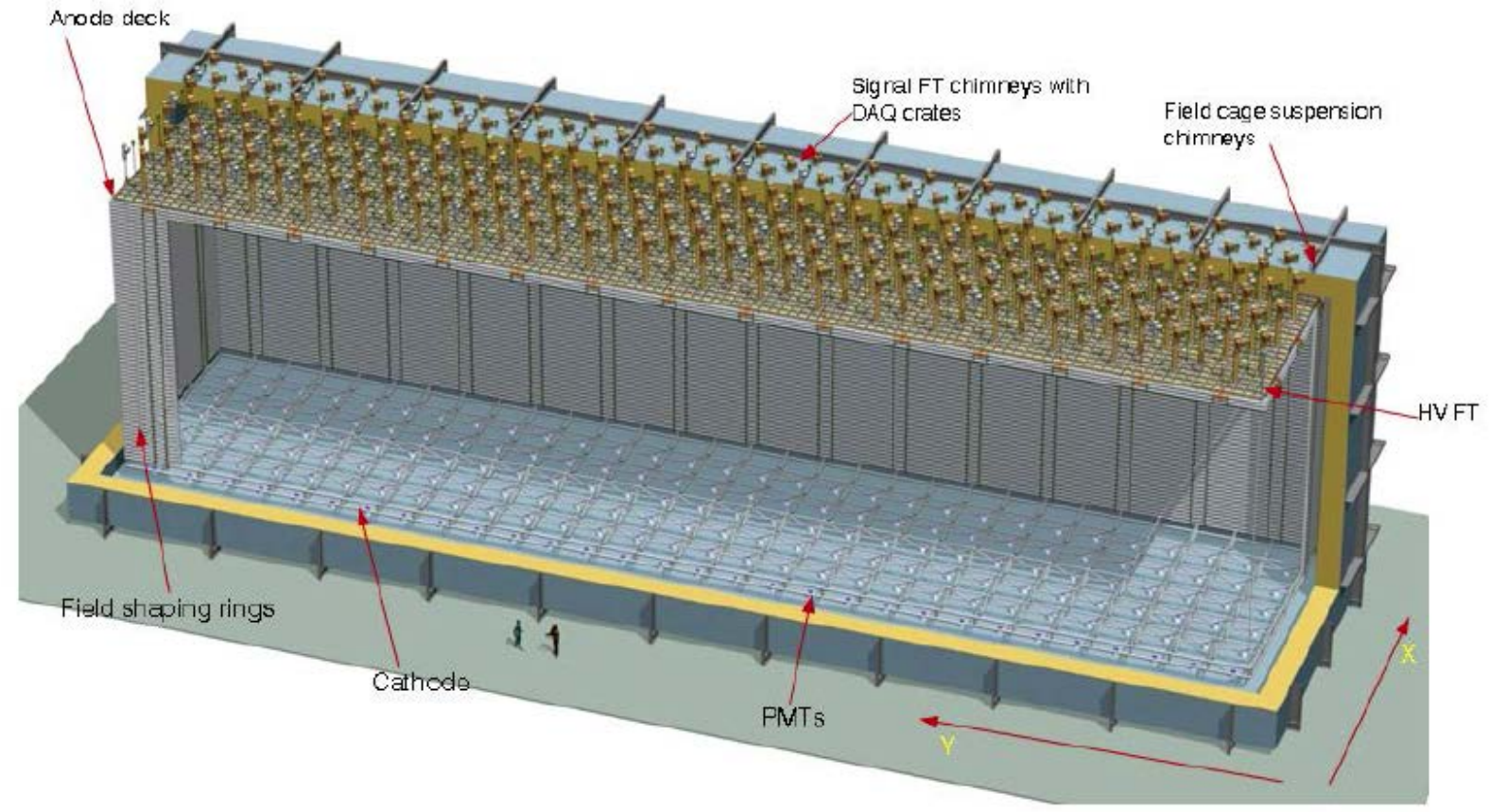}
\end{dunefigure} 

Each \dword{dpmod} is contained in a cryostat with the same internal
dimensions as the \dword{spmod}, but with some variations in cryostat
penetrations.  The \dword{dpmod} consists of a single drift volume
with a vertical drift direction.

The drift volume is enclosed on the top by an array of \num{80}
\dwords{crp}, on the bottom by cathode planes and on the perimeter by
a \dwords{fc}. It is \dpmaxdrift in height with a \dpnominaldriftfield
gradient. The cathode is held at a potential of
\dptargetdriftvoltneg{}.  The \dword{dp} \dword{pds} consists of
\dpnumpmtch \dwords{pmt} at the bottom of the drift volume and is
integrated with the cathode planes.  The \dword{lar} level in
\dword{dpmod} is within the \dword{crp}, just above the collection
grid and below the anode readout plane. A gradient of
\SI{2}{\kilo\volt/\centi\meter} in this region is used to extract the
drift electrons from the liquid. A \dword{lem} with a gradient of
\SI{33}{\kilo\volt/\centi\meter} causes charge multiplication and
amplification of the charge that is then collected on the anode, 
which consists of two perpendicular 
readout strips.

The \dword{dp} \dword{dss} consists of a set of stainless steel
cables that are suspended from feedthroughs on top of the
cryostat. The cables can be extended to the floor of the cryostat where
they are used to lift components to design height. In the case of
\dwords{crp}, there are three cables per panel with active height
control in order to position the panel precisely with respect to the
\dword{lar} surface.

The cryogenic \dword{fe} electronics is installed in the
\dwords{sftchimney} on the roof of the cryostat to process the
\dword{lartpc} signals. Each \dword{sftchimney} is coupled to a \dword{utca}
crate to digitize the signals. These crates are connected via optical
fiber links to the \dword{daq} back end. Arrangement of equipment on
top of the cryostat is similar to the \dword{spmod}.

\section{DUNE Far Detector Consortia}
\label{sec:fdconsortia}

A total of eleven \dword{fd} consortia have been formed to cover 
the subsystems required for the two detector types currently under
consideration.  In particular, three consortia (SP-APA, SP-TPC
Electronics and SP-Photon Detection) pursue subsystems specific to
the single-phase design and another three consortia (DP-CRP, DP-TPC
Electronics, and DP-Photon Detection) pursue designs for \dword{dp}
specific subsystems.  An additional five consortia (HV System, \dword{daq},
\dword{cisc}, Calibration, and Computing)
have responsibility for subsystems common to both detector
technologies.  Figure~\ref{fig:DUNE_consortia} shows the consortia 
associated with the \dword{fd} construction effort along with their 
current leadership teams.  
\begin{dunefigure}[DUNE consortia]{fig:DUNE_consortia}
  {\dword{dune} consortia organization. CL refers to consortium leader
    and TL refers to technical lead.}
  \includegraphics[width=0.99\textwidth]{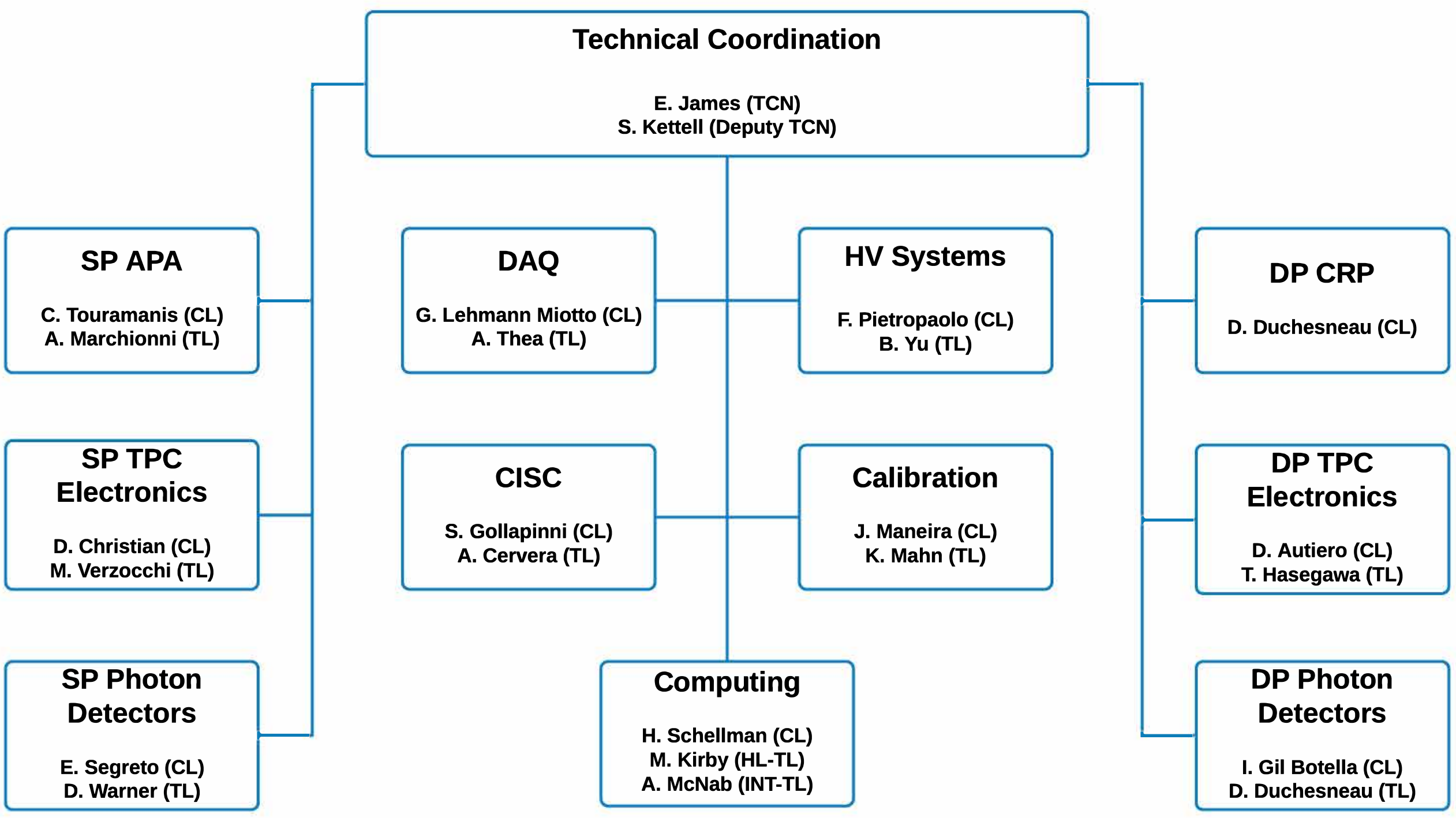}
\end{dunefigure}

\section{Work Breakdown Structure (WBS)}
\label{sec:fdsp-coord-wbs}

The complete scope of the \dword{dune} construction project is captured in a 
\dword{wbs} to explain the distribution of deliverables between 
the consortia.  In combination with interface documentation, the 
\dword{wbs} is used to validate that all necessary scope is covered.  The 
\dword{wbs} is also used as a framework for building \dword{dune} 
detector cost estimates.

The highest-level layers of the \dword{dune} \dword{wbs} are summarized 
in Figure~\ref{fig:WBS_level2}.  At level 1 the \dword{wbs} is broken down into 
six elements corresponding to the five \dword{dune} detector modules (four 
\dword{fd} and one \dword{nd}) and \dword{tc}.  The scope documented
here is fully contained within the \dword{tc}, first \dword{fd} module 
(\dword{sp}), and second \dword{fd} module (\dword{dp}) level 1 elements.   
\begin{dunefigure}[DUNE WBS at level 2]{fig:WBS_level2}
  {High level \dword{dune} \dword{wbs} to level 2.}
  \includegraphics[width=0.75\textwidth]{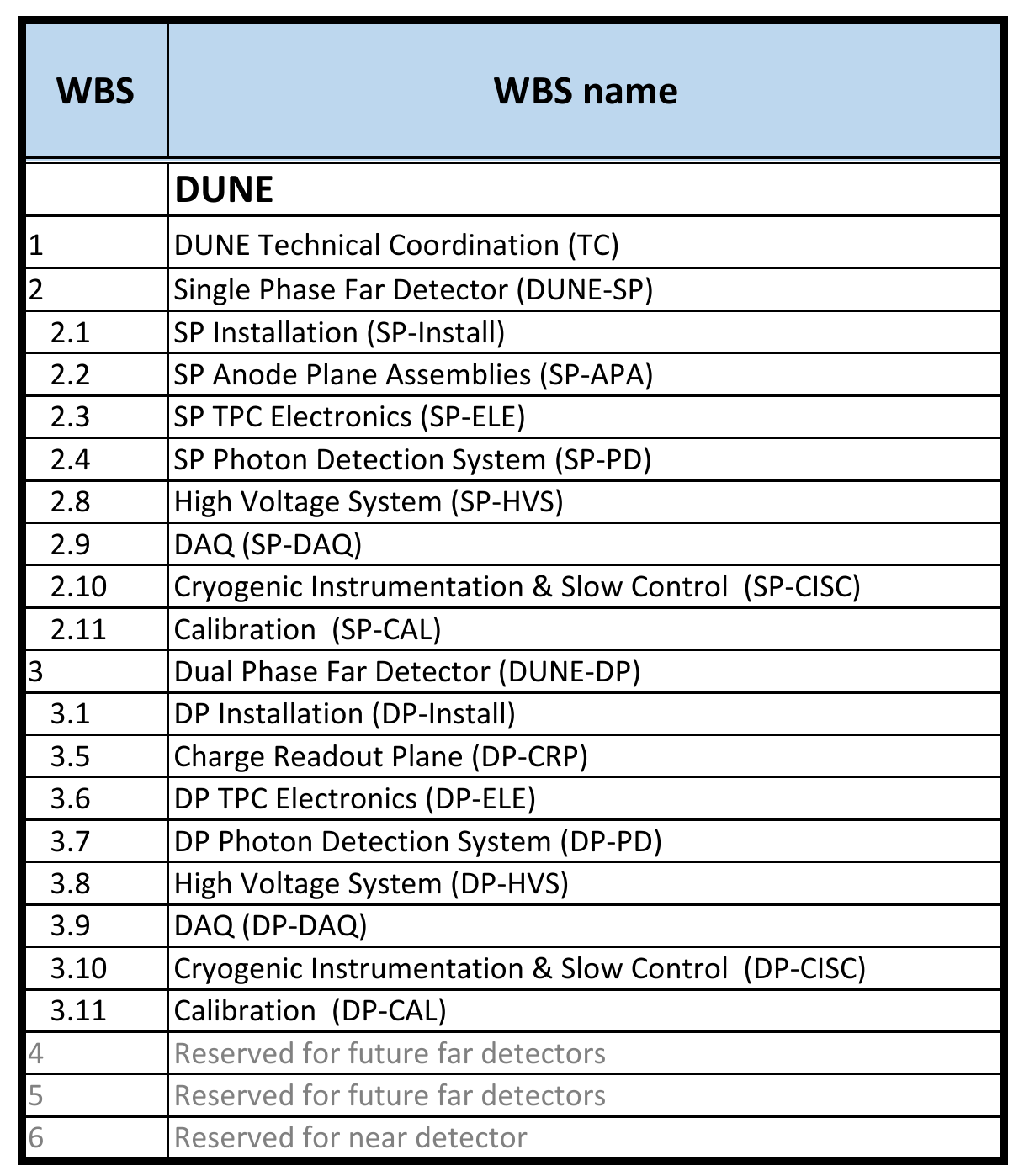}
\end{dunefigure}

For the \dword{fd} module elements at level 1, the \dword{wbs} breaks 
down at level 2 into elements encompassing the deliverables provided by 
each consortium to that \dword{detmodule} along with an element containing 
common deliverables associated with the required detector installation 
and integration effort.  Since each consortium takes responsibility 
for a particular subsystem, this breakdown effectively corresponds to 
a division of deliverables across subsystems. 

The level 3 breakdown of the level 2 subsystem \dword{wbs} elements follow 
a common format that separates required activities into groupings defined 
roughly by their sequence in time.  A total of six elements are used:     
\begin{enumerate}
  \item management,
  \item physics and simulation,
  \item design, engineering, and R\&D,
  \item production setup,
  \item production, and
  \item integration and installation.
\end{enumerate}
The groupings at level 3 allow for the convenient separation of costs
 including those associated with one-time and recurring activities in the
 case where two identical detector modules are constructed.

Lower levels within these \dword{wbs} elements are determined by 
the responsible consortia and generally correspond at level 4 to 
the different primary detector components from which each subsystem
is assembled.  This level 4 structure is repeated under each of the 
level 3 items to ensure that the full cost of each primary detector 
components can be rolled up over the sequence of activities defining 
their design, production, and installation.       

\section{DUNE Design Maturity}

The \dword{dune} project builds on significant development in previous
large \dword{lartpc} detectors (\dword{icarus} and \dword{microboone})
and on substantial development from \dword{lbne} and \dword{lbno}. One
of the most important elements that has significantly advanced the
project development is the successful construction of the
\dword{protodune} detectors and successful operation of
\dword{pdsp}. These detectors use full-size \dword{dune} components
and processes. The construction of \dword{protodune} has established
teams, production lines, \dword{qa} and \dword{qc} processes,
installation, operation, and performance of the final \dword{dune}
detectors.

Based on the success of \dword{protodune}, \dword{dune} has reached
advanced technical maturity, approaching (80\%). The designs from \dword{protodune}
 have significantly advanced for the \dword{dune} \dword{fd}. Most
subsystems completed \dword{pdr} or 60\% reviews on design
modifications beyond \dword{protodune} in advance of the
\dword{tdr}. The overall level of design maturity is now
$\sim\,$90\%. The breakdown of the design maturity level for the
\dword{spmod} by subsystem is provided in
Table~\ref{tab:designmaturity}. The table shows the \dword{dune} design maturity at the time of 
    \dword{pdsp} and at the time of writing of this \dword{tdr}, 
    along with the estimated design effort or weight of each subsystem.
    \begin{dunetable}
    [DUNE-SP design maturity at the time of ProtoDUNE and at the time of the TDR]
    {p{0.35\linewidth}p{0.1\linewidth}p{0.12\linewidth}p{0.07\linewidth} 
    } {tab:designmaturity}
    {\dword{spmod} design maturity}
  System & Weight & \dword{protodune} & \dword{dune}   \\ \toprowrule
  DSS & 10\% & 75\% &  85\% \\ \colhline
  APA & 30\% & 85\% &  95\% \\ \colhline
  TPC Electronics  & 20\% & 80\% &  90\% \\ \colhline
  PDS & 10\% & 50\% &  65\% \\ \colhline
  HVS & 15\% & 80\% &  95\% \\ \colhline
  DAQ & 10\% & 60\% &  80\% \\ \colhline
  CISC & 5\% & 80\% &  90\% \\ \colhline \colhline
  Total& 100\% & 76\% & 90\% \\ 
\end{dunetable}
In particular for the \dword{sp} design: 

\begin{itemize}
\item The \dword{apa} conceptual design was developed in 2010, prototyped
first at 40\% scale, and again in the \dword{35t}. The version deployed in
\dword{pdsp} is close to that for the \dword{spmod} (85\%). 

\item The TPC electronics 
 low-noise system design, including feedthroughs, cables, and
grounding, was successfully prototyped at large scale in
\dword{microboone} and \dword{pdsp}, and is 90\% mature. 
\begin{itemize}  
\item The \dword{fe} chip has gone through eight iterations and was successfully
demonstrated in \dword{microboone} and \dword{pdsp} (90\%). 

\item The \dword{femb} has gone through a similar number of iterations and was
successfully demonstrated in \dword{pdsp} (80\%). We have gained important
knowledge from the \dword{femb}
development cycles, in particular about techniques of the power distribution and
signal routing for the low-noise design. New \dword{femb} prototypes for the \dword{spmod} that use new custom \dwords{asic} differ from the \dword{pdsp} version only in the physical layout of the power distribution and the interconnections between 
two \dwords{asic}. From a logical point of view there are very few differences between
\dwords{femb} accommodating different \dwords{asic}. This has been demonstrated by the
recent \dword{femb} design with \dwords{coldadc} showing nice performance from the lab
test results.

\item The \dword{adc} chip
has evolved from a previous version used in \dword{cmos} 180~nm
technology that was tested to $-50^\circ$C. (70\%). 

\item Key elements of
the \dword{coldata} chip have been prototyped (70\%). 
\end{itemize}  
\item The \dword{hv}
design has evolved from \dword{icarus}, \dword{microboone}, and the
\dword{35t}.  It has been prototyped in subsequent runs of the
\dword{35t} and demonstrated in \dword{pdsp} (80\%). 

\item The \dword{pds}
\dword{xarapu} design has been prototyped at small scale and in
\dword{pdsp} (20\%). The mechanical design has been extensively
developed using the \dword{35t} detector and \dword{pdsp} (85\%). 

\item The
\dword{daq} \dword{artdaq} back end has been developed in several
experiments, including the \dword{35t} and \dword{pdsp}. The
\dword{daq} \dword{felix} \dword{fe} has been developed by
\dword{atlas} and prototyped in \dword{pdsp}.
\end{itemize}

The design maturity of the \dword{dp} detector technology is also
quite advanced. It builds on working noble liquid \dwords{tpc} for
dark matter and neutrinoless double beta decay experiments. A
significant benchmark is the operation of the \dword{wa105} at
\dshort{cern}. The successful construction of \dword{pddp} and tests in
the cold box at \dshort{cern} provide invaluable experience. A critical
test will be operation and analysis of \dword{pddp} at
\dshort{cern}.


\cleardoublepage

\chapter{Integration Engineering}
\label{sec:fdsp-coord-integ-sysengr}

The \dword{dune} \dword{fd} consists of \dwords{spmod} and
\dwords{dpmod}, housed inside cryostats, which in turn are housed
inside the \dword{lbnf} \dword{fscf}.  This nested structure is
mirrored with detector integration in a similar layered manner.  This
chapter explains the method of integration for the \dwords{detmodule}.
The integration of the modules is carried out by the
\dword{tc} engineering support team working within the broader framework of the \dword{jpo} central engineering team.

Integration engineering for \dword{dune} focuses on configuring the
mechanical and electrical systems of each \dword{detmodule} and managing
the interfaces within them. This includes verifying that subassemblies
and their interfaces are built conforming to the approved design,
e.g., \dword{apa} or \dword{pds}. The second major focus
is assuring that the \dwords{detmodule} can be integrated and
installed into their final configuration. And the third major focus is
integrating necessary services provided by \dword{cf} 
with the \dwords{detmodule}.

To this end, the \dword{jpo} engineering team maintains
subsystem component documentation in order to manage the detector
configuration.
The consortia provide engineering data for their detector subsystems to the \dword{jpo} team for incorporation within the global configuration files.

This process, used successfully for \dword{protodune}, 
has been enhanced with the addition of engineering
and design staff and development of interface documents as described
in Section~\ref{sec:fdsp-coord-interface}.

\section{Mechanical Integration Models}
\label{sec:fdsp-coord-integ-models}

The \dword{sp} and \dword{dp} \dwords{detmodule} are large and made of many
intricate components. Fortunately, for the most part, the
components are repetitive and not overly complex
geometrically. Thus, \threed mechanical modeling techniques are well suited
to represent the \dwords{detmodule} and manage their configuration.

At the same time, \threed modelling techniques vary in the way items are
represented and in the way the techniques are carried out. Thus, a set
of \twod integration drawings must be generated; these drawings must be
clear and unambiguous across the collaboration. Such \twod drawings are
the basis for the \threed model accuracy, as well as the basis for the engineering
design of all components.

The consortia choose their mechanical modeling software.
Their model files are transferred via STEP files\footnote{\url{http://www.npd-solutions.com/step.html} ``Standard for the Exchange of Product Model Data'', ISO Standard 10303.}, which the \dword{jpo} engineering team integrates into overall models.  Navisworks\footnote{Navisworks\texttrademark \url{https://www.autodesk.com/products/navisworks/overview}.} software allows for visualization by
the entire collaboration.

\subsection{Static Models}
\label{sec:fdsp-coord-integ-static}

\Dword{tc} engineering support team generates and maintains \threed detector
integration models as well as  \twod integration drawings of the detector.
These models represent the detector at Normal Temperature and Pressure
(\SI{101.3}{kPa} and $20^\circ$C). These models are static because they represent all
components at their design dimensions and locations. They do not represent
effects of gravity, tolerances, cold temperature, and installation and
assembly clearances. Such effects are modeled in the envelope and assembly
models, as described in Section~\ref{sec:fdsp-coord-integ-envelope}.

The \threed models are assembled by combining component models from
various consortia  then shared with the consortia. The \twod
integration drawings, which are generated from the \threed models and
disseminated, show the interfaces to the level of detail necessary to
ensure proper fit and function. Any issues that arise are communicated
to the consortia, and a resolution method is determined.

The \dword{jpo} engineering team will not change any
consortia component models.  The consortia must resolve any issues
using agreed-upon methods  and provide updated models for
reintegration. Using this process, models are kept synchronized with
integration occurring in only one direction: only the consortia modify
their models, and the \dword{jpo} engineering team
integrates and disseminates them.  The \dword{tc} engineering support team works with the consortia as needed to keep their subsystem models current. 
The  \dword{jpo} engineering team defines points in the design 
process where current models are combined and identified as the official
current integration model.

The level of detail in a model is managed actively. When models are
combined and incorporated into global models of facilities, too much
detail leads to very large file sizes.  The \dword{jpo} 
engineering team must ensure the appropriate level of detail at each
stage of model integration.

Several examples from the \dword{spmod} are presented
here. Figure~\ref{fig:dune-sp_overall} shows the overall model of the
\dword{spmod} with one wall of the cryostat removed to make the
interior components visible. The detector has 25 rows, all of which
follow the same construction. At the ends of rows 1 and 25,
\dwords{ewfc} are installed to close the detection volume.  As mentioned
earlier, this model does not include all the details of the detector
components. The components are simplified to keep the overall model
complexity to a manageable level.
\begin{dunefigure}[Overall model of the SP module]{fig:dune-sp_overall}
  {Overall model of the \dword{spmod} showing three of the 25 rows,
    simplified cryostat, \dword{dss} and temporary cryostat opening.}
  \includegraphics[width=0.85\textwidth]{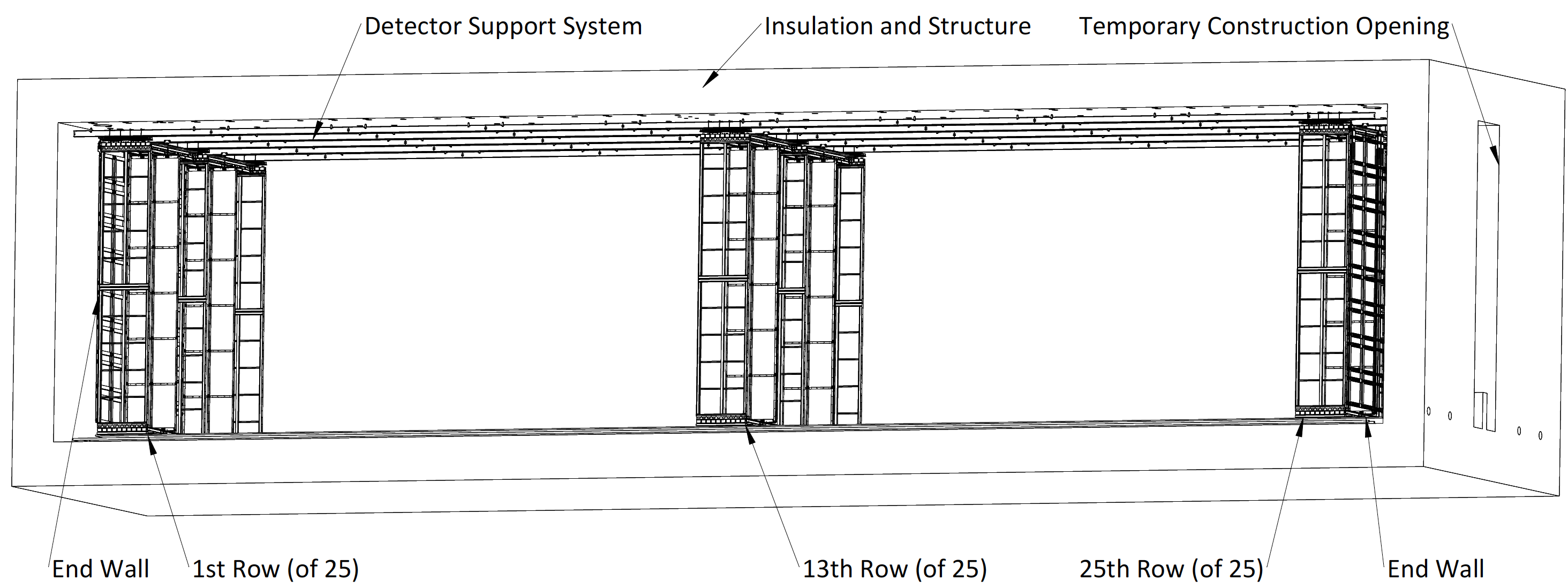}
\end{dunefigure}

Figure~\ref{fig:dune-sp_row} shows the model of one row of the
detector module. Each row is constructed from six \dword{apa}s, four
\dword{cpa}s and eight \dwords{fc} and \dwords{gp}. A total of 25 rows
comprise one \dword{spmod}.
\begin{dunefigure}[Model of one row of the SP module]{fig:dune-sp_row}
  {Model of one row of \dword{spmod} showing overall arrangement and dimensions.}
  \includegraphics[width=0.85\textwidth]{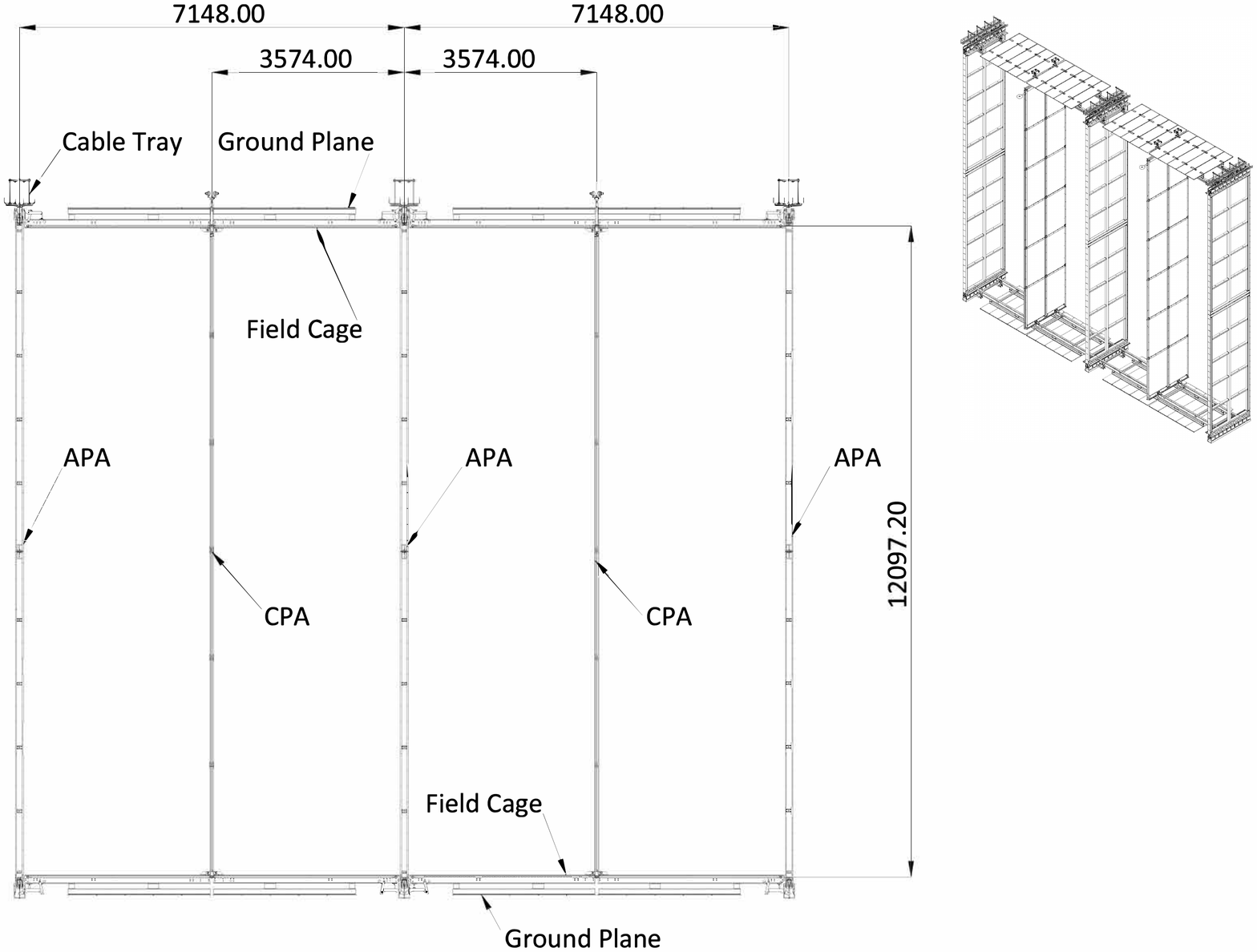}
\end{dunefigure}

Figure~\ref{fig:dune-sp_transverse} shows a cross section of the
\dword{spmod} in the transverse direction and the overall dimensions.
Figure~\ref{fig:dune-sp_long} shows a cross section of the
\dword{spmod} in the longitudinal direction and the overall
dimensions. In both figures, the cryostat structure and insulation are
shown as cross-hatched areas.
\begin{dunefigure}[Section view of the SP module in the
    transverse direction]{fig:dune-sp_transverse}
  {Section view of the \dword{spmod} in the transverse
    direction.}
  \includegraphics[width=0.65\textwidth]{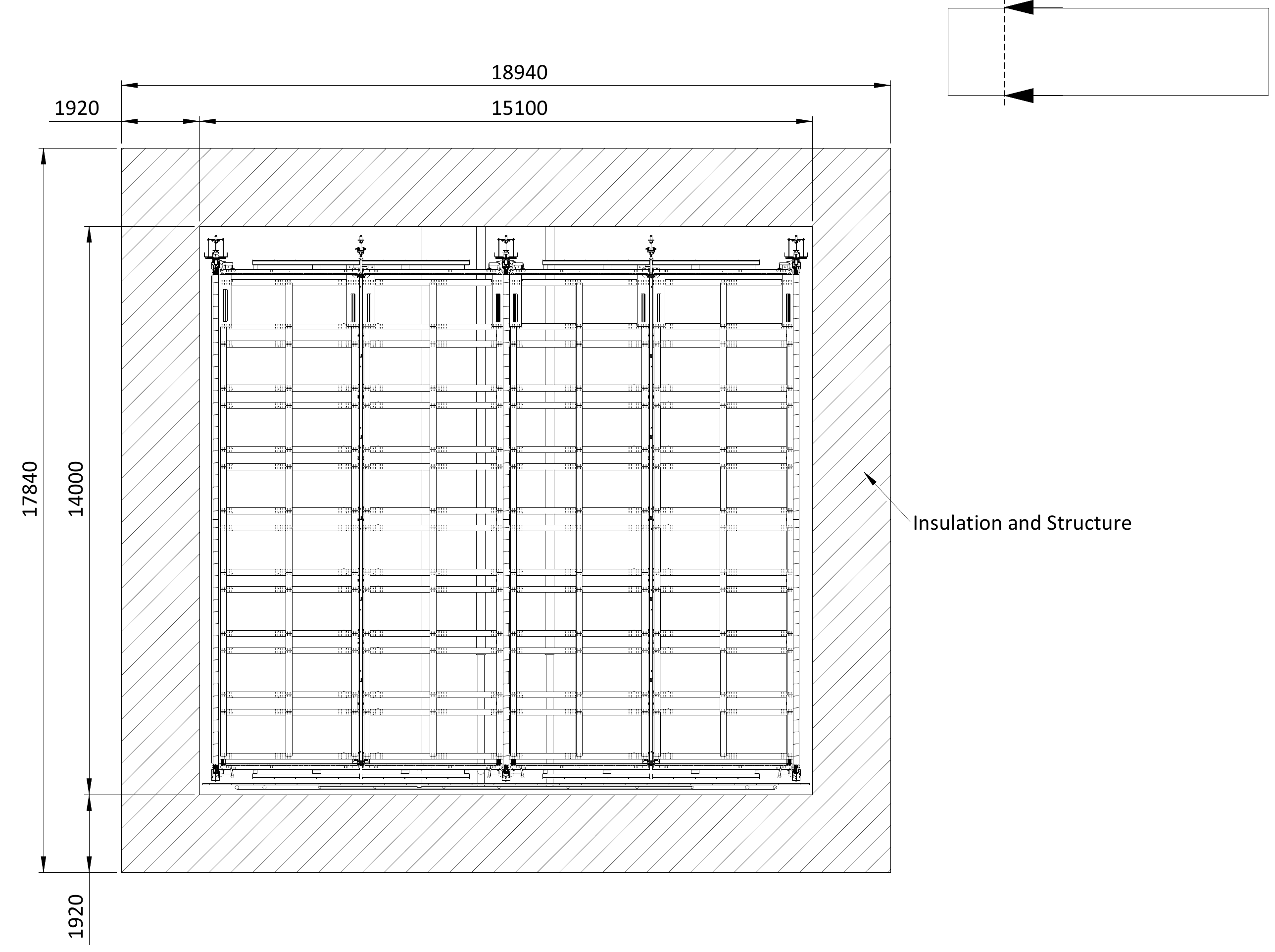}
\end{dunefigure}
\begin{dunefigure}[Overall model of the SP module in the
    longitudinal direction]{fig:dune-sp_long}
  {Overall model of the \dword{spmod} in the longitudinal direction.}
  \includegraphics[width=0.85\textwidth]{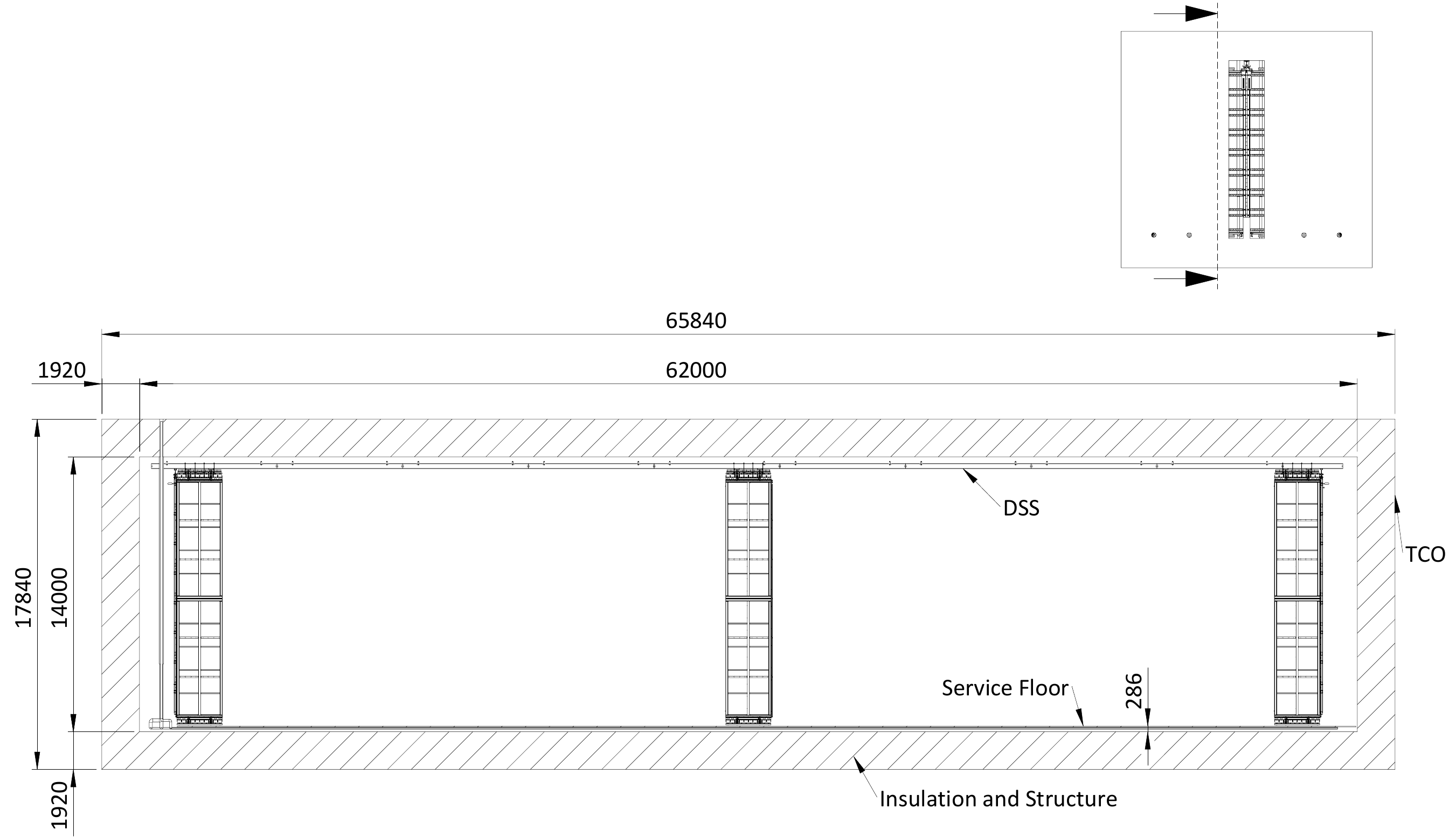}
\end{dunefigure}


\subsection{Envelope and Assembly Models}
\label{sec:fdsp-coord-integ-envelope}

Static models represent the \dword{detmodule} and its components using their exact
design dimensions. Such exact dimensions are needed so that the
detailed component drawings and model remain completely compatible at all
times.

For installation and operation, however, other envelope models are
needed. Envelope models are developed to address issues that affect
installation and operation:
\begin{enumerate}
 \item effects on the detector caused by distortion of the cryostat
   and detector support structure due to gravity;
 \item effects on the detector caused by distortion of the cryostat
   and detector support structure due to loads on the cryostat during
   detector filling and operation;
 \item effects on the detector caused by thermal contraction during
   detector filling and operation;
 \item effects of component and
   assembly tolerances;
 \item clearances needed for installation and envelopes needed for
   access and tooling;
 \item reference models and drawings needed for installation stages
   and to control assembly; and
 \item reference models and drawings needed for alignment and survey.
\end{enumerate}

The models and drawings described above are generated from static
models. Models are also generated to represent combined effects of the
above. In all cases, as with static models, \twod drawings are created
and provide the basis for the installation drawings.

Generating envelope models and drawings are the responsibility of the
\dword{jpo} engineering team in coordination with consortia.

\subsection{Integration and Interface Drawings}
\label{sec:fdsp-coord-integ-drawings}

Within each \dword{detmodule}, components from various consortia are
assembled and installed. In addition, components that are the
responsibility of \dword{integoff} are assembled and installed in
parallel. The interfaces among components are developed and managed
through models and drawings as described in
Section~\ref{sec:fdsp-coord-integ-models}. Many such interfaces must
be controlled to ensure that the detector will fit together. The
following section shows some of the interfaces, control drawing,
dimensions, and configurations.

Figure~\ref{fig:dune-apa_interfaces_top} shows the interfaces for the
top \dword{apa}s in the upper corner of the cryostat. It also shows the position
of the cable penetration for the \dword{apa}. Interfaces with cryostat
corrugations and \dword{lar} fill lines are also shown. The reference plane,
defined as the plane of the \dword{apa} yokes, is explained in the
alignment section (Section~\ref{sec:fdsp-coord-integ-survey}).
\begin{dunefigure}[Interface between upper APA,
   FC, cable trays and DSS]{fig:dune-apa_interfaces_top}
  {\dword{spmod} interface between upper \dword{apa}, \dword{fc}, cable
    trays, and \dword{dss}.}
  \includegraphics[width=0.8\textwidth]{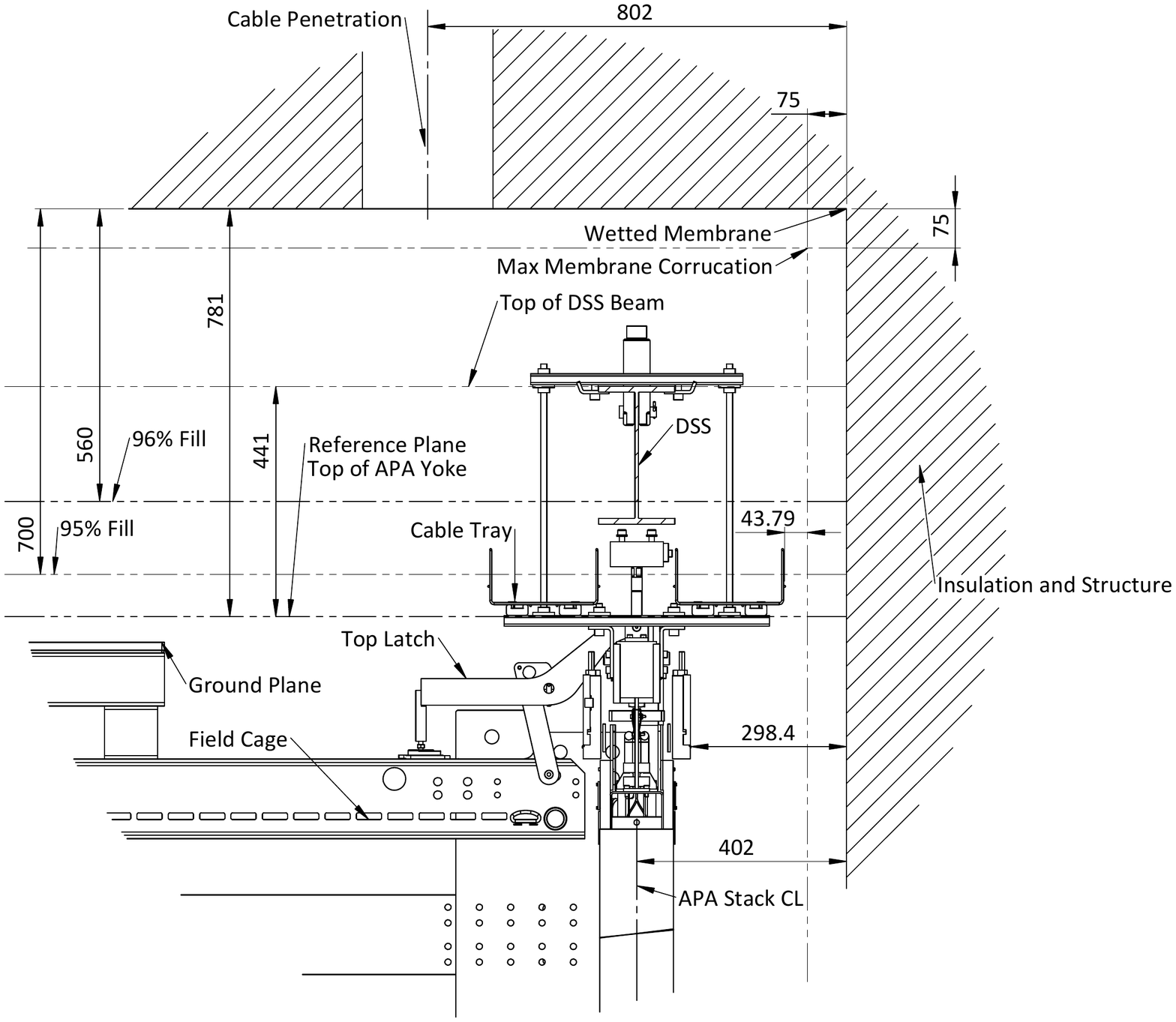}
\end{dunefigure}

Figure~\ref{fig:dune-apa_interfaces_bottom} shows the interfaces for
the bottom \dword{apa}s in the lower corner of the cryostat. In both figures,
the connection latch between the \dwords{fc} and \dword{apa}s is also
shown.
\begin{dunefigure}[Interface between lower APA, FC
    and service floor]{fig:dune-apa_interfaces_bottom}
  {\dword{spmod} interface between lower \dword{apa}, \dword{fc}, and
    service floor.}
  \includegraphics[width=0.7\textwidth]{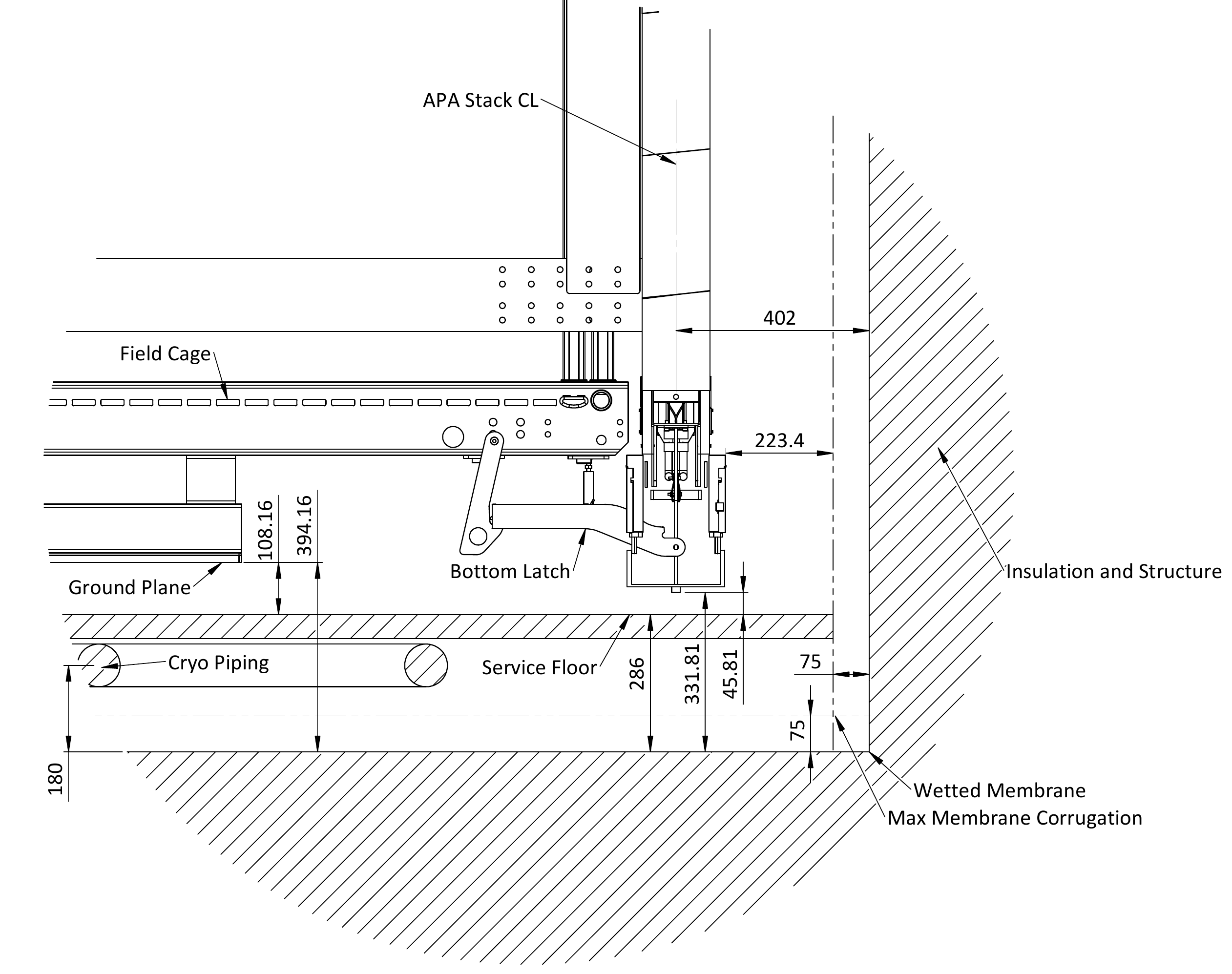}
\end{dunefigure}

Figure~\ref{fig:dune-apa_interfaces_mid} shows the interfaces for the
top of the central row of \dword{apa}s with other components. In this case, a
double latch connection is used. Similarly,
Figure~\ref{fig:dune-cpa_interfaces} shows the interfaces for top of the
\dword{cpa}s with other components.
\begin{dunefigure}[Interface between APA, upper FC
    and GP]{fig:dune-apa_interfaces_mid}
  {\dword{spmod} interface between \dword{apa}, upper \dword{fc}, and \dword{gp}}
  \includegraphics[width=0.8\textwidth]{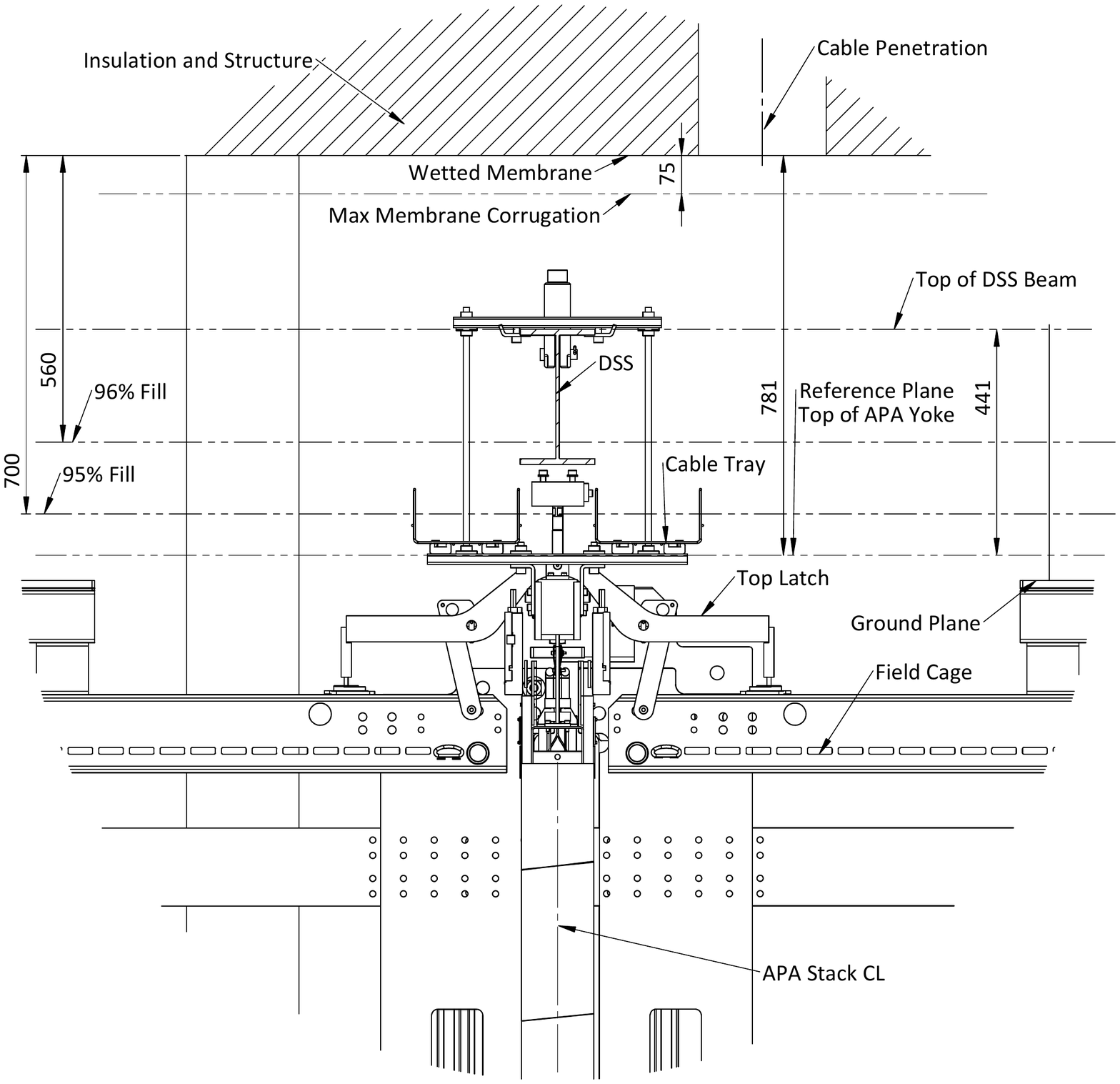}
\end{dunefigure}
\begin{dunefigure}[Interface between CPA, upper FC and GP]
    {fig:dune-cpa_interfaces}
  {\dword{spmod} interface between \dword{cpa}, upper \dword{fc}, and \dword{gp}.}
  \includegraphics[width=0.8\textwidth]{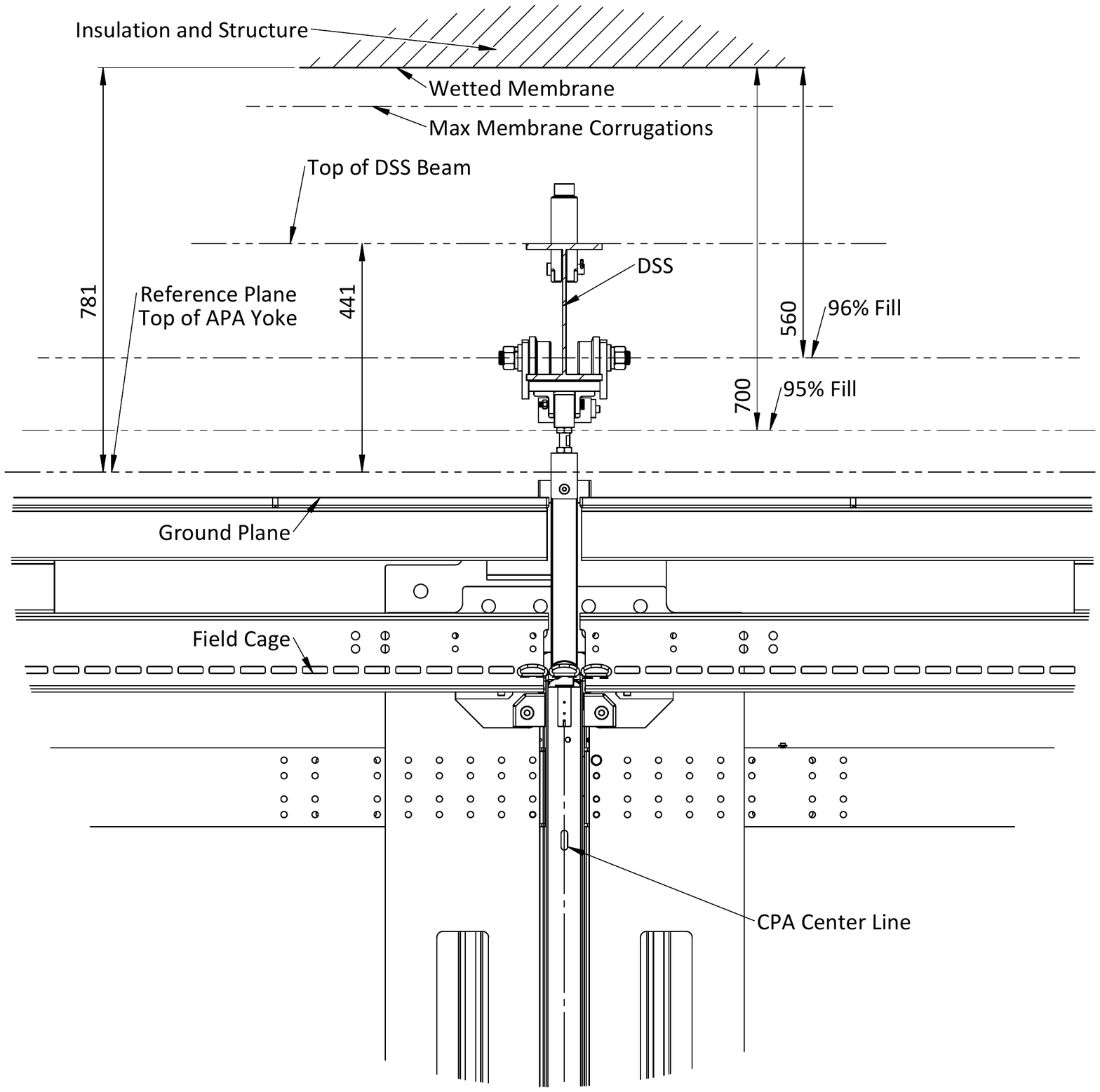}
\end{dunefigure}

These integration drawings are derived directly from the overall
integration model. The overall integration model is assembled from
component models developed by the consortia. Interfaces are controlled
by \dword{tc}, and consortia maintain their model files to be
compatible with the interfaces. During the design phase, models are
assembled and checked continuously. At the time of final design, all
interfaces will be finalized.

Component tolerances and installation clearances are managed through
additional models as described in
Section~\ref{sec:fdsp-coord-integ-envelope}.
Figure~\ref{fig:dune-apa_envelope} shows the \dword{apa}s and
\dword{cpa}s as well as their relative positions that show how they are
constrained within the detector.
\begin{dunefigure}[Envelope dimensions and
    installation clearances for APAs and CPAs (warm)]
    {fig:dune-apa_envelope} {\dword{spmod} graphical
    representation of envelope dimensions and installation clearances
    for \dword{apa}s and \dword{cpa}s in the warm state.}
  \includegraphics[width=0.95\textwidth]{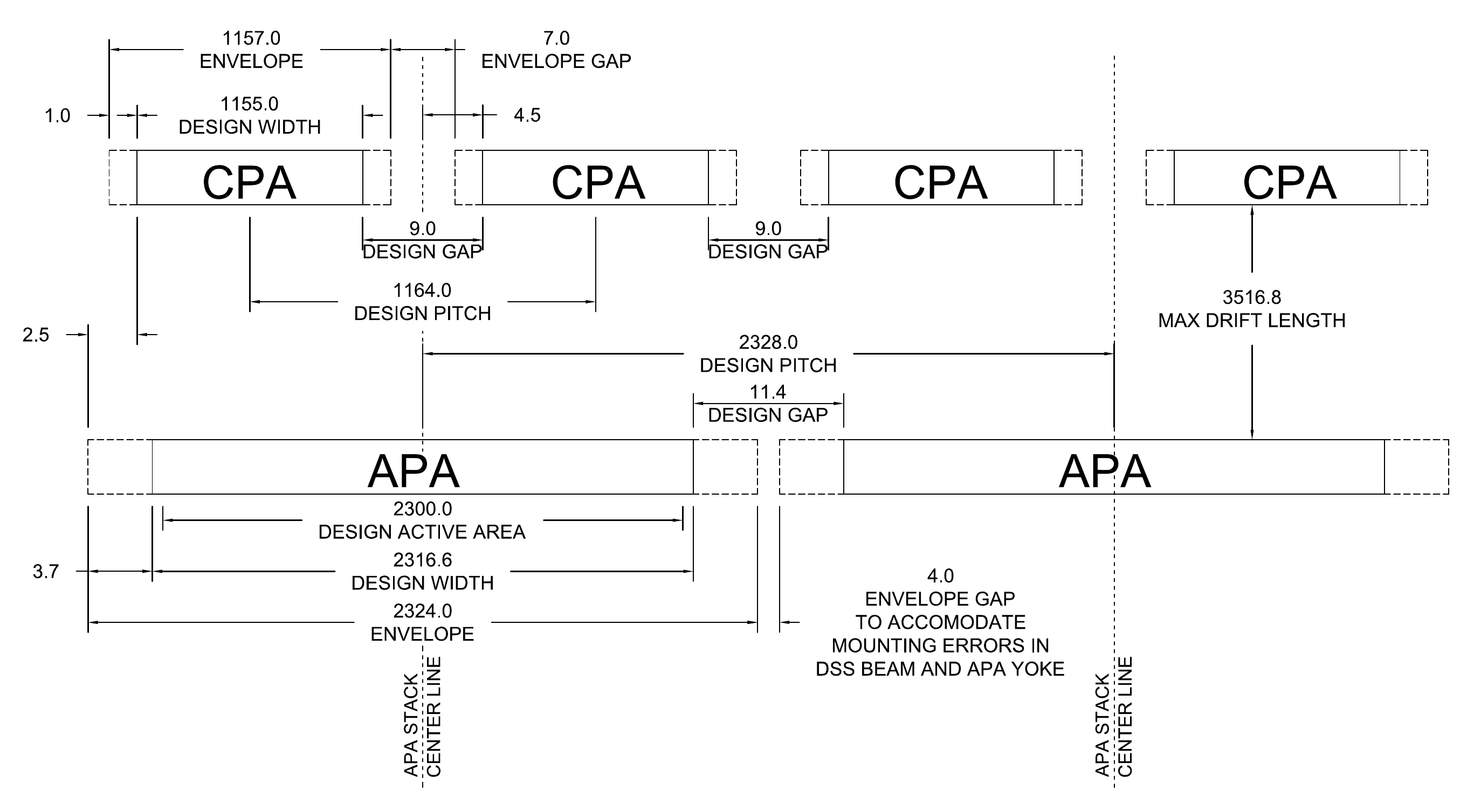}
\end{dunefigure}

Figure~\ref{fig:dune-apa_envelope} shows design dimensions. Component
tolerances and assembly tolerances for the upper and lower
\dword{apa} and \dword{cpa} stacks have been analyzed and are
represented as envelope dimensions. An envelope gap has been defined
to account for tolerances in the support system position and among
components. Taking all of these into account, pitch distances for
\dword{apa}s and \dword{cpa}s have been defined in the warm state.

This figure also shows the design drift distance in the warm
state. The drift distance is defined as the perpendicular distance
between the surface of \dword{cpa}s and the collection wire plane of
\dword{apa}s.

\dword{apa}s and \dword{cpa}s are supported in groups of two or three
on \dword{dss} beams. Fifty beams are arranged into five parallel rows
with 10 beams in each row.  In the cold state, the relative positions
between groups of \dword{apa}s that are supported on different beams
change due to thermal contraction of the beams. Relative positions
within each group supported on the same beam are relatively constant
since \dword{apa} frames and \dword{dss} are both made from stainless
steel.  The effect is that the gap in the active area between some
\dword{apa}s increases.  As can be seen in
Figure~\ref{fig:dune-apa_aa_cold}, in the warm state, the gap in the
active areas between adjacent \dword{apa}s is 28 mm (dotted line). In
the cold state, nine of the 24 gaps increase to approximately 45~mm.
\begin{dunefigure}[APAs active area gap in cold state]{fig:dune-apa_aa_cold} 
    {\dword{spmod} gap in active areas between adjacent \dword{apa}s
      in the cold state. Dashed line warm state, dots cold state.}
    \includegraphics[width=0.8\textwidth]{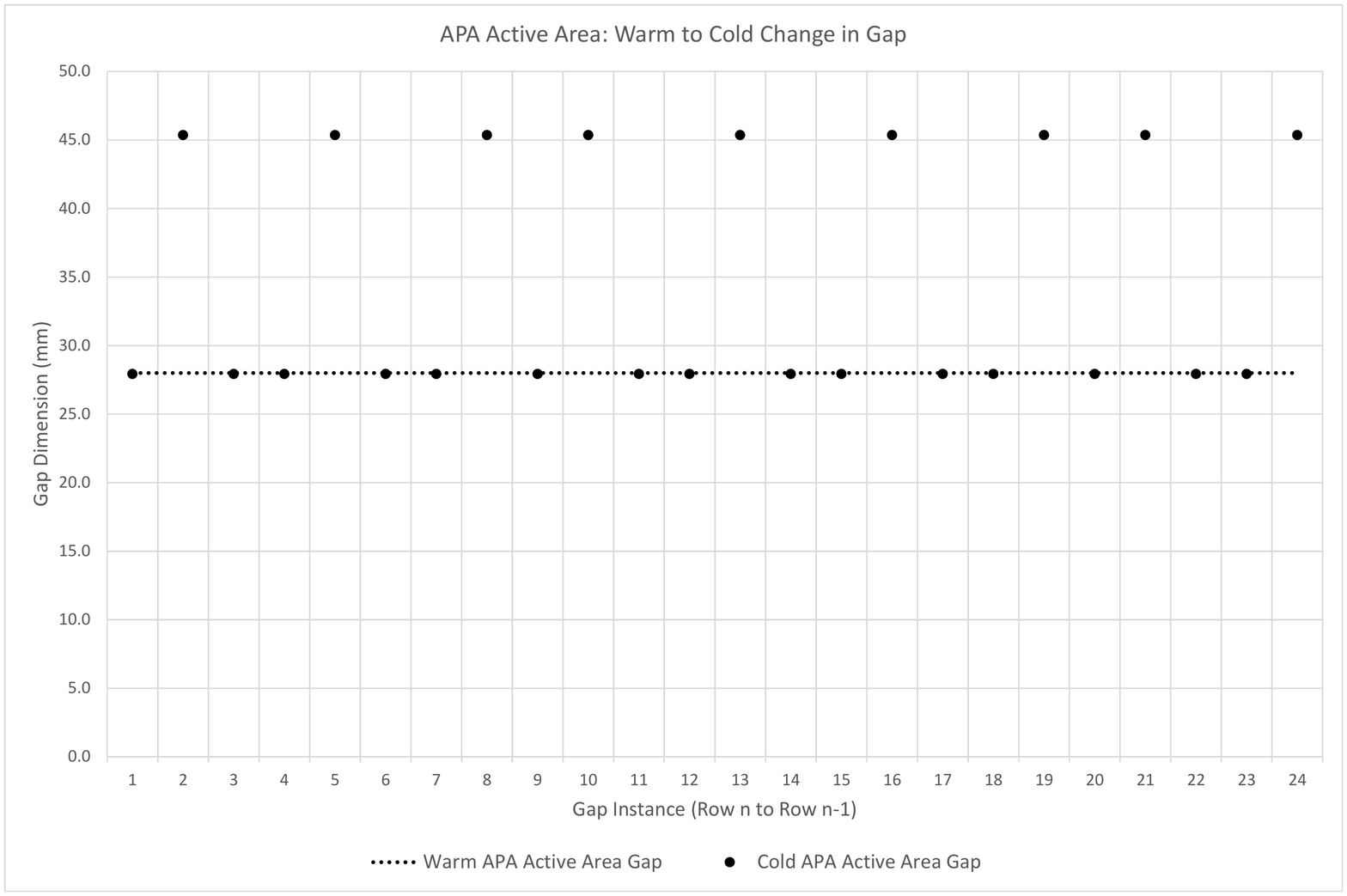}
\end{dunefigure}

The effects of gravity and buoyancy are not represented in the above
analysis. Such effects are under study and will be shown in the models
as design progresses.

Finally, Figure~\ref{fig:dune-floorpipes} shows the interface of the
cryostat service floor with other components.  Before installing the
\dword{detmodule}, a set of cryogenic distribution pipes are installed
on the floor of the cryostat. These as well as the corrugations of the
cryostat membrane would impede movement, hence the need for
a temporary service floor. It will be installed, and later removed, in sections. 
\begin{dunefigure}[SP module interface of cryogenics, service floor and detector]{fig:dune-floorpipes}
{\dword{spmod} interface of cryogenic distribution pipes, service
  floor, and detector.}
\includegraphics[width=0.8\textwidth]{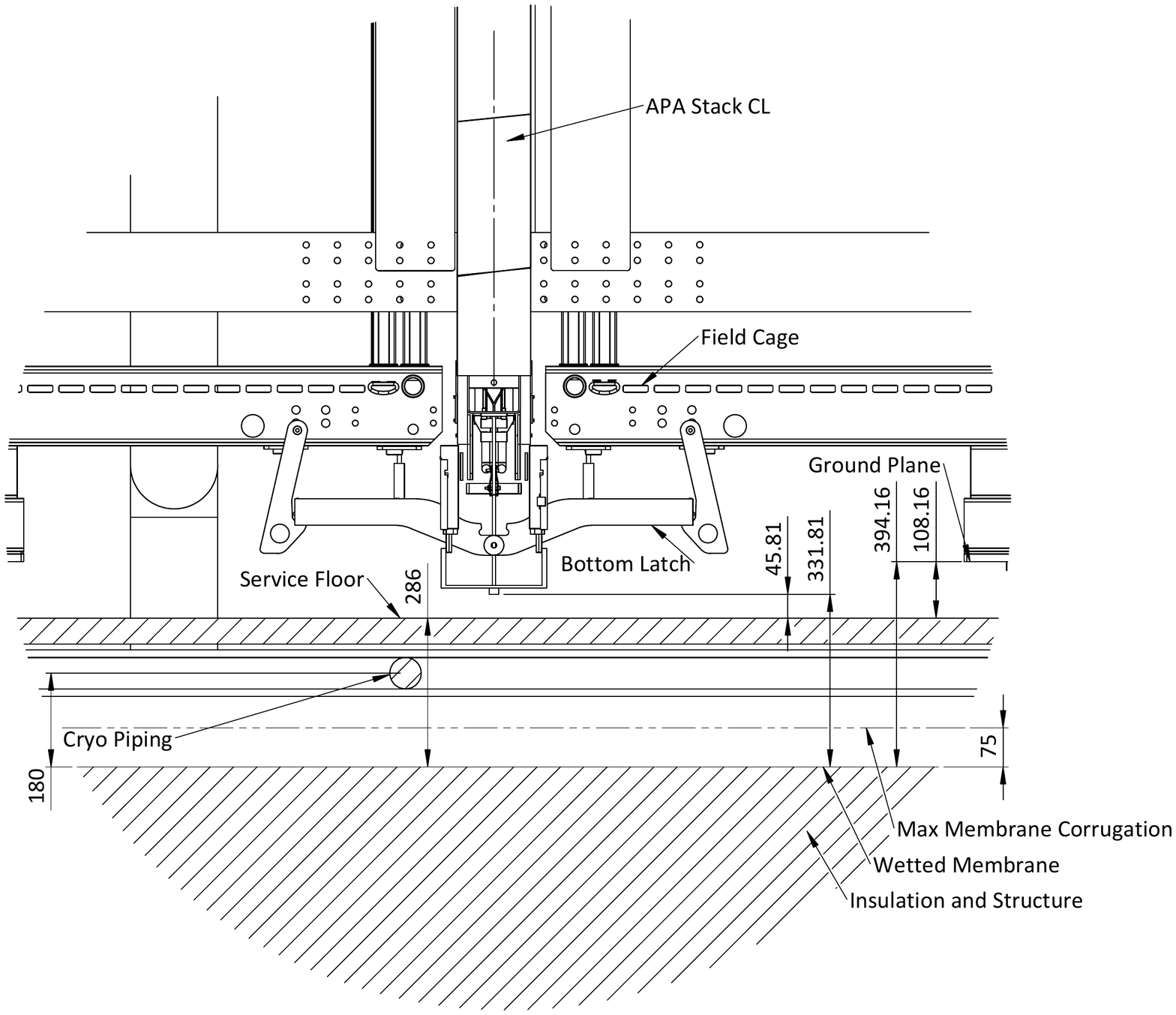}
\end{dunefigure}

\subsection{Detector Survey and Alignment}
\label{sec:fdsp-coord-integ-survey}

The requirement for detector placement within the cryostat is driven
by overall mechanical assembly needs rather than physics.
Interfacing parts must be assembled properly and function as intended.

In this section, reference frames for the detector are defined so that the
overall survey and alignment can be done within the cavern reference
frame.

For the \dword{spmod}, we define a flat and horizontal reference plane
coplanar with the upper \dword{apa} yoke plane; i.e., 75 yoke planes
define this plane. This reference plane is set at exactly \SI{781}{mm}
below the theoretical plane of the cryostat top membrane.

The detector reference plane is coplanar with the upper \dword{apa}
yoke planes because all features, including the active area, are
referenced to this plane. Once this reference plane is defined and
established through survey, all vertical distances within the detector
are referenced and established relative to it.
Figure~\ref{fig:dune-apa_interfaces_mid} shows the reference plane in
relation to the cryostat top membrane and \dword{dss}.

During installation, the height of the \dword{dss} beams is set in
accordance with this relationship. Adjustments are made in the
\dword{dss} to ensure that all the beams are in the correct plane. The
combined effects of gravity, buoyancy, temperature, and \dword{lar}
mass after fill are calculated, and further adjustments are made to
compensate.  This will ensure that the \dword{tpc} position remains as
close as possible to nominal after fill.

The transverse position of the detector is constrained to the center
of the cryostat. Thus, the mid-plane of the middle row \dword{apa} is
coplanar with the vertical mid-plane of the cryostat. This
relationship is verified when the \dword{dss} beams and central row of
\dword{apa}s are installed. The outer rows are similarly aligned and
surveyed with the offset as shown in Figure~\ref{fig:dune-sp_row}.
\begin{dunefigure}[Position of longitudinal
    reference point for the SP module]{fig:dss_feedthru}
  {Position of feedthrough determining the longitudinal reference point of the \dword{spmod}.}
  \includegraphics[width=0.9\textwidth]{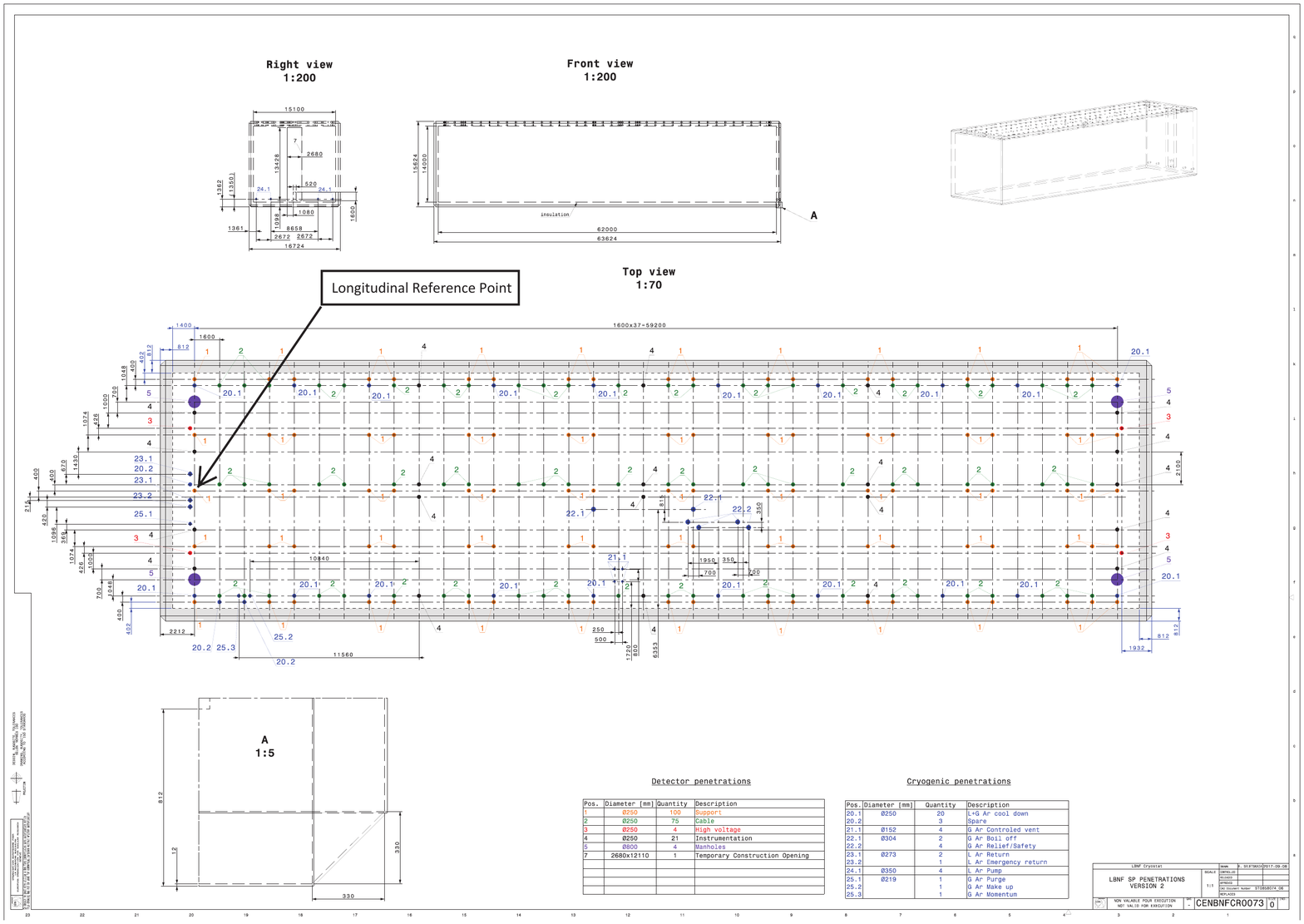}
\end{dunefigure}

The longitudinal reference point of the detector within the cryostat
is defined by the position of the single feedthrough of the central row
farthest from the cryostat opening. This feedthrough position
is shown in Figure~\ref{fig:dss_feedthru}.


\section{Electrical Integration}
\label{sec:fdsp-Integ-electrical}

\subsection{Electrical System Block Drawings, Schematics, Layouts and Wiring Diagrams}
\label{sec:fdsp-coord-electrical}

The \dword{integoff} is responsible for the AC power distribution supplied to
the experiment and to detector electronics racks.  This is described
in Sections~\ref{sec:fdsp-coord-faci-grounding}
and~\ref{sec:fdsp-coord-faci-power}.  Specific guidance has been
provided regarding the use of DC power supplies and cable shield
treatment~\cite{bib:cernedms2095958}. 
They are the same as were
followed at \dword{protodune} and developed during extensive testing
of the \dword{apa} wire readout at \dword{bnl} and \dword{protodune}.
Guidelines are included for the treatment of DC power supplies and the
use and connection of shielded cables.  All systems are reviewed for
compliance to these guidelines during the design review procedure.
Any deviation from the guidelines must be noted and approved by system
engineering.

The \dword{jpo} engineering team will review all electrical systems to ensure that they
follow safe design practice and will pass \dwords{orr} as discussed in
Chapter~\ref{vl:tc-review}.  Review by the \dword{jpo} engineering team will include vetting
of all power and ground paths, adherence to national electrical
standards or equivalents for all commercial equipment, and adherence
to the \dword{dune} electrical design rules.

The electrical design of each subsystem is described by a set of
documents that includes a system-level block diagram and a wiring
diagram that includes a complete description of all power and ground
connections.  Depending on what is being described, a complete set of
schematics, board production files, and wiring diagrams will be
reviewed and archived.  All designs are subject to electrical safety
review, as described above, before production proceeds. The safety
reviews proceed in conjunction with the overall review process as
described in Chapter~\ref{vl:tc-review}.

Consortia will produce system-level block diagrams. The \dword{tcoord}
will ensure that these diagrams are produced and reviewed.  In some
cases, multiple diagrams may be required, e.g., the \dword{cisc}
consortium is responsible for several types of systems (such as
temperature readouts, purity monitors, cameras, pressure sensors),
each requiring a separate diagram. A system-level block diagram should
show the conceptual blocks required in the design along with
connections to other conceptual block elements, both inside and
outside the given consortium.
Figure~\ref{fig:electrical_blockdiagram_example} shows an example of a
system-level block diagram.
\begin{dunefigure}[Sample system-level block diagram]{fig:electrical_blockdiagram_example}
  {Electrical system-level block diagram example (\dword{spmod} TPC electronics).}
 \includegraphics[width=0.85\textwidth]{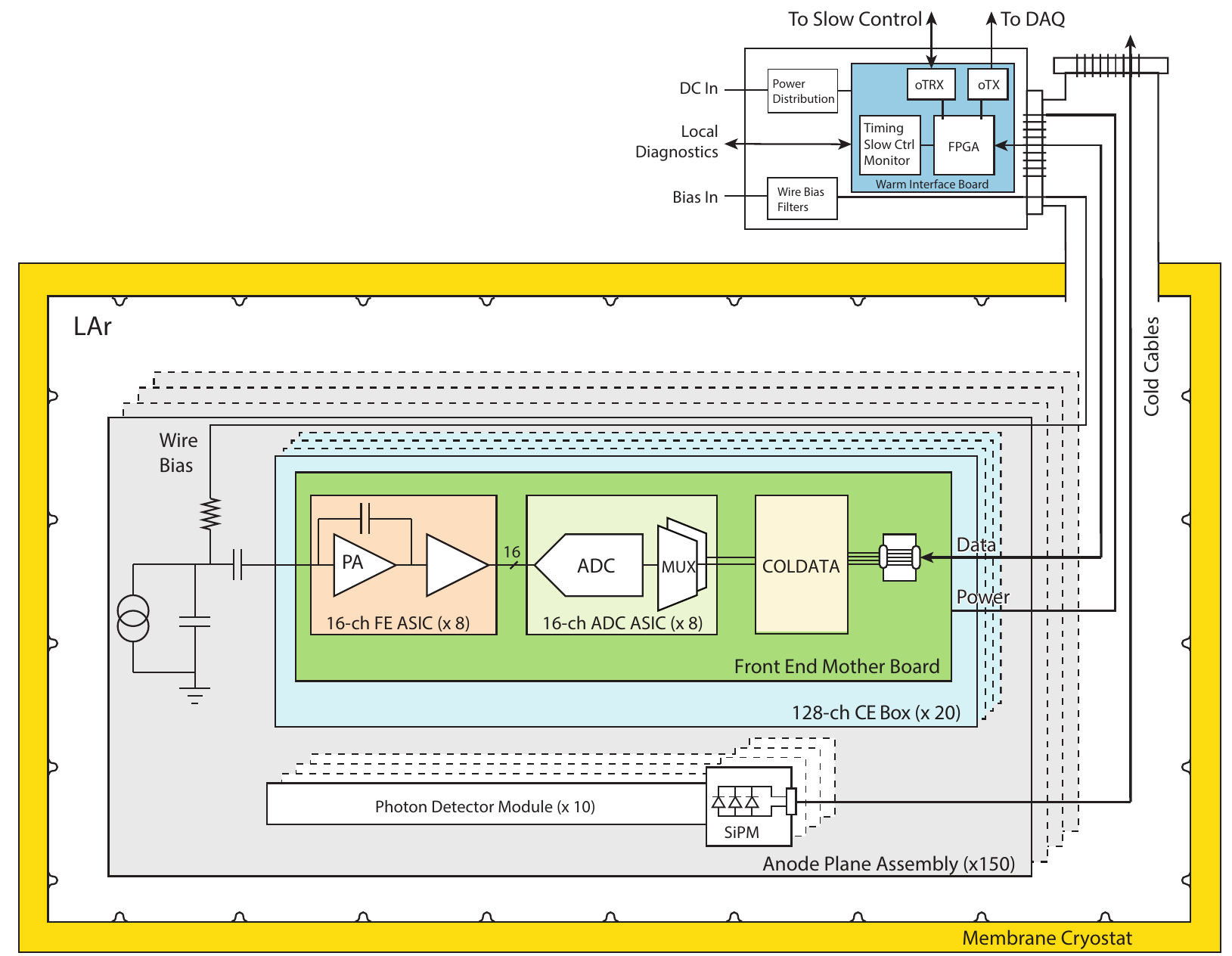}
\end{dunefigure}

All consortia must provide an electrical wiring diagram that 
represents the power and ground distribution within the system being
described.  The paths of power and ground distribution wiring between
circuit elements are specified along with wire types and sizes.  Power
elements like power supplies, fuses (or other protective circuit
elements), power connectors, and pin and wire ampacity are documented.

Electrical schematics show very specifically how individual
components are connected.  Usually, a schematic will represent a
\dword{pcb} design.  Schematics call out specific
parts that are used in the design and include all interconnections.
In the case of a \dword{pcb}, layout files, manufacturing
specifications, and bills of materials document
the design and allow a safety review of any custom boards or
modules.

Wiring diagrams include all wire and cable connections that run
between \dwords{pcb} or electronics modules.  Wires and
cables are described within the diagram and include identification of
\dword{awg}, wire color, cable specification, and
cable connectors and pinouts.

\subsection{Electrical Integration Documentation}
\label{sec:fdsp-coord-integ-electrical}

Interfaces that occur between subsystems of different consortia will
be documented and formally agreed upon between the technical leads of
the coordinating consortia and must be verified by the \dword{jpo} engineering 
team.  Much of the documentation required to describe a subsystem can
be used for the interface documents.

Documentation required for the interface between two electrical
subsystems includes a block diagram that identifies all connections
between the subsystems.  This block diagram must exist in the formal
interface document between consortia.  For each connection, additional
documentation must fully describe the interface details. This
additional detailed information can exist outside of the primary
interface document between the consortia, but that document must point
to it.

A signal cable that runs between \dwords{pcb} belonging to different
consortia is an example of an integration interface.  For a signal
cable, interface documentation includes the connector specifications,
pinouts at each end of the cable, and the pinout of the board
connectors.  Documentation would also describe relevant electrical
signal characteristics that may include signal levels, function,
protocol, bandwidth and timing information.  If different subsystems
refer to a given signal by different names, documentation of signal
name cross reference must be provided.

Consortium technical leads and the \dword{jpo} engineering team sign off
and approve the detailed documentation information on integration
interfaces not included in the primary consortium-to-consortium
interface document.

The \dword{jpo} engineering team will provide unique names and labels
for all racks, crates, boards, power supplies, cables, and any other
electrical type equipment.  A database will be created to track these
devices.

\section{Configuration and Drawing Storage and Dissemination}
\label{sec:fdsp-coord-integ-modelplan}

The consortia and \dword{jpo} engineering team create and share
drawings, models, schematics, production data, and all other
engineering documents. In addition, the \dword{jpo} engineering team
generates and shares all interface drawings and documentation.

Folders have been set up to allow uploading and sharing documents
with appropriate protection. The structure of the folders has been set
up to suit each consortium. The consortia do not necessarily have
similar folder structures or files  and will adapt the structure to fit
their needs.

The folders and files reside on the \dword{edms}. This system and
similar structures were used for \dword{protodune} and are being
used by \dword{lbnf}.

The following shows a high-level outline of the file structure. The
first section is for technical coordination files. The second
section is generic, intended for a consortium. Each consortium will have one such
folder.
\begin{enumerate}
 \item \dword{tc}  
 \begin{enumerate}
  \item mechanical drawings and files (controlled by \dword{dune} lead mechanical engineer)
  
  \begin{enumerate}
    \item \dword{fd} general drawings for illustration (controlled by \dword{tcoord}),
    \item \threed model files of internal detector for periodic upload to global model,
    \item \twod interface drawing files,
    \item alignment and survey files,
    \item \dword{ashriver} installation test facility files,
    \item \dword{qa}/\dword{qc} files,
    \item safety analysis and documentation, and 
    \item design reviews;
  \end{enumerate}
  
  \item electrical and electronics (controlled by lead electrical engineer of \dword{dune})
  \begin{enumerate}
    \item infrastructure requirements for grounding,
    \item consortium interface drawings,
    \item detector electronics grounding guidelines,
    \item detector safety system,
    \item \dword{qa}/\dword{qc} files,
    \item safety analysis and documentation, and 
      \item design reviews;
  \end{enumerate}
  
 \end{enumerate}

 \item consortium files (one per consortium, controlled by consortium technical leads)   

 \begin{enumerate}
   \item \threed model files,
   \item \twod part drawing files,
   \item production files,
   \item general grounding diagrams,
   \item system level block diagrams,
   \item system level wiring diagrams,
   \item software and firmware plans,
   \item custom components, such as \dwords{asic} (one folder per component),
   \item \dword{pcb} components (one folder per component),
   \item cable components (one folder per component), and 
   \item power supply components (one folder per component).
 \end{enumerate}

\end{enumerate}

An image of the \dword{edms} file structure is shown in
Figure~\ref{fig:config_structure}.
\begin{dunefigure}[\dshort{edms} file structure]{fig:config_structure}
  {\dword{edms} file structure.}
  \includegraphics[width=0.4\textwidth]{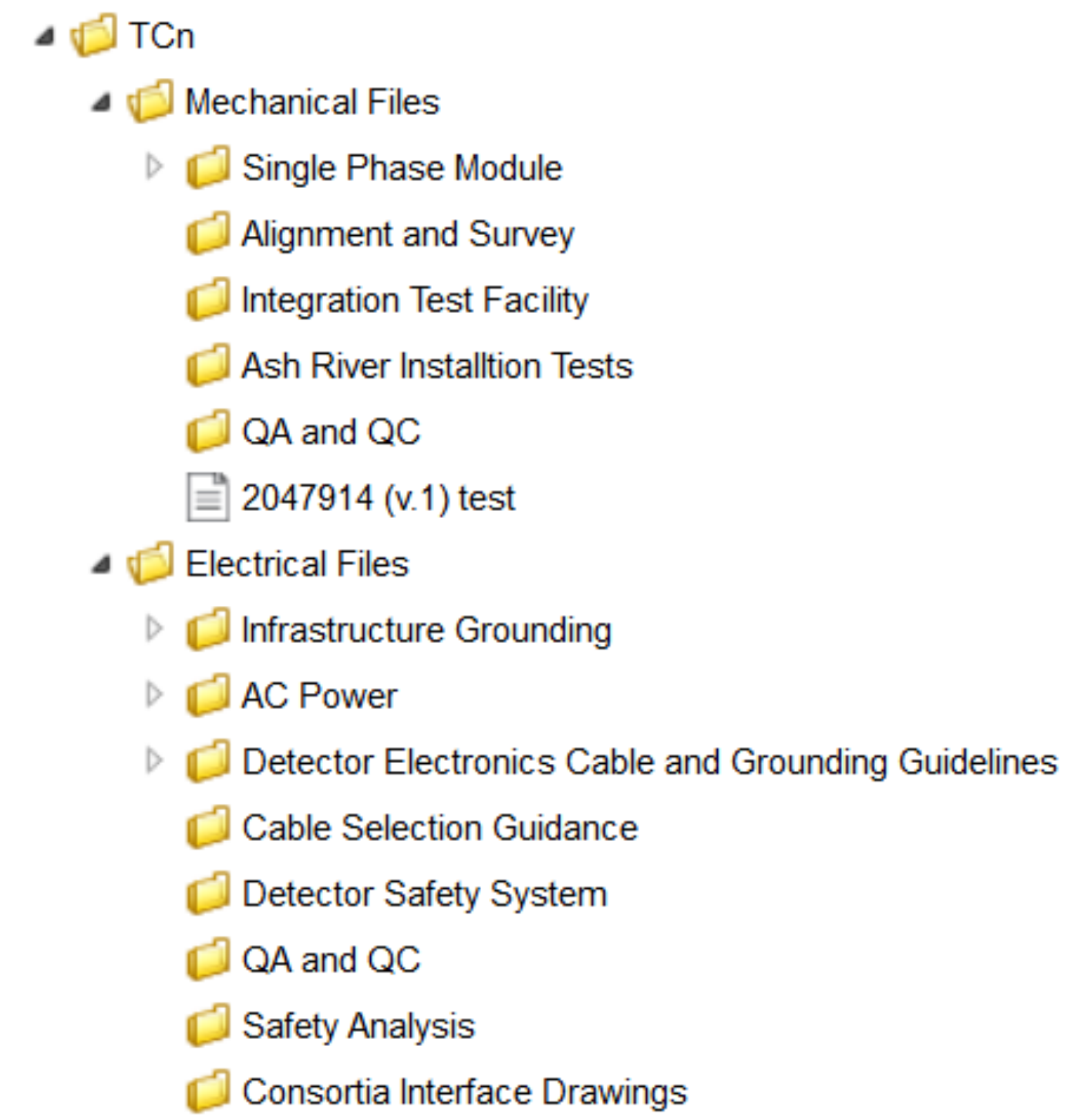}
\end{dunefigure}

\section{Organization of Interfaces and Interface Documents}
\label{sec:fdsp-coord-integ-interface}

An integration mechanism has been developed to manage and create an
overall model of interfaces both within a \dword{detmodule} and
between a \dword{detmodule} and facilities. The mechanism defines
integration nodes, which can be thought of as focus areas, as
explained below.  The \dword{jpo} engineering team carries out and
manages interfaces between the nodes. The integration nodes comprise the following:
\begin{itemize}
\item {\bf detector:} This consists of all \dword{tpc} elements within
  the \dword{lar}. Almost all consortia are involved in this
  integration task, which is mostly mechanical. Consortium engineering
  teams work directly with the \dword{tc} engineering team.  The
  primary interface is with \dword{dss} through hangers. Examples of
  interfaces within this node include  \dword{fc} connections to both
  \dword{cpa}s and \dword{apa}s, \dword{ce} and \dword{pd} cable
  routing within the cryostat, and location of calibration, and
  cryogenics instrumentation.
\item {\bf \dword{dss}:} This consists of all detector support elements,
  cable trays, and feedthroughs.
\item {\bf detector electronics:} This consists of all racks, cooling,
  power, cable trays, cable distribution on top of the cryostats, rack
  protection (smoke detectors, hardware power trip), rack component
  build, and the interface to the \dword{ddss}.
\item {\bf \dword{daq} and electronics:} This includes electronics on
  top of the cryostat, in the \dword{daq} room and in the surface
  rooms. This also includes the fiber optic distribution from the
  surface to the \dword{daq} room, and the fiber optic distribution
  from the \dwords{detmodule} to the \dword{daq} room. It also includes the
  layout and cooling of the \dword{daq} room.
\end{itemize}

\begin{dunefigure}[Integration nodes]{fig:integration_nodes}
  {Overall integration nodes and interfaces.}
  \includegraphics[width=0.7\textwidth]{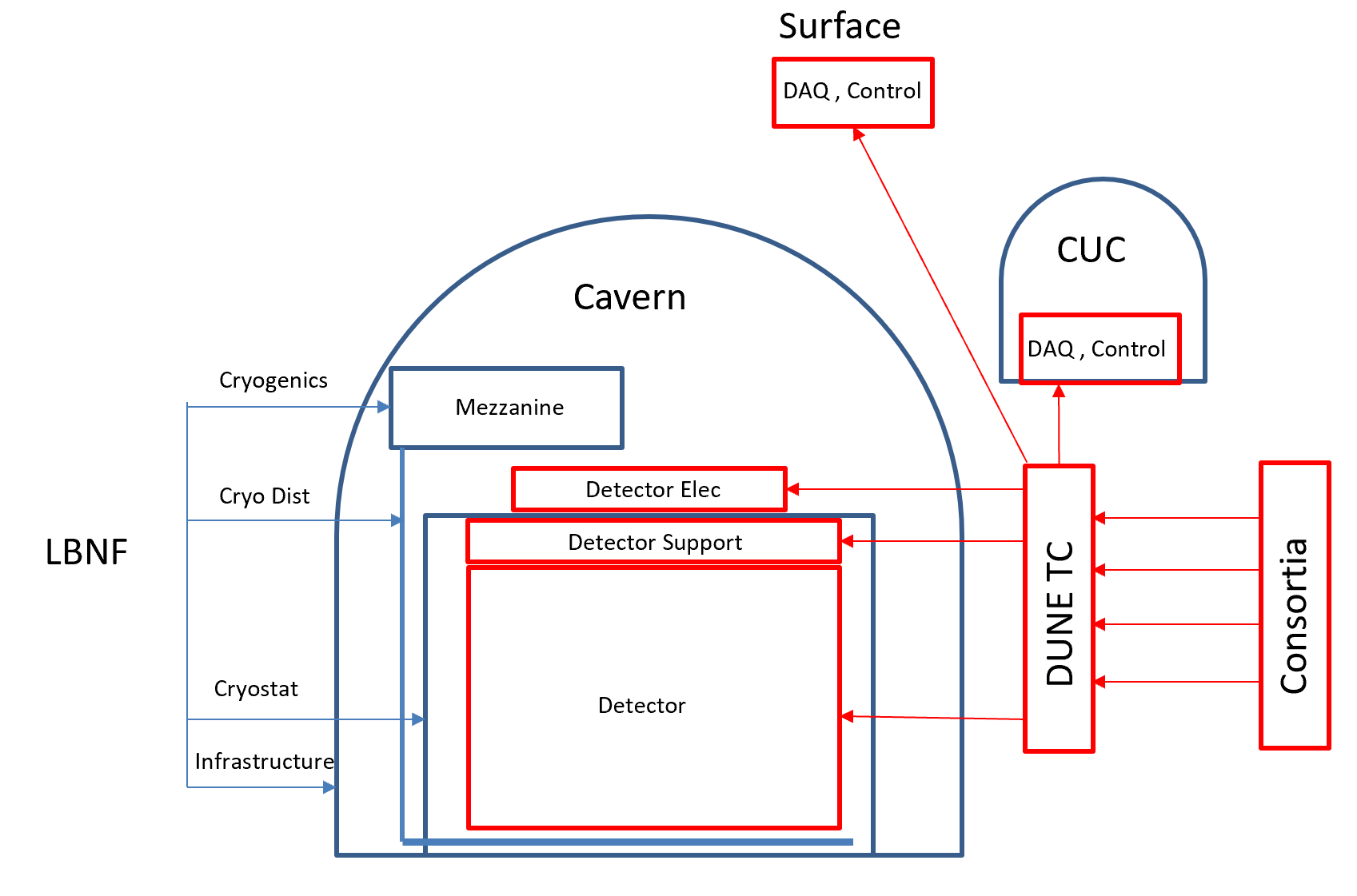}
\end{dunefigure}
Figure~\ref{fig:integration_nodes} shows the interfaces between the
detector and facilities. In this figure, within the cavern, items
provided by \dword{lbnf} are on the left and the items provided by
\dword{dune} are on the right. In addition, the \dword{jpo} engineering
team integrates (ensures that the interfaces are appropriately defined
and managed) the \dword{daq} room in the \dword{cuc} and surface
control and network rooms. Interfaces with \dword{lbnf}
are managed at the boundaries of each integration node. As an example,
the interfaces between \dword{lbnf} and \dword{dune} for the
underground \dword{daq} and control rooms are power, cooling water,
data fibers, and cable penetrations at the room
boundaries. \dword{dune} is responsible for implementing power,
cooling water, data, and signal cables, as well as integrating the
racks.

Interface documents are developed and maintained to manage the
interfaces between consortia and between each consortium and the
\dword{tcoord}. A single document covers the interface between any two
systems, so any one system may have several interface documents. If no
interface exists, the interface document is not provided.  The
interface documents are managed by the appropriate consortium
technical leads and by \dword{tc} project engineers.

The content of interface documents varies depending on the type
of interface. However, the documents are intended to have
a common structure: 

\begin{enumerate}
 \item Definition: defines the interfacing systems.
 \item Hardware: defines the interfacing hardware components,
   electrical and mechanical, in general terms. As an
   example, the \dword{apa} frame needs to support the \dword{pd}
   mounting brackets.
 \item Design: describes the dependencies in design
   methodology, sequence, and standards. As with the
   previous example, the design of the \dword{pd} mounting
   brackets depends on the side tubes chosen for the \dword{apa}.
 \item Production: details responsibilities for component production
   and overall assembly, which may be shared among interfacing
   systems.
 \item Testing: details responsibilities for testing the required
   equipment. Like production, testing is a shared responsibility.
 \item Integration: defines the integration of systems into installable units
   before insertion into the cryostat. 
   It also defines the location, methodology, tooling, and environment for integration.
 \item Installation: defines installation tasks and responsibilities, once
   installable units are assembled, and  defines  
   special transportation or installation tools or fixtures. 
 \item Commissioning: defines overall responsibilities for
   commissioning tasks  and sets parameters. 
 \item Data format, control and error codes: Communications protocols,
   responses and necessary actions are defined.
 \item Appendices: includes technical figures and interfaces 
   in as much detail as necessary. These must include block diagrams that
   show interconnections and detailed documentation of each connection.
\end{enumerate}

The interface documents will be developed and modified during the
technical design period. At the time of this writing, not all
documents have been fully developed. Once the technical design is
finished, the interface documents will be placed under revision
control. A summary of the interface documents is provided in
Section~\ref{sec:fdsp-coord-interface}.

\section{Engineering Change Control}
\label{sec:fdsp-change}

Changes in design and fabrication requirements follow revision
processes for design and fabrication documents, per individual
consortium practice, while ensuring appropriate levels of
verification, review, and approval by the consortium design authority.

Either a consortium or \dword{tc} may initiate and request
changes in design and fabrication requirements that involve interfaces among detector subsystems and between detector subsystems and the facilities.  The
\dword{jpo} engineering team directs requests for changes through the
\dword{tb}. Any change that affects cost or schedule must be approved
by the \dword{exb}.

When shop or site work must be performed before a given 
design
document under configuration management can be formally revised and re-issued, 
an \dword{ecr} must be developed, approved, and distributed. 
Inter-discipline reviews will be
performed when the \dword{ecr} subject matter may impact other
subsystems. The design authority will indicate if it is a one-time
change or if the change is to be incorporated into the design
documents.

\section{Value Engineering}
\label{sec:fdsp-coord-ve}

Value engineering is the process of arriving at cost-effective
solutions to the technical challenges of building the \dword{dune}
detector. \dword{dune} value engineering builds on significant
developments in \dword{lar} detectors dating to the early 1970s,
especially the large \dwords{lartpc}: \dword{icarus} and
\dword{microboone}. Prototyping by both \dword{lbne} and \dword{lbno} has
significantly advanced the value engineering process, leading to
construction of the \dword{protodune} detectors. These detectors validate
\dword{dune} designs and confirm that the necessary performance is
met. Any significant departure from current designs must account for
the success of \dword{protodune}  and may require testing in a second
run of \dword{protodune} to validate that the change does not degrade performance. 

The value engineering process is executed at both the consortium and
\dword{tc} level. For example, the \dword{apa} size has been optimized
over the last 10--15 years, using input from dimensions of the Ross
Shaft, shipping container size, and availability of high-quality long
stainless steel tubes from reliable vendors.  At this point, any
significant change to the \dword{apa} design would likely lead to new
and significant re-engineering costs.

Value engineering is ongoing at all stages of design and will continue
through the fabrication, assembly, and installation phases. In
particular, during the fabrication and assembly stages, when labor costs
are relatively higher, this process can result in significant cost savings. The
consortia and \dword{tc} are actively engaged and have the necessary
experience for this ongoing process.

\cleardoublepage

\chapter{Reviews}
\label{vl:tc-review}

The \dword{integoff} and \dword{tc} review all stages of detector development
and work with each consortium to arrange reviews of the design
(\dword{cdrev}, \dword{pdr} and \dword{fdr}), production (\dword{prr}
and \dword{ppr}), installation (\dword{irr}), and operation
(\dword{orr}) of their system. The reviews are organized by the
\dword{jpo} \dword{ro}. In parallel, the \dword{ro}  reviews all
stages of \dword{lbnf} cryostat and cryogenics development.  These
reviews provide information to the \dword{tb}, \dword{exb}, and \dword{efig}
in evaluating technical decisions. A timeline for the review process
is shown in Figure~\ref{fig:review_timeline}.
\begin{dunefigure}[DUNE review process]{fig:review_timeline}
  {\dword{dune} review process and timeline}
  \includegraphics[width=0.75\textwidth]{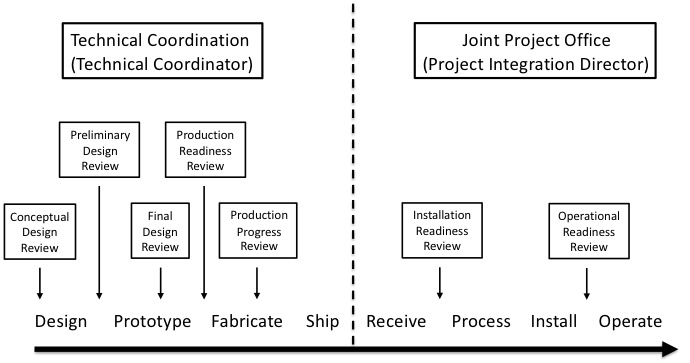}
\end{dunefigure}
Review reports are tracked by the \dword{jpo} \dword{ro} and \dword{tc} and provide
guidance on key issues that require engineering oversight by the
\dword{jpo} engineering team. The \dword{ro}  maintains a
calendar of \dword{dune} reviews. The calendar of reviews is developed
by the \dword{ro}  in consultation with the \dword{ipd},
\dword{tcoord}, \dword{lbnf} project manager, consortia leads, and
\dword{lbnf} subsystem managers.

\Dword{tc} works with consortia leaders to prepare for reviews of all detector
designs.  As part of the \dword{tdr} development \dwords{pdr} were
arranged for many subsystems in advance of the \dword{tdr}. Some
remaining subsystem \dwords{pdr} will occur after the \dword{tdr}. All
subsystems will undergo \dwords{fdr}.  All major technology decisions
will be reviewed before down-select.  \Dword{tc} may form task forces
as needed to address specific issues that require more in depth
review.

\Dword{tc} works with consortia leaders to prepare for review of
detector component production processes.  Production of detector
elements begins only after successful \dwords{prr}. Regular production
progress reviews will be held once production starts depending on the
length of the production process. The \dwords{prr} will typically
include a review of the production of \textit{Module 0}, the first
module produced at the facility. \Dword{tc} will work with consortia
leaders on all production reviews.

The \dword{integoff} works with consortium leaders to prepare for reviews of
detector installation processes 
and later of the installed detector components to
ensure that they are ready for operations.  \Dword{tc} coordinates
technical documents for the \dword{lbnc} \dword{tdr} review.

The review process is an important part of the \dword{dune} \dword{qa}
process, as described in Section~\ref{sec:verification}, for
design and production.

The review process has been in place since 2016 with various reviews
of \dword{protodune} components and has continued into the first
\dword{dune} reviews in 2018--19. Past and scheduled reviews are in
the \dword{dune} Indico at https://indico.fnal.gov/category/586.
Review reports are currently maintained in~\citedocdb{1584}.

\section{Design Reviews}

The \dword{dune} design review process is described
in~\citedocdb{9564}. An updated mandate for the \dword{ro}  is under
development at~\cite{bib:cernedms2173197}. 
Design reviews
for \dword{protodune} were held for each major system. Because the
schedule was extremely tight for \dword{protodune}, only a single
design review was held for each major system. Similarly, a single
\dword{prr} was held for each major system.

The successful operation of \dword{protodune} means \dword{dune} is at
a very advanced state of design. The strategy going forward has been
to hold \dwords{cdrev} for systems with significant changes from
\dword{protodune}, including the \dword{dss}, \dword{pds},
\dword{daq}, and calibration. These are systems that require changes
due either to the size difference between \dword{protodune} and \dword{dune}
or the fact that the former is in a test beam and the latter is underground. All
systems will go through \dwords{pdr} to review design changes from
\dword{protodune} and \dwords{fdr} after the \dword{tdr}.

\Dword{tc} has established an engineering safety committee with
mechanical and electrical engineering experts from collaborating
institutions to develop processes and procedures to evaluate
engineering designs using accepted international safety standards. The
current status of international code equivalencies is discussed
further in Section~\ref{sec:esh_codes}. The codes and standards to
which each system is designed will be reviewed as part of the
\dword{pdr} and \dword{fdr}.

\section{Production Reviews}

Once the designs are finished, production reviews will be held before
significant funds are authorized for large production runs. These
reviews are closely coordinated with the \dword{qa} team. The
expectation is that a \textit{Module 0} be produced and presented as
part of the \dword{prr}. The \textit{Module 0} is the first article
from the production line and provides a useful indication of the
validity of production processes, time estimates, and quality of the
product.

Once production has started, the \dword{ro}  will schedule \dwords{ppr} as
appropriate to monitor production schedule and quality. These reviews
will consist of site visits and will include membership from the
\dword{esh} and \dword{qa} teams.

The \dword{prr} process was exercized during \dword{pdsp}
construction. Because the schedule was extremely tight for
\dword{protodune}, only a single \dword{prr} was held for each major
system, and no \dwords{ppr} were held.

\section{Installation Reviews}

\Dwords{irr} are planned to verify that equipment and procedures are in
place prior to installation of detector components. These will
review \dword{qc} results to verify that as-built detector
components can be successfully installed and operated. A critical part
of these reviews is to establish and verify the \dword{ha} for the
installation activities and mitigate any identified safety risks.  These reviews
will include safety personnel from \dword{surf} and \dword{fnal} as
appropriate.

\section{Operations Reviews}

The \dword{ro}  will conduct an \dword{orr} on subsystems before they are operated.  \Dwords{orr} serve as the final safety check after the 
equipment is installed. These reviews will include safety personnel from \dword{surf} and
\dword{fnal}, as appropriate. 

\section{Review Tracking}

Tracking and controlling review recommendations is part of the review
process. Review committees assess recommendations from earlier
reviews. The \dword{ro}  assures that consortia respond to review
recommendations and works with the consortia to make sure the
responses are appropriately documented and implemented. Reports from
\dword{dune} reviews are maintained in~\citedocdb{1584} along with the
list of recommendations. The \dword{ro}  reports to the \dword{ipd},
\dword{tcoord}, and \dword{lbnf} project manager on recommendations
and progress towards completing actions on these recommendations.

\section{Lessons Learned}
\label{sec:fdsp-coord-lessons}

A detailed list of lessons learned from construction and operation of
\dword{pdsp} is in~\citedocdb{8255}. These lessons have driven planning for
\dword{dune} and have led to design changes in \dword{dune}. Lessons
learned will continue to be updated throughout the design review process
and into production. The methodologies are described in
Section~\ref{sec:quality_improvement}. 

\section{Reporting}
\label{sec:fdsp-coord-reporting}

The \dword{dune} project has published regular monthly reports since
the final design and construction of \dword{protodune} began in
earnest in summer 2016. \Dword{tc} currently plans to continue to compile and
publish these reports. The \dword{dune} project provides
regular reports to the \dword{lbnc} at reviews several times a
year. The \dword{jpo} \dword{ro}  produces reports from design,
production, installation, and operations reviews.

\cleardoublepage

\chapter{Quality Assurance}
\label{vl:tc-QA}

\section{Overview of DUNE Quality Assurance}

\dshort{dune} \dword{tc} monitors technical contributions from
collaborating institutions and provides centralized project
coordination functions. One part of this project coordination is
standardizing \dfirst{qa}/\dfirst{qc} practices, one facet
of which is to assist consortia in defining and implementing
\dword{qa}/\dword{qc} plans that maintain uniform, high
standards across the entire detector construction
effort. Figure~\ref{fig:fnal_qa} shows how \dword{dune} \dword{tc}
derives its \dword{qa} program from the principles of the \fnal \dword{qa} program:
requirements are flowed down through the \dword{lbnf-dune}
\dword{qa} program into the \dword{qc} plans developed for consortium fabrication of
detector components and integration and installation of the detector.
\begin{dunefigure}[\fnal QA]{fig:fnal_qa}
  {Flow-down of \fnal \dword{qa} to consortia}
  \includegraphics[width=0.85\textwidth]{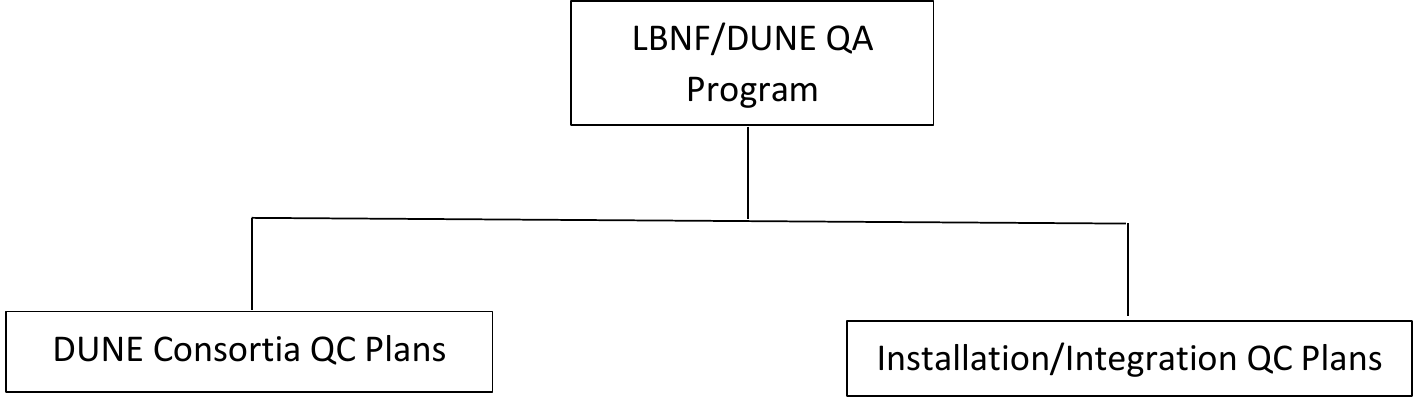}
\end{dunefigure}
The \dword{qa} effort includes design, production readiness, and
progress reviews as appropriate for the \dword{dune} detector
subsystems, as was done for \dword{pdsp} under \dword{tc}
oversight. Installation and operations reviews fall under \dword{integoff}
oversight as is discussed in Chapter~\ref{vl:tc-review}.

\subsection{Purpose}

The primary objective of the \dword{lbnf-dune} \dword{qa} program is
to assure quality in the construction of the \dword{lbnf} facility and
\dword{dune} experiment while providing protection of
\dword{lbnf-dune} personnel, the public and the environment. The
\dword{qa} plan aligns \dword{lbnf-dune} \dword{qa} activities, which
are spread around the world, with the principles of the \fnal Quality
Assurance Manual. The manual identifies the \fnal Integrated Quality
Assurance Program features that serve as the basis for the
\dword{lbnf-dune} \dword{qa} plan.

The \dword{lbnf-dune} \dword{qa} plan outlines the \dword{qa}
requirements for all \dword{lbnf-dune} collaborators and
subcontractors and describes how the requirements will be met.
\Dword{qa} criteria can be satisfied using a graded approach. This
\dword{qa} plan is implemented by the development of quality plans,
procedures, and guides by the consortia to accommodate those specific
quality requirements.

\subsection{Scope}

The \dword{lbnf-dune} \dword{qa} plan provides \Dword{qa}
requirements applicable to all consortia, encompassing all activities
performed from research and development (R\&D) through fabrication and
component commissioning, building on the success of
ProtoDUNE. Consortia are responsible for providing their deliverables,
whether subsystems, components, or services in accordance with
applicable agreements. All parties are responsible for
implementing a quality plan that meet the requirements of the
\dword{lbnf-dune} \dword{qa} plan. Oversight of the work of
the consortia will be the responsibility of the \dword{dune}
\dword{tcoord} and \dword{lbnf-dune} \dword{qa}
manager.

\subsection{Graded Approach}

A key element of the \dword{lbnf-dune} \dword{qa} plan is the
concept of graded approach; that is, applying a level of analysis,
controls, and documentation commensurate with the potential for an
environmental, safety, health, or quality impact. The graded approach
seeks to tailor the kinds and extent of quality controls applied in
the process of fulfilling requirements. Application of the graded
approach entails
\begin{itemize}
  \item identifying activities that present significant \dword{esh}
    and/or quality risk,
  \item defining the activity,
  \item evaluating risk and control choice, and 
  \item documenting and approving the application of the graded
    approach.
\end{itemize}

\section{Quality Assurance Program}

The \dword{lbnf-dune} Systems Engineering teams maintain a
\dword{lbnf-dune} \dword{cmp}~\cite{bib:docdb82}, which identifies the \dword{lbnf} project
Configuration Items Data List (CIDL) and Interface Control matrices
that provide the tier structure for the flow down of \dword{qa} plans,
with the \dword{lbnf-dune} \dword{qa} plan as the top tier.

With the
assistance of the \dword{lbnf-dune} \dword{qa} manager, the consortia will develop specific \dword{qa} plans  for
component or system \dword{qa}. Due to the limited scope of
work of some consortia, they may elect to work under the
\dword{lbnf-dune} \dword{qa} plan for their scope of work. In
case of conflict between sets of \dword{qa} requirements, \dword{dune}
\dword{tc} will provide resolution.

With many institutions carrying responsibility for various aspects of
the project, institutional \dword{qa} plans will be reviewed by
\dword{dune} \dword{tc} to ensure compliance with the
\dword{lbnf-dune} \dword{qa} plan. Using a graded approach,
supplements to institutions existing plans will be implemented for
their \dword{dune} scope of work, if necessary.

Overall \dword{qa} supervision, including all activities described
above, is the responsibility of the \dword{dune} \dword{tcoord}.

\subsection{Responsibility for Project Management}

The \dword{dune} consortium leaders manage their projects and are
responsible for achieving performance goals. The
\dword{lbnf-dune} \dword{qa} manager is responsible for
ensuring that a quality system is established, implemented, and
maintained in accordance with requirements. The
\dword{lbnf-dune} \dword{qa} manager reports to the
\dword{dune} \dword{tcoord} and provides oversight and support to
consortium leaders to ensure a consistent quality program.

\dword{dune} consortium leaders are responsible for quality within
their project and report \dword{qa} issues to the \dword{dune}
\dword{tcoord} and \dword{lbnf-dune} \dword{qa}
manager. \dword{dune} consortium leaders designate \dword{qa}
representatives within their organization and delegate, as appropriate, 
work defined in the \dword{lbnf-dune} \dword{qa} plan, as
shown in Fig.~\ref{fig:dune_qa}.
\begin{dunefigure}[DUNE QA organization]{fig:dune_qa}
  {\dword{dune} \dword{qa} organization.}
  \includegraphics[width=0.75\textwidth]{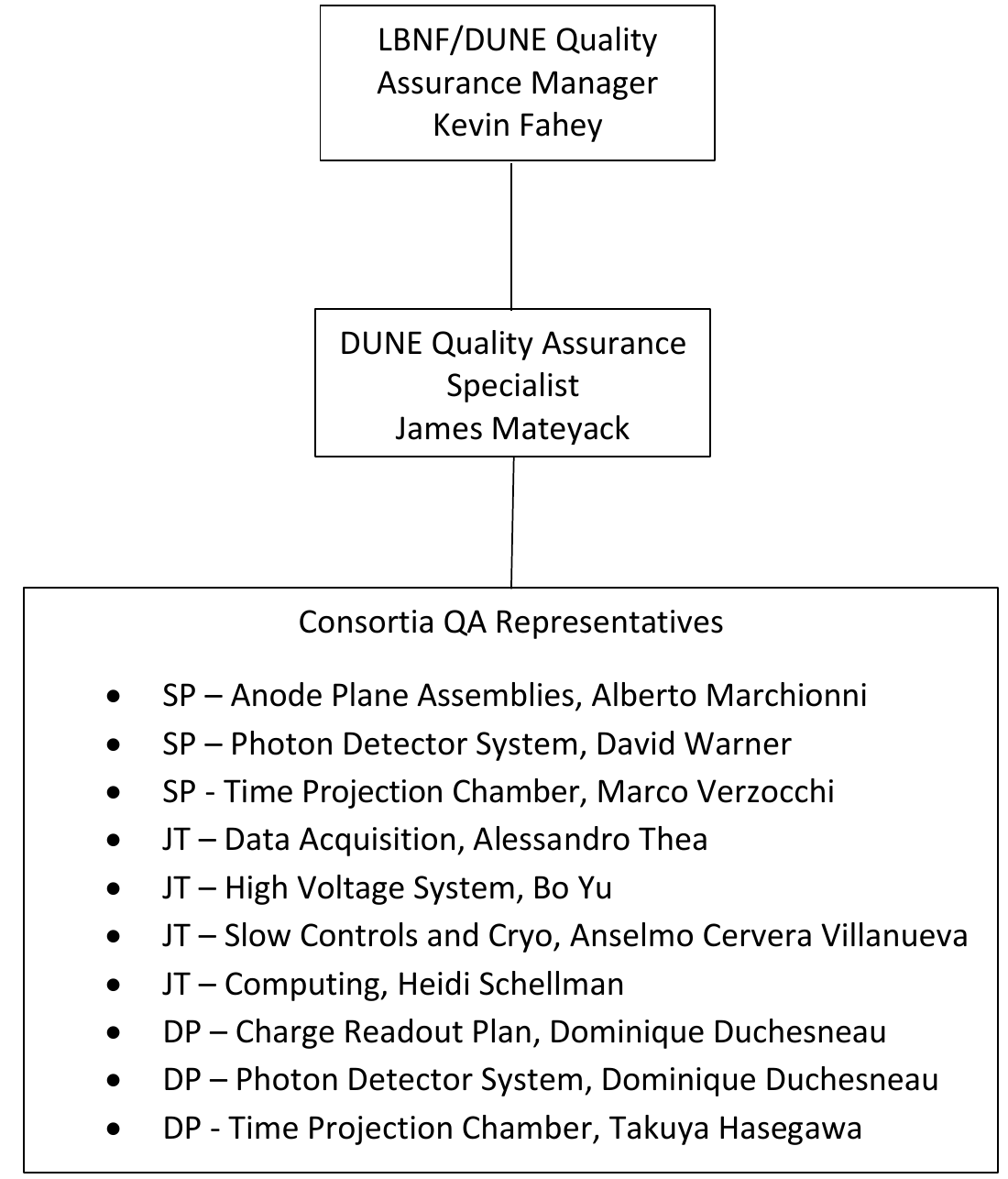}
\end{dunefigure}
The \dword{dune} consortium leaders retain overall responsibility for
\dword{qa} even though they have designated a \dword{qa}
representative.

\subsection{Levels of Authority and Interface}

The \dword{dune} Management Plan, the \dword{lbnf-dune} PMP
and the \dword{lbnf-dune} \dword{qa} plan define the
responsibility, authority and interrelation of personnel who manage,
perform, and verify work that affects quality. The \dword{qa} plan
defines the \dword{qa} roles and responsibilities of the \dword{dune}
project.

All consortium members are responsible for the quality of the work that
they do and for using guidance and assistance that is available. Each
has the authority to stop work and report adverse conditions that
affect quality of \dword{dune} products to their respective
\dword{dune} consortium leader and the \dword{lbnf-dune}
\dword{qa} manager. The consortium leader responsible for \dword{dune}
components or systems is required to determine and document their
acceptance criteria. \dword{dune} personnel at each level are
responsible for evaluation of quality through self-assessments;
however, independent quality assessments may also be requested by
project management.  The \dword{lbnf-dune} \dword{qa} manager is
responsible for development, implementation, assessment, and
improvement of the \dword{qa} program.

The \dword{lbnf-dune} \dword{qa} manager is responsible for
periodically reporting on the performance of the quality system to the
\dword{dune} \dword{tcoord} for review and as a basis for improving
the quality system. The \dword{dune} \dword{tcoord} may call for
\dword{qa} plan readiness assessments as the project nears major
milestones. The \dword{dune} \dword{tcoord}, consortium leaders, and
\dword{lbnf-dune} \dword{qa} manager are all responsible for
providing the resources needed to conduct the project successfully,
including those required to manage, perform, and verify work that
affects quality.

\subsection{Quality Assurance Organization}

\dword{lbnf-dune} \dword{qa} manager may request personnel
from the \dword{dune} project to act on behalf of the
\dword{lbnf-dune} \dword{qa} manager to perform quality
assurance functions, based on need, in accordance with the graded
approach described above. The requested personnel must possess
qualifications or receive the appropriate training required to perform
these functions.

The \dword{dune} \dword{qa} Specialist will be responsible for the
following activities:
\begin{itemize}
	\item Cooperatively develop, monitor, and control \dword{dune}
          \dword{qa} procedures to assure compliance with \dword{dune}
          standards and applicable laws;
     \item Provide assistance for \dword{qa}/\dword{qc} matters in project
       plans, including strategizing technical solutions and
       alternatives on \dword{qa}/\dword{qc} matters and assist in developing
       testing plans with project team members;
	   \item Participate in audits, site inspections, accident
             investigations, and monitor trend analysis to identify
             areas of concern and implement improvements;
	\item Interact with all stakeholders on \dword{qa} issues;
      \item Provide guidance and interpretation on routine and complex
        \dword{qa} matters and problems; and
	\item Participate in reviews at collaborating institutions.
\end{itemize}

Each consortium selects a \dword{qa} representative. 
Each consortium \dword{qa} representative is responsible for
overall coordination of quality requirements to assure they meet
consortium objectives.  The \dword{qa} representative is expected to 
\begin{itemize}
  \item oversee the consortium fabrication facilities for quality
    performance;
  \item interface with the \dword{dune} \dword{qa} specialist on
    consortia \dword{qa} related matters;
  \item monitor the status of all required testing based on the
    consortium \dword{qc} plan;
  \item make sufficient fabrication facility visits to determine
    adequacy of \dword{qc} system performance: check certifications
    of materials and equipment delivered to the facility, spot check
    workmanship, observe testing procedures; and
  \item make or arrange \dwords{ppr} during the fabrication cycle in
    coordination with \dword{tc} and the \dword{jpo} \dword{ro} .
\end{itemize}

Figure~\ref{fig:dune_qa} shows the interface between the
\dword{lbnf-dune} \dword{qa} organization and the consortium \dword{qa}
representatives.  These interfaces will remain when the equipment is
shipped to the far site for installation, but the names may change for
the consortium \dword{qa} representatives.  The consortium \dword{qa}
representatives remain responsible for \dword{qa}/\dword{qc} during the
detector installation process.

\section{Personnel Training and Qualification}

The \dword{dune} consortium leaders are responsible for identifying the
resources to ensure that their team members are adequately trained and
qualified to perform their assigned work. Before allowing personnel to
work independently, they must 
ensure that their team
members have the necessary experience, knowledge, skills and
abilities. Personnel qualifications are based on the following
factors:
\begin{itemize}
 \item previous experience, education, and training;
 \item performance demonstrations or tests to verify previously
   acquired skills;
 \item completion of training or qualification programs; and 
 \item on-the-job training.
\end{itemize}

All \dword{dune} consortium leaders are responsible for ensuring that
their training and qualification requirements are fulfilled, including
periodic re-training to maintain proficiency and qualifications.

\section{Quality Improvement and Lessons Learned}
\label{sec:quality_improvement}

Lessons learned have been developed and utilized by the consortia in
the development of the latest designs.  A lessons learned program
guideline and worksheet has been developed for the use by the
consortia~\citedocdb{8921}.  The project is now at a stage where it is
utilizing the lessons learned gathered from \dword{protodune}. The
lessons learned program remains active although the project is in the
design phase. Lessons learned are being collected during the
performance of activities associated with \dword{iceberg},
\dword{ashriver} \dword{apa} Installation Test Assembly, \dword{cern}
\coldbox facility, and developments toward \dword{protodune2}.

All \dword{dune} consortium members participate in quality improvement
activities that identify opportunities for improvement. They can
respond to the discovery of quality-related issues and follow up on
any required actions. This quality-improvement process requires that
any failures and non-conformance be identified and reported to the
appropriate consortium leader, and that root causes be identified and
corrected. All consortium members are encouraged to identify problems
or potential quality improvements and may do so without fear of
reprisal or recrimination. Items, services, and processes that do not
conform to specified requirements must be identified and controlled
to prevent their unintended use. Inspection and test reports or
similar tools will be used to implement this requirement. Each
consortium leader is responsible for reporting non-conformance to the
\dword{lbnf-dune} \dword{qa} manager, 
who will periodically report these non-conformance to
\dword{dune} \dword{tc}.

\dword{dune} consortium members will perform root cause analysis and
corrective and preventive actions for conditions that do not meet
defined requirements. Consortium leaders may perform root cause
analysis and corrective and preventive actions under their own
procedures or \fnal procedures.  This problem identification, analysis
and resolution process for quality consists of the following steps:
\begin{enumerate}
  \item identify problem;
  \item understand the process;
  \item grade the process and identify Root Cause Analysis (RCA)
    method;
  \item identify possible causes;
  \item collect and analyze data;
  \item communicate lessons learned and document RCA; and
  \item implement corrective and preventative action procedure.
\end{enumerate}


To promote continuous improvement, \dword{dune} \dword{tc} will develop a
lesson learned program based on the \fnal Office of Project Support
Services Lessons Learned Program. This program provides a systematic
approach to identify and analyze relevant information for both good
and adverse work practices that can influence project execution. Where
appropriate, improvement actions are taken to either promote the
repeated application of a positive lesson learned or prevent
recurrence of a negative lesson learned. Lessons learned will be
gathered throughout the project life cycle. As part of the transition
to operations a lessons learned report will be submitted.

In addition, the \dword{lbnf-dune} \dword{qa} manager will
periodically publish a best practices and lessons learned
report. Lessons learned from the \dword{dune} project will be screened
for applicability to other organizations. The \dword{dune} project
will periodically check external lessons learned sources for
applicability to the \dword{dune} project. Sources of lessons learned
include the \dword{doe} Lessons Learned List Server, the \fnal \dword{esh}
Lessons Learned Database, and \dword{dune} team members who
participate in peer reviews of other projects. Reviews of the
\dword{dune} project serve as input to quality improvement.

\section{Documents and Records}

Engineering and technical documents (including drawings) are 
    prepared by \dword{dune} personnel to define the design, manufacture, 
    construction, and installation of their equipment. 
 Ultimately, before these documents are put into effect
they are reviewed and signed by the \dword{dune} consortium leader or
designee. The
\dword{dune} project manages all documents under the document control
systems: \dword{edms} and \docdb, as identified in the \dword{dune}
\dword{cmp}~\cite{bib:docdb82}.  The system to control document preparation, approval,
issuance to users and revision is also described there. Consortium
leaders will use the graded approach described in this plan to
determine work in their scope that requires  \dword{lbnf-dune}
\dword{qa} manager review and signature. This person also reviews project documents that
contain quality requirements. 

Records are prepared and maintained to document how decisions are
made, for instance, decisions on how to arrive at a design, how to
record the processes followed to manufacture components and the means
and methods of cost and schedule change
control. \dword{lbnf-dune} will follow the guidelines for
storing and maintaining records for the project in accordance with
\fnal Records Management\footnote{\url{http://ccd.fnal.gov/records}.}. The
\dword{dune} \dword{tcoord}, \dword{lbnf-dune} \dword{qa}
manager and consortium leaders are responsible for identifying the
information to be preserved. In addition to the technical, cost, and
schedule baseline and all changes to it, records must be preserved as
evidence that a decision or an action was taken, and to provide
the justification for the decision or action.

\section{Work Processes}

\dword{dune} team members are responsible for the quality of their
work, and consortium leaders are responsible for procuring the
resources and support systems to enable their staff to complete their
work with high quality. All \dword{dune} work will be performed using
methods that promote successful completion of tasks, conformance to
\dword{dune} requirements, and compliance with the
\dword{lbnf-dune} \dword{ieshp}. Work processes consist of a
series of actions planned and carried out by qualified personnel using
approved procedures, instructions, and equipment, and under administrative,
technical, and environmental controls, to achieve a high-quality
result.

\subsection{Fabrication Work Processes}

Fabrication work on the \dword{dune} project will be performed to
established technical standards and administrative controls using
approved instructions and procedures. Fabrication work processes with
\dword{qa} inspections and tests will be documented on travelers that
are retained with the hardware item or 
electronically. Items, including consumables, will be identified and
controlled to ensure their proper use and to  prevent the use of
incorrect, unaccepted, or unidentified items. Each consortium will define
a system of controls to ensure that its items are handled, stored,
shipped, cleaned, and preserved to prevent them from deteriorating,
being damaged, or becoming lost. Equipment used for process monitoring
or data collection will be calibrated and maintained. 

Work must be performed safely, in a manner that ensures adequate
protection for employees, the public, and the environment. Consortium
members and the \dword{dune} \dword{tcoord} must exercise a degree of
care commensurate with the work and the associated hazards. See the
\dword{lbnf-dune} \dword{ieshp}~\cite{bib:docdb291} for more details on
\dword{lbnf-dune} integrated safety management systems. 

\subsection{Change-Controlled Work Processes}
\label{sec:change-control}

Changes to design and fabrication requirements should follow the
normal revision process for design and fabrication documents ensuring
the appropriate level of verification, review and approval by the
relevant consortium and \dword{tc}. The change control process flow for
\dword{dune}, as currently envisioned, can be found in~\cite{bib:docdb82}. 
See Figure~\ref{fig:change_control}. 
A Change Control Board (CCB)
led by the \dword{tcoord} advises the \dword{tcoord} on changes.
\begin{dunefigure}[\dshort{lbnf-dune} change control]{fig:change_control}
  {\dword{lbnf-dune} preliminary change control flow chart.}%
  \includegraphics[width=0.95\textwidth]{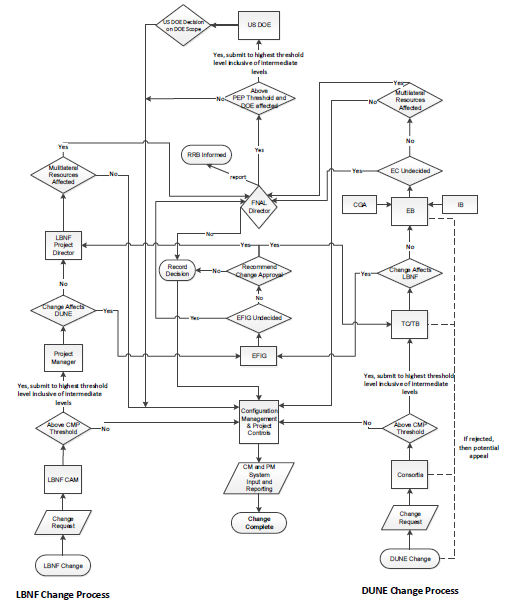}
\end{dunefigure}

When shop or site work must be performed before the associated 
design
document can be formally revised and re-issued, such changes are
accomplished through the development, approval, and distribution of an
\dword{ecr}. This applies to 
designs that are
under configuration management. Interdisciplinary reviews are
performed when the \dword{ecr} subject matter may impact other
subsystems. The \dword{ecr} requestor  indicates whether it is a one-time
change or if the change is to be incorporated into the design
documents. Refer to Section~\ref{sec:fdsp-change} for more information on this
process.  This design change control process is applicable to design
changes that may occur during fabrication and for which the change must be
expedited to avoid schedule delays. These types of changes do not
affect technical scope, cost, or schedule. For changes that affect these aspects, 
the change control process is defined in
Sections~\ref{sec:dune_changecontrol} and~\ref{sec:tc_change control}.

\section{Design}

The \dword{dune} design process provides appropriate control of design
inputs and design products. The primary design inputs are the
\dword{dune} scientific and engineering requirements (e.g., physics
requirements, detector requirements, specifications, drawings,
engineering reports) as discussed in
Section~\ref{sec:fdsp-coord-requirements}.

The basis of the design process requires sound engineering judgment
and practices, adherence to scientific principles, and use of
applicable orders, codes, and standards. This basis of the design
process naturally incorporates environment, health, and safety
concerns.

\subsection{Design Process}

The \dword{lbnf-dune} Systems Engineering website
documentation defines the scope of design work for any given
scientific/engineering work group. From this source, work groups
will begin preliminary design of \dword{dune} by breaking their work
down into sets of engineering drawings, specifications and
reports. This is the design output.

Throughout the design process, engineers and designers work with
consortium leaders and the \dword{lbnf-dune} \dword{qa} manager to
determine \dword{qa} inspection criteria of fabricated products and
installations. Close coordination must be made with \dword{dune}
scientists to assure the engineering satisfies the scientific
requirements of the experiment. Configuration management as documented
in the \dword{lbnf-dune} \dword{cmp}~\cite{bib:docdb82} will be
systematically implemented for \dword{dune}. Final Design work sets
the final \dword{qa} parameters for the parts, assemblies and
installations. Design during final design and production is confined
to change-controlled changes, as above, and to minor changes necessary
to facilitate production, drawing error correction, material
substitutions and similar functional areas.

\subsection{Design Verification and Validation}
\label{sec:verification}

Design is verified and validated to an extent commensurate with its
importance to safety, complexity of design, degree of standardization,
state of the art, and similarity to proven design
approaches. Acceptable verification methods include but are not
limited to any one or a combination of (1) design reviews, (2)
alternative calculations, and (3) prototype, qualification testing,
and/or (4) comparison of the new design with a similar proven design
if available. Verification work must be completed before approval and
implementation of the design.

Design reviews will verify and validate that the following criteria
are met at the appropriate milestone:
\begin{itemize}
 \item adherence to requirements,
 \item technical adequacy of the design,
 \item adequacy of work instructions,
 \item thoroughness of specifications,
 \item test results,
 \item adequacy of technical reports,
 \item adequacy of design calculations and drawings,
 \item reliability and maintainability, and
 \item calibration program for measurement and test equipment.
\end{itemize}
The \dword{dune} Review Plan as discussed in
Chapter~\ref{vl:tc-review} describes the design reviews recommended
for 
particular consortia.

Wherever the design method involves the use of
computer software to make engineering calculations or static dynamic
models of the structure, system, or a component's functionality, the
software must be verified to demonstrate that the software produces
valid results. The verification must be documented in a formal
report of validation that is maintained in records that are accessible
for inspection. However, exemptions may be made for commercially
available software that is widely used and for codes with an extensive
history of refinement and use by multiple institutions. Exemptions
affecting systems or components will be identified to the
\dword{lbnf-dune} Systems Engineering team.

Critical software and firmware computer codes, especially those codes
that are involved in controlling \dword{dune} \dword{daq}, are also subject to reviews for verification and
validation. Some items to be considered during computer code review
are
\begin{itemize}
\item adequacy of code testing scheme,
\item code release control and configuration management,
  \item output data verification against code configuration,
  \item verification that code meets applicable standards, 
  \item verification of code compatibility to other systems that use
    the data,
  \item verification that code meets applicable hardware requirements,
  \item adequacy of code maintenance plans, and
  \item adequacy of code and data backup systems.
\end{itemize}

Validation ensures that any given design product conforms to \dword{dune}
requirements.
During reviews, validation of conformity to requirements
follows verification that the engineering design or computer code
meets all criteria. Engineering designs and computer codes are
validated, preferably before procurement, manufacture, or
construction; but no later than acceptance and use of the item; this
is to ensure the design or computer code
\begin{itemize}
 \item meets \dword{dune} requirements,
 \item contains or makes reference to acceptance criteria, and
 \item identifies all characteristics crucial to the safe and proper
   use of the equipment or system and its associated interfaces.
\end{itemize}

Each inspection, test, or review will feed the \dword{qa} evaluation
process, which is a comparison of results with acceptance criteria, to
determine acceptance or rejection. Rejection identifies the need for
Quality Improvement based on Section~\ref{sec:quality_improvement}. In
some cases, the outcome of the Quality Improvement process may be to
request one or more changes to the design requirements.

\dword{qa} reporting formality escalates as the significance of the
inspection, test or review nonconformance increases. Higher levels of
management must be aware of and participate in the correction of the
most significant nonconformance. Section~\ref{sec:quality_improvement}
identifies the required course of action when nonconformance is
encountered.

\section{Procurement}

\subsection{Procurement Controls}

Procurement controls will be implemented to ensure that
purchased items and services meet \dword{dune} requirements and comply
with the \dword{lbnf-dune} \dword{qa} plan.  The consortium members
requesting procurement of items and services are responsible for
providing all documentation that adequately describes the item or
service being procured so that the supplier can understand the 
requirements for consortium acceptance. Development of this documentation
may be achieved through the involvement of consortium leaders and
established review and approval systems. The following factors will be
considered for review and approval of this documentation:
\begin{itemize}
 \item inclusion of technical performance requirements
 \item identification of required codes and standards, laws and
   regulations;
 \item inclusion of acceptance criteria, including requirements for
   receiving inspection and/or source inspection;
 \item \dword{dune} requirements for vendor qualifications and
   certifications;
 \item \dword{dune} intention to perform acceptance sampling in lieu
   of full inspection and test item acceptance.
\end{itemize}
NOTE: For vendor qualification and acceptance of purchased items or
material by consortium members this may be performed under their own
institution requirements.

Previously accepted suppliers will be monitored to ensure that they
continue supplying acceptable items and services. Source surveillance
is the recommended method to ensure that items are free of damage and
that specified requirements are met. Supplier deliveries will be
verified against previously established acceptance criteria.

Unacceptable supplier items or services will be documented. Records
of supplier performance, Inspection Test Records (ITR), and
contract-required submittals, are kept for future procurement
consideration.

Inspections will be conducted to detect counterfeit and/or suspect
parts. For work funded by \dword{doe}, when counterfeit or suspect parts
are found, they will be identified, segregated, and disposed of in
accordance with the \fnal Quality Assurance Manual Chapter 12020
Suspect/Counterfeit Items (S/CI) Program. \dword{dune} consortia may
use their own institutional procedure for counterfeit/suspect parts.

\subsection{Inspection and Acceptance Testing}

Inspection and testing of electrical, mechanical, and structural
components, and associated services and processes by consortium members
will be conducted using acceptance and performance criteria. ITR
forms, travelers, and a traveler database are the primary tools used
to organize this activity. Inspections will be conducted in accordance
with the graded approach.

Once equipment is received at \dword{sdwf}, the equipment is 
inspected for shipping damage and accuracy against the bill of lading. 
Consortium members perform any additional inspection and testing that 
is required by their design documents either at \dword{sdwf} or 
underground in the clean room as the equipment is prepared for 
installation.

Equipment used for all inspections and tests will be calibrated and
maintained. Calibration will be controlled by a system or systems
making appropriate use of qualified calibration service
providers. Consortium leaders must ensure that equipment requiring
calibration have their calibration status identified on the item or
container, are traceable back to the calibration documentation, and are
tracked to ensure the equipment is calibrated at the required
interval. The \dword{lbnf-dune} \dword{qa} manager 
oversees and supports the \dword{dune} calibration programs.

\section{Assessments}

\subsection{Management Assessments}

Management assessments may be performed by the consortia or \dword{tc}
to evaluate their own management processes (self-assessment) and their
implementation in order to identify noteworthy practices, uncover issues,
identify corrective actions, verify meeting of deliverables, and
ensure that work being performed is satisfactory and done according to the 
requirements. The performance of management assessments is a critical
assurance activity.

\dword{dune} \dword{tc} will monitor progress
of objectives and goals in the consortia to assess whether work is
performed and resources are allocated to meet those objectives and
goals. \dword{dune} \dword{tc}  is responsible for monitoring the
resolution of items identified from assessments, assigning
responsibility for resolution, identifying appropriate timeframes for
resolution, ensuring actions are finalized with appropriate objective
evidence, and documented.

The \dword{lbnf-dune} \dword{qa} manager monitors adequacy of
assessments and progress of corrective actions, and sponsors or conducts
periodic assessments of the effectiveness of implementation of the \dword{qa}
program throughout the \dword{dune} project.

\subsection{Independent Assessments}

The \dword{lbnf-dune} \dword{qa} manager will plan reviews as
independent assessments to assist the \dword{dune} \dword{tcoord} in
identifying opportunities for quality or performance-based improvement,
and to ensure compliance with specified requirements. Independent
assessments of the \dword{dune} projects can be requested by
\dword{dune} management. Independent assessments typically focus on
quality or \dword{esh} management systems, self-assessment programs, or
other organizational functions identified by management. The
\dword{dune} project uses a formal process for assigning
responsibility in response to recommendations from independent
assessments. These recommendations are tracked to closure.

Personnel conducting independent assessments must be technically
qualified and knowledgeable in the areas assessed. A qualified lead
assessor (auditor), who is a Subject Matter Expert (SME) in the
technical area of assessment, is required. The team may include other
SMEs to evaluate the adequacy and effectiveness of activities only if
they are not responsible for the work being assessed.

The \fnal director appoints an independent Long Baseline Neutrino
Committee (LBNC) to advise it and \dword{dune} Management. The role of
this standing committee is described in the \dword{lbnf-dune}
project management plan (PMP). The \dword{doe} and other funding agencies perform external
assessments that provide an objective view of performance and thus
contribute to the independent assessment process. Since such
assessments are not under the control of \dword{dune}, they are not
necessarily considered a part of the independent assessment
criterion. However, \dword{dune} management considers external
assessment results when determining the scope and schedule of
independent assessments.

\section{\dshort{dune} Quality Control}
  
The \dword{dune} consortia are a geographically diverse group of institutions,
collaborating across three continents to fabricate a single integrated
system. As such, careful planning and control of component
fabrication, assembly and testing must be maintained. Each of the
major system components will be fabricated in accordance with
documented procedures and drawings.  These procedures will detail the
inspection and test requirements for each component to ensure that they
meet the requisite specifications prior shipment to \dword{surf}.
Each consortium is responsible for developing the procedures, \dword{qc}
plans, and test plans.

Required inspections and tests during fabrication and installation are
defined in the consortium \dword{tdr} sections. These inspections and tests may
be performed from receipt of material, during fabrication and final
acceptance. For critical equipment source inspection may be performed
in addition. The inspections and tests will be documented on \dword{qc} plans,
travelers, test reports, as applicable.  The procedures will define
the documentation method.

All data from the fabrication and \dword{qc} processes will be maintained in a
database that allows the information to be accessed 
during installation. Each consortium will define the requirements for
the information to be stored in the database. The consortium \dword{tdr}
sections contain information on how they are utilizing the database.

In the case of a nonconformance, the nonconformance will be documented
by the fabrication facility and a recommended disposition will be
provided and forwarded to the consortium technical lead.  If the
disposition is to scrap or rework so that the item meets the requirements
of the specification, the notification to the consortium lead will be 
for information only, and work can continue on performing the
disposition.  If the disposition is ``use as is'' or ``repair'' (where the
item is repaired to meet the functionality but remains in
noncompliance with the specification requirements), work on the item
stops until the disposition has been approved by the design
authority, the consortium technical lead and the \dword{tcoord}.  The \dword{tcoord}
approves the disposition to ensure it does not affect any
other \dword{dune} components during integration. The item will be
re-inspected or tested after completion of the disposition to ensure
that it meets the requirements.

\subsection{\dshort{apa} Quality Control}

Flatness of the \dword{apa} support frame is a key feature and is
defined as the minimum distance between the two parallel planes that 
contain all the points on the surface of the \dword{apa}. After
assembly of the \dword{apa} frame, a laser survey is performed on
the bare frames before  delivery to an \dword{apa}
production site. Three sets of data are compiled into a map that shows
the amount of bow, twist, and fold in the frame. A visual file is also
created for each \dword{apa} from measured data. During \dword{apa}
wiring at the production sites, a final frame survey is completed
after installation of all electrical components, and the as-built
plane-to-plane separations are measured to verify that the
distance between adjacent wire planes meets the tolerances.

Another check performed at the \dword{apa} production site before the
frame is transferred to a winder is necessary to confirm sufficient
electrical contact between the mesh sub-panels and the \dword{apa}
support frame. A resistance measurement is taken immediately after
mesh panel installation for all 20 panels before wiring begins.

All components require inspection and \dword{qc} checks before use on an
\dword{apa}. Most of these tests will be performed at locations other than the
\dword{apa} production sites by institutions within the consortium before the
hardware is shipped for use in \dword{apa} construction. This distributed
model for component production and \dword{qc} is key to enabling the efficient
assembly of \dword{apa}s at the production sites. The critical path components
are the support frames (one per \dword{apa}), grounding mesh panels (20 per
\dword{apa}), and wire carrier boards (204 per \dword{apa}).

The tension of every wire will be measured during production to ensure
that wires have a low probability of breaking or moving excessively in the
detector. Every channel on the completed \dwords{apa} will also be tested for
continuity across the \dword{apa} and isolation from other channels.

A cold testing facility sized for DUNE \dword{apa}s exists at the 
Physical Sciences Laboratory (PSL) at the University of Wisconsin
 that can be
used for such tests. Throughout the construction project, it is
anticipated that 10\% of the produced \dwords{apa} will be shipped to PSL for
cold cycling. This amounts to about one \dword{apa} per year per production site
during the project. 
All \dwords{apa} will still be cold
tested during integration at \dword{surf} and before installation in the \dword{spmod}. 
All active detector components are shipped to the \dword{sdwf} 
before final transport to \dword{surf}. During the storage period, the wire
tensions are measured on all \dwords{apa} to ensure that the transport has
not damaged the wires.

After unpacking an \dword{apa}, a visual inspection is performed, and
wire continuity and tension are measured again. Definite
guidance for the acceptable tension values will be available to inform
decisions on the quality of the \dword{apa}. Clear pass/fail criteria will be
provided as well as clear procedures to deal with individual wires
lying outside the acceptable values. In addition, a continuity test
and a leakage current test is performed on all channels and the data
recorded in the database.

\subsection{\dshort{hv} Quality Control}

The \dword{hv} consortium has developed a comprehensive \dword{qc}
plan for the production, shipping, and installation of the
\dword{spmod} \dword{hv} components. Inventory tagging and tracking
each component is crucial. Documentation in the form of printed
checklists is maintained. Travelers will have been replaced by a system
of tags (``cattle tags'') attached to the units with bar codes that 
key to electronic \dword{qc} data. 

Power supplies used in a \dword{spmod} are tested before
installation. Output voltages and currents must be checked on a known
load. The feedthrough and filters are tested at the same time,
with the planned power supply. The feedthrough must be verified to
hold the required voltage in \dword{tpc}-quality \dword{lar} for several days.

The \dword{qc} tests of the \dwords{hvdb} 
require that all
individual resistors and varistors are submitted to a warm and cold
(87 K) current-voltage measurement. This forms the basis for selecting
components that meet specifications; all electrical components must
pass visual inspection for mechanical damage and all measurement values
(resistance, clamping voltage) must be within 2$\sigma$ of the mean for the
entire sample both in warm and cold tests.

The \dword{qc} process for mechanical components starts at the production
factories by attaching a cattle tag with a unique code to each
production element. A file linked to each code contains the individual
measurements and properties contained in the \dword{qc} checklists for that
element.

\subsection{\dshort{tpc} Electronics Quality Control}

All \dwords{asic} will be tested in \dword{ln} before they are mounted
on the \dwords{femb}; cryogenic testing of the \dwords{femb} is also
planned.  Capacitors and resistors will be
cryogenically tested on a sample basis of a few components from each
reel. Some other components installed on the \dwords{femb}, e.g., 
voltage regulators and crystals, will be qualified in \dword{ln}
before being mounted on the \dwords{femb}.

The \dwords{pcb} for the \dwords{femb} will be tested by the
vendor for electrical continuity and shorts. A visual inspection of
the boards is then performed before installing the discrete components
and the \dwords{asic}. This inspection is repeated after
installation and before the functionality test, which for \dword{dune}
will be performed in \dword{ln}.  After assembly, each \dword{femb}
is tested in \dword{ln} using the current \dword{cts}.

Checks will be performed on all cables during production at room
temperature, before installation and connection to the
\dwords{femb}. These tests  involve continuity and resistance
measurements on the low voltage power and the bias voltage cables, and
bit-error rate measurements on the clock/control and data readout
cables. Connectors will be visually inspected to ensure that they show
no sign of damage. Further tests will take place when the \dword{apa}s
are tested in the \coldbox{}es at \dword{surf} prior to installation
inside the cryostat.

On each cryostat penetration there are two flanges for the \dword{ce}
and one for the \dword{pd} system. The crossing tubes with their spool
pieces are fabricated by industry and tested by vendors to be leak and
pressure proof. The flanges are assembled at consortium institutions
responsible for the \dword{tpc} electronics and \dword{pd} system; the
flanges must undergo both electrical and mechanical tests to ensure
their functionality. Electrical tests comprise checking all the
signals and voltages to ensure they are passed properly between the
two sides of the flange and that no shorts exist. Mechanical tests
involve checking that the flange itself is leak and pressure proof.

\subsection{\dshort{pd} Quality Control}

Materials certification is required (in the \dword{fnal} materials test
stand and other facilities) to ensure materials' compliance with
cleanliness requirements. Cryogenic testing of all materials to be
immersed in \dword{lar} is done to ensure satisfactory performance through repeated
and long-term exposure to \dword{lar}. Special attention will be paid to
cryogenic behavior of fused silica and plastic materials (such as
filter plates and wavelength-shifters), \dwords{sipm}, cables, and
connectors. Testing will be conducted both on small-scale test
assemblies (such as the small test cryostat at Colorado State University (CSU)) and full-scale
prototypes (such as the full-scale CDDF cryostat at CSU). Mechanical
interface testing, beginning with simple mechanical go/no-go gauge
tests, followed by installation into the \dword{pdsp2} system, and
finally full-scale interface testing of the \dword{pd} system into the final
pre-production \dword{tpc} system models will be done; as well as full-system readout tests of the
\dword{pd} readout electronics, including trigger generation and timing,
and tests for electrical interference between the \dword{tpc} and \dword{pd}
signals.

Prior to the start of fabrication, a manufacturing and \dword{qc} plan will be
developed detailing the key manufacturing, inspection, and test
steps. The fabrication, inspection, and testing of the components will
be performed in accordance with documented procedures. This work will
be documented on travelers and applicable test or inspection
reports. Records of the fabrication, inspection and testing will be
maintained. When a component has been identified as being in
noncompliance to the design, the nonconforming condition will be
documented, evaluated and dispositioned as one of (1) use-as-is (does not meet
design but can meet functionality as it is), (2) rework (bring into
compliance with design), (3) repair (will be brought to meet functionality
but will not meet design), or (4) scrap. For products with a disposition
of use-as-is or repair, the nonconformance documentation is
submitted to the design authority for approval. All \dword{qc} data (from
assembly and pre- and post-installation into the \dword{apa}) will be directly
stored to the \dword{dune} database for ready access. 

Monthly summaries of key performance metrics (to be defined) will be
generated and inspected to check for quality trends. Based on the
\dword{pdsp} model, we expect to conduct the following production
testing prior to shipping from assembly site. 
\begin{itemize}
\item dimensional checks of
critical components and completed assemblies to ensure that
system interfaces ar satisfactory;
\item  post-assembly cryogenic checkouts of \dword{sipm} mounting
\dwords{pcb} (prior to assembly into \dword{pd} modules);
\item module dimensional
tolerances using go/no- go gauge set; and 
\item warm scan of complete module
using motor-driven \dword{led} scanner (or UV \dword{led} array).  
\end{itemize}
Following shipping
to the USA reception and checkout facility but prior to storage at
\dword{sdwf}, we will conduct mechanical inspection, a warm scan (using identical scanner to
initial scan), and cryogenic testing of completed modules (at the CSU CDDF or
similar facility).

Following delivery to the underground integration cleanroom, prior to and
during integration and Installation, we will 
\begin{itemize}
\item conduct  a warm scan (using identical
scanner to initial scan);
\item  complete visual inspection of module against
a standard set of inspection points, with photographic records kept
for each module;
\item conduct end-to-end cable continuity and short circuit tests
of assembled cables; 
\item perform a \dword{fe} electronics functionality
check  
\item perform installation \dword{qc} \dword{pd} system pre-installation testing, following the model established for \dword{pdsp}.
\end{itemize}

Prior to installation in the \dword{apa}, the \dword{pd} modules will undergo a warm
scan in a scanner identical to the one at the \dword{pd} module assembly
facility and we will compare the results. The module will also 
undergo a complete visual inspection for defects, and a set of
photographs of selected critical optical surfaces will be taken and entered
into the \dword{qc} record database. Following installation into the \dword{apa} and
cabling, an immediate check for electrical continuity to the \dwords{sipm}
will be conducted. Following the mounting of the \dword{tpc} \dword{ce}
and the \dwords{pd}, the entire \dword{apa} will undergo a cold system
test in a gaseous argon \coldbox, similar to that performed for
\dword{pdsp}. During this test and prior to installation, the \dword{pd} system 
will undergo a final
integrated system check, checking dark and
\dword{led}-stimulated \dword{sipm} performance for all channels, checking for
electrical interference with the \dword{ce}, and confirming
compliance with the detector grounding scheme.

\subsection{Calibration Quality Control}

The manufacturer and the institutions in charge of devices will
conduct  series of tests to ensure the equipment can perform its
intended function as part of \dword{qc}. \dword{qc}
also includes post-fabrication tests and tests run after shipping and
installation. The overall strategy for the calibration devices is to
test the systems for correct and safe operation first in dedicated test
stands, then at \dword{pdsp2}, then as appropriate near \dword{surf},
and finally underground. Electronics and racks associated with each full
system will be tested before transporting them underground. Each calibration
system undergoes specific tests.
\begin{itemize}
\item Ionization Laser System: For assembly and operation of the laser
  and feedthrough interface, 
  the test is carried out on a mock-up
  flange for each of the full hardware sets (periscope, feedthrough,
  laser, power supply, and electronics). All operational parts (UV
  laser, red alignment laser, trigger photodiode, attenuator,
  diaphragm, movement motors, and encoders) are tested for
  functionality before being transported underground.
\item Photoelectron Laser System: The crucial test is to measure the
  light transmission of all fibers at 266 nm. A suitable transmission
  acceptance threshold will be established based on studies during the
  development phase. Studies to estimate the number of photoelectrons
  emitted as a function of intensity (based on distance of fiber
  output to the metallic tab) will also be undertaken.
\item Laser Beam Location System: For the \dword{lbls}, to ensure uniformity
  across all clusters, the main test is to verify 
   that the PIN diodes are all functional and that their light
  detection efficiency is within a specified range. For the mirror-based system, the reflectivity
  of all mirrors will  be tested prior to assembly.
\item Pulsed Neutron Source System: The first test will be safe
  operation of the system in a member institution radiation-safe
  facility. Then the system will be validated at \dword{pdsp2}. The
  same procedure will be carried out for any subsequent devices before
  the devices are transported to \dword{surf} and underground. System
  operation will be tested with shielding assembled to confirm safe
  operating conditions and sufficient neutron yields using an external
  dosimeter as well as with the installed neutron monitor.
\end{itemize}

\subsection{\dshort{daq} Quality Control}

SP-DAQ-6 Data verification: The \dword{daq} must check integrity of data at
every data transfer step. It only deletes data from the local
storage after confirmation that data have been correctly recorded to
permanent storage. Data integrity checking is fundamental to ensure
data quality. The high overall experiment uptime goal requires that the \dword{daq} 
be stringently designed for reliability, fault tolerance, and
redundancy  -- criteria that aim to reduce overall downtime. The \dword{daq}
monitors the quality of the detector data and of its own operational
status, performs automated error detection, and has recovery
capabilities.

The \dword{eb} subsystem provides bookkeeping
functionality for the raw data. This  includes the documenting of
simple mappings, such as which trigger is stored in which raw data
file, as well as more sophisticated quality checks. For example, it
will know which time windows and geographic regions of the detector
are requested for each trigger, and in the unlikely event that some
fraction of the requested detector data cannot be stored in the event
record, it will document that mismatch.

Data Quality Monitoring: While the \dword{daqccm} contains an element of
monitoring, here \dword{dqm}
refers to a subsystem that quickly analyzes the data in order to
determine the general quality of the detector and \dword{daq} operation. This
 allows operators to promptly detect and respond to any
unexpected changes and assures high exposure times for later physics
analyses.

A \dword{daq} \dword{dqm} will be developed (including
necessary infrastructure, visualization, and algorithms) that 
processes a subset of detector data in order to provide prompt feedback
to the detector operators. This system will be designed to allow it to
evolve as the detector and its data are understood during
commissioning and early operation, and to cope with any evolution of
detector conditions.

While the hardware design will be done at the institutions working in
this area, the production of prototypes and final \dwords{pcb} will be
outsourced to companies, allowing for early identification of those
companies that can guarantee a high-quality card production.

\subsection{\dshort{cisc} Quality Control}

The manufacturer and the institution in charge of device assembly will
conduct a series of tests to ensure the equipment can perform its
intended function as part of \dword{qc}. \dword{qc} also includes post fabrication
tests and tests run after shipping and installation. For complex
systems, the entire system will be tested before shipping. Additional
\dword{qc} procedures can be performed underground after installation.

Slow Controls Hardware Networking and computing systems will be
purchased commercially, requiring \dword{qc}. However, the new servers
are tested after delivery to confirm they suffered no damage during
shipping. The new system is allowed to burn in overnight or for a few
days, running a diagnostics suite on a loop in order to validate the
manufacturer's \dword{qa} process.

\section{ProtoDUNE to DUNE QA Approach}

The approach to \dfirst{qa}/\dfirst{qc} for \dword{dune} is going to be
very similar to the activities and oversight that was performed for
\dword{protodune}.  For \dword{protodune}, the major
\dword{qa}/\dword{qc} activities included review of the consortium
\dword{tdr} \dword{qa}/\dword{qc} sections; assisting the consortia in
development and review of \dword{qc} plans (production and
installation), fabrication, inspection and test procedures,
installation plans and documentation; and the performance of
\dwords{prr} at eleven consortium fabrication facilities.  There was
also \dword{qc} participation in the \dword{protodune} design reviews.
The \dwords{prr} looked at the following criteria:
\begin{itemize}
  \item final \dword{qa} plans for institutions not adopting the
    \dword{lbnf-dune} \dword{qa} Plan;
  \item final production drawings, specifications, and manufacturing
    and test procedures;
  \item final safety documents (i.e., hazard analysis documentation);
  \item component \dword{qc} plan (i.e., travelers, test reports,
    software verification and validation documents, supplier
    documentation);
  \item final procurement documents per institution practice; and
  \item completion and evaluation of prototypes, and review of production
    process and \dword{qc} results.
\end{itemize}
The reviews ensured the facilities
were prepared for production and that any kinks in the processes had been
identified and mitigations performed. 
The positive outcome of
these reviews was the amount of equipment received at \dword{cern} with little
to no damage.

\dwords{ppr} will be performed at the fabrication facilities for the
\dword{dune} detector components. This type of review has been added due to the
larger number of components required and the increased number of
fabrication facilities. The goal of these reviews is to ensure
consistent fabrication processes between the facilities. If an issue
is identified at one facility, it can be communicated to 
others to prevent recurrence. 

For \dword{protodune}, installation was performed at \dword{cern} under the
guidance of \dword{cern} policies and procedures. Installation of the
\dword{dune} \dwords{detmodule} at \dword{surf} will fall under similar procedures in the
\dword{lbnf-dune} \dword{qa} plan.

\cleardoublepage

\chapter{Environment, Safety, and Health}
\label{vl:tc-ESH}

\dword{lbnf-dune} is committed to protecting the health and safety of
staff, the community, and the environment, as stated in the
\dword{lbnf-dune} integrated \dword{esh} plan, as well as to ensuring a
safe work environment for \dword{dune} workers at all institutions and
protecting the public from hazards associated with constructing and
operating \dword{dune}.  Accidents and injuries are preventable, and
the \dword{esh} team will work with the global \dword{lbnf-dune}
project and collaboration to establish an injury-free workplace.
All work will be performed so as to preserve  the quality of the environment and
prevent property damage.

The \dword{lbnf-dune} \dword{esh} program complies with applicable
standards and local, state, federal, and international legal
requirements through the \dword{fnal} Work Smart set of standards and the
contract between \dword{fra} and the \dword{doe}
Office of Science (FRA-DOE). \fnal, as the host laboratory,
established the \dword{sdsd} to provide facility support.
\dword{sdsd} is responsible for support of \dword{lbnf-dune}
operations at \dword{surf}.

The \dword{lbnf-dune} \dword{esh} program strives to prevent
injuries or illness and seeks to continually improve safety and health
management.  To the maximum practical extent, all hazards must be
eliminated or minimized through substitution, or engineering or
administrative controls.  Where engineering or administrative controls
are not feasible, workers will use \dword{ppe}.

The \dword{lbnf-dune} \dword{esh} management system is
designed to work hand-in-hand with the \dword{surf} emergency
management systems to protect the public, workers, and the environment;
ensure compliance with the FRA-DOE contract and \fnal Work Smart
standards; and improve the \dword{dune} ability to meet or
exceed stakeholder expectations and execute the
scientific mission.  \dword{dune} uses a set of criteria to plan, direct,
control, coordinate, assure, and improve how \dword{esh} policies,
objectives, processes, and procedures are established, implemented,
monitored and achieved.

The \dword{lbnf} facilities at \dword{surf} are subject to
the requirements of the \dword{doe} Worker Safety and Health Program,
Title 10, Code Federal Regulations, Part 851 (10 CFR 851). These
requirements are promulgated through the \fnal Director Policy
Manual\footnote{\fnal Director's Policy Manual is:
  http://www.fnal.gov/directorate/Policy\_Manual.html}, and the \fnal
\dword{esh} Manual\footnote{\fnal \dword{esh} Manual is:
  http://esh.fnal.gov/xms/ESHQ-Manuals/feshm} (\dword{feshm}), which
align with the \dword{surf} \dword{esh} manual.

\section{\dshort{lbnf-dune} \dshort{esh} Management and Oversight}

The \dword{tcoord} and \dword{ipd} have responsibility for
implementation of the \dword{dune} \dword{esh} program for the construction and installation activities, respectively.  The
\dword{lbnf-dune} \dword{esh} manager reports to the
\dword{tcoord} and \dword{ipd} and is responsible for providing
\dword{esh} support and oversight for development and implementation of the 
\dword{lbnf-dune} \dword{esh} program. Figure~\ref{fig:dune_esh} shows
the \dword{lbnf-dune} \dword{esh} organization.

The \dword{dune} \dword{esh} coordinator reports to the
\dword{lbnf-dune} \dword{esh} manager and has primary responsibility
for \dword{esh} support and oversight of the \dword{dune} \dword{esh}
program for activities at collaborating insitutions as shown in
Figure~\ref{fig:dune_esh_construction}.  The far and near site
\dword{esh} coordinators are responsible for providing daily field support and
oversight for all installation activities at the \dword{surf}
and \dword{fnal} sites, as shown in Figure~\ref{fig:dune_esh_installation}.

Additional \dword{esh} subject matter experts (SMEs) are available to
provide supplemental support to the project through the \fnal
\dword{esh} Section. The \fnal \dword{esh} Section, FRA-DOE, CERN and
\dword{surf} will provide supplemental \dword{esh} oversight to
validate implementation of the \dword{lbnf-dune} \dword{esh}
program. FRA-DOE maintains a daily oversight presence at the far and
near sites.

The \dword{lbnf-dune} \dword{esh} plan defines the \dword{esh}
requirements applicable to installation activities at the \dword{surf}
site. Regular \dword{esh} walkthroughs will be conducted by
\dword{lbnf-dune} \dword{esh} management personnel. All
findings will be documented in the \fnal Predictive Solutions
database system.

\section{National Environmental Protection Act Compliance}

In compliance with the National Environmental Protection Act (NEPA) and in
accordance with \dword{doe} Policy 451.1, the
\dword{lbnf-dune} project performed an assessment  of 
environmental impacts that are possible during the construction and operation of the
project.  
This assessment~\cite{bib:docdb122}   identifies the
potential environmental impacts and the safety and health hazards
that could occur or be present during the design, construction, and operating phases of
\dword{lbnf-dune}.  The environmental assessment presented an
analysis of the potential environmental consequences of the facility
and compared them to the consequences of a No Action Alternative. 
It also included a detailed analysis of all potential environmental,
safety, and health hazards associated with construction and operation
of the facility.  The environmental assessment has been completed and
a finding of no significant impact (FONSI)~\cite{bib:docdb122} issued in September 2015.

\section{Codes/Standards Equivalencies}
\label{sec:esh_codes}

\dword{dune} will rely on significant contributions from international
partners. In many cases, an international partner will contribute
equipment for installation at \dword{fnal} or \dword{surf}, built
following international standards. \dword{fnal} has established a
process under the international agreement with \dword{cern}, detailed
in \dword{feshm}~\cite{FNAL:FESHM2000} Chapter 2110, to establish code equivalency between
USA and international engineering design codes and standards. This
process allows the laboratory to accept in-kind contributions from
international partners or purchase equipment designed using
international standards while ensuring an equivalent level
of safety.

At the time of this writing, \dword{fnal} has completed the following code
equivalencies:
\begin{itemize}
 \item pressure vessels designed using EN13445;
 \item structures designed using EN 1990, EN 1991, EN1993, EN 1999 (a
   subset of the Eurocodes), and EN 14620;
 \item CE-marked pressure piping systems designed using PED 97/23 EN 13480;
 \item CE-marked relief valves designed using PED 2014/68/EU EN ISO 4126;
 \item CE-marked electrical equipment for measurement and control; and
   laboratory use designed using IEC 61010-1 and IEC 61010-2-030.
\end{itemize}

As necessary, the laboratory code equivalency process will be followed
to establish equivalency to other international codes and
standards. The current list of completed code equivalencies can be
found in~\cite{bib:eshdocdb3303}. 

\section{\dshort{esh} Requirements at Collaborating Laboratories and Institutions}

All work performed at collaborating institutions will be completed
following that institution's \dword{esh} policies and
programs. Equipment and operating procedures provided by the
collaborating institution will conform to the \dword{dune} project
\dword{esh} and integrated safety management policies and
procedures. The \dword{esh} organization at collaborating institutions
provides \dword{esh} oversight for work activities carried
out at their facilities.

\section{\dshort{lbnf-dune} \dshort{esh} Program at \dshort{surf}}

\subsection{Site and Facility Access}

All \dword{dune} workers requiring access to the \dword{surf} site
register through the \fnal Global Services Office to receive the
necessary user training and a \fnal identification number that can be
used to apply for a \dword{surf} identification badge, through the
\dword{surf} Administrative Services Office. This is coordinated by
\dword{sdsd}. The \dword{surf} identification badge allows access to
the \dword{surf} site as part of the \dword{surf} Site Access Control
Program. The \fnal Global Services Office has extensive experience
with international collaborators.

\dword{surf} underground access will require that working groups
obtain a \dword{tap} for each daily access to the
underground areas.  All personnel within each working group must be
individually listed on the \dword{tap}, per the \dword{surf} Site
Access Control Program. All personnel are required to ``brass in and
out'' via the brass board located at the entrance to the Ross Cage prior to accessing the underground facilities.

\subsection{\dshort{esh} Training}

All personal performing work onsite at \dword{surf} are required to
attend \dword{surf} \dword{esh} Site Orientation prior to their work
at the site.  This includes \dword{surf} Surface and Underground
training classes, as well as associated Cultural Heritage
training. Arrangements will be made for all workers to complete this
training. In addition, unescorted-access training will be provided to
personnel for each underground working level (\dword{4850l} and 4910L) at
which they will perform work.  The \dword{lbnf}/\dword{dune}
\dword{esh} management team will present a project-specific
introductory \dword{esh} presentation.

\subsection{Personnel Protective Equipment}

Personal protective equipment (PPE) is not a substitute for
engineering and administrative controls. These controls will be
implemented, to the extent feasible, to mitigate the hazard so that
the need for \dword{ppe} is reduced or eliminated.

Personnel must wear the following \dword{ppe} when on site the \dword{surf} site.
\begin{itemize}
\item At a minimum, all workers personnel shall wear
  steel-toed boots, long pants and shirts with at least four-inch sleeves when
  performing non-office work at \dword{surf}. 
  \item All personnel entering the work site at the Ross (or Yates) Dry
    shall wear hard hats (brim facing forward), gloves,
    safety glasses with rigid side-shields and a reflective high
    visibility (e.g., orange) shirt, coat, or vest (minimum ANSI Class 2).
    Exceptions to these minimum requirements will be approved by the
    \dword{lbnf-dune} \dword{esh} manager and noted in the
    activity-specific \dword{ha} maintained by the \dword{dune} \dword{esh} coordinator.
  \item When working underground, all personnel will carry an Ocenco M20.2 self rescuer
    and a hard hat cap lamp.  Ocenco 7.5 emergency breathing apparatus devices
    will be stored underground for additional emergency support.
  \item Hard hats must meet the ANSI Z89.1 standard as defined by 29
    CFR 1926.100 and bear the Z89-.1 designation. 
   \item Eye protection must meet the requirement of 29 CFR
      1926.102. Safety glasses must be ANSI approved and be marked
      with the ANSI marking Z87.1 designation.
    \item Hearing protection must be appropriate to the work environment, as
      defined in the activity-based \dword{ha}.
    \item Workers will don any specialized \dword{ppe} required for specific work tasks as
      defined in the activity-based \dword{ha}. 
\end{itemize}

\subsection{Work Planning and Controls}

The goal of the work planning and controls process is to determine how to do the work
safely, correctly, and efficiently. All work activities are subject to the work
planning and controls process which includes the development of \dword{ha}
documentation. The first steps are to initiate careful thought about the work,
determine the scope of the job, identify the potential hazards associated with it,
and determine mitigation strategies. We ensure that all affected employees understand
the full work plan and what is expected of them and that they have all the
appropriate materials to do the job properly. The Work Planning and \dword{ha} program is
documented in Chapter 2060 in the \dword{feshm} manual~\cite{feshm}. 
 All work planning documentation is reviewed and
approved by  line management supervision, the \dword{dune} \dword{esh} coordinator, and the \dword{dune}
\dword{irr} or \dword{orr} committees prior to the start of work activities.

A work planning meeting will be held 
before each shift. 
The meeting 
is led by the shift supervisor, supported by the \dword{dune} \dword{esh} coordinator, and
attended by all personnel working on site during that shift. The meeting is intended to inform the workers of potential safety hazards and
hazard mitigations relating to the various work activities, ensure
that employees have the necessary \dword{esh} training and \dword{ppe}, answer any
questions relating to the work activities, and authorize the work
activities for that shift. The meeting is expected to last approximately 15 minutes, but 
may vary depending on the activities planned for the shift. 

A Safety Data Sheet (SDS) will be available for all chemicals and
hazardous materials that are used on site. All chemicals and hazardous
materials brought to the \dword{surf} site must be reviewed and approved by the
\dword{dune} \dword{esh} coordinator and the \dword{surf} \dword{esh}
Department before arriving at site.  SDS documentation will be
submitted to the \dword{dune} \dword{esh} coordinator prior to the
material arriving on site.

\subsection{Emergency Management}

Any injuries, accidents, or spills are to be reported immediately to the
\dword{lbnf-dune} \dword{esh} manager, from either the \dword{surf} Emergency
Response Coordinator or the installation manager, through the \dword{dune} far site
\dword{esh} coordinators and the \dword{dune} \dword{esh} coordinator. 
Any personnel that
experience any injury will be sent (or transported, if needed) to the Black Hills
Medical Clinic or Regional Health Lead-Deadwood hospital. 
The supervisor  completes an initial incident
investigation report and submits the report to the \dword{lbnf-dune}
\dword{esh} manager within 24 hours.

The \dword{surf} \dword{esh}
Manual\footnote{\url{http://sanfordlab.org/esh}} maintains the Emergency
Management and Emergency Response Plan 
(ERP)\footnote{\url{http://sanfordlab.org/esh/manual/32-emergency-response-plan-erp-policy}}
for the site. All personnel will receive ERP training and the ERP
flowchart for emergency notification process is posted at all
telephones. For all emergencies at the \dword{surf} site, personnel
contact Emergency Response personnel by using any building
phone, dialing the hoist operator, or by calling 911 from any outside
line (e.g., cell phone).

Emergency Response Groups that are not part of \dword{surf} but are recognized
outside resources are the Lawrence County Emergency Manager, the Lead and
Deadwood Fire Departments, Lawrence County Search and Rescue, Black
Hills Life Flight, and the Rapid City Fire Department HAZMAT team. The
\dword{surf} ERP is distributed to these resources to facilitate outside
emergency response.

\dword{sdsta} will maintain an emergency response incident command
system and an emergency response team (\dword{ert}) on all shifts that can access the
underground sites with normal surface fire department response
times. This team provides multiple response capabilities for both
surface and underground emergencies but specializes in underground
rescue through MSHA Metal/Non-Metal Mine Rescue training.  
The \dword{ert} has a defined training schedule and conducts regular
walkthroughs in areas of response. The team conducts  emergency drills 
on both the surface and underground sites, in which all personnel on
site are required to participate. \dword{ert} personnel includes a
minimum of one emergency medical technician (EMT) and/or paramedic per
shift. 

\dword{surf} implements a guide program for both the surface and
underground areas. The guide program has an established training
program.  Visitors and other untrained personnel must be escorted by a trained
guide when on site at \dword{surf}. 
In addition, a minimum of one guide is stationed on each 
level underground at all times where work is occurring. This guide provides
supplemental emergency support to unescorted, access-trained
personnel. Guides are trained as first responders to help in a medical
emergency until the ERT arrives.

In the case of an underground emergency such as fire or \dword{odh},
evacuation to the surface takes place through the Ross or Yates Shafts.
If full evacuation is not  possible, the refuge chamber on the \dword{4850l}
 can shelter up to 144 persons for 98 hours.

\subsection{Fire Protection, \dshort{odh} and Life Safety}

The \dword{dune} installation team members are required to police their work areas
frequently and maintain good housekeeping. Teams generating common garbage and other
waste must dispose of it at frequent, regular intervals. 
Containers will be provided for the
collection and separation of waste, trash, oily or used rags, and other
refuse.  Containers used for garbage and other oily, flammable, or
hazardous wastes, (e.g., caustics, acids, or harmful dusts) will be equipped with covers.  Chemical agents or
substances that might react to create a hazardous condition, must
be stored and disposed of separately, as assessed by the
\dword{lbnf-dune} \dword{esh} coordinator.

The \dword{lbnf-dune} \dword{esh} coordinator collects SDS documentation for chemicals and hazardous materials and determines proper storage cabinets for them.  The documentation is made readily available with the materials in the cabinets for the workers. 


All open flame, welding, cutting, or grinding work activities require completion and approval of a \dword{fnal} ``hot'' work permit.  The \dword{dune}
\dword{esh} coordinator coordinates the issuance of the permit.
The team completing the work will be responsible for
providing all the required materials, personnel, and \dword{ppe} 
to conduct the hot work. All hot work permits must be
provided to the \dword{surf} \dword{esh} Department.

Cables installed for \dword{dune} are chosen to be
consistent with current \fnal standards for cable insulation and must
comply with recognized standards concerning cable fire resistance. 
This reduces the probability of a fire starting and of adverse health effects due to
combustion products of cable insulation materials.

Fire and life safety requirements for \dword{lbnf-dune} areas were
analyzed in the \dword{lbnf-dune} Far Site Fire and Life Safety
  Assessment~\citedocdb{14245}. ARUP provided code analysis, fire
modeling, egress calculations, and the design of the fire protection
features for \dword{lbnf} \dword{fscf}.  Additionally, the ARUP
consultant, SRK Consulting, modeled additional fire scenarios and the potential
spread of toxic fumes and heat in the drifts used by
\dword{lbnf-dune} for evacuation, verifying that the system design and evacuation 
    processes will be safe.   All caverns will be equipped with
fire detection and suppression systems, with both visual and audible
notification.  All fire alarms and system supervisory signals will be
monitored in the \dword{surf} Incident Command Center.  The
\dword{surf} \dword{ert} will respond with additional support from the
Lead and Deadwood Fire Departments and the county's emergency management
department.

\dword{odh} requirements were assessed through the \dword{lbnf-dune}
\dword{odh} analysis. The caverns have been classified as \dword{odh} 1 and
the drifts are classified as  \dword{odh} 0. The caverns will be equipped with
an \dword{odh} monitoring and alarm system, with independent visual and
audible notification systems.  All \dword{odh} alarms and system
supervisory signals will be monitored in the \dword{surf} Incident
Command Center.  Each occupant entering an  \dword{odh} area will receive  \dword{odh}
training and carry Ocenco M20.3 escape packs.

Emergency conditions from smoke or \dword{odh} incidents underground are
primarily mitigated by the large ventillation rate in the \dword{surf}
underground area.

The facility emergency management plan will be reviewed and updated as
necessary during construction, installation, and operation activities
based on the egress strategy defined in the ARUP Fire and Safety
Report and the  \dword{surf} Emergency Management
  Plan\footnote{\url{http://sanfordlab.org/ehs/manual/31-emergency-management-policy}}.

Radon levels are presently monitored in the occupied underground
facilities. This monitoring program will extend to the \dword{lbnf}
underground areas in coordination with \dword{fnal} and \dword{sdsta}
\dword{esh} personnel.

\subsection{Earthquake Design Standards}

For surface and underground structures, the design standard is the
latest edition of the International Building Code (IBC) (2018), wherein
Chapter 16  is Structural Design and Section 1613 is Earthquake
Loads. The Lead seismic region, according to the American Society of
Civil Engineers (ASCE 7), is between the lowest risk (the same as that 
for \fnal) and the next level up. Both are minimal risks.

\subsection{Material handling and Equipment Operation}

All overhead cranes, gantry cranes, fork lifts, and motorized equipment,
e.g., trains and carts, will be operated only by trained
operators. Other equipment, e.g., scissor lifts, pallet jacks, hand
tools, and shop equipment, will be operated only by personnel trained
for the particular piece of equipment. 

Hoisting and rigging operations will be evaluated and planned.  A
member of the trained rigging team must identify the hazards and
determine the controls necessary to maintain an acceptable level of
risk.  A Hoisting and Rigging Lift Plan is required for complex and
critical lifts. This plan must be documented using the \fnal Hoisting
and Rigging Lift Plan or similar plan accepted by \fnal. The \dword{ha}
 documentation will include the development of critical lift
plans for specific phases of installation. 

All equipment operating in the underground facility will be diesel or
electric powered. Diesel is allowed due to the large
ventillation rate in the underground area.  There will be no gasoline or propane powered
equipment in the underground facility.

\subsection{Stop Work Authority}

If a worker identifies any unanticipated or unsafe conditions 
or non-compliant
practices occurring in their work activity, 
the trained worker is empowered and expected to stop the activity
 and notify their supervisor and \dword{dune} \dword{esh} coordinator of
this action. All workers on the \dword{dune} project have the
authority to stop work in any situation that presents an imminent
threat to safety, health, or the environment. Work may not resume
until the circumstances are investigated and the deficiencies corrected,
including the concurrence of the \dword{dune} \dword{ipd}
and \dword{lbnf-dune} \dword{esh} manager.


\subsection{Operational Readiness}

The \dword{dune} review process consists of design, production,
installation, and operation reviews as described in
Chapter~\ref{vl:tc-review}. These reviews include lifting fixture
load testing and work planning and controls documentation. The
\dword{jpo} \dword{ro}  is involved at all stages of the review
process. All major stakeholders (including \dword{fnal}, \dword{surf} and
\dword{dune} collaborating institutions) will be involved as
appropriate. The \dword{ro}  will complete both system and process
readiness reviews to authorize installation activities at
\dword{surf}.  Operational readiness reviews will be completed prior
 to the operation of detector components.

\subsection{Lessons Learned}

The \dword{lbnf} project is currently working with \dword{sdsta} and the 
engineering consultant ARUP to implement \dword{esh} procedures and
protocols for training, emergency management, fire
protection, and life safety. The \fnal \dword{esh} Section, \dword{doe}, and
\dword{lbnf} \dword{esh} have completed a series of assessments of
critical \dword{sdsta} \dword{esh} programs including underground access,
emergency management, electrical safety, rigging, and fire
protection. The findings and lessons learned identified in these
\dword{esh} program assessments are tracked within the \fnal issues management
database, iTrack.

\dword{fnal} completed a review to
identify critical lessons learned from the previous underground
neutrino project NuMI/MINOS  in May 2009. The findings from this
exercise were documented in a report entitled Executive Summary of
Major NuMI Lessons Learned.  We are using the \dword{dune} lessons learned from
\dword{protodune}
to further develop and enhance
the \dword{dune} engineering review and work planning and controls
processes.

Lessons learned are disseminated in areas of applicability and
flowed-down for appropriate implementation. Any action items
associated with lessons learned are tracked in iTrack. Lessons learned
are reviewed and evaluated by both \dword{fnal} and \dword{lbnf-dune} management.

\cleardoublepage

\appendix
\chapter{Project Document Summary}
\label{ch:tc-sp-project}
\section{Interface Documents}
\label{sec:fdsp-coord-interface}

A set of interface documents defines the scope of each subsystem and
with progressively more detail defines the detailed interfaces between
subsystems. There are three sets of interface documents. One set of
documents includes all of the consortia-to-consortia interfaces. A second
set includes the interfaces between the consortia and the facilities
(provided either by \dword{tc}, \dword{lbnf} or the \dword{integoff}). The
third set is between the consortia and the installation team. All
three sets are managed by \dword{jpo} engineering team.

The \dword{dune} interface documents are actively maintained in the
\dword{edms}, but a copy has been archived in DocDB that captures the
status of these documents at the time of this \dword{tdr}.  A matrix
with links to the interface documents in DocDB between consortia for
the \dword{spmod} are shown in
Table~\ref{tab:interface_sp_consortia}. 
The interface documents for the \dword{dpmod} are in preparation as of this writing. 
\begin{dunetable}
  [\dshort{spmod} inter-consortium interface document matrix]
  {|p{0.08\linewidth}|rp{0.08\linewidth}||rp{0.08\linewidth}||rp{0.08\linewidth}||rp{0.08\linewidth}|rp{0.08\linewidth}||rp{0.08\linewidth}|rp{0.08\linewidth}|||}
  {tab:interface_sp_consortia}
  {\dword{spmod} consortium-to-consortium interface document matrix. All entries point to documents in the DUNE DocDB.}
       & PDS  & TPC Elec   & HV   & DAQ  & CISC & CAL  & COMP \\ \toprowrule
  APA  & \cite{bib:docdb6667} &  \cite{bib:docdb6670} & \cite{bib:docdb6673} & \cite{bib:docdb6676} &  \cite{bib:docdb6679} &  \cite{bib:docdb7048} &  \cite{bib:docdb7102} \\ \colhline
  PDS  &      &  \cite{bib:docdb6718} &  \cite{bib:docdb6721} & \cite{bib:docdb6727} &  \cite{bib:docdb6730} &  \cite{bib:docdb7051} & \cite{bib:docdb7105} \\ \colhline
  TPC Elec   &      &      & \cite{bib:docdb6739} & \cite{bib:docdb6742} & \cite{bib:docdb6745} & \cite{bib:docdb7054} & \cite{bib:docdb7108} \\ \colhline
  HV   &      &      &      & \cite{bib:docdb6736} & \cite{bib:docdb6787} & \cite{bib:docdb7066} & \cite{bib:docdb7120} \\ \colhline
  DAQ  &      &      &      &      & \cite{bib:docdb6790} & \cite{bib:docdb7069} & \cite{bib:docdb7123} \\ \colhline
  CISC &      &      &      &      &      & \cite{bib:docdb7072} & \cite{bib:docdb7126} \\ \colhline
  CAL  &      &      &      &      &      &      & \cite{bib:docdb6868} \\ 
\end{dunetable}

A matrix with links to the interface documents in DocDB between each consortium and the facility, installation, and DUNE physics for the \dword{spmod} are
shown in Table~\ref{tab:interface_sp_tc}.
\begin{dunetable}
  [\dshort{spmod} consortium-TC interface document matrix]
  {|p{0.15\linewidth}||rp{0.08\linewidth}||rp{0.08\linewidth}||rp{0.08\linewidth}||rp{0.08\linewidth}||rp{0.08\linewidth}||rp{0.08\linewidth}||rp{0.08\linewidth}||rp{0.08\linewidth}|}
  {tab:interface_sp_tc}
 {\dword{spmod} consortium-to-TC interface document matrix. All entries point to documents in the DUNE DocDB.}
                &  APA & PDS  & TPC Elec   & HV   & DAQ  & CISC & CAL  & COMP \\ \toprowrule
  Facility      & \cite{bib:docdb6967} & \cite{bib:docdb6970} & \cite{bib:docdb6973} & \cite{bib:docdb6985} & \cite{bib:docdb6988} & \cite{bib:docdb6991} & \cite{bib:docdb6829} & \cite{bib:docdb6841} \\ \colhline
  Installation  & \cite{bib:docdb6994} & \cite{bib:docdb6997} & \cite{bib:docdb7000} & \cite{bib:docdb7012} & \cite{bib:docdb7015} & \cite{bib:docdb7018} & \cite{bib:docdb6847} & \cite{bib:docdb6853} \\ \colhline
  Physics       & \cite{bib:docdb7075} & \cite{bib:docdb7078} & \cite{bib:docdb7081} & \cite{bib:docdb7093} & \cite{bib:docdb7096} & \cite{bib:docdb7099} & \cite{bib:docdb6865} &   \cite{bib:docdb6871}   \\ 
\end{dunetable}

%

\section{Schedule Milestones}
\label{sec:fdsp-coord-controls}

A series of tiered milestones have been developed for the \dword{dune}
project. The spokespersons and host laboratory director are
responsible for the tier 0 milestones. Three tier 0 milestones have
been defined and the dates set:
\begin{enumerate}
\item Start main cavern excavation \hspace{2.58in} 2020
\item Start \dword{detmodule}~1 installation \hspace{2.1in} 2024
\item Start operations of \dwords{detmodule} \#1--2 with beam \hspace{0.8in} 2028
\end{enumerate}
These dates will be revisited after the U.S. \dword{lbnf} project is
reviewed. The \dword{tcoord}, \dword{ipd} and \dword{lbnf} project
manager hold the tier 1 milestones.  The consortia hold tier 2
milestones. Table~\ref{tab:DUNE_schedule} provides a high level version of the
\dword{dune} milestones from the \dword{lbnf-dune} schedule.

\begin{dunetable}
[\dshort{dune} schedule milestones]
{p{0.75\textwidth}p{0.18\textwidth}}
{tab:DUNE_schedule}
{\dword{dune} schedule milestones for first two far detector modules. Key DUNE dates and milestones, defined for planning purposes in this TDR, are shown in orange.  Dates will be finalized following establishment of the international project baseline.}
Milestone & Date   \\ \toprowrule
Final design reviews  & 2020 \\ \colhline
Start of APA production & August 2020 \\ \colhline
Start photosensor procurement & July 2021 \\ \colhline
Start TPC electronics procurement  & December 2021 \\ \colhline
Production readiness reviews  &  2022    \\ \colhline
\rowcolor{dunepeach} South Dakota Logistics Warehouse available& \sdlwavailable      \\ \colhline
Start of ASIC/FEMB production   & May 2022   \\ \colhline
Start of DAQ server procurement &September 2022    \\ \colhline
\rowcolor{dunepeach} Beneficial occupancy of cavern 1 and \dshort{cuc}& \cucbenocc      \\ \colhline
Finish assembly of initial PD modules (80)      &March 2023    \\ \colhline
\rowcolor{dunepeach} \dshort{cuc} \dshort{daq} room available& \accesscuccountrm      \\ \colhline
Start of DAQ installation&      May 2023   \\ \colhline
Start of FC production for \dshort{detmodule} \#1       &September 2023   \\ \colhline
Start of CPA production for \dshort{detmodule} \#1&     December 2023   \\ \colhline
\rowcolor{dunepeach} Top of \dshort{detmodule} \#1 cryostat accessible& \accesstopfirstcryo      \\ \colhline
Start TPC electronics installation on top of \dshort{detmodule} \#1     & April 2024   \\ \colhline
Start FEMB installation on APAs for \dshort{detmodule} \#1 &    August 2024    \\ \colhline
\rowcolor{dunepeach}Start of \dshort{detmodule} \#1 \dshort{tpc} installation& \startfirsttpcinstall      \\ \colhline
\rowcolor{dunepeach} Top of \dshort{detmodule} \#2 cryostat accessible& \accesstopsecondcryo      \\ \colhline 
Complete FEMB installation on APAs for \dshort{detmodule} \#1   &March 2025    \\ \colhline
End DAQ installation    &May 2025    \\ \colhline
\rowcolor{dunepeach} End of \dshort{detmodule} \#1 \dshort{tpc} installation& \firsttpcinstallend      \\ \colhline 
\rowcolor{dunepeach}Start of \dshort{detmodule} \#2 \dshort{tpc} installation& \startsecondtpcinstall      \\ \colhline
End of FC production for \dshort{detmodule} \#1 &January 2026     \\ \colhline
End of APA production for \dshort{detmodule} \#1        &April 2026    \\ \colhline
\rowcolor{dunepeach} End \dshort{detmodule} \#2 \dshort{tpc} installation& \secondtpcinstallend      \\  \colhline
\rowcolor{dunepeach}Start detector module \#1 operations & July 2026 \\
\end{dunetable}

To monitor progress, \dword{jpo} scheduling team will maintain the
\dword{lbnf-dune} schedule that links all consortium schedules and
contains milestones for each consortia.  The schedules will go under
change control after each consortium agrees to the milestone dates,
the \dword{tdr} is approved, and the \dword{lbnf} project is baselined.

To ensure that the \dword{dune} detector remains on schedule,
\dword{tc} will monitor schedule status from each
consortium and organize reviews of schedules and risks as appropriate.
As schedule problems arise, \dword{tc} will work with the affected
consortium to resolve the problems. If problems cannot be solved, the
\dword{tcoord} will take the issue to the \dword{tb} and \dword{exb}.

A monthly report with input from all the consortia will be published by
\dword{tc} and provided to the \dword{lbnc}. This will
include updates on consortium and \dword{tc} technical progress
against the schedule.

\section{Requirements}
\label{sec:fdsp-coord-requirements}

The scientific goals of \dword{dune} as described in 
Volume~\volnumberexec:~\voltitleexec of this \dword{tdr}  include
\begin{itemize}
\item a comprehensive program of neutrino oscillation measurements
  including the search for \dword{cpv};
\item measurement of $\nu_{e}$ flux from a core-collapse supernova within our
  galaxy, should one occur during \dword{dune} operations; and 
\item a search for baryon number violation.
\end{itemize}
These goals motivate a number of key detector requirements: drift
field, electron lifetime, system noise, photon detector light yield
and time resolution. The \dword{exb} has approved a list of high-level
detector specifications, including those listed above. These will be
maintained in \dword{edms}, and the high-level requirements with
significant impact on physics (applying to both \dword{sp} and
\dword{dp} \dwords{detmodule} are highlighted in Table~\ref{tab:dunephysicsreqs}.
\begin{dunetable}
  [DUNE physics-related specifications owned by EB]
  {p{0.025\textwidth}p{0.06\textwidth}p{0.2\textwidth}p{0.35\textwidth}p{0.15\textwidth}p{0.1\textwidth}}
  {tab:dunephysicsreqs}
  {\dword{dune} physics-related specifications owned by \dword{exb}}
  ID & System & Parameter & Physics Requirement Driver & Requirement & Goal \\ \toprowrule
  1   & HVS    & Minimum drift field &  Limit recombination, diffusion and space charge impacts on particle ID. Establish adequate \dword{s/n} for tracking. & >\SI{250}{V/cm} & \spmaxfield \\ \colhline
  2   & TPC Elec     & System noise & The noise specification is driven by pattern recognition and two-track separation.  & <\SI{1000}{enc} & ALARA \\ \colhline
  3   & PDS    & Light yield  & The light yield shall be sufficient to measure time of events with visible energy above 200 MeV.  Goal is 10\% energy measurement for visible energy of 10 MeV.  & >\SI{0.5}{pe/MeV} & >\SI{5}{pe/MeV}  \\ \colhline
  4   & PDS    & Time resolution  & The time resolution of the photon detection system shall be sufficient to assign a unique event time.  & $<\,\SI{1}{\micro\second}$ & $<\,\SI{100}{\nano\second}$  \\ \colhline
  5   & all    & liquid argon purity & The LAr purity shall be sufficient to enable drift e- lifetime of 3 (10)ms & $<$\,\SI{100}{ppt} & $<$\,\SI{30}{ppt} \\ 
\end{dunetable}
Eleven other significant specifications for the \dword{spmod}
owned by the \dword{exb} are listed in Table~\ref{tab:dunephysicsspecs}
along with another twelve high-level engineering specifications.
\begin{dunetable}
  [DUNE high-level system specifications owned by the EB]
  {p{0.025\textwidth}p{0.06\textwidth}p{0.2\textwidth}p{0.35\textwidth}p{0.15\textwidth}p{0.1\textwidth}}
  {tab:dunephysicsspecs}
  {\dword{dune} high-level system specifications owned by \dword{exb}}
  ID & System & Parameter & Physics Requirement Driver & Requirement & Goal \\ \toprowrule
  6   & APA & Gaps between APAs  & minimize events lost due to vertex in gaps between APAs (15mm on same support beam, 30mm on adjacent beams) & <\SI{30}{mm} & <\SI{15}{mm} \\ \colhline
  7   & DSS & Drift field uniformity & tolerance on drift field due to component location & $<\,\SI{1}{\%}$  &   \\ \colhline
  8   & APA & wire angles  & 0$^\circ$ collection, $\pm$35.7$^\circ$ induction &  &  \\ \colhline
  9   & APA & wire spacing  & \SI{4.669}{mm} for U,V; \SI{4.790}{mm} for X,G &  &  \\ \colhline
  10  & APA & wire position tolerance  & & $\pm\,\SI{0.5}{mm}$  &  \\ \colhline
  11  & HVS & Drift field uniformity & tolerance on drift field due to HVS system & $<\,\SI{1}{\%}$  &  \\ \colhline
  12  & HVS & Cathode power supply ripple & very small compared to intrinsic electronics noise & $<\,\SI{100}{enc}$ &   \\ \colhline
  13  & TPC Elec & Front end peaking time  & optimize vertex resolution & \SI{1}{\micro\second} &  \\ \colhline
  14  & TPC Elec & Signal saturation  & largest signals occur with multiple protons in the primary vertex & 500k $e^-$ &  \\ \colhline
  15  & cryo & LAr N$_2$ contamination  & optical attenuation length in liquid argon with 50~ppm of N$_2$ contamination is roughly 3~m & $<\,\SI{25}{ppm}$ &  \\ \colhline
29  & all & Detector uptime  &  risk of missing a supernova burst & $<\,\SI{98}{\%}$ & $<\,\SI{99}{\%}$  \\  \colhline
30  & all & Individual detector module uptime  &  meet physics goals in timely fashion & $<\,\SI{90}{\%}$ & $<\,\SI{95}{\%}$  \\ 
\end{dunetable}
The high level \dword{dune} requirements that drive the \dword{lbnf} design are
maintained in~\citedocdb{112} and under change control. These are owned by
the \dword{dune} \dword{tcoord} and \dword{lbnf} project manager.

Lower level detector specifications are held by the consortia and
described in the \dword{dune} \dword{tdr} 
Volumes~\volnumbersp{}, \voltitlesp{}, and~\volnumberdp{}, \voltitledp{}. A complete list of detector specifications is
provided in Section~\ref{sec:fdsp-app-requirements}.

\section{Full DUNE Requirements}
\label{sec:fdsp-app-requirements}

\subsection{Single-phase}
\label{sec:tc-req-sp}

\begin{footnotesize}
\begin{longtable}{p{0.12\textwidth}p{0.18\textwidth}p{0.17\textwidth}p{0.25\textwidth}p{0.16\textwidth}}
\caption{Specifications for SP-FD \fixmehl{ref \texttt{tab:spec:SP-FD}}} \\
  \rowcolor{dunesky}
       Label & Description  & Specification \newline (Goal) & Rationale & Validation \\  \colhline

   \newtag{SP-FD-1}{ spec:min-drift-field }  & Minimum drift field  &  $>$\,\SI{250}{ V/cm} \newline ($>\,\SI{500}{ V/cm}$) &  Lessens impacts of $e^-$-Ar recombination, $e^-$ lifetime, $e^-$ diffusion and space charge. &  ProtoDUNE \\ \colhline

  \newtag{SP-FD-2}{ spec:system-noise }  & System noise  &  $<\,\SI{1000}\,e^-$ &  Provides $>$5:1 S/N on induction planes for  pattern recognition and two-track separation. &  ProtoDUNE and simulation \\ \colhline

  \newtag{SP-FD-3}{ spec:light-yield }  & Light yield  &  $>\,\SI{20}{PE/MeV}$ (avg), $>\,\SI{0.5}{PE/MeV}$ (min) &  Gives PDS energy resolution comparable to that of the TPC for 5-7 MeV SN $\nu$s, and allows tagging of $>\,\SI{99}{\%}$ of nucleon decay backgrounds with light at all points in detector. &  Supernova and nucleon decay events in the FD with full simulation and reconstruction. \\ \colhline
    
   \newtag{SP-FD-4}{ spec:time-resolution-pds }  & Time resolution  &  $<\,\SI{1}{\micro\second}$ \newline ($<\,\SI{100}{\nano\second}$) &  Enables \SI{1}{mm} position resolution for \SI{10}{MeV} SNB candidate events for instantaneous rate $<\,\SI{1}{m^{-3}ms^{-1}}$. &   \\ \colhline
    
   \newtag{SP-FD-5}{ spec:lar-purity }  & Liquid argon purity  &  $<$\,\SI{100}{ppt} \newline ($<\,\SI{30}{ppt}$) &  Provides $>$5:1 S/N on induction planes for  pattern recognition and two-track separation. &  Purity monitors and cosmic ray tracks \\ \colhline

  \newtag{SP-FD-6}{ spec:apa-gaps }  & Gaps between APAs   &  $<\,\SI{15}{mm}$ between APAs on same support beam; $<\,\SI{30}{mm}$ between APAs on different support beams &  Maintains fiducial volume.  Simplified contruction. &  ProtoDUNE \\ \colhline

  \newtag{SP-FD-7}{ spec:misalignment-field-uniformity }  & Drift field uniformity due to component alignment  &  $<\,1\,$\% throughout volume &  Maintains APA, CPA,  FC orientation and shape. &  ProtoDUNE \\ \colhline

  \newtag{SP-FD-8}{ spec:apa-wire-angles }  & APA wire angles  &  \SI{0}{\degree} for collection wires, $\pm\,$\SI{35.7}{\degree} for induction wires &  Minimize inter-APA dead space. &  Engineering calculation \\ \colhline

  \newtag{SP-FD-9}{ spec:apa-wire-spacing }  & APA wire spacing  &  \SI{4.669}{mm} for U,V; \SI{4.790}{mm} for X,G &  Enables 100\% efficient MIP detection, \SI{1.5}{cm} $yz$ vertex resolution. &  Simulation \\ \colhline

  \newtag{SP-FD-10}{ spec:apa-wire-pos-tolerance }  & APA wire position tolerance  &  $\pm\,\SI{0.5}{mm}$ &  Interplane electron transparency; $dE/dx$, range, and MCS calibration. &  ProtoDUNE and simulation \\ \colhline

  \newtag{SP-FD-11}{ spec:hvs-field-uniformity }  & Drift field uniformity due to HVS  &  $<\,\SI{1}{\%}$ throughout volume &  High reconstruction efficiency. &  ProtoDUNE and simulation \\ \colhline

  \newtag{SP-FD-12}{ spec:hv-ps-ripple }  & Cathode HV power supply ripple contribution to system noise  &  $<\,\SI{100}e^-$ &  Maximize live time; maintain high S/N. &  Engineering calculation, in situ measurement,   ProtoDUNE \\ \colhline

  \newtag{SP-FD-13}{ spec:fe-peak-time }  & Front-end peaking time  &  \SI{1}{\micro\second} &  Vertex resolution; optimized for \SI{5}{mm} wire spacing. &  ProtoDUNE and simulation \\ \colhline
    
   \newtag{SP-FD-14}{ spec:sp-signal-saturation }  & Signal saturation level  &  \num{500000} $e^-$ \newline (Adjustable so as to see saturation in less than \SI{10}{\%} of beam-produced events) &  Maintain calorimetric performance for multi-proton final state. &  Simulation \\ \colhline

  \newtag{SP-FD-15}{ spec:lar-n-contamination }  & LAr nitrogen contamination  &  $<\,\SI{25}{ppm}$ &  Maintain \SI{0.5}{PE/MeV} PDS sensitivity required for triggering proton decay near cathode. &  In situ measurement \\ \colhline

  \newtag{SP-FD-16}{ spec: }  & (downtime requirement replaced with 29 and 30)  &   &   &   \\ \colhline
    
   \newtag{SP-FD-17}{ spec:cathode-resistivity }  & Cathode resistivity  &  $>\,\SI{1}{\mega\ohm/square}$ \newline ($>\,\SI{1}{\giga\ohm/square}$) &  Detector damage prevention. &  ProtoDUNE \\ \colhline

  \newtag{SP-FD-18}{ spec:cryo-monitor-devices }  & Cryogenic monitoring devices  &   &  Constrain uncertainties on detection efficiency, fiducial volume. &  ProtoDUNE \\ \colhline

  \newtag{SP-FD-19}{ spec:adc-sampling-freq }  & ADC sampling frequency  &  $\sim\,\SI{2}{\mega\hertz}$ &  Match \SI{1}{\micro\second} shaping time. &  Nyquist requirement and design choice \\ \colhline

  \newtag{SP-FD-20}{ spec:adc-number-of-bits }  & Number of ADC bits  &  \num{12} bits &  ADC noise contribution negligible (low end); match signal saturation specification (high end). &  Engineering calculation and design choice \\ \colhline

  \newtag{SP-FD-21}{ spec:ce-power-consumption }  & Cold electronics power consumption   &  $<\,\SI{50}{ mW/channel} $ &  No bubbles in LAr to reduce HV discharge risk. &  Bench test \\ \colhline

  \newtag{SP-FD-22}{ spec:data-rate-to-tape }  & Data rate to tape  &  $<\,\SI{30}{PB/year}$ &  Cost.  Bandwidth. &  ProtoDUNE \\ \colhline

  \newtag{SP-FD-23}{ spec:sn-trigger }  & Supernova trigger  &  $>\,\SI{95}{\%}$ efficiency for a SNB producing at least 60 interactions with a neutrino energy >10 MeV in 12 kt of active detector mass during the first 10 seconds of the burst. &  $>\,$95\% efficiency for SNB within 20 kpc &  Simulation and bench tests \\ \colhline

  \newtag{SP-FD-24}{ spec:local-e-fields }  & Local electric fields  &  $<\,\SI{30}{kV/cm}$ &  Maximize live time; maintain high S/N. &  ProtoDUNE \\ \colhline

  \newtag{SP-FD-25}{ spec:non-fe-noise }  & Non-FE noise contributions  &  $<<\,\SI{1000}\,e^- $ &  High S/N for high reconstruction efficiency. &  Engineering calculation and ProtoDUNE \\ \colhline

  \newtag{SP-FD-26}{ spec:lar-impurity-contrib }  & LAr impurity contributions from components  &  $<<\,\SI{30}{ppt} $ &  Maintain HV operating range for high live time fraction. &  ProtoDUNE \\ \colhline

  \newtag{SP-FD-27}{ spec:radiopurity }  & Introduced radioactivity  &  less than that from $^{39}$Ar &  Maintain low radiological backgrounds for SNB searches. &  ProtoDUNE and assays during construction \\ \colhline

  \newtag{SP-FD-28}{ spec:dead-channels }  & Dead channels  &  $<\,\SI{1}{\%}$ &  Minimize the degradation in physics performance over the $>\,20$-year detector operation. &  ProtoDUNE and bench tests \\ \colhline
    
   \newtag{SP-FD-29}{ spec:dp-det-uptime }  & Detector uptime  &  $>\,$98\% \newline ($>\,$99\%) &  Meet physics goals in timely fashion. &  ProtoDUNE \\ \colhline
    
   \newtag{SP-FD-30}{ spec:dp-det-mod-uptime }  & Individual detector module uptime  &  $>\,$90\% \newline ($>\,$95\%) &  Meet physics goals in timely fashion. &  ProtoDUNE \\ \colhline

\label{tab:specs:SP-FD}
\end{longtable}
\end{footnotesize}

\begin{footnotesize}
\begin{longtable}{p{0.12\textwidth}p{0.18\textwidth}p{0.17\textwidth}p{0.25\textwidth}p{0.16\textwidth}}
\caption{APA specifications \fixmehl{ref \texttt{tab:spec:SP-APA}}} \\
  \rowcolor{dunesky}
       Label & Description  & Specification \newline (Goal) & Rationale & Validation \\  \colhline

  \newtag{SP-FD-6}{ spec:apa-gaps }  & Gaps between APAs   &  $<\,\SI{15}{mm}$ between APAs on same support beam; $<\,\SI{30}{mm}$ between APAs on different support beams &  Maintains fiducial volume.  Simplified contruction. &  ProtoDUNE \\ \colhline

  \newtag{SP-FD-7}{ spec:misalignment-field-uniformity }  & Drift field uniformity due to component alignment  &  $<\,1\,$\% throughout volume &  Maintains APA, CPA,  FC orientation and shape. &  ProtoDUNE \\ \colhline

  \newtag{SP-FD-8}{ spec:apa-wire-angles }  & APA wire angles  &  \SI{0}{\degree} for collection wires, $\pm\,$\SI{35.7}{\degree} for induction wires &  Minimize inter-APA dead space. &  Engineering calculation \\ \colhline

  \newtag{SP-FD-9}{ spec:apa-wire-spacing }  & APA wire spacing  &  \SI{4.669}{mm} for U,V; \SI{4.790}{mm} for X,G &  Enables 100\% efficient MIP detection, \SI{1.5}{cm} $yz$ vertex resolution. &  Simulation \\ \colhline

  \newtag{SP-FD-10}{ spec:apa-wire-pos-tolerance }  & APA wire position tolerance  &  $\pm\,\SI{0.5}{mm}$ &  Interplane electron transparency; $dE/dx$, range, and MCS calibration. &  ProtoDUNE and simulation \\ \colhline

  \newtag{SP-APA-1}{ spec:apa-unit-size }  & APA unit size  &  \SI{6.0}{m} tall $\times$ \SI{2.3}{m} wide &  Maximum size allowed for fabrication, transportation, and installation.  &  ProtoDUNE-SP  \\ \colhline

  \newtag{SP-APA-2}{ spec:apa-active-area }  & Active area  &  Maximize total active area. &  Maximize area for data collection  &  ProtoDUNE-SP  \\ \colhline

  \newtag{SP-APA-3}{ spec:apa-wire-tension }  & Wire tension  &  \SI{6}{N} $\pm$ \SI{1}{N} &  Prevent contact beween wires and minimize  break risk &  ProtoDUNE-SP \\ \colhline

  \newtag{SP-APA-4}{ spec:apa-bias-voltage }  & Wire plane bias voltages  &  The setup, including boards, must hold 150\% of max operating voltage. &  Headroom in case adjustments needed &  E-field simulation sets wire bias voltages. ProtoDUNE-SP confirms performance. \\ \colhline

  \newtag{SP-APA-5}{ spec:apa-frame-planarity }  & Frame planarity (twist limit)  &  $<$\SI{5}{mm} &  APA transparency.  Ensures wire plane spacing change of $<$0.5 mm.  &  ProtoDUNE-SP \\ \colhline

  \newtag{SP-APA-6}{ spec:apa-bad-channels }  & Missing/unreadable channels  &  $<$1\%, with a goal of $<$0.5\% &  Reconstruction efficiency &  ProtoDUNE-SP \\ \colhline

\label{tab:specs:SP-APA}
\end{longtable}
\end{footnotesize}
\begin{footnotesize}
\begin{longtable}{p{0.12\textwidth}p{0.18\textwidth}p{0.17\textwidth}p{0.25\textwidth}p{0.16\textwidth}}
\caption{HV specifications \fixmehl{ref \texttt{tab:spec:SP-HV}}} \\
  \rowcolor{dunesky}
       Label & Description  & Specification \newline (Goal) & Rationale & Validation \\  \colhline

   \newtag{SP-FD-1}{ spec:min-drift-field }  & Minimum drift field  &  $>$\,\SI{250}{ V/cm} \newline ($>\,\SI{500}{ V/cm}$) &  Lessens impacts of $e^-$-Ar recombination, $e^-$ lifetime, $e^-$ diffusion and space charge. &  ProtoDUNE \\ \colhline

  \newtag{SP-FD-11}{ spec:hvs-field-uniformity }  & Drift field uniformity due to HVS  &  $<\,\SI{1}{\%}$ throughout volume &  High reconstruction efficiency. &  ProtoDUNE and simulation \\ \colhline

  \newtag{SP-FD-12}{ spec:hv-ps-ripple }  & Cathode HV power supply ripple contribution to system noise  &  $<\,\SI{100}e^-$ &  Maximize live time; maintain high S/N. &  Engineering calculation, in situ measurement,   ProtoDUNE \\ \colhline
    
   \newtag{SP-FD-17}{ spec:cathode-resistivity }  & Cathode resistivity  &  $>\,\SI{1}{\mega\ohm/square}$ \newline ($>\,\SI{1}{\giga\ohm/square}$) &  Detector damage prevention. &  ProtoDUNE \\ \colhline

  \newtag{SP-FD-24}{ spec:local-e-fields }  & Local electric fields  &  $<\,\SI{30}{kV/cm}$ &  Maximize live time; maintain high S/N. &  ProtoDUNE \\ \colhline
    
   \newtag{SP-FD-29}{ spec:dp-det-uptime }  & Detector uptime  &  $>\,$98\% \newline ($>\,$99\%) &  Meet physics goals in timely fashion. &  ProtoDUNE \\ \colhline
    
   \newtag{SP-FD-30}{ spec:dp-det-mod-uptime }  & Individual detector module uptime  &  $>\,$90\% \newline ($>\,$95\%) &  Meet physics goals in timely fashion. &  ProtoDUNE \\ \colhline

  \newtag{SP-HV-1}{ spec:power-supply-stability }  & Maximize power supply stability  &  $>\,\SI{95}{\%}$ uptime &  Collect data over long period with high uptime. &  ProtoDUNE \\ \colhline
    
   \newtag{SP-HV-2}{ spec:hv-connection-redundancy }  & Provide redundancy in all \dshort{hv} connections.  &  Two-fold \newline (Four-fold) &  Avoid interrupting data collection or causing accesses to the interior of the detector. &  Assembly QC \\ \colhline

\label{tab:specs:SP-HV}
\end{longtable}
\end{footnotesize}
\begin{footnotesize}
\begin{longtable}{p{0.12\textwidth}p{0.18\textwidth}p{0.17\textwidth}p{0.25\textwidth}p{0.16\textwidth}}
\caption{TPC electronics specifications \fixmehl{ref \texttt{tab:spec:SP-ELEC}}} \\
  \rowcolor{dunesky}
       Label & Description  & Specification \newline (Goal) & Rationale & Validation \\  \colhline

  \newtag{SP-FD-2}{ spec:system-noise }  & System noise  &  $<\,\SI{1000}\,e^-$ &  Provides $>$5:1 S/N on induction planes for  pattern recognition and two-track separation. &  ProtoDUNE and simulation \\ \colhline

  \newtag{SP-FD-13}{ spec:fe-peak-time }  & Front-end peaking time  &  \SI{1}{\micro\second} &  Vertex resolution; optimized for \SI{5}{mm} wire spacing. &  ProtoDUNE and simulation \\ \colhline
    
   \newtag{SP-FD-14}{ spec:sp-signal-saturation }  & Signal saturation level  &  \num{500000} $e^-$ \newline (Adjustable so as to see saturation in less than \SI{10}{\%} of beam-produced events) &  Maintain calorimetric performance for multi-proton final state. &  Simulation \\ \colhline

  \newtag{SP-FD-19}{ spec:adc-sampling-freq }  & ADC sampling frequency  &  $\sim\,\SI{2}{\mega\hertz}$ &  Match \SI{1}{\micro\second} shaping time. &  Nyquist requirement and design choice \\ \colhline

  \newtag{SP-FD-20}{ spec:adc-number-of-bits }  & Number of ADC bits  &  \num{12} bits &  ADC noise contribution negligible (low end); match signal saturation specification (high end). &  Engineering calculation and design choice \\ \colhline

  \newtag{SP-FD-21}{ spec:ce-power-consumption }  & Cold electronics power consumption   &  $<\,\SI{50}{ mW/channel} $ &  No bubbles in LAr to reduce HV discharge risk. &  Bench test \\ \colhline

  \newtag{SP-FD-25}{ spec:non-fe-noise }  & Non-FE noise contributions  &  $<<\,\SI{1000}\,e^- $ &  High S/N for high reconstruction efficiency. &  Engineering calculation and ProtoDUNE \\ \colhline

  \newtag{SP-FD-28}{ spec:dead-channels }  & Dead channels  &  $<\,\SI{1}{\%}$ &  Minimize the degradation in physics performance over the $>\,20$-year detector operation. &  ProtoDUNE and bench tests \\ \colhline

  \newtag{SP-ELEC-1}{ spec:num-FE-baselines }  & Number of baselines in the front-end amplifier  &  \num{2} &  Use a single type of amplifier for both induction and collection wires &  ProtoDUNE \\ \colhline
    
   \newtag{SP-ELEC-2}{ spec:gain-FE-amplifier }  & Gain of the front-end amplifier  &  $\sim\SI{20}{mV/fC}$ \newline (Adjustable in the range \SIrange{5}{25}{mV/fC}) &  The gain of the FE amplifier is obtained from the maximum charge to be observed without saturation and from the operating voltage of the amplifier, that depends on the technology choice. &   \\ \colhline
    
   \newtag{SP-ELEC-3}{ spec:syncronization-CE }  & System synchronization  &  \SI{50}{ns} \newline (\SI{10}{ns}) &  The dispersion of the sampling times on different wires of the APA should be much smaller than the sampling time (500 ns) and give a negligible contribution to the hit resolution. &   \\ \colhline

  \newtag{SP-ELEC-4}{ spec:num-channels-FEMB }  & Number of channels per front-end motherboard  &  \num{128} &  The total number of wires on one side of an APA, 1,280, must be an integer multiple of the number of channels on the FEMBs. &  Design \\ \colhline
    
   \newtag{SP-ELEC-5}{ spec:FEMB-data-link }  & Number of links between the FEMB and the WIB  &  \num{4} at \SI{1.28}{Gbps} \newline (\num{2} at \SI{2.56}{Gbps}) &  Balance between reducing the number of links and reliability and power issues when increasing the data transmission speed. &  ProtoDUNE, Laboratory measurements on bit error rates \\ \colhline

  \newtag{SP-ELEC-6}{ spec:number-FEMB-per-WIB }  & Number of FEMBs per WIB  &  \num{4} &  The total number of FEMB per WIB is a balance between the complexity of the boards, the mechanics inside the WIEC, and the required processing power of the FPGA on the WIB.  &  ProtoDUNE, Design \\ \colhline

  \newtag{SP-ELEC-7}{ spec:WIB-data-link }  & Data transmission speed between the WIB and the DAQ backend  &  \SI{10}{Gbps} &  Balance between cost and reduction of the number of optical fiber links for each WIB. &  ProtoDUNE, Laboratory measurements on bit error rates \\ \colhline

  \newtag{SP-ELEC-8}{ spec:cold-cables-xsec }  & Maximum diameter of conduit enclosing the cold cables while they are routed through the APA frame  &  \SI{6.35}{cm} (2.5") &  Avoid the need for further changes to the APA frame and for routing the cables along the cryostat walls &  Tests on APA frame prototypes \\ \colhline

\label{tab:specs:SP-ELEC}
\end{longtable}
\end{footnotesize} 
\begin{footnotesize}
\begin{longtable}{p{0.12\textwidth}p{0.18\textwidth}p{0.17\textwidth}p{0.25\textwidth}p{0.16\textwidth}}
\caption{PDS specifications \fixmehl{ref \texttt{tab:spec:SP-PDS}}} \\
  \rowcolor{dunesky}
       Label & Description  & Specification \newline (Goal) & Rationale & Validation \\  \colhline

  \newtag{SP-FD-3}{ spec:light-yield }  & Light yield  &  $>\,\SI{20}{PE/MeV}$ (avg), $>\,\SI{0.5}{PE/MeV}$ (min) &  Gives PDS energy resolution comparable to that of the TPC for 5-7 MeV SN $\nu$s, and allows tagging of $>\,\SI{99}{\%}$ of nucleon decay backgrounds with light at all points in detector. &  Supernova and nucleon decay events in the FD with full simulation and reconstruction. \\ \colhline
    
   \newtag{SP-FD-4}{ spec:time-resolution-pds }  & Time resolution  &  $<\,\SI{1}{\micro\second}$ \newline ($<\,\SI{100}{\nano\second}$) &  Enables \SI{1}{mm} position resolution for \SI{10}{MeV} SNB candidate events for instantaneous rate $<\,\SI{1}{m^{-3}ms^{-1}}$. &   \\ \colhline

  \newtag{SP-FD-15}{ spec:lar-n-contamination }  & LAr nitrogen contamination  &  $<\,\SI{25}{ppm}$ &  Maintain \SI{0.5}{PE/MeV} PDS sensitivity required for triggering proton decay near cathode. &  In situ measurement \\ \colhline

  \newtag{SP-PDS-1}{ spec:pds-cleanroom-assbly }  & Clean assembly area  &  Class \num{100000} clean assembly area &  Demonstrated as satisfactory in \dshort{pdsp}, and is the \dshort{dune} assembly area standard. &  \dshort{pdsp} and in \fnal materials test stand \\ \colhline

  \newtag{SP-PDS-2}{ spec:spatial-localization }  & Spatial localization in $y$-$z$ plane  &  $<\,\SI{2.5}{\meter}$ &  Enables accurate matching of \dshort{pd} and \dshort{tpc} signals. &  \dshort{snb} neutrino and \dshort{ndk} simulation in the \dshort{fd} \\ \colhline

  \newtag{SP-PDS-3}{ spec:env-light-exposure }  & Environmental light exposure  &  No exposure to sunlight. All other unfiltered sources: $<\,\num{30}$ minutes integrated across all exposures &  Shown to prevent damage to \dshort{wls} coatings due to UV. &  Studies in \dshort{pdsp}, and at IU \\ \colhline

  \newtag{SP-PDS-4}{ spec:env-humidity-limit }  & Environmental humidity limit  &  $<\,\SI{50}{\%}$ RH at \SI{70}{\degree F} &  Demonstrated to prevent damage to \dshort{wls} coatings due to humidity. &  \dshort{pd} optical coating studies \\ \colhline

  \newtag{SP-PDS-5}{ spec:light-tightness }  & Light-tight cryostat  &  Cryostat light leaks responsible for $<\,\SI{10}{\%}$  of data transferred from PDS to DAQ &  Minimizing false triggers due to cryostat light leaks helps limit the data transfer rate to  \dshort{daq}. &  \dshort{pdsp} and \dshort{iceberg} \\ \colhline

  \newtag{SP-PDS-7}{ spec:mech-deflection }  & Mechanical deflection (static)  &  $<\,\SI{5}{\milli\meter}$ &  Minimize motion of \dshort{pd} modules inside the \dshort{apa} (due to static and dynamic loads) to avoid damaging \dshort{apa}. &  \dshort{pd} \dshort{fea}, \dshort{pdsp}, \dshort{iceberg}; Ash River integration tests and CERN pre-production integration tests pending \\ \colhline

  \newtag{SP-PDS-8}{ spec:apa-install }  & Clearance for installation through APA side tubes  &  $>\,\SI{1}{\milli\meter}$ &  Maintain required clearance to allow \dshort{pd} insertion into \dshort{apa} following wire wrapping. &  \dshort{pd} \dshort{fea}, \dshort{pdsp}, \dshort{iceberg};  Ash River integration tests and CERN pre-production integration tests pending \\ \colhline

  \newtag{SP-PDS-9}{ spec:mech-compatibility }  & No mechanical interference with APA, SP-CE and SP-HV detector elements (clearance)  &  $>\,\SI{1}{\milli\meter}$ &  \dshort{pd} mounting and securing element tolerances must prevent interference with \dshort{apa} and \dshort{ce} cable bundles. &  \dshort{iceberg}, Ash River integration tests, and the CERN pre-production integration tests \\ \colhline

  \newtag{SP-PDS-10}{ spec:pds-cable }  & APA intrusion limit for PD cable routing   &  $<\,\SI{6}{\milli\meter}$ &  \dshort{pd} modules must install into \dshort{apa} frames following wire wrapping.  \dshort{pd} modules must not occlude \dshort{apa} side tubes. &  \dshort{iceberg}, Ash River integration  tests, and the CERN pre-production integration tests \\ \colhline

  \newtag{SP-PDS-11}{ spec:pds-cablemate }  & PD cabling cannot limit upper-lower APA junction gap  &  \SI{0}{\milli\meter} separation mechanically allowed &  \dshort{pd} cable connections must not limit the minimum upper and lower \dshort{apa} separation. &  \dshort{iceberg}, Ash River integration  tests, and the CERN pre-production integration tests \\ \colhline

  \newtag{SP-PDS-12}{ spec:pds-clearance }  & Maintain PD-APA clearance at LAr temperature  &  $>\,\SI{0.5}{\milli\meter}$ &  \dshort{pd} mounting frame and cable harness must accommodate thermal contraction of itself and \dshort{apa} frame. &  Thermal modeling, \dshort{protodune}, \dshort{iceberg}, CERN pre-production integration tests \\ \colhline

  \newtag{SP-PDS-13}{ spec:pds-datarate }  & Data transfer rate from SP-PD to DAQ  &  $<\,\SI{8}{Gbps}$ &  \dshort{pd} data transfer must not exceed \dshort{daq} data throughput capability. &  Maximum bandwidth out of the \dshort{pd} electronics is \SI{80}{Mbps} \\ \colhline

  \newtag{SP-PDS-14}{ spec:pds-signaltonoise }  & Signal-to-noise in SP-PD  &  $>\,\num{4}$ &  Keep data rate within electronics bandwidth limits. &  \dshort{pdsp}, \dshort{iceberg} and \dshort{pdsp2} \\ \colhline

  \newtag{SP-PDS-15}{ spec:pds-darkrate }  & Dark noise rate in SP-PD  &  $<\,\SI{1}{kHz}$ &  Keep data rate within electronics bandwidth limits. &  Pre-production photosensor testing, \dshort{pdsp}, \dshort{iceberg} and \dshort{pdsp2} \\ \colhline

  \newtag{SP-PDS-16}{ spec:pds-dynamicrange }  & Dynamic Range in SP-PD  &  $<\,\SI{20}{\%}$ &  Keep the rate of saturating channels low enough for effective mitigation. &  Pre-production photosensor testing, \dshort{pdsp}, \dshort{iceberg} and \dshort{pdsp2} \\ \colhline

\label{tab:specs:SP-PDS}
\end{longtable}
\end{footnotesize}
\begin{footnotesize}
\begin{longtable}{p{0.12\textwidth}p{0.18\textwidth}p{0.17\textwidth}p{0.25\textwidth}p{0.16\textwidth}}
\caption{Calibration specifications \fixmehl{ref \texttt{tab:spec:SP-CALIB}}} \\
  \rowcolor{dunesky}
       Label & Description  & Specification \newline (Goal) & Rationale & Validation \\  \colhline

   \newtag{SP-FD-1}{ spec:min-drift-field }  & Minimum drift field  &  $>$\,\SI{250}{ V/cm} \newline ($>\,\SI{500}{ V/cm}$) &  Lessens impacts of $e^-$-Ar recombination, $e^-$ lifetime, $e^-$ diffusion and space charge. &  ProtoDUNE \\ \colhline

  \newtag{SP-FD-2}{ spec:system-noise }  & System noise  &  $<\,\SI{1000}\,e^-$ &  Provides $>$5:1 S/N on induction planes for  pattern recognition and two-track separation. &  ProtoDUNE and simulation \\ \colhline
    
   \newtag{SP-FD-5}{ spec:lar-purity }  & Liquid argon purity  &  $<$\,\SI{100}{ppt} \newline ($<\,\SI{30}{ppt}$) &  Provides $>$5:1 S/N on induction planes for  pattern recognition and two-track separation. &  Purity monitors and cosmic ray tracks \\ \colhline

  \newtag{SP-FD-7}{ spec:misalignment-field-uniformity }  & Drift field uniformity due to component alignment  &  $<\,1\,$\% throughout volume &  Maintains APA, CPA,  FC orientation and shape. &  ProtoDUNE \\ \colhline

  \newtag{SP-FD-9}{ spec:apa-wire-spacing }  & APA wire spacing  &  \SI{4.669}{mm} for U,V; \SI{4.790}{mm} for X,G &  Enables 100\% efficient MIP detection, \SI{1.5}{cm} $yz$ vertex resolution. &  Simulation \\ \colhline

  \newtag{SP-FD-11}{ spec:hvs-field-uniformity }  & Drift field uniformity due to HVS  &  $<\,\SI{1}{\%}$ throughout volume &  High reconstruction efficiency. &  ProtoDUNE and simulation \\ \colhline

  \newtag{SP-FD-13}{ spec:fe-peak-time }  & Front-end peaking time  &  \SI{1}{\micro\second} &  Vertex resolution; optimized for \SI{5}{mm} wire spacing. &  ProtoDUNE and simulation \\ \colhline

  \newtag{SP-FD-22}{ spec:data-rate-to-tape }  & Data rate to tape  &  $<\,\SI{30}{PB/year}$ &  Cost.  Bandwidth. &  ProtoDUNE \\ \colhline

  \newtag{SP-FD-23}{ spec:sn-trigger }  & Supernova trigger  &  $>\,\SI{95}{\%}$ efficiency for a SNB producing at least 60 interactions with a neutrino energy >10 MeV in 12 kt of active detector mass during the first 10 seconds of the burst. &  $>\,$95\% efficiency for SNB within 20 kpc &  Simulation and bench tests \\ \colhline

  \newtag{SP-FD-24}{ spec:local-e-fields }  & Local electric fields  &  $<\,\SI{30}{kV/cm}$ &  Maximize live time; maintain high S/N. &  ProtoDUNE \\ \colhline

  \newtag{SP-FD-25}{ spec:non-fe-noise }  & Non-FE noise contributions  &  $<<\,\SI{1000}\,e^- $ &  High S/N for high reconstruction efficiency. &  Engineering calculation and ProtoDUNE \\ \colhline

  \newtag{SP-FD-26}{ spec:lar-impurity-contrib }  & LAr impurity contributions from components  &  $<<\,\SI{30}{ppt} $ &  Maintain HV operating range for high live time fraction. &  ProtoDUNE \\ \colhline

  \newtag{SP-FD-27}{ spec:radiopurity }  & Introduced radioactivity  &  less than that from $^{39}$Ar &  Maintain low radiological backgrounds for SNB searches. &  ProtoDUNE and assays during construction \\ \colhline
    
   \newtag{SP-FD-29}{ spec:dp-det-uptime }  & Detector uptime  &  $>\,$98\% \newline ($>\,$99\%) &  Meet physics goals in timely fashion. &  ProtoDUNE \\ \colhline
    
   \newtag{SP-FD-30}{ spec:dp-det-mod-uptime }  & Individual detector module uptime  &  $>\,$90\% \newline ($>\,$95\%) &  Meet physics goals in timely fashion. &  ProtoDUNE \\ \colhline

  \newtag{SP-CALIB-1}{ spec:efield-calib-precision }  & Ionization laser \efield measurement precision  &  $\SI{1}{\%}$ &  \efield affects energy and position measurements. &  ProtoDUNE and external experiments. \\ \colhline
    
   \newtag{SP-CALIB-2}{ spec:efield-calib-coverage }  & Ionization laser \efield measurement coverage  &  $>\SI{75}{\%}$ \newline ($\SI{100}{\%}$) &  Allowable size of the uncovered detector regions is set by the highest reasonably expected field distortions, \SI{4}{\%}. &  ProtoDUNE \\ \colhline
    
   \newtag{SP-CALIB-3}{ spec:efield-calib-granularity }  & Ionization laser \efield measurement  granularity  &  $~30\times 30\times 30~$\SI{}{\centi\metre\cubed} \newline ($10\times 10\times 10~$\SI{}{\centi\metre\cubed}) &  Minimum measurable region is set by the maximum expected distortion and position reconstruction requirements. &  ProtoDUNE \\ \colhline
    
   \newtag{SP-CALIB-4}{ spec:laser-location-precision }  & Laser beam location precision  &  $~\SI{0.5}{\milli\radian}$ \newline ($<\SI{0.5}{\milli\radian}$) &  The necessary spatial precision does not need to be smaller than the APA wire gap. &  ProtoDUNE \\ \colhline
    
   \newtag{SP-CALIB-5}{ spec:neutron-source-coverage }  & Neutron source coverage  &  $>\SI{75}{\%}$ \newline ($\SI{100}{\%}$) &  Set by the energy resolution requirements at low energy. &  Simulations \\ \colhline
    
   \newtag{SP-CALIB-6}{ spec:data-volume-laser }  & Ionization laser data volume per year (per 10 kt)  &  $>\SI{184}{TB/yr/10 kt}$ \newline ($>\SI{368}{TB/yr/10 kt}$) &  The laser data volume must allow the needed coverage and granularity. &  ProtoDUNE and simulations \\ \colhline
    
   \newtag{SP-CALIB-7}{ spec:data-volume-pns }  & Neutron source data volume per year (per 10 kt)  &  $>\SI{144}{TB/yr/10 kt}$ \newline ($>\SI{288}{TB/yr/10 kt}$) &  The pulsed neutron system must allow the needed coverage and granularity. &  Simulations \\ \colhline

\label{tab:specs:SP-CALIB}
\end{longtable}
\end{footnotesize}
\begin{footnotesize}
\begin{longtable}{p{0.12\textwidth}p{0.18\textwidth}p{0.17\textwidth}p{0.25\textwidth}p{0.16\textwidth}}
\caption{DAQ specifications \fixmehl{ref \texttt{tab:spec:SP-DAQ}}} \\
  \rowcolor{dunesky}
       Label & Description  & Specification \newline (Goal) & Rationale & Validation \\  \colhline

  \newtag{SP-FD-22}{ spec:data-rate-to-tape }  & Data rate to tape  &  $<\,\SI{30}{PB/year}$ &  Cost.  Bandwidth. &  ProtoDUNE \\ \colhline

  \newtag{SP-FD-23}{ spec:sn-trigger }  & Supernova trigger  &  $>\,\SI{95}{\%}$ efficiency for a SNB producing at least 60 interactions with a neutrino energy >10 MeV in 12 kt of active detector mass during the first 10 seconds of the burst. &  $>\,$95\% efficiency for SNB within 20 kpc &  Simulation and bench tests \\ \colhline

  \newtag{SP-DAQ-1}{ spec:DAQ-readout }  & DAQ readout throughput: The DAQ shall be able to accept the continuous data stream from the TPC and Photon detectors.  &  1.5 TB/s per single phase detector module &  Specification from TPC and PDS electronics &  Modular test on ProtoDUNE; overall throughput scales linearly with number of APAs \\ \colhline

  \newtag{SP-DAQ-2}{ spec:DAQ-throughput }  & DAQ storage throughput: The DAQ shall be able to store selected data at an average throughput of 10 Gb/s, with temporary peak throughput of 100 Gb/s.  &  10 Gb/s average storage throughput; 100 Gb/s peak temporary storage throughput per single phase detector module &  Average throughput estimated from physics and calibration requirements; peak throughput allowing for fast storage of SNB data ($\sim 10^4$ seconds to store 120 TB of data).  &  ProtoDUNE demonstrated steady storage at $\sim$ 40 Gb/s for a storage volume of 700 TB. Laboratory tests will allow to demonstrate the performance reach. \\ \colhline

  \newtag{SP-DAQ-3}{ spec:DAQ-readout-window }  & DAQ readout window: The DAQ shall support storing triggered data of one or more APAs with a variable size readout window, from few $\mu$s (calibration) to 100 s (SNB), with a typical readout window for triggered interactions of 5.4 ms.  &  10 $\mu$s < readout window < 100 s &  Storage of the complete dataset for up to 100 s is required by the SNB physics studies; the typical readout window of 5.4 ms is defined by the drift time in the detector; calibration triggers can be configured to readout data much shorter time intervals. &  Implementation techniques to be validated on the ProtoDUNE setup and in test labs. \\ \colhline

  \newtag{SP-DAQ-4}{ spec:trigger-calibration }  & Calibration trigger: The DAQ shall provide the means to distribute time-synchronous commands to the calibration systems, in order to fire them, at a configurable rate and sequence and at configurable intervals in time. Those commands may be distributed during physics data taking or during special calibration data taking sessions. The DAQ shall trigger and acquire data at a fixed, configurable interval after the distribution of the commands, in order to capture the response of the detector to calibration stimuli.  &   &  Calibration is essential to attain required detector performance comprehension. &  Techniques for doing this have been run successfully in MicroBooNE and ProtoDUNE.  \\ \colhline

  \newtag{SP-DAQ-5}{ spec:data-record }  & Data record: Corresponding to every trigger, the DAQ shall form a data record to be transferred to offline together with the metadata necessary for validation and processing.  &   &  Needed for offline analysis. &  Common experimental practice. \\ \colhline

  \newtag{SP-DAQ-6}{ spec:data-verification }  & Data verification: The DAQ shall check integrity of data at every data transfer step. It shall only delete data from the local storage after confirmation that data have been correctly recorded to permanent storage.  &   &  Data integrity checking is fundamental to ensure data quality. &   \\ \colhline

  \newtag{SP-DAQ-7}{ spec:trigger-high-energy }  & High-energy Trigger: The DAQ shall trigger and acquire data on visible energy deposition >100 MeV. Data acquisition may be limited to the area in which activity was detected.  &  $>$\SI{100}{\MeV} &  Driven by DUNE physics mission. &  Physics TDR. 100 MeV is an achievable parameter; lower thresholds are possible. \\ \colhline

  \newtag{SP-DAQ-8}{ spec:trigger-low-energy }  & Low-energy Trigger: The DAQ shall trigger and acquire data on visible energy deposition > 10 MeV of single neutrino interactions. Those triggers will normally be fired using a pre-scaling factor, in order to limit the data volume.  &  $>$\SI{10}{\MeV} &  Driven by DUNE physics mission. &  Physics TDR. 10 MeV is an achievable parameter; lower thresholds are possible. \\ \colhline

  \newtag{SP-DAQ-9}{ spec:daq-deadtime }  & DAQ deadtime: While taking data within the agreed conditions, the DAQ shall be able to trigger and acquire data without introducing any deadtime.  &   &  Driven by DUNE physics mission. &  Zero deadtime is an achievable inter-event deadtime but a small deadtime would not significantly compromise physics sensitivity.  \\ \colhline

\label{tab:specs:SP-DAQ}
\end{longtable}
\end{footnotesize} 
\begin{footnotesize}
\begin{longtable}{p{0.12\textwidth}p{0.18\textwidth}p{0.17\textwidth}p{0.25\textwidth}p{0.16\textwidth}}
\caption{CISC specifications \fixmehl{ref \texttt{tab:spec:SP-CISC}}} \\
  \rowcolor{dunesky}
       Label & Description  & Specification \newline (Goal) & Rationale & Validation \\  \colhline

   \newtag{SP-FD-5}{ spec:lar-purity }  & Liquid argon purity  &  $<$\,\SI{100}{ppt} \newline ($<\,\SI{30}{ppt}$) &  Provides $>$5:1 S/N on induction planes for  pattern recognition and two-track separation. &  Purity monitors and cosmic ray tracks \\ \colhline

  \newtag{SP-FD-15}{ spec:lar-n-contamination }  & LAr nitrogen contamination  &  $<\,\SI{25}{ppm}$ &  Maintain \SI{0.5}{PE/MeV} PDS sensitivity required for triggering proton decay near cathode. &  In situ measurement \\ \colhline

  \newtag{SP-FD-18}{ spec:cryo-monitor-devices }  & Cryogenic monitoring devices  &   &  Constrain uncertainties on detection efficiency, fiducial volume. &  ProtoDUNE \\ \colhline

  \newtag{SP-FD-25}{ spec:non-fe-noise }  & Non-FE noise contributions  &  $<<\,\SI{1000}\,e^- $ &  High S/N for high reconstruction efficiency. &  Engineering calculation and ProtoDUNE \\ \colhline
    
   \newtag{SP-FD-29}{ spec:dp-det-uptime }  & Detector uptime  &  $>\,$98\% \newline ($>\,$99\%) &  Meet physics goals in timely fashion. &  ProtoDUNE \\ \colhline
    
   \newtag{SP-FD-30}{ spec:dp-det-mod-uptime }  & Individual detector module uptime  &  $>\,$90\% \newline ($>\,$95\%) &  Meet physics goals in timely fashion. &  ProtoDUNE \\ \colhline

  \newtag{SP-CISC-1}{ spec:inst-noise }  & Noise from Instrumentation devices  &  $<<\,\SI{1000}\,e^- $ &  Max noise for 5:1 S/N for a MIP passing near cathode; per SBND and DUNE CE &  ProtoDUNE \\ \colhline
    
   \newtag{SP-CISC-2}{ spec:inst-efield }  & Max. E field near instrumentation devices  &  $<\,\SI{30}{kV/cm}$ \newline ($<\,\SI{15}{kV/cm}$) &  Significantly lower than max field of 30 kV/cm per DUNE HV  &  3D electrostatic simulation \\ \colhline
    
   \newtag{SP-CISC-3}{ spec:elec-lifetime-prec }  & Precision in electron lifetime  &  $<\,$1.4\% \newline ($<\,$1\%) &  Required for accurate charge reconstruction per DUNE-FD Task Force report. &  ProtoDUNE-SP and CITF \\ \colhline

  \newtag{SP-CISC-4}{ spec:elec-lifetime-range }  & Range in electron lifetime  &  \SIrange{0.04}{10}{ms} in cryostat, \SIrange{0.04}{30}{ms} inline &  Slightly beyond best values observed so far in other detectors.  &  ProtoDUNE-SP and CITF \\ \colhline
    
   \newtag{SP-CISC-11}{ spec:temp-repro }  & Precision: temperature reproducibility  &  $<\,\SI{5}{mK}$ \newline (\SI{2}{mK}) &  Enables validation of CFD models, which predicts gradients below 15 mK &  ProtoDUNE-SP and CITF \\ \colhline
    
   \newtag{SP-CISC-14}{ spec:temp-stability }  & Temperature stability  &  $<\,\SI{2}{mK}$ at all places and times \newline (Match precision requirement at all places, at all times) &  Measure the temp map with sufficient precision during the entire duration &  ProtoDUNE-SP \\ \colhline
    
   \newtag{SP-CISC-27}{ spec:camera-cold-coverage }  & Cold camera coverage  &  $>\,$80\% of HV surfaces \newline (\num{100}\%) &  Enable detailed inspection of issues near HV surfaces. &  Calculated from location, validated in prototypes. \\ \colhline
    
   \newtag{SP-CISC-51}{ spec:slowcontrol-alarm-rate }  & Slow control alarm rate  &  $<\,$150/day \newline ($<\,$50/day) &  Alarm rate low enough to allow response to every alarm. &  Detector module; depends on experimental conditions \\ \colhline
    
   \newtag{SP-CISC-52}{ spec:slowcontrol-num-vars }  & Total No. of variables  &  $>\,\num{150000}$ \newline (\SIrange{150000}{200000}{}) &  Scaled from ProtoDUNE-SP &  ProtoDUNE-SP and CITF \\ \colhline
    
   \newtag{SP-CISC-54}{ spec:slowcontrol-archive-rate }  & Archiving rate  &  \SI{0.02}{Hz} \newline (Broad range \SI{1}{Hz} to \num{1} per few min.) &  Archiving rate different for each variable, optimized to store important information  &  ProtoDUNE-SP \\ \colhline

\label{tab:specs:SP-CISC}
\end{longtable}
\end{footnotesize} 
\begin{footnotesize}
\begin{longtable}{p{0.12\textwidth}p{0.18\textwidth}p{0.17\textwidth}p{0.25\textwidth}p{0.16\textwidth}}
\caption{Installation specifications \fixmehl{ref \texttt{tab:spec:SP-INST}}} \\
  \rowcolor{dunesky}
       Label & Description  & Specification \newline (Goal) & Rationale & Validation \\  \colhline

  \newtag{SP-INST-1}{ spec:logistics-material-handling }  & Compliance with the SURF Material Handling Specification for all material transported underground  &  SURF Material Handling Specification &  Loads must fit in the shaft be lifted safely. &  Visual and documentation check \\ \colhline

  \newtag{SP-INST-2}{ spec:logistics-shipping-coord }  & Coordination of shipments with CMGC; DUNE to schedule use of Ross Shaft  &  2 wk notice to CMGC &  Both DUNE and CMGC need to use Ross Shaft &  Deliveries will be rejected \\ \colhline

  \newtag{SP-INST-3}{ spec:logistics-materials-buffer }  & Maintain materials buffer at logistics facility in SD   &  $>1$ month &  Prevent schedule delays in case of shipping or customs delays &  Documentatation and progress reporting \\ \colhline

  \newtag{SP-INST-4}{ spec:apa-storage-sd }  & APA stroage at logistics facility in SD  &  700 m$^2$ &  Store APAs during lag between production and installation &  Agree upon space needs \\ \colhline

  \newtag{SP-INST-5}{ spec:cleanroom-specification }  & Installation cleanroom Specificaiton  &  ISO 8 &  Reduce dust (contains U/Th) to prevent induced radiological background in detector &  Monitor air purity \\ \colhline

  \newtag{SP-INST-6}{ spec:cleanroom-uv-filters }  & UV filter in installation cleanrooms for PDS sensor protection  &  filter $<\SI{400}{nm}$ for $>$ 2 wk exp; $<\SI{520}{nm}$ all else &  Prevent damage to PD coatings  &  Visual or spectrographic inspection \\ \colhline

\label{tab:specs:SP-INST}
\end{longtable}
\end{footnotesize}

\subsection{Dual-phase}
\label{sec:tc-req-dp}

\begin{footnotesize}
\begin{longtable}{p{0.12\textwidth}p{0.18\textwidth}p{0.17\textwidth}p{0.25\textwidth}p{0.16\textwidth}}
\caption{Specifications for DP-FD \fixmehl{ref \texttt{tab:spec:DP-FD}}} \\
  \rowcolor{dunesky}
       Label & Description  & Specification \newline (Goal) & Rationale & Validation \\  \colhline

   \newtag{DP-FD-1}{ spec:dp-min-drift-field }  & Minimum drift field  &  $>$\,\SI{250}{ V/cm} \newline ($>\,\SI{500}{ V/cm}$) &  Reduces impacts of $e^-$-Ar recombination, $e^-$ lifetime, $e^-$ diffusion and space charge. &  ProtoDUNE \\ \colhline

  \newtag{DP-FD-2}{ spec:dp-system-noise }  & System noise  &  $<\,\SI{1000}\,e^-$ &  Studies suggest that a minimum of 5:1 S/N on individual strip measurements allows for sufficient reconstruction performance. &  ProtoDUNE and simulation \\ \colhline

  \newtag{DP-FD-3}{ spec:dp-light-yield }  & Light yield  &  $>\,\SI{1}{PE/MeV}$ (at anode), $>\,\SI{5}{PE/MeV}$ (avg  over active volume) &  Enable drift position determination of \dshort{ndk} candidates. Enable \dshort{pds}-based triggering on galactic SNBs. &  Full sim/reco of \dshort{ndk}, \dshort{snb} $\nu$ and radiological events. \\ \colhline
    
   \newtag{DP-FD-4}{ spec:dp-time-resolution-pds }  & Time resolution  &  $<\,\SI{1}{\micro\second}$ \newline ($<\,\SI{100}{\nano\second}$) &  Enables \SI{1}{mm} position resolution for \SI{10}{MeV} SNB candidate events for instantaneous rate $<\,\SI{1}{m^{-3}ms^{-1}}$. &   \\ \colhline

  \newtag{DP-FD-5}{ spec:dp-lar-purity }  & Liquid argon purity  &  $>$\,\SI{5}{ms} &  Directly impacts the number of electrons received at the CRP collection strips and hence the S/N. &  Purity monitors and cosmic ray tracks \\ \colhline

  \newtag{DP-FD-6}{ spec:dp-crp-gaps }  & Gaps between CRPs   &  $<\,\SI{30}{mm}$ between adjacent CRPs &  Required for CRP positioning and shrinkage.  Simplified contruction. &  ProtoDUNE \\ \colhline

  \newtag{DP-FD-7}{ spec:dp-misalignment-field-uniformity }  & Drift field uniformity due to component alignment  &  $<\,1\,$\% throughout volume &  Maintains TPC and  FC orientation and shape. &  ProtoDUNE \\ \colhline

  \newtag{DP-FD-8}{ spec:dp-crp-eff-gain }  & CRP effective gain  &  \num{6} &  Yields S/N of  20 for 6 m drift and 34.6 for 3 m (for E 250 V/cm and 5 ms e lifetime). For 7 ms electron lifetime yields S/N of  12.5 for 12 m drift, 27.3 for 6 m, and 40.5 for 3 m. &   \\ \colhline

  \newtag{DP-FD-9}{ spec:dp-crp-strip-spacing }  & CRP strips spacing  &  $<\,\SI{4.7}{mm}$ &  Enables 100\% efficient MIP detection, \SI{1.5}{cm} $yz$ vertex resolution. &  Simulation \\ \colhline

  \newtag{DP-FD-10}{ spec:dp-crp-planarity }  & CRP planarity  &  $\pm\,\SI{0.5}{mm}$ &  Guarantee the uniformity of the extraction field and the immersion of the extraction grid.  &  ProtoDUNE and simulation \\ \colhline

  \newtag{DP-FD-11}{ spec:dp-hvs-field-uniformity }  & Drift field uniformity due to HVS  &  $<\,\SI{1}{\%}$ throughout volume &  High reconstruction efficiency. &  ProtoDUNE and simulation \\ \colhline

  \newtag{DP-FD-12}{ spec:dp-hv-ps-ripple }  & Cathode HV power supply ripple contribution to system noise  &  $<\,\SI{100}e^-$ &  Maximize live time; maintain high S/N. &  Engineering calculation, in situ measurement,   ProtoDUNE \\ \colhline
    
   \newtag{DP-FD-13}{ spec:dp-fe-peak-time }  & Front-end peaking time  &  \SI{1}{\micro\second} \newline (\SI{1}{\micro\second} achieved in current design) &  Vertex resolution; 1 us matches  3mm pitch and DP S/N ratio. &  ProtoDUNE and simulation \\ \colhline

  \newtag{DP-FD-14}{ spec:dp-signal-saturation }  & Signal saturation level  &  \num{7500000} electrons &  Maintain calorimetric performance for multi-proton final state; takes into account an effective CRP gain of 20 in the DP signal dynamics. &  Simulation \\ \colhline

  \newtag{DP-FD-15}{ spec:dp-lar-n-contamination }  & LAr nitrogen contamination  &  $<\,\SI{3}{ppm}$ &  Higher contaminations signifcantly affect the no. of photons that reach the \dshort{pmt}. &  In situ measurement \\ \colhline

  \newtag{DP-FD-16}{ spec: }  & (downtime requirement replaced with 29 and 30)  &   &   &   \\ \colhline
    
   \newtag{DP-FD-17}{ spec:dp-cathode-resistivity }  & Cathode resistivity  &  $>\,\SI{1}{\mega\ohm/square}$ \newline ($>\,\SI{1}{\giga\ohm/square}$) &  Detector damage prevention. &  ProtoDUNE \\ \colhline

  \newtag{DP-FD-18}{ spec:dp-cryo-monitor-devices }  & Cryogenic monitoring devices  &   &  Constrain uncertainties on detection efficiency, fiducial volume. &  ProtoDUNE \\ \colhline

  \newtag{DP-FD-19}{ spec:dp-adc-sampling-freq }  & ADC sampling frequency  &  $\sim\,\SI{2.5}{\mega\hertz}$ &  Match \SI{1}{\micro\second} shaping time. &  Nyquist requirement and design choice \\ \colhline

  \newtag{DP-FD-20}{ spec:dp-adc-number-of-bits }  & Number of ADC bits  &  \num{12} bits &  ADC noise contribution negligible (low end); match signal saturation specification (high end). &  Engineering calculation and design choice \\ \colhline

  \newtag{DP-FD-21}{ spec:dp-ce-power-consumption }  & TPC analog cold FE electronics power consumption   &  $<\,\SI{50}{ mW/channel} $ &  No bubbles in LAr to reduce HV discharge risk. &  Bench test \\ \colhline

  \newtag{DP-FD-22}{ spec:dp-data-rate-to-tape }  & Data rate to tape  &  $<\,\SI{30}{PB/year}$ &  Cost.  Bandwidth. &  ProtoDUNE \\ \colhline

  \newtag{DP-FD-23}{ spec:dp-sn-trigger }  & Supernova trigger  &  $>\,\SI{95}{\%}$ efficiency for a SNB producing at least 60 interactions with a neutrino energy >10 MeV in 12 kt of active detector mass during the first 10 seconds of the burst. &  $>\,$95\% efficiency for SNB within 100 kpc &  Simulation and bench tests \\ \colhline

  \newtag{DP-FD-24}{ spec:dp-local-e-fields }  & Local electric fields  &  $<\,\SI{30}{kV/cm}$ &  Maximize live time; maintain high S/N. &  ProtoDUNE \\ \colhline

  \newtag{DP-FD-25}{ spec:de-non-fe-noise }  & Non-FE noise contributions  &  $<<\,\SI{1000}\,e^- $ &  High S/N for high reconstruction efficiency. &  Engineering calculation and ProtoDUNE \\ \colhline

  \newtag{DP-FD-26}{ spec:dp-lar-impurity-contrib }  & LAr impurity contributions from components  &  $<<\,\SI{30}{ppt} $ &   &  ProtoDUNE \\ \colhline

  \newtag{DP-FD-27}{ spec:dp-radiopurity }  & Introduced radioactivity  &  less than that from $^{39}$Ar &  Maintain low radiological backgrounds for SNB searches. &  ProtoDUNE and assays during construction \\ \colhline

  \newtag{DP-FD-28}{ spec:dp-dead-channels }  & Dead channels  &  $<\,\SI{1}{\%}$ &  Contingency for possible efficiency loss for $>\,$20 year operation.  All DP electronics are accessible. &  ProtoDUNE \\ \colhline
    
   \newtag{DP-FD-29}{ spec:dp-det-uptime }  & Detector uptime  &  $>\,$98\% \newline ($>\,$99\%) &  Meet physics goals in timely fashion. &  ProtoDUNE \\ \colhline
    
   \newtag{DP-FD-30}{ spec:dp-det-mod-uptime }  & Individual detector module uptime  &  $>\,$90\% \newline ($>\,$95\%) &  Meet physics goals in timely fashion. &  ProtoDUNE \\ \colhline

\label{tab:specs:DP-FD}
\end{longtable}
\end{footnotesize}

\begin{footnotesize}
\begin{longtable}{p{0.12\textwidth}p{0.18\textwidth}p{0.17\textwidth}p{0.25\textwidth}p{0.16\textwidth}}
\caption{DP CRP specifications \fixmehl{ref \texttt{tab:spec:DP-CRP}}} \\
  \rowcolor{dunesky}
       Label & Description  & Specification \newline (Goal) & Rationale & Validation \\  \colhline

   \newtag{DP-FD-1}{ spec:dp-min-drift-field }  & Minimum drift field  &  $>$\,\SI{250}{ V/cm} \newline ($>\,\SI{500}{ V/cm}$) &  Reduces impacts of $e^-$-Ar recombination, $e^-$ lifetime, $e^-$ diffusion and space charge. &  ProtoDUNE \\ \colhline

  \newtag{DP-FD-2}{ spec:dp-system-noise }  & System noise  &  $<\,\SI{1000}\,e^-$ &  Studies suggest that a minimum of 5:1 S/N on individual strip measurements allows for sufficient reconstruction performance. &  ProtoDUNE and simulation \\ \colhline

  \newtag{DP-FD-3}{ spec:dp-light-yield }  & Light yield  &  $>\,\SI{1}{PE/MeV}$ (at anode), $>\,\SI{5}{PE/MeV}$ (avg  over active volume) &  Enable drift position determination of \dshort{ndk} candidates. Enable \dshort{pds}-based triggering on galactic SNBs. &  Full sim/reco of \dshort{ndk}, \dshort{snb} $\nu$ and radiological events. \\ \colhline
    
   \newtag{DP-FD-4}{ spec:dp-time-resolution-pds }  & Time resolution  &  $<\,\SI{1}{\micro\second}$ \newline ($<\,\SI{100}{\nano\second}$) &  Enables \SI{1}{mm} position resolution for \SI{10}{MeV} SNB candidate events for instantaneous rate $<\,\SI{1}{m^{-3}ms^{-1}}$. &   \\ \colhline

  \newtag{DP-FD-5}{ spec:dp-lar-purity }  & Liquid argon purity  &  $>$\,\SI{5}{ms} &  Directly impacts the number of electrons received at the CRP collection strips and hence the S/N. &  Purity monitors and cosmic ray tracks \\ \colhline

  \newtag{DP-FD-6}{ spec:dp-crp-gaps }  & Gaps between CRPs   &  $<\,\SI{30}{mm}$ between adjacent CRPs &  Required for CRP positioning and shrinkage.  Simplified contruction. &  ProtoDUNE \\ \colhline

  \newtag{DP-FD-8}{ spec:dp-crp-eff-gain }  & CRP effective gain  &  \num{6} &  Yields S/N of  20 for 6 m drift and 34.6 for 3 m (for E 250 V/cm and 5 ms e lifetime). For 7 ms electron lifetime yields S/N of  12.5 for 12 m drift, 27.3 for 6 m, and 40.5 for 3 m. &   \\ \colhline

  \newtag{DP-FD-9}{ spec:dp-crp-strip-spacing }  & CRP strips spacing  &  $<\,\SI{4.7}{mm}$ &  Enables 100\% efficient MIP detection, \SI{1.5}{cm} $yz$ vertex resolution. &  Simulation \\ \colhline

  \newtag{DP-FD-10}{ spec:dp-crp-planarity }  & CRP planarity  &  $\pm\,\SI{0.5}{mm}$ &  Guarantee the uniformity of the extraction field and the immersion of the extraction grid.  &  ProtoDUNE and simulation \\ \colhline

  \newtag{DP-CRP-1}{ spec:crp-vert-precision }  & CRP vertical positioning precision  &  $<\,\SI{1}{\milli\meter}$ &  The extraction grid  must remain below the liquid argon surface &  Obtained by design \\ \colhline

\label{tab:specs:DP-CRP}
\end{longtable}
\end{footnotesize}
\begin{footnotesize}
\begin{longtable}{p{0.12\textwidth}p{0.18\textwidth}p{0.17\textwidth}p{0.25\textwidth}p{0.16\textwidth}}
\caption{DP HV specifications \fixmehl{ref \texttt{tab:spec:DP-HV}}} \\
  \rowcolor{dunesky}
       Label & Description  & Specification \newline (Goal) & Rationale & Validation \\  \colhline

   \newtag{DP-FD-1}{ spec:dp-min-drift-field }  & Minimum drift field  &  $>$\,\SI{250}{ V/cm} \newline ($>\,\SI{500}{ V/cm}$) &  Reduces impacts of $e^-$-Ar recombination, $e^-$ lifetime, $e^-$ diffusion and space charge. &  ProtoDUNE \\ \colhline

  \newtag{DP-FD-11}{ spec:dp-hvs-field-uniformity }  & Drift field uniformity due to HVS  &  $<\,\SI{1}{\%}$ throughout volume &  High reconstruction efficiency. &  ProtoDUNE and simulation \\ \colhline

  \newtag{DP-FD-12}{ spec:dp-hv-ps-ripple }  & Cathode HV power supply ripple contribution to system noise  &  $<\,\SI{100}e^-$ &  Maximize live time; maintain high S/N. &  Engineering calculation, in situ measurement,   ProtoDUNE \\ \colhline
    
   \newtag{DP-FD-17}{ spec:dp-cathode-resistivity }  & Cathode resistivity  &  $>\,\SI{1}{\mega\ohm/square}$ \newline ($>\,\SI{1}{\giga\ohm/square}$) &  Detector damage prevention. &  ProtoDUNE \\ \colhline

  \newtag{DP-FD-24}{ spec:dp-local-e-fields }  & Local electric fields  &  $<\,\SI{30}{kV/cm}$ &  Maximize live time; maintain high S/N. &  ProtoDUNE \\ \colhline

   \newtag{DP-HV-1}{ spec:hvdb-redundancy }  & Provide redundancy in HV distribution  &  $>\,\num{2}$ HVDB chain \newline (\num{12} HVDB chains) &  Ensure the HV connections to the detector &  ProtoDUNE and calculations \\ \colhline

\label{tab:specs:DP-HV}
\end{longtable}
\end{footnotesize}
\begin{footnotesize}
\begin{longtable}{p{0.12\textwidth}p{0.18\textwidth}p{0.17\textwidth}p{0.25\textwidth}p{0.16\textwidth}}
\caption{DP TPC electronics specifications \fixmehl{ref \texttt{tab:spec:DP-ELEC}}} \\
  \rowcolor{dunesky}
       Label & Description  & Specification \newline (Goal) & Rationale & Validation \\  \colhline

  \newtag{DP-FD-2}{ spec:dp-system-noise }  & System noise  &  $<\,\SI{1000}\,e^-$ &  Studies suggest that a minimum of 5:1 S/N on individual strip measurements allows for sufficient reconstruction performance. &  ProtoDUNE and simulation \\ \colhline
    
   \newtag{DP-FD-4}{ spec:dp-time-resolution-pds }  & Time resolution  &  $<\,\SI{1}{\micro\second}$ \newline ($<\,\SI{100}{\nano\second}$) &  Enables \SI{1}{mm} position resolution for \SI{10}{MeV} SNB candidate events for instantaneous rate $<\,\SI{1}{m^{-3}ms^{-1}}$. &   \\ \colhline
    
   \newtag{DP-FD-13}{ spec:dp-fe-peak-time }  & Front-end peaking time  &  \SI{1}{\micro\second} \newline (\SI{1}{\micro\second} achieved in current design) &  Vertex resolution; 1 us matches  3mm pitch and DP S/N ratio. &  ProtoDUNE and simulation \\ \colhline

  \newtag{DP-FD-14}{ spec:dp-signal-saturation }  & Signal saturation level  &  \num{7500000} electrons &  Maintain calorimetric performance for multi-proton final state; takes into account an effective CRP gain of 20 in the DP signal dynamics. &  Simulation \\ \colhline

  \newtag{DP-FD-19}{ spec:dp-adc-sampling-freq }  & ADC sampling frequency  &  $\sim\,\SI{2.5}{\mega\hertz}$ &  Match \SI{1}{\micro\second} shaping time. &  Nyquist requirement and design choice \\ \colhline

  \newtag{DP-FD-20}{ spec:dp-adc-number-of-bits }  & Number of ADC bits  &  \num{12} bits &  ADC noise contribution negligible (low end); match signal saturation specification (high end). &  Engineering calculation and design choice \\ \colhline

  \newtag{DP-FD-21}{ spec:dp-ce-power-consumption }  & TPC analog cold FE electronics power consumption   &  $<\,\SI{50}{ mW/channel} $ &  No bubbles in LAr to reduce HV discharge risk. &  Bench test \\ \colhline

\label{tab:specs:DP-ELEC}
\end{longtable}
\end{footnotesize}
\begin{footnotesize}
\begin{longtable}{p{0.12\textwidth}p{0.18\textwidth}p{0.17\textwidth}p{0.25\textwidth}p{0.16\textwidth}}
\caption{DP PDS specifications \fixmehl{ref \texttt{tab:spec:DP-PDS}}} \\
  \rowcolor{dunesky}
       Label & Description  & Specification \newline (Goal) & Rationale & Validation \\  \colhline

  \newtag{DP-FD-3}{ spec:dp-light-yield }  & Light yield  &  $>\,\SI{1}{PE/MeV}$ (at anode), $>\,\SI{5}{PE/MeV}$ (avg  over active volume) &  Enable drift position determination of \dshort{ndk} candidates. Enable \dshort{pds}-based triggering on galactic SNBs. &  Full sim/reco of \dshort{ndk}, \dshort{snb} $\nu$ and radiological events. \\ \colhline
    
   \newtag{DP-FD-4}{ spec:dp-time-resolution-pds }  & Time resolution  &  $<\,\SI{1}{\micro\second}$ \newline ($<\,\SI{100}{\nano\second}$) &  Enables \SI{1}{mm} position resolution for \SI{10}{MeV} SNB candidate events for instantaneous rate $<\,\SI{1}{m^{-3}ms^{-1}}$. &   \\ \colhline

  \newtag{DP-FD-15}{ spec:dp-lar-n-contamination }  & LAr nitrogen contamination  &  $<\,\SI{3}{ppm}$ &  Higher contaminations signifcantly affect the no. of photons that reach the \dshort{pmt}. &  In situ measurement \\ \colhline

  \newtag{DP-PDS-1}{ spec:hit-relative-timing }  & Relative timing accuracy among hits  &  $<\,\SI{100}{ns RMS}$ &  Enable effective clustering of \dshort{pmt} signals based on relative hit timing information. &  Full sim/reco of \dshort{ndk}, \dshort{snb} $\nu$ and radiological events. \\ \colhline

  \newtag{DP-PDS-2}{ spec:hit-snr }  & Hit signal-to-noise ratio  &  $>\,\num{5}$ &  Efficiently reconstruct single-\phel hits while rate of electronics noise hits remains manageable. &  Single-\phel and baseline noise \dshort{rms} measurements in $3\times1\times1$ prototype. \\ \colhline

  \newtag{DP-PDS-3}{ spec:pmt-dark-rate }  & PMT dark count rate  &  $<\,\SI{100}{kHz}$ &  Avoid effects on clustering algorithm and \dshort{pds}-based calorimetry. &  Characterization of \dshorts{pmt} at cryo temps prior to installation. \\ \colhline

  \newtag{DP-PDS-4}{ spec:pds-analog-range }  & Analog range per channel  &  $>\,\SI{100}{PE/(ch\times 6\nano\second)}$ &  Minimize hit saturation for scintillation-based calorimetry over \SI{6}{ns} prompt light period, esp. for beam events. &  Full sim/reco of beam $\nu$ interactions.  \\ \colhline

\label{tab:specs:DP-PDS}
\end{longtable}
\end{footnotesize}
\begin{footnotesize}
\begin{longtable}{p{0.12\textwidth}p{0.18\textwidth}p{0.17\textwidth}p{0.25\textwidth}p{0.16\textwidth}}
\caption{DP calibration specifications \fixmehl{ref \texttt{tab:spec:DP-CALIB}}} \\
  \rowcolor{dunesky}
       Label & Description  & Specification \newline (Goal) & Rationale & Validation \\  \colhline

   \newtag{DP-FD-1}{ spec:dp-min-drift-field }  & Minimum drift field  &  $>$\,\SI{250}{ V/cm} \newline ($>\,\SI{500}{ V/cm}$) &  Reduces impacts of $e^-$-Ar recombination, $e^-$ lifetime, $e^-$ diffusion and space charge. &  ProtoDUNE \\ \colhline

  \newtag{DP-FD-2}{ spec:dp-system-noise }  & System noise  &  $<\,\SI{1000}\,e^-$ &  Studies suggest that a minimum of 5:1 S/N on individual strip measurements allows for sufficient reconstruction performance. &  ProtoDUNE and simulation \\ \colhline

  \newtag{DP-FD-5}{ spec:dp-lar-purity }  & Liquid argon purity  &  $>$\,\SI{5}{ms} &  Directly impacts the number of electrons received at the CRP collection strips and hence the S/N. &  Purity monitors and cosmic ray tracks \\ \colhline

  \newtag{DP-FD-7}{ spec:dp-misalignment-field-uniformity }  & Drift field uniformity due to component alignment  &  $<\,1\,$\% throughout volume &  Maintains TPC and  FC orientation and shape. &  ProtoDUNE \\ \colhline

  \newtag{DP-FD-9}{ spec:dp-crp-strip-spacing }  & CRP strips spacing  &  $<\,\SI{4.7}{mm}$ &  Enables 100\% efficient MIP detection, \SI{1.5}{cm} $yz$ vertex resolution. &  Simulation \\ \colhline

  \newtag{DP-FD-11}{ spec:dp-hvs-field-uniformity }  & Drift field uniformity due to HVS  &  $<\,\SI{1}{\%}$ throughout volume &  High reconstruction efficiency. &  ProtoDUNE and simulation \\ \colhline
    
   \newtag{DP-FD-13}{ spec:dp-fe-peak-time }  & Front-end peaking time  &  \SI{1}{\micro\second} \newline (\SI{1}{\micro\second} achieved in current design) &  Vertex resolution; 1 us matches  3mm pitch and DP S/N ratio. &  ProtoDUNE and simulation \\ \colhline

  \newtag{DP-FD-22}{ spec:dp-data-rate-to-tape }  & Data rate to tape  &  $<\,\SI{30}{PB/year}$ &  Cost.  Bandwidth. &  ProtoDUNE \\ \colhline

  \newtag{DP-FD-23}{ spec:dp-sn-trigger }  & Supernova trigger  &  $>\,\SI{95}{\%}$ efficiency for a SNB producing at least 60 interactions with a neutrino energy >10 MeV in 12 kt of active detector mass during the first 10 seconds of the burst. &  $>\,$95\% efficiency for SNB within 100 kpc &  Simulation and bench tests \\ \colhline

  \newtag{DP-FD-24}{ spec:dp-local-e-fields }  & Local electric fields  &  $<\,\SI{30}{kV/cm}$ &  Maximize live time; maintain high S/N. &  ProtoDUNE \\ \colhline

  \newtag{DP-FD-25}{ spec:de-non-fe-noise }  & Non-FE noise contributions  &  $<<\,\SI{1000}\,e^- $ &  High S/N for high reconstruction efficiency. &  Engineering calculation and ProtoDUNE \\ \colhline

  \newtag{DP-FD-26}{ spec:dp-lar-impurity-contrib }  & LAr impurity contributions from components  &  $<<\,\SI{30}{ppt} $ &   &  ProtoDUNE \\ \colhline

  \newtag{DP-FD-27}{ spec:dp-radiopurity }  & Introduced radioactivity  &  less than that from $^{39}$Ar &  Maintain low radiological backgrounds for SNB searches. &  ProtoDUNE and assays during construction \\ \colhline
    
   \newtag{DP-FD-29}{ spec:dp-det-uptime }  & Detector uptime  &  $>\,$98\% \newline ($>\,$99\%) &  Meet physics goals in timely fashion. &  ProtoDUNE \\ \colhline
    
   \newtag{DP-FD-30}{ spec:dp-det-mod-uptime }  & Individual detector module uptime  &  $>\,$90\% \newline ($>\,$95\%) &  Meet physics goals in timely fashion. &  ProtoDUNE \\ \colhline

  \newtag{DP-CALIB-1}{ spec:efield-calib-precision }  & Ionization laser \efield measurement precision  &  \SI{1}{\%} &  \efield affects energy and position measurements. &  ProtoDUNE and external experiments. \\ \colhline
    
   \newtag{DP-CALIB-2}{ spec:efield-calib-coverage }  & Ionization laser \efield measurement coverage  &  $>\,\SI{93}{\%}$ \newline (\SI{100}{\%}) &  Allowable size of the uncovered detector regions is set by the highest reasonably expected field distortions, \SI{15}{\%}. &  ProtoDUNE \\ \colhline
    
   \newtag{DP-CALIB-3}{ spec:efield-calib-granularity }  & Ionization laser \efield measurement  granularity  &  \SI{10 x 10 x 10}{\centi\metre} \newline (\SI{10 x 10 x 10}{\centi\metre}) &  Minimum measurable region is set by the maximum expected distortion and position reconstruction requirements. &  ProtoDUNE \\ \colhline
    
   \newtag{DP-CALIB-4}{ spec:laser-position-precision }  & Laser beam position precision  &  \SI{0.5}{\milli\radian} \newline ($<\,\SI{0.5}{\milli\radian}$) &  The necessary spatial precision does not need to be smaller than the CRP strip spacing. &  ProtoDUNE \\ \colhline
    
   \newtag{DP-CALIB-5}{ spec:neutron-source-coverage }  & Neutron source coverage  &  $>\,\SI{75}{\%}$ \newline (\SI{100}{\%}) &  Set by the energy resolution requirements at low energy. &  Simulations \\ \colhline
    
   \newtag{DP-CALIB-6}{ spec:data-volume-laser }  & Ionization laser DAQ rate per year (per 10 kt)  &  $>\,\SI{37}{TB/yr/10 kt}$ \newline ($>\,\SI{74}{TB/yr/10 kt}$) &  The laser data volume must allow the needed coverage and granularity. &  ProtoDUNE and simulations \\ \colhline
    
   \newtag{DP-CALIB-7}{ spec:data-volume-pns }  & Neutron source DAQ rate per year (per 10 kt)  &  $>\,\SI{170}{TB/yr/10 kt}$ \newline ($>\,\SI{340}{TB/yr/10 kt}$) &  The pulsed neutron system must allow the needed coverage and granularity. &  Simulations \\ \colhline

\label{tab:specs:DP-CALIB}
\end{longtable}
\end{footnotesize}
\begin{footnotesize}
\begin{longtable}{p{0.12\textwidth}p{0.18\textwidth}p{0.17\textwidth}p{0.25\textwidth}p{0.16\textwidth}}
\caption{DP DAQ specifications \fixmehl{ref \texttt{tab:spec:DP-DAQ}}} \\
  \rowcolor{dunesky}
       Label & Description  & Specification \newline (Goal) & Rationale & Validation \\  \colhline

  \newtag{DP-FD-22}{ spec:dp-data-rate-to-tape }  & Data rate to tape  &  $<\,\SI{30}{PB/year}$ &  Cost.  Bandwidth. &  ProtoDUNE \\ \colhline

  \newtag{DP-FD-23}{ spec:dp-sn-trigger }  & Supernova trigger  &  $>\,\SI{95}{\%}$ efficiency for a SNB producing at least 60 interactions with a neutrino energy >10 MeV in 12 kt of active detector mass during the first 10 seconds of the burst. &  $>\,$95\% efficiency for SNB within 100 kpc &  Simulation and bench tests \\ \colhline

  \newtag{DP-DAQ-1}{ spec:DAQ-readout }  & DAQ readout throughput: The DAQ shall be able to accept the continuous, compressed data stream from the TPC and Photon detectors.  &  \SI{65}{GB/s} per dual phase detector module &  Specification from TPC and PDS electronics &  Modular test on test bench; overall throughput scales linearly with the number of electronics crates.  \\ \colhline

  \newtag{DP-DAQ-2}{ spec:DAQ-throughput }  & DAQ storage throughput: The DAQ shall be able to store selected data at an average throughput of 2 Gb/s, with temporary peak throughput of 100 Gb/s.  &  \SI{2}{GB/s} average storage throughput; \SI{100}{GB/s} peak temporary storage throughput per dual phase detector module &  Average throughput estimated from physics and calibration requirements; peak throughput allowing for fast storage of SNB data.  &  ProtoDUNE demonstrated steady storage at $\sim$ 40 Gb/s for a storage volume of 700 TB. Laboratory tests will allow to demonstrate the performance reach. \\ \colhline

  \newtag{DP-DAQ-3}{ spec:DAQ-readout-window }  & DAQ readout window: The DAQ shall support storing triggered data of one or more CRPs with a variable size readout window, from few $\mu$s (calibration) to 100 s (SNB), with a typical readout window for triggered interactions of 7.5 ms.  &  \SI{10}{\micro\second} < readout window < \SI{100}{s} &  Storage of the complete dataset for up to  \SI{100}{s} is required by the SNB physics studies; the typical readout window of  \SI{7.5}{ms} is defined by the drift time in the detector; calibration triggers can be configured to readout data much shorter time intervals. &  Implementation techniques to be validated on the ProtoDUNE setup and in test labs. \\ \colhline

  \newtag{DP-DAQ-4}{ spec:trigger-calibration }  & Calibration trigger: The DAQ shall provide the means to distribute time-synchronous commands to the calibration systems, in order to fire them, at a configurable rate and sequence and at configurable intervals in time. Those commands may be distributed during physics data taking or during special calibration data taking sessions. The DAQ shall trigger and acquire data at a fixed, configurable interval after the distribution of the commands, in order to capture the response of the detector to calibration stimuli.  &   &  Calibration is essential to attain required detector performance comprehension. &  Techniques for doing this have been run successfully in MicroBooNE and ProtoDUNE.  \\ \colhline

  \newtag{DP-DAQ-5}{ spec:data-record }  & Data record: Corresponding to every trigger, the DAQ shall form a data record to be transferred to offline together with the metadata necessary for validation and processing.  &   &  Needed for offline analysis. &  Common experimental practice. \\ \colhline

  \newtag{DP-DAQ-6}{ spec:data-verification }  & Data verification: The DAQ shall check integrity of data at every data transfer step. It shall only delete data from the local storage after confirmation that data have been correctly recorded to permanent storage.  &   &  Data integrity checking is fundamental to ensure data quality. &   \\ \colhline

  \newtag{DP-DAQ-7}{ spec:trigger-high-energy }  & High-energy Trigger: The DAQ shall trigger and acquire data on visible energy deposition $>\,$\SI{100}{\MeV}. Data acquisition may be limited to the area in which activity was detected.  &  $>\,$\SI{100}{\MeV} &  Driven by DUNE physics mission. &  Physics TDR. \SI{100}{\MeV} is an achievable parameter; lower thresholds are possible. \\ \colhline

  \newtag{DP-DAQ-8}{ spec:trigger-low-energy }  & Low-energy Trigger: The DAQ shall trigger and acquire data on visible energy deposition $>\,$\SI{10}{\MeV} of single neutrino interactions. Those triggers will normally be fired using a pre-scaling factor, in order to limit the data volume.  &  $>\,$\SI{10}{\MeV} &  Driven by DUNE physics mission. &  Physics TDR. \SI{10}{\MeV} is an achievable parameter; lower thresholds are possible. \\ \colhline

  \newtag{DP-DAQ-9}{ spec:daq-deadtime }  & DAQ deadtime: While taking data within the agreed conditions, the DAQ shall be able to trigger and acquire data without introducing any deadtime.  &   &  Driven by DUNE physics mission. &  Zero deadtime is an achievable inter-event deadtime but a small deadtime would not significantly compromise physics sensitivity.  \\ \colhline

\label{tab:specs:DP-DAQ}
\end{longtable}
\end{footnotesize}
\begin{footnotesize}
\begin{longtable}{p{0.12\textwidth}p{0.18\textwidth}p{0.17\textwidth}p{0.25\textwidth}p{0.16\textwidth}}
\caption{DP CISC specifications \fixmehl{ref \texttt{tab:spec:DP-CISC}}} \\
  \rowcolor{dunesky}
       Label & Description  & Specification \newline (Goal) & Rationale & Validation \\  \colhline

  \newtag{DP-FD-5}{ spec:dp-lar-purity }  & Liquid argon purity  &  $>$\,\SI{5}{ms} &  Directly impacts the number of electrons received at the CRP collection strips and hence the S/N. &  Purity monitors and cosmic ray tracks \\ \colhline

  \newtag{DP-FD-15}{ spec:dp-lar-n-contamination }  & LAr nitrogen contamination  &  $<\,\SI{3}{ppm}$ &  Higher contaminations signifcantly affect the no. of photons that reach the \dshort{pmt}. &  In situ measurement \\ \colhline

  \newtag{DP-FD-18}{ spec:dp-cryo-monitor-devices }  & Cryogenic monitoring devices  &   &  Constrain uncertainties on detection efficiency, fiducial volume. &  ProtoDUNE \\ \colhline

  \newtag{DP-FD-25}{ spec:de-non-fe-noise }  & Non-FE noise contributions  &  $<<\,\SI{1000}\,e^- $ &  High S/N for high reconstruction efficiency. &  Engineering calculation and ProtoDUNE \\ \colhline
    
   \newtag{DP-FD-29}{ spec:dp-det-uptime }  & Detector uptime  &  $>\,$98\% \newline ($>\,$99\%) &  Meet physics goals in timely fashion. &  ProtoDUNE \\ \colhline
    
   \newtag{DP-FD-30}{ spec:dp-det-mod-uptime }  & Individual detector module uptime  &  $>\,$90\% \newline ($>\,$95\%) &  Meet physics goals in timely fashion. &  ProtoDUNE \\ \colhline

  \newtag{DP-CISC-1}{ spec:inst-noise }  & Noise from Instrumentation devices  &  $<<\,\SI{1000}\,e^- $ &  Max noise for 5:1 S/N for a MIP passing near cathode; per SBND and DUNE CE &  ProtoDUNE \\ \colhline
    
   \newtag{DP-CISC-2}{ spec:inst-efield }  & Max. E field near instrumentation devices  &  $<\,\SI{30}{kV/cm}$ \newline ($<\,\SI{15}{kV/cm}$) &  Significantly lower than max field of 30 kV/cm per DUNE HV  &  3D electrostatic simulation \\ \colhline
    
   \newtag{DP-CISC-3}{ spec:elec-lifetime-prec }  & Precision in electron lifetime  &  $<\,$2.3\% \newline ($<\,$1\%) &  Required to accurately reconstruct charge per DUNE-FD Task Force report. &  ProtoDUNE and CITF \\ \colhline

  \newtag{DP-CISC-4}{ spec:elec-lifetime-range }  & Range in electron lifetime  &  \SIrange{0.04}{10}{ms} in cryostat, \SIrange{0.04}{30}{ms} inline &  Slightly more than best values so far observed in other detectors. &  ProtoDUNE and CITF \\ \colhline
    
   \newtag{DP-CISC-11}{ spec:temp-repro }  & Precision: temperature reproducibility  &  $<\,\SI{5}{mK}$ \newline (\SI{2}{mK}) &  Allows validating CFD models that predict gradients less than 15 mK. &  ProtoDUNE and CITF \\ \colhline
    
   \newtag{DP-CISC-14}{ spec:temp-stability }  & Temperature stability  &  $<\,\SI{2}{mK}$ at all places and times \newline (Match precision requirement at all places, at all times) &  Measures temperature map with sufficient precision for the duration of thermometer operations. &  ProtoDUNE \\ \colhline
    
   \newtag{DP-CISC-27}{ spec:camera-cold-coverage }  & Cold camera coverage  &  $>\,$80\% of HV surfaces \newline (\num{100}\%) &  Enables detailed inspection of issues near HV surfaces. &  Calculated from location, validated in prototypes. \\ \colhline
    
   \newtag{DP-CISC-51}{ spec:slowcontrol-alarm-rate }  & Slow control alarm rate  &  $<\,$150/day \newline ($<\,$50/day) &  Keeps rate low enough to allow response to every alarm. &  Detector module; depends on experimental conditions \\ \colhline
    
   \newtag{DP-CISC-52}{ spec:slowcontrol-num-vars }  & Total No. of variables  &  $>\,\num{150000}$ \newline (\SIrange{150000}{200000}{}) &  Scaled from ProtoDUNE &  ProtoDUNE and CITF \\ \colhline
    
   \newtag{DP-CISC-54}{ spec:slowcontrol-archive-rate }  & Archiving rate  &  \SI{0.02}{Hz} \newline (Broad range \SI{1}{Hz} to \num{1} per few min.) &  Archiving rate differs by variable, optimized to store important information &  ProtoDUNE \\ \colhline

\label{tab:specs:DP-CISC}
\end{longtable}
\end{footnotesize}

\section{Risks}
\label{sec:fdsp-coord-risks}

\dword{dune} initiated a risk registry in 2018 (available
in~\citedocdb{6443}). This document includes consortium risks and
\dword{tc} risks. It includes a summary of the most significant
overall \dword{dune} risks.  This registry has been updated for the
\dword{tdr} and the full listing can be found in
Appendix~\ref{sec:fdsp-app-risk}.We  expect to update it
approximately yearly. The previous update occurred in early 2018
before \dword{protodune} was completed and the most recent update is for the
\dword{tdr}. Another update is planned for 2020. 
\dword{lbnf} and
\dword{dune}-U.S. would like \dword{dune} to update and expand this risk
register to allow a \dword{mc} analysis of cost and schedule risks to
the U.S. project resulting from international \dword{dune}
risks. This request is under consideration as it may be useful for
other national projects as well.  Successfully operating
\dword{protodune} retired many \dword{dune} risks in
\dword{dune}. This includes most risks associated with the technical
design, production processes, \dword{qa}, integration, and
installation. Residual risks remain relating to design and production
modifications associated with scaling to \dword{dune}, mitigations to
known installation and performance issues in \dword{protodune},
underground installation at \dword{surf}, and organizational growth.

The highest technical risks include development of a system to
deliver \SI{600}{kV} to the \dword{dp} cathode; general delivery of the
required \dword{hv}; cathode and \dword{fc} discharge to the cryostat
membrane; noise levels, particularly for the \dword{ce}; 
number of dead channels; lifetime of components surpassing \dunelifetime{}; 
\dword{qc} of all components; verification of improved \dword{lem}
performance; verification of new cold  \dword{adc} and  \dword{coldata} performance;
argon purity; electron drift lifetime; \phel light yield;
incomplete calibration plan; and incomplete connection of design to
physics. Other significant risks include insufficient funding, optimistic
production schedules, incomplete plans for integration, testing and installation. 

One update to the risk registry since 2018 has been for \dword{tc}, to add 
some risks  associated with  DUNE  integration and  installation  
(see \spchinstall{}, Table~9.2)

In addition to installation-related risks, \dword{tc} is developing its
own set of overall project risks not captured by conortia.  Key risks
for \dword{tc} to manage include the following:
\begin{enumerate}
\item Consortia leave too much scope unaccounted for and too much falls
  to  the \dword{comfund}.
\item Insufficient organizational systems are put into place to
  ensure that this complex, international mega-science project,
  including \dword{tc}, \dword{fnal} as host laboratory, \dword{surf}, \dword{doe}, and all international
  partners continue to work together successfully to ensure that 
  appropriate processes and services are provided for the success of
  the project.
\item Inability of \dword{tc} to obtain sufficient personnel resources to
  ensure that \dword{tc} can oversee and coordinate all project tasks.  While the USA, 
  as host country, has a special responsibility to \dword{tc}, personnel resources should
  be directed to \dword{tc} from each collaborating country. 
\end{enumerate}

The consortia have provided preliminary versions of risk analyses that
have been collected on the \dword{tc} webpage (\citedocdb{6443}). These have
been developed into an overall risk register that will be monitored
and maintained by \dword{tc} in coordination with the consortia. This
full set of risks can be found in 
Section~\ref{sec:fdsp-app-risk}.

\section{Full DUNE Risks}
\label{sec:fdsp-app-risk}

\subsection{Single-phase}
\label{sec:tc-risks-sp}

\begin{footnotesize}
\begin{longtable}{P{0.18\textwidth}P{0.20\textwidth}P{0.32\textwidth}P{0.02\textwidth}P{0.02\textwidth}P{0.02\textwidth}} 
\caption[APA risks]{APA risks (P=probability, C=cost, S=schedule) The risk probability, after taking into account the planned mitigation activities, is ranked as 
L (low $<\,$\SI{10}{\%}), 
M (medium \SIrange{10}{25}{\%}), or 
H (high $>\,$\SI{25}{\%}). 
The cost and schedule impacts are ranked as 
L (cost increase $<\,$\SI{5}{\%}, schedule delay $<\,$\num{2} months), 
M (\SIrange{5}{25}{\%} and 2--6 months, respectively) and 
H ($>\,$\SI{20}{\%} and $>\,$2 months, respectively). \fixmehl{ref \texttt{tab:risks:DP-FD-CISC}}} \\
\rowcolor{dunesky}
ID & Risk & Mitigation & P & C & S  \\  \colhline
RT-SP-APA-01 & Loss of key personnel & Implement succession planning and formal project documentation & L & L & M \\  \colhline
RT-SP-APA-02 & Delay in finalisation of APA frame design & Close oversight on prototypes and interface issues & L & L & M \\  \colhline
RT-SP-APA-03 & One additional pre-production APA may be necessary & Close oversight on approval of designs, commissioning of tooling and assembly procedures & L & L & L \\  \colhline
RT-SP-APA-04 & APA winder construction takes longer than planned & Detailed plan to stand up new winding machines at each facility & M & L & M \\  \colhline
RT-SP-APA-05 & Poor quality of APA frames and/or inaccuracy in the machining of holes and slots & Clearly specified requirements and seek out backup vendors & L & L & M \\  \colhline
RT-SP-APA-06 & Insufficient scientific manpower at APA assembly factories & Get institutional commitments for requests of necessary personnel in research grants & M & M & L \\  \colhline
RT-SP-APA-07 & APA production quality does not meet requirements & Close oversight on assembly procedures & L & M & M \\  \colhline
RT-SP-APA-08 & Materials shortage at factory & Develop and execute a supply chain management plan & M & L & L \\  \colhline
RT-SP-APA-09 & Failure of a winding machine - Drive chain parts failure & Regular maintenance and availability of spare parts & L & L & L \\  \colhline
RT-SP-APA-10 & APA assembly takes longer time than planned  & Estimates based on protoDUNE. Formal training of every tech/operator & L & M & M \\  \colhline
RT-SP-APA-11 & Loss of one APA due to an accident & Define handling procedures supported by engineering notes & M & L & L \\  \colhline
RT-SP-APA-12 & APA transport box inadequate & Construction and test of prototype transport boxes & L & L & M \\  \colhline
RO-SP-APA-01 & Reduction of the APA assembly time & Improvements in the winding head and wire tension mesurements & M & M & M \\  
 &  &  &  &  &  \\  \colhline

\label{tab:risks:SP-FD-APA}
\end{longtable}
\end{footnotesize}

\begin{footnotesize}
\begin{longtable}{P{0.18\textwidth}P{0.20\textwidth}P{0.32\textwidth}P{0.02\textwidth}P{0.02\textwidth}P{0.02\textwidth}} 
\caption[HV risks]{HV risks (P=probability, C=cost, S=schedule) The risk probability, after taking into account the planned mitigation activities, is ranked as 
L (low $<\,$\SI{10}{\%}), 
M (medium \SIrange{10}{25}{\%}), or 
H (high $>\,$\SI{25}{\%}). 
The cost and schedule impacts are ranked as 
L (cost increase $<\,$\SI{5}{\%}, schedule delay $<\,$\num{2} months), 
M (\SIrange{5}{25}{\%} and 2--6 months, respectively) and 
H ($>\,$\SI{20}{\%} and $>\,$2 months, respectively).  \fixmehl{ref \texttt{tab:risks:SP-FD-HV}}} \\
\rowcolor{dunesky}
ID & Risk & Mitigation & P & C & S  \\  \colhline
RT-SP-HV-01 & Open circuit on the field cage divider chain & Component selection and cold tests. Varistor protection. & L & L & L \\  \colhline
RT-SP-HV-02 & Damage to the resistive Kapton film on CPA & Careful visual inspection of panel surfaces.  Replace panel if scratches are deep and long  & L & L & L \\  \colhline
RT-SP-HV-03 & Sole source for Kapton resistive surface; and may go out of production & Another potential source of resistive Kapton identified. Possible early purchase if single source. & M & L & L \\  \colhline
RT-SP-HV-04 & Detector components are damaged during shipment to the far site  & Spare parts at  LW. FC/CPA modules can be swapped and replaced from factories in a few days. & L & L & L \\  \colhline
RT-SP-HV-05 & Damages (scratches, bending) to aluminum profiles of Field Cage modules & Require sufficent spare profiles for substitution. Alternate: local coating with epoxy resin. & L & L & L \\  \colhline
RT-SP-HV-06 & Electric field uniformity is not adequate for muon momentum reconstruction  & Redundant components; rigorous screening. Structure based on CFD. Calibration can map E-field. & L & L & L \\  \colhline
RT-SP-HV-07 & Electric field is below goal during stable operations & Improve the protoDUNE SP HVS design to reduce surface E-field and eliminate exterior insulators. & M & L & L \\  \colhline
RT-SP-HV-08 & Damage to CE in event of discharge  & HVS was designed to reduce discharge to a safe level. Higher resistivity cathode could optimize. & L & L & L \\  \colhline
RT-SP-HV-09 & Free hanging frames can swing in the fluid flow  & Designed for flow using fluid model; Deformation can be calibrated by lasers or cosmic rays. & L & L & L \\  \colhline
RT-SP-HV-10 & FRP/ Polyethene/ laminated Kapton component lifetime is less than expected & Positive experience in other detectors. Gain experience with LAr TPC's; exchangeable feedthrough. & L & L & L \\  \colhline
RT-SP-HV-11 & International funding level for SP HVS too low & Cost reduction through design optimization. Effort to increase international collaboration. & M & M & M \\  \colhline
RT-SP-HV-12 & Underground installation is more labor intensive or slower than expected & SWF contingency, full-scale trial before installation. Estimates based on ProtoDUNE experience. & L & L & L \\  \colhline

\label{tab:risks:SP-FD-HV}
\end{longtable}
\end{footnotesize}

\begin{footnotesize}
\begin{longtable}{P{0.18\textwidth}P{0.20\textwidth}P{0.32\textwidth}P{0.02\textwidth}P{0.02\textwidth}P{0.02\textwidth}} 
\caption[TPC electronics risks]{TPC electronics risks (P=probability, C=cost, S=schedule) The risk probability, after taking into account the planned mitigation activities, is ranked as 
L (low $<\,$\SI{10}{\%}), 
M (medium \SIrange{10}{25}{\%}), or 
H (high $>\,$\SI{25}{\%}). 
The cost and schedule impacts are ranked as 
L (cost increase $<\,$\SI{5}{\%}, schedule delay $<\,$\num{2} months), 
M (\SIrange{5}{25}{\%} and 2--6 months, respectively) and 
H ($>\,$\SI{20}{\%} and $>\,$2 months, respectively). \fixmehl{ref \texttt{tab:risks:SP-FD-TPCELEC}}} \\
\rowcolor{dunesky}
ID & Risk & Mitigation & P & C & S  \\  \colhline
RT-SP-TPC-001 & Cold ASIC(s) not meeting specifications & Multiple designs, use of appropriate design rules for operation in LAr & H  & M & L \\  \colhline
RT-SP-TPC-002 & Delay in the availability of ASICs and FEMBs & Increase pool of spares for long lead items, multiple QC sites for ASICs, appropriate measures against ESD, monitoring of yields & M & L & L \\  \colhline
RT-SP-TPC-003 & Damage to the FEMBs / cold cables during or after integration with the APAs & Redesign of the FEMB/cable connection, use of CE boxes, ESD protections, early integration tests & M & L & L \\  \colhline
RT-SP-TPC-004 & Cold cables cannot be run through the APAs frames & Redesign of APA frames, integration tests at Ash River and at CERN, further reduction of cable plant & L & L & L \\  \colhline
RT-SP-TPC-005 & Delay and/or damage to the TPC electronics components on the top of the cryostat & Sufficient spares, early production and installation, ESD protection measures & L & L & L \\  \colhline
RT-SP-TPC-006 & Interfaces between TPC electronics and other consortia not adequately defined & Early integration tests, second run of ProtoDUNE-SP with pre-production components & M & L & L \\  \colhline
RT-SP-TPC-007 & Insufficient number of spares & Early start of production, close monitoring of usage of components, larger stocks of components with long lead times & M & L & L \\  \colhline
RT-SP-TPC-008 & Loss of key personnel & Distributed development of ASICs, increase involved of university groups, training of younger personnel & H & L & M \\  \colhline
RT-SP-TPC-009 & Excessive noise observed during detector commissioning & Enforce grounding rules, early integration tests, second run of ProtoDUNE-SP with pre-production components, cold box testing at SURF & L & L & M \\  \colhline
RT-SP-TPC-010 & Lifetime of components in the LAr & Design rules for cryogenic operation of ASICs, measurement of lifetime of components, reliability studies & L & n/a & n/a \\  \colhline
RT-SP-TPC-011 & Lifetime of components on the top of the cryostat & Use of filters on power supplies, stockpiling of components that may become obsolete, design rules to minimize parts that need to be redesigned / refabricated & L & M & L \\  \colhline

\label{tab:risks:SP-FD-TPCELEC}
\end{longtable}
\end{footnotesize}

\begin{footnotesize}
\begin{longtable}{P{0.18\textwidth}P{0.20\textwidth}P{0.32\textwidth}P{0.02\textwidth}P{0.02\textwidth}P{0.02\textwidth}} 
\caption[PD system risks]{PD system risks (P=probability, C=cost, S=schedule) The risk probability, after taking into account the planned mitigation activities, is ranked as 
L (low $<\,$\SI{10}{\%}), 
M (medium \SIrange{10}{25}{\%}), or 
H (high $>\,$\SI{25}{\%}). 
The cost and schedule impacts are ranked as 
L (cost increase $<\,$\SI{5}{\%}, schedule delay $<\,$\num{2} months), 
M (\SIrange{5}{25}{\%} and 2--6 months, respectively) and 
H ($>\,$\SI{20}{\%} and $>\,$2 months, respectively). \fixmehl{ref \texttt{tab:risks:SP-FD-PD}}} \\
\rowcolor{dunesky}
ID & Risk & Mitigation & P & C & S  \\  \colhline
RT-SP-PD -01 & Additional photosensors and engineering required to ensure PD modules collect enough light to meet system physics performance specifications. & Extensive validation of \dshort{xarapu} design to demonstrate they meet specification. & L & M & L \\  \colhline
RT-SP-PD-02 & Improvements to active ganging/front end electronics required to meet the specified 1~$\mu$s time resolution. & Extensive validation of photosensor ganging/front end electronics design to demonstrate they meet specification. & L & L & L \\  \colhline
RT-SP-PD-03 & Evolutions in the design of the photon detectors due to validation testing experience require modifications of the TPC elements at a late time. & Extensive validation of \dshort{xarapu} design to demonstrate they meet specification and control of PD/APA interface. & L & L & L \\  \colhline
RT-SP-PD-04 & Cabling for PD and CE within the \dshort{apa} frame or during the 2-APA assembly/installation procedure require additional engineering/development/testing. & Validation of PD/APA/CE cable routing in prototypes at Ash River. & L & L & L \\  \colhline
RT-SP-PD-05 & Experience with validation prototypes shows that the mechanical design of the PD is not adequate to meet system specifications. & Early validation of \dshort{xarapu} prototypes and system interfaces to catch problems ASAP. & L & L & L \\  \colhline
RT-SP-PD-06 & pTB WLS filter coating not sufficiently stable, contaminates \dshort{lar}. & Mechanical acceleration of coating wear.  Long-term tests of coating stability. & L & L & L \\  \colhline
RT-SP-PD-07 & Photosensors fail due to multiple cold cycles or extended cryogen exposure. & Execute testing program for cryogenic operation of photosensors including mutiple cryogenic immersion cycles. & L & L & L \\  \colhline
RT-SP-PD-08 & SiPM active ganging cold amplifiers fail or degrade detector performance. & Validation testing if photosensor ganging in multiple test beds. & L & L & L \\  \colhline
RT-SP-PD-09 & Previously undetected electro-mechanical interference discovered during integration. & Validation of electromechanical designin Ash River tests and at \dshort{pdsp2}. & L & L & L \\  \colhline
RT-SP-PD-10 & Design weaknesses manifest during module logistics-handling. & Validation of shipping packaging and handling prior to shipping.  Inspection of modules shipped to site immediately upon receipt. & L & L & L \\  \colhline
RT-SP-PD-11 & PD/CE signal crosstalk. & Validation in \dshort{pdsp}, \dshort{iceberg} and \dshort{pdsp2}. & L & L & L \\  \colhline
RT-SP-PD-12 & Lifetime of \dshort{pd} components outside cryostat. & Specification of environmental controls to mitigate detector aging. & L & L & L \\  \colhline

\label{tab:risks:SP-FD-PD}
\end{longtable}
\end{footnotesize}

\begin{footnotesize}
\begin{longtable}{P{0.18\textwidth}P{0.20\textwidth}P{0.32\textwidth}P{0.02\textwidth}P{0.02\textwidth}P{0.02\textwidth}} 
\caption[Calibration risks]{Calibration risks (P=probability, C=cost, S=schedule) The risk probability, after taking into account the planned mitigation activities, is ranked as 
L (low $<\,$\SI{10}{\%}), 
M (medium \SIrange{10}{25}{\%}), or 
H (high $>\,$\SI{25}{\%}). 
The cost and schedule impacts are ranked as 
L (cost increase $<\,$\SI{5}{\%}, schedule delay $<\,$\num{2} months), 
M (\SIrange{5}{25}{\%} and 2--6 months, respectively) and 
H ($>\,$\SI{20}{\%} and $>\,$2 months, respectively).  \fixmehl{ref \texttt{tab:risks:SP-FD-CAL}}} \\
\rowcolor{dunesky}
ID & Risk & Mitigation & P & C & S  \\  \colhline
RT-SP-CAL-01 & Inadequate baseline design & Early detection allows R\&D of alternative designs accommodated through multipurpose ports & L & M & M \\  \colhline
RT-SP-CAL-02 & Inadequate engineering or production quality & Dedicated small scale tests and full prototyping at ProtoDUNE; pre-installation QC & L & M & M \\  \colhline
RT-SP-CAL-03 & Laser impact on PDS & Mirror movement control to avoid direct hits; turn laser off in case of PDS saturation & L & L & L \\  \colhline
RT-SP-CAL-04 & Laser beam location system stops working & QC at installation time, redundancy in available targets, including passive, alternative methods & L & L & L \\  \colhline
RT-SP-CAL-05 & Laser beam misaligned & Additional (visible) laser for alignment purposes & M & L & L \\  \colhline
RT-SP-CAL-06 & The neutron anti-resonance is much less pronounced & Dedicated measurements at LANL and test at ProtoDUNE & L & L & L \\  \colhline
RT-SP-CAL-07 & Neutron activation of the moderator and cryostat & Neutron activation studies and simulations & L & L & L \\  \colhline
RT-SP-CAL-08 & Neutron yield not high enough & Simulations and tests at ProtoDUNE; alternative, movable design & L & M & M \\  \colhline
RO-SP-CAL-09 & Laser beam is stable at longer distances than designed & tests at ProtoDUNE & M & H & L \\  \colhline

\label{tab:risks:SP-FD-CAL}
\end{longtable}
\end{footnotesize}

\begin{footnotesize}
\begin{longtable}{P{0.18\textwidth}P{0.20\textwidth}P{0.32\textwidth}P{0.02\textwidth}P{0.02\textwidth}P{0.02\textwidth}} 
\caption[DAQ risks]{DAQ risks (P=probability, C=cost, S=schedule) The risk probability, after taking into account the planned mitigation activities, is ranked as 
L (low $<\,$\SI{10}{\%}), 
M (medium \SIrange{10}{25}{\%}), or 
H (high $>\,$\SI{25}{\%}). 
The cost and schedule impacts are ranked as 
L (cost increase $<\,$\SI{5}{\%}, schedule delay $<\,$\num{2} months), 
M (\SIrange{5}{25}{\%} and 2--6 months, respectively) and 
H ($>\,$\SI{20}{\%} and $>\,$2 months, respectively).  \fixmehl{ref \texttt{tab:risks:SP-FD-DAQ}}} \\
\rowcolor{dunesky}
ID & Risk & Mitigation & P & C & S  \\  \colhline
RT-SP-DAQ-01 & Detector noise specs not met & ProtoDUNE experience with noise levels and provisions for data processing redundancy in DAQ system; ensure enough headroom of bandwidth to FNAL. & L & L & L \\  \colhline
RT-SP-DAQ-02 & Externally-driven schedule change & Provisions for standalone testing and commissioning of production DAQ components, and schedule adjustment & L & L & L \\  \colhline
RT-SP-DAQ-03 & Lack of expert personnel & Resource-loaded plan for DAQ backed by institutional commitments, and schedule adjustment using float & L & L & H \\  \colhline
RT-SP-DAQ-04 & Power/space requirements exceed CUC capacity & Sufficient bandwidth to surface and move module 3/4 components to an expanded surface facility & L & L & L \\  \colhline
RT-SP-DAQ-05 & Excess fake trigger rate from instrumental effects & ProtoDUNE performance experience, and provisions for increase in event builder and high level filter capacity, as needed; headroom in data link to FNAL. & L & L & L \\  \colhline
RT-SP-DAQ-06 & Calibration requirements exceed acceptable data rate & Provisions for increase in event builder and high level filter capacity, as neeed; headroom in data link to FNAL. & L & L & L \\  \colhline
RT-SP-DAQ-07 & Cost/performance of hardware/computing excessive & Have prototyping and pre-production phases, reduce performance using margin or identify additional funds & L & L & L \\  \colhline
RT-SP-DAQ-08 & PDTS fails to scale for DUNE requirements & Hardware upgrade & L & L & L \\  \colhline
RT-SP-DAQ-09 & WAN network & Extensive QA and development of failure mode recovery and automation, improved network connectivity, and personnel presence at SURF as last resort. & L & M & M \\  \colhline
RT-SP-DAQ-10 & Infrastructure & Design with redundancy, prior to construction, and improve power/cooling system. & M & M & L \\  \colhline
RT-SP-DAQ-11 & Custom electronics manifacturing issues & Diversify the manifacturers used for production; run an early pre-production and apply stringent QA criteria. & L & M & M \\  \colhline

\label{tab:risks:SP-FD-DAQ}
\end{longtable}
\end{footnotesize}

\begin{footnotesize}
\begin{longtable}{P{0.18\textwidth}P{0.20\textwidth}P{0.32\textwidth}P{0.02\textwidth}P{0.02\textwidth}P{0.02\textwidth}} 
\caption[CISC risks]{CISC risks (P=probability, C=cost, S=schedule) The risk probability, after taking into account the planned mitigation activities, is ranked as 
L (low $<\,$\SI{10}{\%}), 
M (medium \SIrange{10}{25}{\%}), or 
H (high $>\,$\SI{25}{\%}). 
The cost and schedule impacts are ranked as 
L (cost increase $<\,$\SI{5}{\%}, schedule delay $<\,$\num{2} months), 
M (\SIrange{5}{25}{\%} and 2--6 months, respectively) and 
H ($>\,$\SI{20}{\%} and $>\,$2 months, respectively).  \fixmehl{ref \texttt{tab:risks:SP-FD-CISC}}} \\
\rowcolor{dunesky}
ID & Risk & Mitigation & P & C & S  \\  \colhline
RT-SP-CISC-01 & Baseline design from ProtoDUNEs for an instrumentation device is not adequate for DUNE far detectors & Focus on early problem discovery in ProtoDUNE so any needed redesigns can start as soon as possible. & L & M & L \\  \colhline
RT-SP-CISC-02 & Swinging of long instrumentation devices (T-gradient monitors or PrM system) & Add additional intermediate constraints to prevent swinging. & L & L & L \\  \colhline
RT-SP-CISC-03 & High E-fields near instrumentation devices cause dielectric breakdowns in \dshort{lar} & CISC systems placed as far from cathode and FC as possible. & L & L & L \\  \colhline
RT-SP-CISC-04 & Light pollution from purity monitors and camera light emitting system & Use PrM lamp and camera lights outside PDS trigger window; cover PrM cathode to reduce light leakage. & L & L & L \\  \colhline
RT-SP-CISC-05 & Temperature sensors can induce noise in cold electronics & Check for noise before filling and remediate, repeat after filling. Filter or ground noisy sensors. & L & L  & L \\  \colhline
RT-SP-CISC-06 & Disagreement between lab and \em{in situ} calibrations for ProtoDUNE-SP dynamic T-gradient monitor & Investigate and improve both methods, particularly laboratory calibration. & M & L & L \\  \colhline
RT-SP-CISC-07 & Purity monitor electronics induce noise in TPC and PDS electronics. & Operate lamp outside TPC+PDS trigger window. Surround and ground light source with Faraday cage. & L & L & L \\  \colhline
RT-SP-CISC-08 & Discrepancies between measured temperature map and CFD simulations in ProtoDUNE-SP & Improve simulations with additional measurements inputs; use fraction of sensors to predict others   & L & L & L \\  \colhline
RT-SP-CISC-09 & Difficulty correlating purity and temperature in ProtoDUNE-SP impairs understanding cryo system. & Identify causes of discrepancy, modify design. Calibrate PrM differences, correlate with RTDs. & L & L & L \\  \colhline
RT-SP-CISC-10 & Cold camera R\&D fails to produce prototype meeting specifications \& safety requirements & Improve insulation and heaters. Use cameras in ullage or inspection cameras instead. & M & M & L \\  \colhline
RT-SP-CISC-11 & HV discharge caused by inspection cameras & Study E-field in and on housing and anchoring system. Test in HV facility. & L & L & L \\  \colhline
RT-SP-CISC-12 & HV discharge destroying the cameras & Ensure sufficient redundancy of cold cameras. Warm cameras are replaceable. & L & M & L \\  \colhline
RT-SP-CISC-13 & Insufficient light for cameras to acquire useful images & Test cameras with illumination similar to actual detector. & L & L & L \\  \colhline
RT-SP-CISC-14 & Cameras may induce noise in cold electronics & Continued R\&D work with grounding and shielding in realistic conditions. & L & L & L \\  \colhline
RT-SP-CISC-15 & Light attenuation in long optic fibers for purity monitors  & Test the max.\ length of usable fiber, optimize the depth of bottom PrM, number of fibers. & L & L & L \\  \colhline
RT-SP-CISC-16 & Longevity of purity monitors & Optimize PrM operation to avoid long running in low purity. Technique to protect/recover cathode. & L & L & L \\  \colhline
RT-SP-CISC-17 & Longevity: Gas analyzers and level meters may fail. & Plan for future replacement in case of failure or loss of sensitivity.  & M & M & L \\  \colhline
RT-SP-CISC-18 & Problems in interfacing  hardware devices (e.g. power supplies) with slow controls & Involve slow control experts in choice of hardware needing control/monitoring.
 & L & L & L \\  \colhline

\label{tab:risks:SP-FD-CISC}
\end{longtable}
\end{footnotesize}

\begin{footnotesize}
\begin{longtable}{P{0.18\textwidth}P{0.20\textwidth}P{0.32\textwidth}P{0.02\textwidth}P{0.02\textwidth}P{0.02\textwidth}} 
\caption[SP module installation risks]{SP module installation risks (P=probability, C=cost, S=schedule) The risk probability, after taking into account the planned mitigation activities, is ranked as 
L (low $<\,$\SI{10}{\%}), 
M (medium \SIrange{10}{25}{\%}), or 
H (high $>\,$\SI{25}{\%}). 
The cost and schedule impacts are ranked as 
L (cost increase $<\,$\SI{5}{\%}, schedule delay $<\,$\num{2} months), 
M (\SIrange{5}{25}{\%} and 2--6 months, respectively) and 
H ($>\,$\SI{20}{\%} and $>\,$2 months, respectively).  \fixmehl{ref \texttt{tab:risks:SP-FD-INST}}} \\
\rowcolor{dunesky}
ID & Risk & Mitigation & P & C & S  \\  \colhline
RT-INST-01 & Personnel injury & Follow established safety plans. & M & L & H \\  \colhline
RT-INST-02 & Shipping delays & Plan one month buffer to store  materials locally. Provide logistics manual. & H & L & L \\  \colhline
RT-INST-03 & Missing components cause delays & Use detailed inventory system to verify availability of  necessary components.  & H & L & L \\  \colhline
RT-INST-04 & Import, export, visa issues  & Dedicated \dshort{fnal} \dshort{sdsd}division will expedite import/export and visa-related issues. & H & M & M \\  \colhline
RT-INST-05 & Lack of available labor  & Hire early and use Ash River setup to train \dshort{jpo} crew. & L & L & L \\  \colhline
RT-INST-06 & Parts do not fit together & Generate \threed model, create interface drawings, and prototype detector assembly. & H & L & L \\  \colhline
RT-INST-07 & Cryostat damage & Use cryostat false floor and temporary protection. & L & L & M \\  \colhline
RT-INST-08 & Weather closes SURF & Plan for \dshort{surf} weather closures & H & L & L \\  \colhline
RT-INST-09 & Detector failure during \cooldown & Cold test individual components then cold test \dshort{apa} assemblies immediately before installation. & L & H & H \\  \colhline

\label{tab:risks:SP-FD-INST}
\end{longtable}
\end{footnotesize}
 

\subsection{Dual-phase}
\label{sec:tc-risks-dp}

For each risk, the risk probability, after taking into account the planned mitigation activities, is ranked as 
L (low $<\,$\SI{10}{\%}), 
M (medium \SIrange{10}{25}{\%}), or 
H (high $>\,$\SI{25}{\%}). 
The cost and schedule impacts are ranked as 
L (cost increase $<\,$\SI{5}{\%}, schedule delay $<\,$\num{2} months), 
M (\SIrange{5}{25}{\%} and 2--6 months, respectively) and 
H ($>\,$\SI{20}{\%} and $>\,$2 months, respectively).

\begin{footnotesize}
\begin{longtable}{P{0.18\textwidth}P{0.20\textwidth}P{0.32\textwidth}P{0.02\textwidth}P{0.02\textwidth}P{0.02\textwidth}} 
\caption[DP CRP Risks]{Risks for DP-FD-CRP (P=probability, C=cost, S=schedule) More information at \dshort{riskprob}. \fixmehl{ref \texttt{tab:risks:DP-FD-CRP}}} \\
\rowcolor{dunesky}
ID & Risk & Mitigation & P & C & S  \\  \colhline
RT-DP-CRP-01 & Poor quality of G10  frames and/or inaccuracy in the hole machining & Clearly specified requirements, followup and seek out backup vendors & L & L  & M \\  \colhline
RT-DP-CRP-02 & LEM production takes longer than expected & Define a production schedule allowing enough contingencies to limit the assembly impact   & L & L  & M \\  \colhline
RT-DP-CRP-03 & One of the CRP assembly site not ready on time & Close oversight on construction of tooling and preparation of assembly sites & L & M & H  \\  \colhline
RT-DP-CRP-04 & Materials shortage at production site & Develop and execute a supply chain management & M & L  & L \\  \colhline
RT-DP-CRP-05 & Failure of extraction grid winding machine & Regular maintenance and availability of spare parts & L & L  & L \\  \colhline
RT-DP-CRP-06 & CRP assembly takes longer time than planned & Estimates based on ProtoDUNE-DP. Formal training of every tech/operator at each site. & L & M & M \\  \colhline

\label{tab:risks:DP-FD-CRP}
\end{longtable}
\end{footnotesize}

\begin{footnotesize}
\begin{longtable}{P{0.18\textwidth}P{0.20\textwidth}P{0.32\textwidth}P{0.02\textwidth}P{0.02\textwidth}P{0.02\textwidth}} 
\caption[DP HV risks]{Risks for DP-FD-HV (P=probability, C=cost, S=schedule) More information at \dshort{riskprob}. \fixmehl{ref \texttt{tab:risks:DP-FD-HV}}} \\
\rowcolor{dunesky}
ID & Risk & Mitigation & P & C & S  \\  \colhline
RT-DP-HV-01 & Broken resistors or varistors on voltage divider boards & Redundancy of resistors, varistors, and \dshorts{hvdb}.  & L & L & L \\  \colhline
RT-DP-HV-02 & \efield uniformity is not adequate for muon momentum reconstruction & Regularly map out field using a laser calibration sysem. & L & L & L \\  \colhline
RT-DP-HV-03 & \efield is below specification during stable operations & Improve purity by more aggressive filtering. & M & M & L \\  \colhline
RT-DP-HV-04 & Space charge from positive ions distorting the \efield beyond expectation & Minimize insulators facing cryostat wall ground. & M & M & L \\  \colhline
RT-DP-HV-05 & Damage to \dshort{ce} in event of discharge & Minimize the energy released in a short time using highly resistive connections. & L & L & L \\  \colhline
RT-DP-HV-06 & Energy stored in FC (in DP) is suddenly discharged & Delay energy discharge by connecting neighboring Al profiles with resistive sheaths.  & L & L & L \\  \colhline
RT-DP-HV-07 & Detector components are damaged during shipment to the far site & Make sufficient spares and increase the number of shipping boxes.  & L & L & L \\  \colhline
RT-DP-HV-08 & Damages (scratches, bending) to aluminum profiles of Field Cage modules & Make sufficient spares and increase the number of shipping boxes.  & L & L & L \\  \colhline
RT-DP-HV-09 & Bubbles from heat in PMTs or resistors cause HV discharge & A large area of cathode consists of high resistance rods, delaying the energy release.   & L & L & L \\  \colhline
RT-DP-HV-10 & Free hanging frames can swing in the fluid flow &  & L & L & L \\  \colhline
RT-DP-HV-11 & FRP/ Polyethene/ laminated Kapton component lifetime is less than expected &  & L & L & L \\  \colhline
RT-DP-HV-12 & Lack of collaboration effort on this HV system & Continue recruiting collaborators. & L & L & L \\  \colhline
RT-DP-HV-13 & International funding level for DP HVC too low & Employ cost saving measures and  recruit collaborators. & L & L & L \\  \colhline
RT-DP-HV-14 & Underground installation is more labor intensive or slower than expected & Increase labor contingency and refine labor cost estimates. Further improve installation procedure. & L & L & L \\  \colhline

\label{tab:risks:DP-FD-HV}
\end{longtable}
\end{footnotesize}

\begin{footnotesize}
\begin{longtable}{P{0.18\textwidth}P{0.20\textwidth}P{0.32\textwidth}P{0.02\textwidth}P{0.02\textwidth}P{0.02\textwidth}} 
\caption[DP TPC electronics risks]{Risks for DP-FD-TPC (P=probability, C=cost, S=schedule) More information at \dshort{riskprob}. \fixmehl{ref \texttt{tab:risks:DP-FD-TPC}}} \\
\rowcolor{dunesky}
ID & Risk & Mitigation & P & C & S  \\  \colhline
RT-DP-TPC-01 & Component obsolescence over the experiment lifetime & Monitor component stocks and procure an adequate number of spares at the time of production & L & M & L \\  \colhline
RT-DP-TPC-02 & Modification to the LRO FE electronics due to evolution in design of PD design & A strict and timely following of the evolution of DP PDS & L & L & M \\  \colhline
RT-DP-TPC-03 & Damage to electronics due to HV discharges or other causes & FE analog electronics is protected with TVS diodes. Electronics can be easily replaced. & L & L & L \\  \colhline
RT-DP-TPC-04 & Problems with FE card extraction due to insufficient overhead clearance & Addressed by imposing a clearance requirement on \dshort{lbnf} & L & L & L \\  \colhline
RT-DP-TPC-05 & Overpressure in the \dshorts{sftchimney} & The \dshorts{sftchimney} are equipped with overpressure release valves & L & L & L \\  \colhline
RT-DP-TPC-06 & Leak of nitrogen inside the \dshort{dpmod} via cold flange & Monitor chimney pressure for leaks and switch to argon cooling in case of a leak & L & L & L \\  \colhline
RT-DP-TPC-07 & Data flow increase due to inefficient compression caused by higher noise & Have a sufficiently large (a factor of \num{5}) margin in the available bandwidth & L & L & L \\  \colhline
RT-DP-TPC-08 & Damage to \dshort{utca} crates due to presence of water on the roof of the cryostat & \dshort{lbnf} requirement that the cryostat top remains dry & L & L & L \\  \colhline
RT-DP-TPC-09 & Clogging ventilation system of \dshort{utca} crates due to bad air quality & \dshort{lbnf} requirement that the air quality is comparable to a standard industrial environment & L & L & L \\  \colhline

\label{tab:risks:DP-FD-TPC}
\end{longtable}
\end{footnotesize}

\begin{footnotesize}
\begin{longtable}{P{0.18\textwidth}P{0.20\textwidth}P{0.32\textwidth}P{0.02\textwidth}P{0.02\textwidth}P{0.02\textwidth}} 
\caption[DP PDS risks]{Risks for DP-FD-PDS (P=probability, C=cost, S=schedule) More information at \dshort{riskprob}. \fixmehl{ref \texttt{tab:risks:DP-FD-PDS}}} \\
\rowcolor{dunesky}
ID & Risk & Mitigation & P & C & S  \\  \colhline
RT-DP-PDS-01 & Insufficient light yield due to inefficient PDS design & Increase PMT photo-cathode coverage and/or WLS reflector foils coverage. & L & M & L \\  \colhline
RT-DP-PDS-02 & Poor coating quality for \dshort{tpb} coated surfaces and \dshort{lar} contamination by \dshort{tpb} & Test quality and ageing properties of TPB coating techniques. Elaborate improved techniques if needed. & L & L & L \\  \colhline
RT-DP-PDS-03 & \dshort{pmt} channel loss due to faulty \dshort{pmt} base design & Optimize clustering algorithms. Improve \dshort{pmt} base design from analysis of possible failure modes in \dshort{pddp}. & L & L & L \\  \colhline
RT-DP-PDS-04 & Bad \dshort{pmt} channel due to faulty connection between \dshort{hv}/signal cable and \dshort{pmt} base & Optimize clustering algorithms. Connectivity tests in \lntwo prior to installation. & L & L & L \\  \colhline
RT-DP-PDS-05 & \dshort{pmt} signal saturation & Tuning of \dshort{pmt} gain. In worst case, redesign front-end to adjust to analog input range of ADC. & M & L & L \\  \colhline
RT-DP-PDS-06 & Excessive electronics noise to distinguish \dshort{lar} scintillation light & Measurement of noise levels during commissioning prior to \lar filling. Modifications to grounding, shielding, or power distribution schemes. & M & L & L \\  \colhline
RT-DP-PDS-07 & Availability of resources for work at the installation/integration site less than planned & Move people temporarily from institutions involved in the \dshort{pds} consortium to the integration/installation site. & L & L & L \\  \colhline
RT-DP-PDS-08 & Damage of \dshorts{pmt} during shipment to the experiment site & Special packaging to avoid possible \dshort{pmt} damage during shipment. Contingency of 10\% spare  \dshorts{pmt}. & L & L & L \\  \colhline
RT-DP-PDS-09 & Damage of optical fibers during installation & Fibers will be last DP-PDS item to be installed. Detailed documentation for all DP-PDS installation tasks.   & L & L & L \\  \colhline
RT-DP-PDS-10 & Excessive exposure to ambient light of \dshort{tpb} coated surfaces, resulting in degraded performance & \dshort{tpb} coated surfaces temporarily covered until cryostat closing. Detailed installation procedure to minimize exposure to ambient light. & L & L & L \\  \colhline
RT-DP-PDS-11 & \dshort{pmt} implosion during \dshort{lar} filling & No mitigation necessary, considering \SI{7}{bar} pressure rating of \dshorts{pmt} and experience with same/similar \dshorts{pmt} in other large liquid detectors. & L & L & L \\  \colhline
RT-DP-PDS-12 & Insufficient light yield due to poor \dshort{lar} purity & Procurement of \dshort{lar} from the manufacturer will require less than 3 ppM in nitrogen. & M & L & L \\  \colhline
RT-DP-PDS-13 & \dshort{pmt} channel or \dshort{pds} sector loss due to failures in \dshort{hv}/signal rack & Ease of maintenance outside cryostat and availability of spares for all components of at least one \dshort{hv}/signal rack. & L & L & L \\  \colhline
RT-DP-PDS-14 & Unstable response of the photon detection system over the lifetime of the experiment & Channel-level instabilities corrected via light calibration system. Detector-level instabilities corrected via cosmic-ray muon calibration data. & L & L & L \\  \colhline
RT-DP-PDS-15 & Bubbles from heat in \dshorts{pmt} or resistors cause \dshort{hv} discharge of the cathode & Verify the power density of the \dshort{pmt} bases are within specifications. Monitor and interlock \dshort{pmt} power supply currents. & L & L & L \\  \colhline
RT-DP-PDS-16 & Reflector/\dshort{wls} panel assemblies together with the \dshort{fc} walls can swing in the fluid flow & Allow appropriate open areas within/between reflector/\dshort{wls} panel assemblies to minimize drag. & L & L & L \\  \colhline

\label{tab:risks:DP-FD-PDS}
\end{longtable}
\end{footnotesize}

\begin{footnotesize}
\begin{longtable}{P{0.18\textwidth}P{0.20\textwidth}P{0.32\textwidth}P{0.02\textwidth}P{0.02\textwidth}P{0.02\textwidth}} 
\caption[DP calibration risks]{Risks for DP-FD-CAL (P=probability, C=cost, S=schedule) More information at \dshort{riskprob}. \fixmehl{ref \texttt{tab:risks:DP-FD-CAL}}} \\
\rowcolor{dunesky}
ID & Risk & Mitigation & P & C & S  \\  \colhline
RT-DP-CAL-01 & Inadequate baseline design & Early detection allows R\&D of alternative designs accommodated through multipurpose ports. & L & M & M \\  \colhline
RT-DP-CAL-02 & Inadequate engineering or production quality & Dedicated small-scale tests and full prototyping at ProtoDUNE; pre-installation QC. & L & M & M \\  \colhline
RT-DP-CAL-03 & Laser impact on PDS & Mirror movement control to minimize direct hits; intelock to keep laser off while PMTs are on. & L & L & L \\  \colhline
RT-DP-CAL-04 & Laser positioning system stops working & QC at installation time, redundancy in available targets, including passive, alternative methods. & L & L & L \\  \colhline
RT-DP-CAL-05 & Laser beam misaligned & Additional (visible) laser for alignment purposes. & M & L & L \\  \colhline
RT-DP-CAL-06 & The neutron anti-resonance is much less pronounced & Dedicated measurements at LANL and test at ProtoDUNE. & L & L & L \\  \colhline
RT-DP-CAL-07 & Neutron activation of the moderator and cryostat & Neutron activation studies and simulations. & L & L & L \\  \colhline
RT-DP-CAL-08 & Neutron yield not high enough & Simulations and tests at ProtoDUNE; alternative, movable design & L & M & M \\  \colhline
RO-DP-CAL-09 & Laser beam is stable at longer distances than designed & tests at ProtoDUNE & M & H & L \\  \colhline

\label{tab:risks:DP-FD-CAL}
\end{longtable}
\end{footnotesize}

\begin{footnotesize}
\begin{longtable}{P{0.18\textwidth}P{0.20\textwidth}P{0.32\textwidth}P{0.02\textwidth}P{0.02\textwidth}P{0.02\textwidth}} 
\caption[DP DAQ Risks]{Risks for DP-FD-DAQ (P=probability, C=cost, S=schedule) More information at \dshort{riskprob}. \fixmehl{ref \texttt{tab:risks:DP-FD-DAQ}}} \\
\rowcolor{dunesky}
ID & Risk & Mitigation & P & C & S  \\  \colhline
RT-DP-DAQ-01 & Detector noise specs not met & ProtoDUNE experience with noise levels and provisions for data processing redundancy in DAQ system; ensure enough headroom of bandwidth to FNAL. & L & L & L \\  \colhline
RT-DP-DAQ-02 & Externally-driven schedule change & Provisions for standalone testing and commissioning of production DAQ components, and schedule adjustment & L & L & L \\  \colhline
RT-DP-DAQ-03 & Lack of expert personnel & Resource-loaded plan for DAQ backed by institutional commitments, and schedule adjustment using float & L & L & H \\  \colhline
RT-DP-DAQ-04 & Power/space requirements exceed CUC capacity & Sufficient bandwidth to surface and move module 3/4 components to an expanded surface facility & L & L & L \\  \colhline
RT-DP-DAQ-05 & Excess fake trigger rate from instrumental effects & ProtoDUNE performance experience, and provisions for increase in event builder and high level filter capacity, as needed; headroom in data link to FNAL. & L & L & L \\  \colhline
RT-DP-DAQ-06 & Calibration requirements exceed acceptable data rate & Provisions for increase in event builder and high level filter capacity, as neeed; headroom in data link to FNAL. & L & L & L \\  \colhline
RT-DP-DAQ-07 & Cost/performance of hardware/computing excessive & Have prototyping and pre-production phases, reduce performance using margin or identify additional funds & L & L & L \\  \colhline
RT-DP-DAQ-08 & PDTS fails to scale for DUNE requirements & Hardware upgrade & L & L & L \\  \colhline
RT-DP-DAQ-09 & WAN network & Extensive QA and development of failure mode recovery and automation, improved network connectivity, and personnel presence at SURF as last resort. & L & M & M \\  \colhline
RT-DP-DAQ-10 & Infrastructure & Design with redundancy, prior to construction, and improve power/cooling system. & M & M & L \\  \colhline
RT-DP-DAQ-11 & Custom electronics manifacturing issues & Diversify the manifacturers used for production; run an early pre-production and apply stringent QA criteria. & L & M & M \\  \colhline

\label{tab:risks:DP-FD-DAQ}
\end{longtable}
\end{footnotesize}

\begin{footnotesize}
\begin{longtable}{P{0.18\textwidth}P{0.20\textwidth}P{0.32\textwidth}P{0.02\textwidth}P{0.02\textwidth}P{0.02\textwidth}} 
\caption[DP CISC risks]{Risks for DP-FD-CISC (P=probability, C=cost, S=schedule) More information at \dshort{riskprob}. \fixmehl{ref \texttt{tab:risks:DP-FD-CISC}}} \\
\rowcolor{dunesky}
ID & Risk & Mitigation & P & C & S  \\  \colhline
RT-DP-CISC-001 & Baseline design from ProtoDUNEs for an instrumentation device is not adequate for DUNE far detectors & Focus on early problem discovery in ProtoDUNE so any needed redesigns can start as soon as possible. & L & M & L \\  \colhline
RT-DP-CISC-002 & Swinging of long instrumentation devices (T-gradient monitors or PrM system) & Add additional intermediate constraints to prevent swinging. & L & L & L \\  \colhline
RT-DP-CISC-003 & High E-fields near instrumentation devices cause dielectric breakdowns in \dshort{lar} & CISC systems shielded and placed as far from cathode and FC as possible. & L & L & L \\  \colhline
RT-DP-CISC-004 & Light pollution from purity monitors and camera light emitting system & Use PrM lamp and camera lights outside PDS trigger window; cover PrM cathode to reduce light leakage. & L & L & L \\  \colhline
RT-DP-CISC-005 & Temperature sensors can induce noise in cold electronics & Check for noise before filling and remediate, repeat after filling. Filter or ground noisy sensors. & L & L  & L \\  \colhline
RT-DP-CISC-006 & Disagreement between lab and \em{in situ} calibrations for ProtoDUNE-SP dynamic T-gradient monitor & Investigate and improve both methods, particularly laboratory calibration. & M & L & L \\  \colhline
RT-DP-CISC-007 & Purity monitor electronics induce noise in TPC and PDS electronics. & Operate lamp outside TPC+PDS trigger window. Surround and ground light source with Faraday cage. & L & L & L \\  \colhline
RT-DP-CISC-008 & Discrepancies between measured temperature map and CFD simulations in ProtoDUNE-SP & Improve simulations with additional measurements inputs; use fraction of sensors to predict others   & L & L & L \\  \colhline
RT-DP-CISC-009 & Difficulty correlating purity and temperature in ProtoDUNE-SP impairs understanding cryo system. & Identify causes of discrepancy, modify design. Calibrate PrM differences, correlate with RTDs. & L & L & L \\  \colhline
RT-DP-CISC-010 & Cold camera R\&D fails to produce prototype meeting specifications \& safety requirements & Improve insulation and heaters. Use cameras in ullage or inspection cameras instead. & M & M & L \\  \colhline
RT-DP-CISC-011 & HV discharge caused by inspection cameras & Study E-field in and on housing and anchoring system. Test in HV facility. & L & L & L \\  \colhline
RT-DP-CISC-012 & HV discharge destroying the cameras & Ensure sufficient redundancy of cold cameras. Warm cameras are replaceable. & L & M & L \\  \colhline
RT-DP-CISC-013 & Insufficient light for cameras to acquire useful images & Test cameras with illumination similar to actual detector. & L & L & L \\  \colhline
RT-DP-CISC-014 & Cameras may induce noise in cold electronics & Continued R\&D work with grounding and shielding in realistic conditions. & L & L & L \\  \colhline
RT-DP-CISC-015 & Light attenuation in long optic fibers for purity monitors  & Test the max.\ length of usable fiber, optimize the depth of bottom PrM, number of fibers. & L & L & L \\  \colhline
RT-DP-CISC-016 & Longevity of purity monitors & Optimize PrM operation to avoid long running in low purity. Technique to protect/recover cathode. & L & L & L \\  \colhline
RT-DP-CISC-017 & Longevity: Gas analyzers and level meters may fail. & Plan for future replacement in case of failure or loss of sensitivity.  & M & M & L \\  \colhline
RT-DP-CISC-018 & Problems in interfacing  hardware devices (e.g. power supplies) with slow controls & Involve slow control experts in choice of hardware needing control/monitoring.
 & L & L & L \\  \colhline

\label{tab:risks:DP-FD-CISC}
\end{longtable}
\end{footnotesize}

\begin{footnotesize}
\begin{longtable}{P{0.18\textwidth}P{0.20\textwidth}P{0.32\textwidth}P{0.02\textwidth}P{0.02\textwidth}P{0.02\textwidth}} 
\caption[DP module installation risks]{Risks for DP-FD-INST (P=probability, C=cost, S=schedule) More information at \dshort{riskprob}. \fixmehl{ref \texttt{tab:risks:DP-FD-INST}}} \\
\rowcolor{dunesky}
ID & Risk & Mitigation & P & C & S  \\  \colhline
RT-DPINST-01 & Personnel injury & Follow the safety rules in force. & L & L & H \\  \colhline
RT-DPINST-02 & Cryostat damage during installation & Use temporary protection for the corrugated membrane. & L & L & M \\  \colhline
RT-DPINST-03 & Detector components damage during transport/installation & Handling should be done by trained people, following the handling instructions provided by the consortia and in presence of a technical expert. & L & M & M \\  \colhline
RT-DPINST-04 & Detector components failure during test & Only trained people should test equipments. Spare components must be available at the warehouse facility. & L & L & M \\  \colhline
RT-DPINST-05 & Components interferences & 3D model, survey at the construction sites, and full scale assembly tests. & M & L & L \\  \colhline
RT-DPINST-06 & Shipping delays/missing parts & Buffer material and detector components at the warehouse facility and use inventory tools to follow the fundamental items. & L & L & L \\  \colhline
RT-DPINST-07 & Lack of specialised/trained manpower & Hire and train personnel. Plan with all the underground stakeholders enough in advance the required personnel. & L & L & L \\  \colhline

\label{tab:risks:DP-FD-INST}
\end{longtable}
\end{footnotesize}

\section{Hazard Analysis Report (HAR)}
\label{sec:fdsp-har}

A key element of an effective \dword{esh} program is the hazard
identification process. Hazard identification allows production of a
list of hazards within a facility, so these hazards can be screened
and managed through a suitable set of controls.

The \dword{lbnf-dune} project completed a \dword{har}
to ensure that identified hazards are mitigated early in 
the design process.  The focus of the report is on process hazards,
not activity hazards that are typically covered in a job hazard
analysis.  The \dword{har} has been completed, identifying
hazards anticipated in the project's construction and operational
phases.

The hazard \dword{har} looks at the consequences of a hazard
to establish a pre-mitigation risk category. Proposed mitigation is
applied to hazards of concern to reduce risk and then establishes a
post-mitigation risk category.

As the \dword{dune} design matures, the \dword{har} will be
updated to ensure that all hazards are properly identified and
controlled through design and safety management system programs.  In
addition, some sections of the \dword{har} are used to meet
the safety requirements as defined in 10 CFR 851 and \dword{doe} Order
420.2C, Safety of Accelerator Facilities.  Table~\ref{tab:hazards}
summarizes these hazards.  The sections following the table describe
in more detail the hazards that are most applicable to \dword{dune}
activities and the design and operational controls used to mitigate
these hazards. The results of these evaluations confirm that the
potential risks from construction, operations, and maintenance are
acceptable. Individual activity-based \dword{ha} will be
developed for each work \dword{lbnf-dune} activity at
\dword{surf}.

\begin{longtable}{|p{0.35\textwidth}|p{0.28\textwidth}|p{0.28\textwidth}|}
  \caption[List of identified hazards]{List of identified hazards}
  \label{tab:hazards} \\  \toprowrule
  \rowtitlestyle   HA-1 (Construction) & HA-2 (Natural Phenomena) & HA-3 (Environmental)   \\ \toprowrule
  Site Clearing, Excavation, Mining, Tunneling (explosives), Vertical/Horizontal Conveyance Systems,
  Confined space, Heavy Equipment, Work at Elevations (steel, roofing), Material Handling (rigging)
  Utility interfaces, (electrical, steam, chilled water), Slips/trips/falls, Weather related conditions
  Scaffolding, Transition to Operations, Radiation Generating Devices &
  Seismic, Flooding, Wind, Lightning, Tornado &
  Construction impacts,
  Storm water discharge (construction and operations), Operations impacts, Soil and groundwater activation/contamination,
  Tritium contamination, Air activation, Cooling water activation (HVAC and Machine),
  Oils/chemical leaks or spills, Discharge/emission points (atmospheric/ground)\\ \colhline
  \rowtitlestyle HA-4 (Waste) & HA-5 (Fire) & HA-6 (Electrical)   \\ \toprowrule
  Construction Phase, Facility maintenance, Experimental Operations, Industrial, Hazardous, Radiological &
  Facility Occupancy Classification, Construction Materials, Storage, Flammable/combustible liquids,
  Flammable gasses, Egress/access, Electrical, Lightning, Welding/cutting/brazing work, Smoking  &
  Facility, Experimental, Job built Equipment, Low Voltage/High Current, High Voltage/High Power,
  Maintenance, Arc flash, Electrical shock, Cable tray overloading/mixed utilities, Exposed 110V,
  Stored energy (capacitors \& inductors), Be in contactors   \\ \colhline
  \rowtitlestyle   HA-7 (Mechanical) & HA-8 (Cryo/ODH) & HA-9 (Confined Space)   \\ \toprowrule
  Construction Tools, Machine Shop Tools, Industrial Vehicles, Drilling, Cutting, Grinding,
  Pressure/Vacuum Vessels and Lines, High Temp Equipment (Bakeouts) &
  Thermal, Cryogenic systems, Pressure, Handling and Storage,
  Liquid argon/nitrogen spill/leak, Use of inert gases (argon, nitrogen, helium), Specialty gases &
  Sumps, Utility Chases        \\ \colhline
  \rowtitlestyle   HA-11 (Chemical) & HA-14 (Laser) & HA-15 (Material Handling)   \\ \toprowrule
  Toxic, Compressed gas, Combustibles, Explosives, Flammable gases, Lead (shielding), Cryogenic &
  Alignment Laser, Testing and Calibration, Magnetic Fields, Calibration \& Testing &
  Overhead cranes/hoists, Fork trucks, Manual material handling, Delivery area distribution,
  Manual movement of materials, Hoisting \& Rigging, Lead, Beryllium Windows,Oils, Solvents, Acids,
  Cryogens, Compressed Gases   \\ \colhline
  \rowtitlestyle   HA-16 (Experimental Ops) &  &    \\ \toprowrule
  Electrical equipment, Water Hazard, Working from heights (scaffolding/lifts), Transportation of hazardous materials,
  Liquid Argon/Nitrogen, Chemicals (Corrosive, Reactive, Flammable), Elevations, Ionizing radiation,
  Ozone production, Slips, trips, falls, Machine tools/hand tools, Stray static magnetic fields, Research gasses (Inert, Flammable) &
  &   \\   \colhline
\end{longtable}

\subsection{Construction Hazards (LBNF-DUNE HA-1)}

The project will use the existing work planning and
control process for the laboratories along with a construction project safety and health
plan to communicate these policies and procedures as required by \dword{doe}
Order 413.3b. The installation and construction hazards
anticipated for the \dword{lbnf-dune} project include the following:
\begin{itemize}
\item Site clearing;
\item Excavation;
\item Installing vertical/horizontal conveyance systems;
\item Confined space;
\item Heavy equipment operation;
\item Work at elevation (erecting steel, roofing);
\item Material handling (rigging);
\item Utility interfaces (electrical, chilled water, ICW, natural gas);
\item Slips/trips/falls;
\item Weather related conditions;
\item Scaffolding;
\item Transition to operations; and
\item Devices generating radiation.
\end{itemize}

To reduce risks from construction hazards, \fnal will use engineered
and approved excavation and fall protection systems.  Heavy equipment
will use required safety controls. The \fnal construction safety
oversight program includes periodic evaluation of the construction
site and construction activities, \dword{ha} for all
subcontractor activities and frequent \dword{esh} communications at
the daily tool box meetings of subcontractors.

\subsection{Natural Phenomena (LBNF-DUNE HA-2)}

The \dword{lbnf-dune} design will be governed by the
International Building Code, 2015 edition; \dword{doe} Standard
(STD)-1020, 2016 edition, Natural Phenomena Hazard Analysis and Design
Criteria for \dword{doe} Facilities, guided the design in meeting the
natural phenomena hazard requirements.  The International Building
Code specifies design criteria for wind loading, snow loading, and
seismic events.

\dword{lbnf-dune} was determined to be a low-hazard,
performance category 1 facility according to the \dword{doe}
STD-1021-93. \dword{lbnf-dune} areas will contain small
quantities of activated, radioactive, and hazardous chemical
materials. Should a natural phenomenon hazard cause significant
damage, the impact will be mission-related and will not pose a hazard
to the public or the environment.

\subsection{Environmental Hazards (LBNF-DUNE HA-3)}

Environmental hazards from \dword{dune} include potentially releasing
chemicals to soil, groundwater, surface water, air, or sanitary sewer
systems that could, if not controlled, exceed regulatory limits.

\fnal maintains an environmental management system equivalent to ISO
14001, consisting of programs for protecting the environment, assuring
compliance with applicable environmental regulations and standards,
and avoiding adverse environmental impact through continual
improvement.  These programs are documented in the 8000 and 11000
series of chapters in the \dword{feshm}.  The environmental mitigation
plan also meets federal and state regulations.

\subsection{Waste Hazards (LBNF-DUNE HA-4)}

Waste-related hazards from \dword{dune} include the potential for
releasing waste materials (oils, solvents, chemicals, and radioactive
material) to the environment, injury to personnel, and reactive or
explosive event. Typical initiators will be transportation accidents,
incompatible materials, insufficient packaging or labeling, failure of
packaging, and a natural phenomenon.

During installation and \dword{dune} operation, we anticipate few
hazardous materials will be used. Such materials include paints,
epoxies, solvents, oils, and lead in the form of shielding. No current
or anticipated activities at \dword{dune} would expose workers to
levels of contaminants (dust, mists, or fumes) above regulatory
limits.

The \dword{esh}\&Q section industrial hygiene group and hazard control
technology team manage the program and guide collaborators subject to
waste-related hazards.  Their staff identify workplace hazards, help
identify controls, and monitor implementation. Industrial hygiene
hazards will be evaluated, identified, and mitigated as part of the
work planning and control hazard assessment process.

\subsection{Fire Hazards (LBNF-DUNE HA-5)}

Fire hazards have been evaluated and addressed to comply with
\dword{doe} Order 420.1C, Facility Safety, Chapter II and
\dword{doe}-STD 1066, Fire Protection Design Criteria.  The intent of
these documents is to meet \dword{doe}'s highly protected risk (HPR)
approach to fire protection.  In addition, the National Fire
Protection Association Standard 520, Standard on Subterranean Spaces,
was used in developing the basis for design related to fire
protection/life safety.

The combustible loads and the use of flammable and/or reactive
materials in the \dword{lbnf-dune} facility are controlled
following the International Building Code building occupancy
classification. Certain ancillary buildings outside the main structure
may be classified as higher hazard areas (Use Group H occupancy),
including the gas cylinder and chemical storage rooms, because they
hold more concentrated quantities of flammable or combustible
materials.  The control area concept used in the International
Building Code and National Fire Protection Association standards will
be followed for hazardous chemical use and storage areas to provide
the most flexibility and control of materials by allowing individual inventory
thresholds per control area.  The \dword{lbnf-dune} facility
will be equipped with fire detection systems and alarm systems that
will monitor water flow in case of fire suppression activation, as well
as monitor control valves and detection systems.

Audible/visual alarm notification devices will alert building
occupants.  Manual pull stations for the fire alarms will be installed
at all building exits.  Following National Fire Protection Association
90A, air handling systems will have photoelectric smoke detectors.
Smoke detection will be provided in areas with highly sensitive
electronic equipment.  Combinations of audible and visual alarm
notification devices will be set up throughout the underground
enclosures and service buildings to alert occupants. All fire alarm
signals will report through a centralized system at \dword{surf}.
Fire alarm and supervisory signals will be transmitted to internal and
external emergency responders using the campus reporting system.

While fixed fire protection systems afford an excellent level
of protection, additional strategies that include operational controls
that minimize combustible materials, adequately
fused power supplies, fire safety inspections, and operational
readiness reviews will be used to further reduce fire hazards
within the facility following \dword{doe} highly protected risk methods.

Experimental cabling will meet the requirements of the National Fire
Protection Association 70 and National Electrical Code, 2015 edition.
Preferred cables will be fire resistant, using appropriately
designated cable types for plenum or general-purpose cables.  When
there is a large investment in equipment for experiment power or
computer rack systems, or when equipment is custom-made (as opposed to
off-the-shelf commercial electronics), a device to detect faults or
smoke in the system will be provided.  This device will also shut
down the individual rack or racks when smoke or faults are detected.

\subsection{Electrical Hazards (LBNF-DUNE HA-6)}

\dword{lbnf-dune} will have significant facility-related
systems and subsystems that produce or use high voltage, high current,
or high levels of stored energy, all of which can present electrical
hazards to personnel. Electrical hazards include electric shock and
arc flash from exposed conductors, defective and substandard
equipment, lack of training, or improper procedures.

\fnal has a well-established electrical safety program that
incorporates de-energizing equipment, isolation barriers, \dword{ppe}, 
and training. The cornerstone of the program is
the lockout/tagout following the \dword{feshm} Chapter 2100, \fnal
Energy Control Program (Lockout/Tagout).

Design, installation, and operation of electrical equipment will
comply with the National Electrical code (NFPA 70), applicable parts
of Title 29 Code of Federal Regulations, Parts 1910 and 1926, National
Fire Protection Association 70E, and \fnal electrical safety policies
documented in the \dword{feshm} 9000 series chapters. Equipment
procured from outside vendors or international in-kind partners will
be either certified by a nationally recognized testing laboratory,
conform to international standards previously evaluated and deemed
equivalent to USA standards, or inspected and accepted using \fnal's
electrical equipment inspection policies outlined in \dword{feshm}
9110, Electrical Utilization Equipment Safety.

\subsection{Noise/Vibration/Thermal/Mechanical (LBNF-DUNE HA-7)}

Hazards include overexposure of personnel to noise and vibrations as
specified by the American Conference of Governmental Industrial
Hygienists and US Occupational Safety and Health Administration
(OSHA), which set noise limits to avoid permanent hearing loss, also
known as permanent threshold shift. Vibration of equipment can
contribute to noise levels and could damage or interfere with
sensitive equipment.

\dword{lbnf-dune} will use a wide variety of equipment that
will produce a wide range of noise and vibration. Support equipment,
such as pumps, motors, fans, machine shops, and general HVAC all
contribute to point source and overall ambient noise levels. While
noise will typically be below the ACGIH and OSHA eight-hour time-weighted
average, certain areas with mechanical equipment could exceed that
criterion and will require periodic monitoring, posting, and use of
protective equipment. Ambient background noise is more a concern for
collaborator comfort, stress level, and fatigue.

The detector facilities use a wide variety of noisy equipment. Items such as pumps,
fans, and machine shop devices 
are possible sources of
noise levels that might exceed the \fnal noise action
levels. \dword{feshm} Chapter 4140, Hearing Conservation, details
requirements for reducing noise and protecting personnel exposed to
excessive noise levels. Warning signs are posted wherever hazardous
noise levels may occur, and hearing protection devices are readily
available. Ways to reduce noise and vibration will be incorporated
into the \dword{lbnf-dune} design. These techniques include
using low-noise and low-vibration-producing equipment, especially for fans in
the HVAC equipment, isolating noise-producing equipment by segregating
or enclosing it, and using sound deadening materials on walls and
ceilings.

\subsection{Cryogenic/Oxygen Deficiency Hazard (LBNF-DUNE HA-8)}

The \dword{lbnf-dune} project will use large volumes of liquid
argon, nitrogen, and helium within the \dword{fscf}. Cryogenic
hazards could include \dword{odh} atmospheres due to failure of
the cryogenics systems, thermal (cold burn) hazards from cryogenic
components, and pressure hazards. Initiators could include the failure
or rupture of cryogenics systems from overpressure, failure of
insulating vacuum jackets, mechanical damage or failure, deficient
maintenance, or improper procedures.

Cryogenic liquids and gases are extremely dangerous to humans.
They can destroy tissue and damage materials and equipment past repair
by altering characteristics and properties (e.g., size, strength, and
flexibility) of metals and other materials.

Although cryogens are used extensively at \fnal, quantities that may
be used within a facility are strictly limited. Uses beyond defined
limits require \dword{odh} analyses and using
ventilation, \dword{odh} monitoring, or other controls.

Cryogenics systems are subject to formal project review, which includes
independent reviews by a subpanel of the Cryogenic Safety Subcommittee
following National Fire Protection Association Chapter 5032, Cryogenic
System Review. The members of this panel have relevant knowledge in
appropriate areas. They review the system safety documentation,
\dword{odh} analysis documentation, and the equipment before new
systems are permitted to begin the \cooldown process.

\fnal has developed and successfully deployed \dword{odh} monitoring
systems throughout the laboratory to support its current cryogenic
operations. The systems provide both local and remote alarms when
atmospheres contain less than 19.5\% oxygen by volume.

\fnal has a mature training program to address cryogenic safety
hazards. Key program elements include \dword{odh} training,
pressurized gas safety, and general cryogenic safety.

\subsection{Confined Space Hazards (LBNF-DUNE HA-9)}

Hazards from confined spaces could result in death or injury from
asphyxiation, compressive asphyxiation, smoke inhalation, or impact
with mechanical systems. Initiators would include failure of cryogenics
systems that are releasing liquid, gas, or fire, or failure of mechanical
systems. 

The \fnal confined-space program is outlined in \fnal Environmental
and Safety Manual Chapter 4230, Confined
Spaces. \dword{lbnf-dune} facilities will be incorporated into
this program. The emphasis at the \dword{lbnf-dune} design
phase will be to create the minimum number of confined spaces by
clearly articulating the definition of confined spaces to facility
designers to assure that such spaces have adequate egress, that mechanical
spaces are adequately sized, and, wherever possible, that no confined space
exists at all. During facility operations, the existing campus
confined-space program, along with appropriate labeling of confined
spaces, work planning and control, and entry permits will be used to
control access to these spaces.

\subsection{Chemical/Hazardous Materials Hazards (LBNF-DUNE HA-11)}

The \dword{dune} facility anticipates minimal use of chemical and hazardous
materials. Materials like paints, epoxies, solvents, oils, and lead
shielding may be used during construction and operation of the
facility. Exposure to these materials could result in injury;
exposure could also exceed regulatory limits. Initiators could be
experimental operations, transfer of material, failure of packaging,
improper marking or labeling, a reactive or explosive event, improper
selection of or lack of \dword{ppe}, or a
natural phenomenon.

\fnal maintains a database of hazardous chemicals in compliance with
the requirements imposed by 10 CFR 851 and \dword{doe} orders. In
addition to an inventory of chemicals at the facility, copies of each
manufacturer's safety data sheets (SDS) are maintained. Reviews of
conventional safety measures at the facilities show that using these
chemicals does not warrant special controls other than appropriate
signs, procedures, appropriate use of  \dword{ppe},
and hazard communication training. \dword{dune} will also supply
SDS documentation to the \dword{surf} \dword{esh}
department for all chemicals and hazardous materials that arrive on
site.

The industrial hygiene program, detailed in the \dword{feshm} 4000
series chapters, addresses potential hazards to workers using such
materials. The program identifies how to evaluate workplace hazards
when planning work and the controls necessary to either eliminate or
mitigate these hazards to an acceptable level.

Specific procedures are also in place for safe handling, storing,
transporting, inspecting, and disposing of hazardous materials. These
are contained in the \dword{feshm} 8000 and 10000 series chapters,
Environmental Protection and Material Handling and Transportation,
which describe how to comply with the standards set by the Code of
Federal Regulations, Occupational Safety and Health Standards, Hazard
Communication, Title 29 CFR, Part 1910.1200.

\subsection{Lasers \& Other Non-Ionizing Radiation Hazards (LBNF-DUNE HA-14)}
\label{sec:tc-esh-lasers}

Production and delivery of Class 3B and Class 4, near-infrared, UV,
and visible lasers must be completely contained in transport pipes or
designated enclosures for the Class 3B and Class 4 lasers, thus
creating a laser controlled area. (This will be in accordance with
\fnal \dword{feshm} chapter 4260.)  Establishing the laser controlled
area prevents areas around it from exceeding the maximum permissible
exposure as set by the \fnal laser safety officer.

\subsection{Material Handling Hazards (LBNF-DUNE HA-15)}

\dword{dune} will require a significant amount of manual and mechanical
material handling during the construction, installation, and operations
phases.  
These activities present hazards that include serious injury or
death to equipment operators and bystanders, damage to equipment and
structures, and interruption of the program.  Additional material
handling hazards from forklift and tow cart operations include injury
to the operator or personnel in the area and contact with equipment or
structures. Cranes and hoists will be used during fabrication,
testing, removal, and installation of equipment. The error precursors
associated with this type of work may include irregularly shaped loads,
awkward load attachments, limited space, obscured sight lines, and
poor communication.  The material or equipment being moved will
typically be one of a kind, expensive, or of considerable programmatic
value, and without dedicated lifting points or an obvious center
of gravity.

Lessons learned from across the \dword{doe} complex and OSHA have been
evaluated and incorporated into the \fnal material handling programs
documented in the \dword{feshm} 10000 series chapters.  The laboratory
limits personnel with access to mechanical material handling equipment
like cranes and forklifts to those who have successfully completed the
laboratory's training programs and demonstrated competence in
operating this equipment.

\subsection{Experimental Operations (LBNF-DUNE HA-16)}
Experimental activity undertaken at \dword{lbnf-dune} will be
fully reviewed under the \dword{orc} process
and by other experts as needed (e.g., representatives from electrical
safety, fire safety, environmental compliance, industrial hygiene,
cryogenic safety, and industrial safety), to identify and manage the
hazards of each experimental operation. The shift leader will ensure
that all safety reviews take place for each activity and that any
issues are appropriately addressed. The \dword{orc}  process will document
these reviews, covering the necessary controls and management approval
to proceed.

Typically, the \dword{orc}  process evaluates the scope of the proposed
experimental activity and identifies the hazards and controls to
mitigate them. The process ensures that collaborators are properly
trained, that qualified, hazardous material is kept to a minimum, that
engineering controls are deployed as a preferred mitigation, and that
\dword{ppe} is appropriate for the hazard.

\cleardoublepage


\cleardoublepage
\printglossaries

\cleardoublepage
\cleardoublepage
\renewcommand{\bibname}{References}
\bibliographystyle{utphys} 
\bibliography{common/tdr-citedb}

\end{document}